\documentclass[a4paper,10pt]{book}
\usepackage{amsmath,amssymb,amsfonts,bbm,amsthm,  amscd}
\usepackage{graphicx}
\usepackage[latin1]{inputenc}
\usepackage{psfrag}
\usepackage{epsfig,epic,eepic}
\usepackage{subfigure}

\allowdisplaybreaks

\newcommand{\f}{\frac}
\newcommand{\tr}{\mathrm{tr}}

\newcommand{\pd}{\partial}

\newcommand{\su}{\mathfrak{su}}
\renewcommand{\sl}{{\mathfrak sl}}

\renewcommand{\u}{{\mathfrak u}}

\newcommand{\spin}{\mathfrak{spin}}

\def\vphi{\varphi}
\def\la{\langle}
\def\ra{\rangle}
\def\q{(d_{j_0}, \Theta)}

\newcommand{\N}{\mathbb{N}}
\newcommand{\Z}{\mathbb{Z}}

\newcommand{\R}{\mathbb{R}}

\newcommand{\C}{\mathbb{C}}

\def\ss{{\mathcal S}}
\def\ppp{{\mathcal P}}
\def\vtheta{\vartheta}

\newcommand{\sixj}[2]{\left\{\begin{array}{ccc} #1 \\ #2 \end{array}\right\}}

\newcommand{\cF}{{\mathcal F}}

\newcommand{\cI}{{\mathcal I}}

\newcommand{\cL}{{\mathcal L}}
\newcommand{\cH}{{\mathcal H}}

\newcommand{\cO}{{\mathcal O}}
\newcommand{\tcO}{\widetilde{{\cal O}}}

\newcommand{\cM}{{\mathcal M}}
\newcommand{\cN}{{\mathcal N}}

\newcommand{\cC}{{\mathcal C}}
\newcommand{\cS}{{\mathcal S}}
\newcommand{\SU}{\mathrm{SU}}
\newcommand{\SL}{\mathrm{SL}}
\newcommand{\GL}{\mathrm{GL}}
\newcommand{\SO}{\mathrm{SO}}
\newcommand{\U}{\mathrm{U}}

\newcommand{\Spin}{\mathrm{Spin}}

\newcommand{\vJ}{\vec{J}}
\newcommand{\vV}{\vec{V}}
\newcommand{\vK}{\vec{K}}
\def\bv{{\bf v}}
\newcommand{\w}{\wedge}
\def\tDelta{{\widetilde{\Delta}}}

\newcommand{\id}{\mathbb{I}}

\def\mm{\mathcal{M}}
\def\eps{\epsilon}
\def\om{\omega}
\def\Om{\Omega}
\def\tl{\widetilde}

\def\tz{\tl{z}}
\def\tX{\tl{X}}
\def\arr{\rightarrow}

\def\nn{\nonumber}

\def\hh{{\cal H}}
\def\Hs{{H_S}}
\def\Hp{{H_p}}
\def\tpsi{\tilde{\psi}}
\def\tphi{\tilde{\phi}}
\def\tvphi{\tilde{\vphi}}
\def\th{\tilde{h}}
\def\tk{\tilde{k}}
\def\tj{\tilde{j}}
\def\tbeta{\tilde{\beta}}
\def\trho{\tilde{\rho}}

\def\sj{$\{6j\}$}

\newcommand{\be}{\begin{equation}}
\newcommand{\ee}{\end{equation}}
\newcommand{\bes}{\begin{eqnarray}}
\newcommand{\ees}{\end{eqnarray}}
\newcommand{\bea}{\begin{eqnarray}}
\newcommand{\eea}{\end{eqnarray}}

\newcommand{\Ref}[1]{(\ref{#1})}

\newcommand{\mat}[2]{\left(\begin{array}{#1}#2
\end{array}\right)}

\def\pp{\partial}

\def\dag{^\dagger}

\newtheorem{theorem}{Theorem}[section]

\newtheorem{proposition}[theorem]{Proposition}

\newcommand{\equa} [1] {\begin{equation} #1\end{equation}}

\newcommand{\tabl} [2] {\begin{array} {#1} #2 \end{array}}
\newcommand{\bin} [2] {\left (\begin{array}{c}#2\\#1\end{array} \right ) }

\def\ie{{i.e. \/}}
\def\eg{{e.g. \/}}

\begin{document}
%

\thispagestyle{empty}
\sloppy\hbadness=5000\vbadness=100
{\center 

\fbox{%
\raisebox{.8cm}{\parbox{11cm}{%
\large \sc \center Ecole Normale Sup\'erieure de Lyon\\ Laboratoire de Physique}

}}\\
\vspace{1cm}
{\huge \sc Th\`ese\\} 
\vspace{1.5cm}
Pr\'esent\'ee par\\
\vspace{1cm}
{\large \sc Ma\"it\'e Dupuis} \\
\vspace{1cm}
Pour obtenir le grade de \\
\vspace{.5cm}
{\large \sc Docteur de l'Ecole Normale Sup\'erieure de Lyon}\\
\vspace{1cm}
Ecole doctorale : Physique et Astrophysique de Lyon\\
\vspace{1.2cm}
{\huge  \sc Spin foam models for quantum gravity \\ \vspace{.2cm} \sc and  semi-classical limit}\\

\vspace{0.9cm}

{\Large Mod\`eles de mousses de spin pour la gravit\'e quantique et leur r\'egime semi-classique}\\

\vspace{2cm}

Soutenue le 16 d\'ecembre 2010, devant la commission d'examen compos\'ee de\\
\begin{center}
\begin{tabular}[c]{ll}
{Mr John  \sc Barrett} & Membre\\
{Mr Laurent \sc  Freidel} & Membre/Rapporteur\\
{Mr Etera \sc Livine} & Membre\\
{Mr Jean-Michel  \sc Maillet } & Membre \\
{Mr Alejandro \sc  Perez} & Membre/Rapporteur
\end{tabular}
\end{center}
}


 \newpage
\strut
\newpage

\section*{Remerciements}

Merci tout d'abord \`a Etera, mon directeur de th\`ese qui m'a fait d\'ecouvrir le monde de la recherche et la gravit\'e quantique!
 
\noindent
Merci \`a tous les membres du laboratoire de physique de l'ENS (je pense en particulier \`a mes diff\'erents co-bureaux!) qui m'ont si gentillement accueillie puis support\'ee pendant ces trois ann\'ees!

\noindent
Merci aussi \`a tous les chercheurs qui forment la (encore petite mais si dynamique, jeune et sympathique) communaut\'e de la gravit\'e quantique \`a boucles! Et un merci particulier \`a tous ceux qui m'ont invit\'ee dans leur groupe et qui se sont tous montr\'es tr\`es ouverts \`a la discussion et \`a la collaboration: Florian \`a Sydney; Carlo, Alejandro et Simone \`a Marseille; Thomas \`a Erlangen; Laurent et Lee \`a Waterloo; Bianca et Daniele \`a Berlin; Renate \`a Utrecht.

\noindent
Merci aussi \`a mes deux rapporteurs, Alejandro et Laurent, pour leur lecture attentive de cette th\`ese, ainsi qu'aux autres membres du jury de ma soutenance, Etera, Jean-Michel et John pour leur int\'er\^et \`a mon travail. 

\noindent
Un merci sp\'ecial \`a Andr\'e qui a une importance toute sp\'eciale pour moi et qui a pris le temps de relire cette th\`ese.

\noindent
Et bien s\^ur un grand merci \`a tous ceux qui m'ont entour\'e pendant ces trois ann\'ees! Flo pour son soutien quotidien!  Ma famille: mes parents, mon fr\`ere Xavier et ma soeur Margot ainsi que tous mes super oncles, tantes, cousins, cousines et grands-parents! Mes amis: ceux de Besan\c on qui me connaissent depuis si longtemps, les amis rencontr\'es en pr\'epa, les amis de l'ENS, les amis de l'athl\'e ainsi que mes diff\'erents entra\^ineurs.... Je ne donne pas vos noms mais merci \`a vous tous pour tous les moments partag\'es ensembles qui ont \'et\'e si importants!

\newpage
\strut
\newpage

\section*{R\'esum\'e}

{\Large\sc Mod\`eles de mousses de spin pour la gravit\'e quantique et leur r\'egime semi-classique}


\vspace{0.3cm}

Les mousses de spin fournissent un formalisme d{'}int\'egrale de chemin pour la gravit\'e quantique qui s{'}inspire de la gravit\'e quantique \`a boucles. Ils d\'ecrivent la structure quantique de l{'}espace-temps et l{'}\'evolution temporelle des \'etats cin\'ematiques de la gravit\'e quantique \`a boucles. La quantification covariante en terme de mousses de spin est bas\'ee sur l{'}\'ecriture de la relativit\'e g\'en\'erale comme une th\'eorie topologique contrainte. \\
Les contraintes, appel\'ees contraintes de simplicit\'e,  introduisent les degr\'es de libert\'es locaux et permettent de passer d{'}une th\'eorie topologique \`a une th\'eorie de la g\'eom\'etrie de l'espace-temps. Elles sont donc essentielles. Cependant leur impl\'ementation est encore mal comprise. Dans cette th\`ese, une mani\`ere originale d{'}imposer les contraintes est propos\'ee: les contraintes de simplicit\'es sont reformul\'ees en utilisant un nouveau formalisme construit \`a partir d'oscillateurs harmoniques et  des \'etats coh\'erents, solutions des contraintes, sont donn\'es. \\
D'autre part, un mod\`ele de mousse de spin pour la gravit\'e quantique est coh\'erent s{'}il peut \^etre reli\'e \`a l'approche canonique \`a boucles et poss\`ede la bonne limite semi-classique.\\
Un lien entre les \'etats cin\'ematiques de la gravit\'e quantique \`a boucles et les \'etats fronti\`eres d'une mousse de spin est ici explicit\'e reliant ainsi clairement l'approche cannonique et l'approche covariante. \\ 
Nous proposons aussi de nouvelles techniques pour calculer le d\'eveloppement asymptotique semi-classique des amplitudes de transition de la gravit\'e quantique. En particulier dans le contexte de la gravit\'e 3d, des outils analytiques n\'ecessaires pour calculer toutes les corrections quantiques des corr\'elations du champs gravitationnel sont present\'es. Des calculs explicites, bas\'es sur des m\'ethodes d{'}approximation et sur l{'}utilisation de relations de r\'ecurrence sur les amplitudes de mousses de spins, ont \'et\'e effectu\'es. Les r\'esultats sont pertinents pour d\'eriver des corrections quantiques \`a la dynamique du champ gravitationnel.  

\smallskip

Mots cl\'e : relativit\'e g\'en\'erale, th\'eorie des champs topologique, gravit\'e quantique, mousses de spin, r\'eseaux de spins. \\
 \\
{\Large\sc Spin foam models for quantum gravity and semi-classical limit}

\vspace{0.3cm}

The spinfoam framework is a proposal for a regularized path integral for quantum gravity. Spinfoams define quantum space-time structures describing the evolution in time of the spin network states for quantum geometry derived from Loop Quantum Gravity (LQG). The construction of this covariant approach is based on the formulation of General Relativity as a topological theory plus the so-called simplicity constraints which introduce local degrees of freedom.\\
The simplicity constraints are essential in turning the non-physical topological theory into 4d gravity. In this PhD manuscript, an original way to impose the simplicity constraints in 4d Euclidean gravity using harmonic oscillators is proposed and new coherent states, solutions of the constraints, are given. \\
A consistent spinfoam model for quantum gravity has to be connected to LQG and must have the right semi-classical limit. \\
An explicit map between the spin network states of LQG and the boundary states of spinfoam models is given connecting the canonical and the covariant approaches. \\
New techniques to compute semiclassical asymptotic expressions for the transition amplitudes of 3d quantum gravity and to extract semi-classical information from a spinfoam model are introduced. Explicit computations based on approximation methods and on the use of recurrence relations on spinfoam amplitudes have been performed. The results are relevant to derive quantum corrections to the dynamics of the gravitational field.

 \smallskip

Keywords: general relativity, topological field theory, quantum gravity, spin networks, spin foams.\\

\vspace{0.4cm}

Laboratoire de Physique - Ecole Normale Sup\'erieure de Lyon - UMR 5672

\tableofcontents

\chapter{Introduction}

One of the deepest questions for physicists is to unravel 
the essential nature of space and time and how the world came into existence. This quest on the meaning of space and  time has a very long history which teaches us that an answer to a question can radically change our view of the world. 
Although a new theory on the nature of space and time can be the end of a journey, it is often the starting point for more questions. A new theory  brings us new frameworks and new technical tools which open new frontiers and enlarge our  perception. It improves our understanding of the notions of space and time and and gives us access to more and more refined scales of observations. 
The historical evolution of space-time theories from the Aristotelician world, the Newton's views to Einstein's breakthrough,  emphasizes  that a final description of our physical world is far from being done. 

\begin{figure}[ht] 
\begin{center}
\includegraphics[width=8cm]{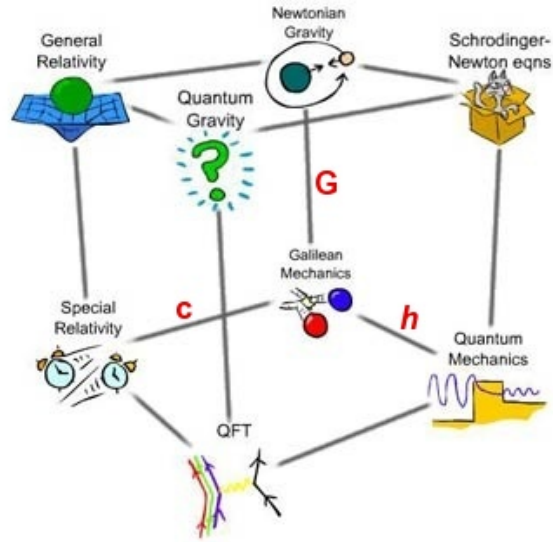}
\end{center}
\caption{Fundamental constants and theories. } \label{cst}
\end{figure}

Three dimensional constants play a fundamental role in the description of our physical world:  the speed of light in vacuum\footnote{Its value was fixed to 299 792 m.s$^{-1}$ in 1983 by the BIPM (Bureau International des Poids et Mesures) and this value does define the meter.} 
$c \sim 3 \times10^9 m.s^{-1}$, Planck's constant $h \sim 6.6 \times 10^{-34} m^2. kg. s^{-1}$, the Newton universal gravitational constant\footnote{The fractional uncertainty in $G$ precisely measured in the 1980's (uncertainty of $0.0128 \%$) is thousands of times larger than those of other fundamental constants. Moreover, recently the value of $G$ has been called into question by new measurements which suggest the uncertainty of $G$ could be much larger than originally thought.} $G \sim 6.7 \times 10^{-34} m^3.kg^{-1}.s^{-1}$. Their order of magnitude is much larger (or smaller) than the meter, kilogram and second, which are the units (from the International System of Units) significant to describe 
 direct everydays life experiments. 
 Each of these three constants is associated to a physical regime and corresponds to a particular scale of observation. They are specific to some theories and the links between each constant and the different theories are summarized in  Fig. \ref{cst}.

Let us present some of the theories appearing in Fig. \ref{cst}. In the general relativity theory, Einstein makes the revolutionary claim that the geometry of space is not fixed but evolves in time. Quantum mechanics tells us that every dynamical quantity is not continuum but quantized. The size of a ``quantum" is characterized by Planck's constant $h$. At the end of the first half of the twentieth century, Freeman Dyson, Richard Feynman among many others consistently combined quantum theory with special relativity to give the so-called quantum field theory in which both constants, $h$ and $c$ appear. A very useful and experimental successful example of a quantum field theory fixed Minkowski background is the standard model. It describes three of  the four  fundamental interactions which govern the myriad of  phenomena in nature: electromagnetism, the strong nuclear force which holds atomic nuclei together and the weak nuclear force which is responsible for radioactive decay; gravity is however neglected. The standard model is therefore an approximation describing particle physics in the lab and in a variety of astrophysical situations when the gravitational force can be considered weak below a given  energy scale which is the so-called Planck scale. It is defined by the combination of the three constants: $c$, $\hbar= h/ 2 \pi$, $G$. The Planck length equals: $l_p=\sqrt{\frac{\hbar G}{c^3}}$ in 4d and its value is then $l_p=1.616252(81) \times 10^{-35}$m that corresponds to an energy $E_p=1.22 \times 10^{28}$ eV. 
$E_p$ can be compared with the energies obtained in our most powerful accelerator, the Large Hardon Collider in Geneva, which are of the order of $10^{12}$ eV.  Below this Planck scale, the gravity effects can be neglected and the standard model is a good approximation, above a theory of quantum gravity is necessary to describe space and time. Since the 1980s, the problem of understanding what happens to general relativity at the extreme short-distance Planck scale $l_p$ has been one of the most pressing questions in theoretical physics. Before going on with quantum gravity, let us point out that the regime where both $G$  and $h$ have to be taken into account can be studied thanks to the so-called Schrodinger-Newton equations. Theoretically, this brings nothing really new, however, some  experimental results obtained in 2001 are very interesting: physicists have observed quantized states of matter under the influence of gravity for the first time. Ultra-cold neutrons moving in a gravitational field do not move smoothly but jump from one height to another, as predicted by quantum theory \cite{ultracoldNeutrons}.

Let us detail the plan of the next three main parts.

\begin{figure}[ht] 
\begin{center}
\includegraphics[width=9.5cm]{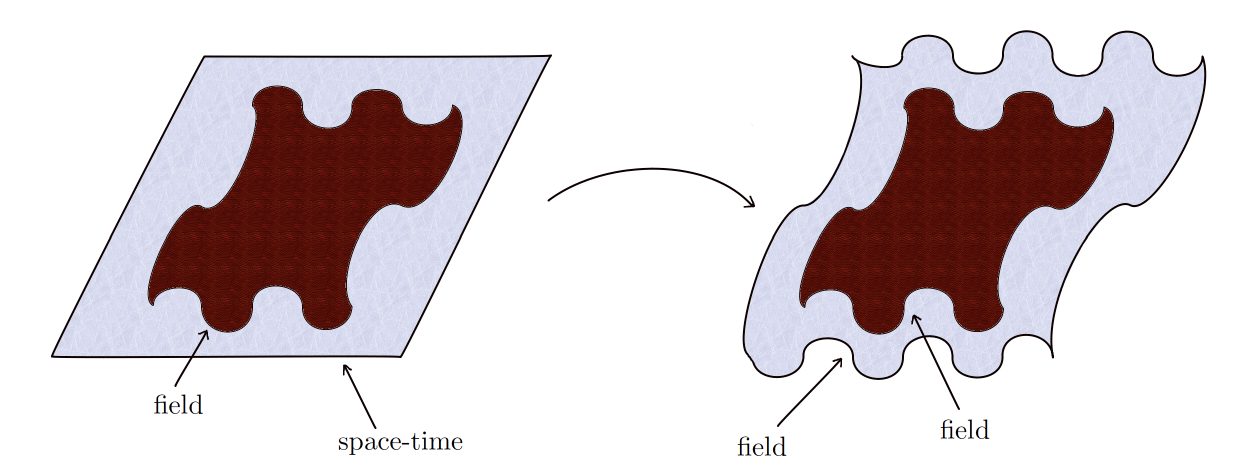}
\end{center}
\caption{Space-time geometry becomes a dynamical object like  the other fields.} \label{GR}
\end{figure}

In the first part, we will give the reasons why we need a theory of quantum gravity. Several serious answers to this challenge have been proposed. These include non-commutative geometry, string theory, loop quantum gravity, emergent gravity,... This  three year work is  exclusively based on the so-called loop quantum gravity theory \cite{book-carlo, book-thomas} which is a theory focusing on the problem of quantizing gravity with no aim to find a unified description of all interactions. Loop quantum gravity is an attempt to define a quantization of gravity paying great attention to the conceptual lessons of general relativity. 
For example, general relativity teaches us   that the degrees of freedom of the gravitational field are encoded in the geometry of spacetime and in this sense, the gravitational interaction is fundamentally different from all  other known forces. The spacetime geometry is fully dynamical and the notion of absolute space on top of which 'things happen' ceases to make sense in gravitational physics (see Fig. \ref{GR}). 
The idea is therefore  to build a theory of quantum gravity which will not be based on a notion of background geometry. In  Part \ref{part2}, we will describe  loop quantum gravity  established as a proposal of background independent and non perturbative quantization of general relativity. This presentation will be done in two steps. We will first recall the Hamiltonian formulation of general relativity starting with the conventional ADM variables and then introduce the variables that are used in the definition of loop quantum gravity. The second step is the quantization: we will give a short review of the canonical approach of quantization of general relativity and then describe the Hilbert space of the quantum states and in particular define the {\it spin network states}. The spin network states are the building block of loop quantum gravity: they provide a basis of the kinematical Hilbert space and diagonalize some geometric operators, such as surface areas. The kinematics of loop quantum gravity is in fact beautifully under control. However, understanding  the  dynamics is still in progress. In the last section of Part \ref{part2}, we will present the {\it spin foam} formalism which can be considered as a possible approach to solve the dynamics of spin networks. Indeed, a spin foam picture emerges when considering the evolution in ``time" of spin networks. But spin foam models can  also be naturally interpreted as a form of  path integral approach to quantum gravity, \textit{the covariant approach}. In this approach, one abandons the canonical approach and seeks a functional integral  description of transition amplitudes between two quantum gravity states. The states are given by 3-geometries relative to 3d hypersurfaces not necessarily connected and the histories interpolating between them are 4-geometries inducing the given 3-geometries on the hypersurfaces. The different steps -- the first order formalism of general relativity, the discretization,  the quantification -- will be concretely illustrated in  Section \ref{PRmodel} where we will give the technical details of the construction and the derivation of  a spin foam amplitude in the case of the Ponzano-Regge model which is a spin foam model for 3d quantum gravity.  3d general relativity has no local degree of freedom: it can be discretized and quantized exactly without losing any of its physical content and consequently offers a simpler laboratory than the physically relevant 4d case.

In  Part \ref{part2} and Part \ref{part3}, new results are presented. The spin foam framework is the starting point of all this work. Two of the main issues for this approach are:
\begin{itemize}
\item to define a consistent 4d model of quantum gravity,
\item to have a low-energy prediction of the theory.
\end{itemize}
The difficulties of the first point come from the fact that unlike the 3d case, 4d general relativity is not a topological theory but a topological theory plus constraints.  The issue is to understand how to impose consistently the constraints -- the so-called {\it simplicity constraints } -- at the quantum level in order to turn the topological theory into a geometrical theory and to introduce local degrees of freedom. The second point is essential to test the theory and to make predictions but it is difficult since the starting point is a non-perturbative formalism. That is why it is challenging  to define a perturbative expansion in the Planck scale $l_p$ in order to check if the correct semi-classical limit arises. The next step is to explore if loop quantum gravity can provide a UV completion of the perturbative quantization in terms of graviton, which unlike the other interactions in the standard model turns out to be non-renormalizable. This could consequently provide  predictions for possible experiments. 
These two different lines have been explored and our results can bring some elements of response to both issues. 

Another important question related to the low-energy limit predictions will be  to understand the behavior of spin foam models under renormalization and to study the coarse-graining of such model. This is fundamental in order to truly define the continuum limit of spinfoam models and their semi-classical regime. In fact, we do not expect a unique spin foam model for 4d gravity but a class of models characterized by a renormalization flow. This question to define a renormalization group flow in a background independent context has still to be seriously tackled \cite{rivasseau}.

Part \ref{part3} will be devoted to the problem of defining  a spin foam model for 4d gravity. We will first recall the definition of the Barrett-Crane model which was the first explicit attempt to solve the simplicity constraints. However,  the  loop quantum gravity point of view   or the spin foam graviton calculations seem to indicate that too many degrees of freedom of the 3d space geometry are frozen in this model. We will then introduce the EPRL-FK models which were proposed to address this issue prior to the inception of this work. In these models, the simplicity constraints are imposed weakly either following a procedure inspired from the Gupta-Bleuler quantization 
or using coherent states. After a review on these two different approaches, we will present some results, -- which we presented in \cite{article4} -- regarding one aspect of the problematic focusing on the  relation between loop quantum gravity and the spin foam formalism. More precisely, we have defined the various ways to map the loop quantum gravity spin networks onto certain subspaces of Lorentz invariant states -- {\it the projected spin networks} -- which satisfy the simplicity constraints and would therefore legitimately implement the dynamics and evolution of spin networks for loop quantum gravity.

 In the last part of  Part \ref{part3} tackling the issue of the definition of a 4d spin foam model for gravity, we will present some results -- we presented in  \cite{article3} -- focusing on the implementation of the simplicity constraints at the quantum level using a new framework introduced in \cite{UN1} and developed more recently in \cite{UN2, UN3}. This framework leads to a better interpretation of the structure of the Hilbert space of $\SU(2)$ {\it intertwiners}\footnote{$\SU(2)$ intertwiners are $\SU(2)$ invariant tensors attached to each vertex of a spin network. They are the basic building blocks of spin network states and glued together, they generate the quantum 3d space. The intertwiner space geometrical interpretation is therefore necessary for a better understanding of loop quantum  gravity at both kinematical and dynamical levels.} 
 from $\U(N)$ representations.  This $\U(N)$ framework, based on the Schwinger representations of the $\su(2)$ Lie algebra in term of a couple of harmonic oscillators,  proposes a closed algebra of geometric observables acting on the space of $\SU(2)$ intertwiners and a new set of coherent states which are covariant under $\U(N)$ transformations. A first part of my work has been to complete the analysis of the $\U(N)$ framework for $\SU(2)$ intertwiners initiated in \cite{UN1, UN2, UN3} in order to have an explicit action of the different newly introduced geometrical observables on the $\U(N)$ coherent states. Then, we have shown how these $\U(N)$ tools can be applied to the analysis of the simplicity constraints for 4d Euclidean gravity and proposed new sets of constraints. Solving them in term of the $\U(N)$ coherent states has yielded weak solutions to all simplicity constraints for arbitrary values of the Immirzi parameter\footnote{The Immirzi parameter, denoted $\gamma$, appears  when adding  to the classical action for gravity  a topological term which is required in order to have a well-defined connection on the boundary. This has no
  effect on the classical equations of motion but this introduces a quantization ambiguity.}.

In Part \ref{part3}, we will present new results concerning the issue of the low-energy interpretation of the theory from the spin foam formalism and show how they are relevant to derive quantum corrections to the classical dynamics of the gravitational field. This work is based on the ``spinfoam graviton" framework proposed by Carlo Rovelli and collaborators who have introduced a technique to study n-point functions within loop quantum gravity \cite{graviton1, graviton2}. The graviton propagator  corresponds to the correlator between excitations of quanta of space or more explicitly to the correlator between geometrical observables such as the areas of elementary surfaces. The two main ingredients to define it are the boundary amplitude that codes the quantum gravity dynamics -- the $\{6j\}$ symbol in the Ponzano-Regge model, the $\{10j\}$ symbol in the Barrett-Crane model -- and a weighted functional of spin networks. This latter  is usually chosen to be a semi-classical state 
 peaked on both the intrinsic and extrinsic geometry of a closed 3d surface, interpreted as the physical boundary of a 4d spacetime region. We have focused on a 3d toy model to explore the properties of the boundary amplitude which is the quantum amplitude associated to a vertex of the spin foam\footnote{ This 3d toy model is  a topological model. We thus know how to quantize it exactly as a spin foam model and in addition we know how to introduce defaults such as particles which introduce local degrees of freedom.}.
The 3d case has the advantage that the semi-classical limit is better understood and all the issues of the 4d case can be addressed in this simplified context. In the Ponzano-Regge model, the vertex amplitude is given by the $\{6j\}$ symbol from the recoupling theory of the representations of $\SU(2)$. We need to understand the corrections to the asymptotical behavior of the $\{6j\}$ symbol in order to compute the higher order quantum corrections to the classical propagator of the graviton. We have developed two methods to study the asymptotic expansion of the $\{6j\}$ symbol: either preforming a brute-force approximation starting from the explicit algebraic formula of the $\{6j\}$ symbol as a sum over some products of factorials \cite{article1}, or using a recursion relation for the $\{6j\}$ symbol derived from the invariance of the $\{6j\}$ symbol under Pachner moves (Biedenharn-Elliott identity) \cite{article2}. The first method has allowed  to investigate the asymptotical behavior of the $\{6j\}$ symbol at next-to-leading order and to compute it analytically. With the second method, we have provided explicit formulas for up to the third order correction beyond the leading order for the particular case of a 'isosceles' $\{6j\}$ symbol and we have in addition shown how the relation recursions can be used to derive Ward-Takahashi-like identities between the expectation values of graviton-like spin foam correlations. \\
Another problem we tackled concerns the second key ingredient regarding the graviton propagator, \ie the boundary state. The original ansatz  is to take as boundary state a Gaussian state with a phase factor in the Hilbert space of boundary spin networks. However, this state has to be physical, that is to be a gauge invariant spin network state that solves the quantum gravity constraints. A criterion to select a physical boundary state had already been proposed in \cite{physical} in the 3d case. We have extended this approach to determine physical boundary state in the 4d case considering the Barrett-Crane model \cite{article5}.

\part{From Loop Quantum Gravity to Spin foam models} \label{part1}

\chapter{Motivations for a quantum theory of gravity}

 {\it Nevertheless, due to the inner atomic movement of electrons, atoms would have to radiate not only electromagnetic but also gravitational energy, if only in tiny amounts. As this is hardly true in Nature, it appears that quantum theory would have to modify not only Maxwellian electrodynamics but also the new theory of gravitation.} \\
 \begin{center}
Albert Einstein,
\\
({\it Preussische Akademie Sitzungsberichte}, 1916)
 \end{center}

\bigskip

 Two of the most exciting developments of  XXth century physics were general relativity and quantum theory. Both have modified our understanding of the physical world in depth. General relativity treats gravity, while the 'standard model' -- the culminating development of quantum theory -- treats the rest of the forces of nature. 
 
 General relativity represents the result of a long line of developments that go all the way back to  Galileo's thought experiments about relativity of the motion,  Mach's arguments  about the nature of space and time and finally to Einstein's magnificent conceptual breakthrough.  It has provided beautiful insights about the nature of the Universe through astrophysics and cosmology and even led to key technological developments such as the Global Positioning System.  
 
Quantum theory has been developed  roughly at the time, triggering new  discoveries  about the fundamental nature of matter and its interactions, through atomic physics, nuclear and particle physics, condensed matter physics. 

Even though both theories had impressive successes in making predictions which were checked with very high precision, they have their own limits. 
 
 For example, central objects for astronomers and theoretical physicists  are black holes. A black hole  is a region of space from which nothing, not even light, can escape. Its defining feature is the appearance of an event horizon -- a boundary in spacetime through which matter and light can only pass inward. Although  the discovery in the early 1970s by   Bekenstein and Hawking among many others of the thermodynamic behavior of black holes (see e.g. \cite{wald})
 -- achieved primarily by classical and semiclassical analyses --  has provided  a better understanding of this notion of black hole horizon, many important issues remain unresolved. Primary among these are the ``black hole information paradox"\footnote{It suggests that physical information could ``disappear" in a black hole, allowing many physical states to evolve into precisely the same state.} and the gravitational singularity in the center of the black hole where the spacetime curvature becomes infinite. 
%
 \\
In the context of cosmology, the Friedmann, Lema\^\i tre, Robertson, Walker  models\footnote{The Friedmann - Lema\^\i tre - Robertson - Walker metric is an exact solution of Einstein's field equations of general relativity; it describes a simply connected, homogeneous, isotropic expanding or contracting universe (see \cite{book-wald, carroll}).} 
 and perturbations thereof  are empirical successes. Indeed, it appears that the rich data that we now have and are likely to accumulate in the near future from astrophysical observations would be adequately described   by these models derived from general relativity and quantum field theory. However, these theories are conceptually incomplete since they assume that the universe begins with a ``Big Bang" at which matter densities and space-time curvature become infinite, \ie a singularity. The appearance of singularities in general relativity is commonly perceived as signaling the breakdown of the theory.

At the atomic scale, the gravitational force can be safely neglected and this is also true for the current high energy experiments. Therefore it might be surprising at first sight that gravitational and quantum effects should  be taken into account both at the same time. However a simple dimensional analysis shows that at the Planck scale $l_P=\sqrt{\frac{\hbar G}{c^3}}\sim 10^{-35}m$, the gravity effects  become important again and cannot be neglected any more. This means that   before reaching the singularity in a black hole one has to enter a regime where gravity is quantized. In particular, one can expect that the notion of singularity is blurred and regularized, solving this singularity  issue \cite{BH-sing}.  A similar reasoning applies in cosmology. The Universe should have cooled down from an extreme  phase where the gravitational degrees of freedom were quantized. The notion of ``Big Bang" should be blurred and regularized  by the quantum effects.  This justifies, at least theoretically, the necessity to study quantum gravity. 
 

 From an experimental point of view, quantum gravity is a theory expected to describe regimes that are so far inaccessible: indeed in the current most  powerful accelerator, the Large Hadron Collider, we are able to obtain energies of the order of a few  TeV  while the Planck energy $E_p$, the energy scale at which quantum gravity effects are believed to become important, is $10^6$ TeV. 
 
 It was therefore thought for a long time that one could only see quantum gravitational effects in the cosmological realm, these effects being moreover extremely weak and hard to measure. This situation has  changed; different experiments both terrestrial (VERITAS) and spatial (FERMI satellite), using extreme high energy particle physics (essentially gamma-rays bursts) are being set up and hope to measure quantum gravitational effects  \cite{amelino}.  A new experimental window is opening,  allowing us to explore the fundamental structure of spacetime. The current situation in this field is therefore exciting, one has to make physical predictions to the outcome of these experiments, and make concrete statements about the physics involved in the quantum gravity regime.


We need a  working theory of gravity that takes quantum effects into account. This theory of quantum gravity should in particular synthesis the generally relativistic principle of background independence and the uncertainty principle of quantum mechanics. However,  it is a non trivial task  to assemble the two theories, general relativity and quantum mechanics,  into a single coherent picture of the world. Indeed, the ``naive" try consisting in applying the standard (perturbative) quantum field theory tools so successful for the standard model, does fail: the perturbative quantization of general relativity is non-renormalizable.  Moreover one can argue that this approach is not really respecting the essence of general relativity since one introduces a background around which one quantizes the gravitational fluctuations. Introducing a background does break a fundamental feature of general relativity, the background invariance. 

Following the lead of Bergmann and Dirac, physicists (Arnowitt, Deser, Misner) have also tried to apply the Dirac quantization rules to general relativity during the mid 60's. Quickly enough it was realized that using the metric formulation, this approach was also bound to fail due to the intrinsic complexity of the formulation.

It was not before the mid 80's that Ashtekar realized that with a clever change of variables the Hamiltonian analysis of general relativity was greatly simplified. The subsequent rediscovery by Rovelli and Smolin of the spin networks introduced in the 70's by Penrose as the basis for the kinematical Hilbert space sparked   the  \textit{loop quantum   gravity} field.  While trying to understand its quantum dynamical aspects, the notion of \textit{spinfoam} was introduced. They can also be viewed as a (discretized) path integral quantization of general relativity. For a 3d space-time case, it was realized those spinfoams were in fact similar to  an old model,  the Ponzano-Regge model, introduced in the 60's \cite{PR}. Since then the spinfoam approach was developed on its own as a way to regularize the general relativity partition function.

\chapter{Hamiltonian formalism of General Relativity}
General relativity is a constrained theory. I will first give the Hamiltonian formulation of general relativity in terms of ADM variables  (see \cite{mtw} for a review) which was used by Arnowitt, Deser, Misner (ADM) among many others to apply  Dirac's quantization program to general relativity. The Dirac's procedure can be schematically divided in three steps \cite{book-dirac, teitelboim, matschull}
\begin{itemize}
\item Find a representation of the phase space variables of the theory as operators in a kinematical Hilbert space $\mathcal{H}_{kin}$ satisfying the standard commutation relations: $\{\;, \;\}\rightarrow -i/\hbar [\;,\;]$.
\item Promote the constraints to (self-adjoint) operators in $\mathcal{H}_{kin}$.
\item Characterize the space of solutions of the constraints and define the corresponding inner product. This would define the so-called physical Hilbert space  $\cH_{phys}$.
\end{itemize}
These steps should then be completed with a physical interpretation of the quantum observables -- the gauge invariant observables are operators commuting with the constraints.
Dirac's procedure applied to the ADM phase space defines  a kinematical Hilbert space which is already ill-defined and on which there is no measure theory. Therefore, we will not apply the quantization program proposed by Dirac to the set of ADM variables but we will see how we can define new variables from the ADM ones to reformulate general relativity in a way more amenable to Dirac's quantization procedure. Indeed, all choices of fundamental variables do not work out as well when quantizing a theory and the key to {\it loop quantum gravity} has precisely been a choice of different variables to describe gravity. From the ADM variables, the definition of the new variables will be done in two steps 
\bes
\textrm{ADM formuation }& \qquad \quad \textrm{  triad formulation } & \qquad \textrm{ $\SU(2)$ Ashtekar-Barbero connection variables } \nn \\
( \pi^{ab}, \,q_{ab}) \qquad & \longrightarrow  \quad \qquad (E^a_i, \,K^j_b) \quad& \longrightarrow  \qquad  (P^a_i 
, \, A^j_b
) \nn
\ees
The first change of variables will introduce a new constraint and the second one is a canonical transformation. In the first section, I will  recall the definition of the ADM variables. Then, in the second section I give  more details on both changes of variables.
Finally, in the last section, I will present the classical framework of  \textit{covariant loop quantum gravity} which is a program of quantizing gravity \textit{\`a la loop} starting from a Lorentz covariant canonical formulation. In fact, this covariant loop quantum gravity theory uses the same techniques and tools as loop quantum gravity but in this case the gauge group is the Lorentz group $\SL(2, \C)$ instead of $\SU(2)$.

\section{Canonical formulation of General Relativity in ADM variables}

 The Hamiltonian formulation of gravity is the basis of any attempt to canonically quantize gravity.  The Hamiltonian treatment of General Relativity is based on a $3+1$ splitting of space-time \cite{bergmann1, bergmann2, bergmann-komar, komar, dirac1}. This splitting allows to coordinatize the phase space $\Gamma$ explicitly. The phase space defined as the space of solutions of the equation of motion is a covariant notion, however the coordinatization we choose for it breaks covariance.  Given a four-geometry -- a globally (hyperbolic) spacetime $\mathcal{M}$ with metric tensor fields $g_{\mu \nu}$ of Lorentzian signature\footnote{$(-+++)$ induces a positive definite metric on spacelike metric. In some cases, we will consider a Riemannian manifold, that is the metric will have a $(++++)$ signature.} $(-+++)$ --  we  consider a one-parameter family of three-dimensional spacelike hypersurfaces $\Sigma_t$ ($t=constant$), with spatial coordinates $x^a\;(a=1,2,3)$ \cite{book-wald}. In the following, we shall use latin letters $a,b,\cdots$ for space indices and greek letters $\mu,  \nu,\cdots $  for space-time indices. Let us consider a nearby pair $(\Sigma_t, \, \Sigma_{t+dt})$ of spacelike hypersurfaces labelled by the time coordinates $t$, and $t+dt$. The foliation allows us to identify the function $t \in \R$ as a (unphysical) time parameter.  We need to define ``moving forward in time" with the parameter time $t$ starting from the surface $\Sigma_t$ and reaching the surface $\Sigma_{t+dt}$. For this, let us choose a vector field $t^\mu$ on $\cM$ satisfying $t^\mu \nabla_\mu t=1$. 
   \begin{figure}[ht]
\begin{center}
 \includegraphics[width=7cm]{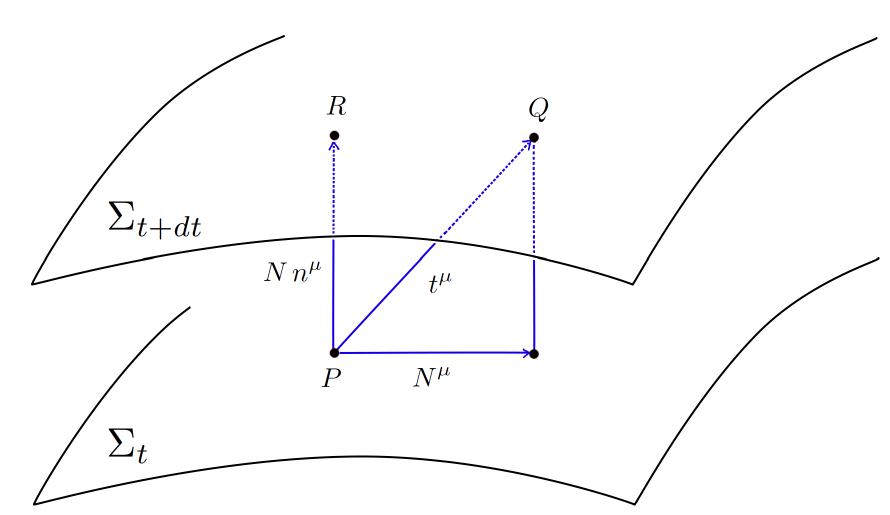}
\end{center}
\caption{Coordinates for two nearby hypersurfaces labelled by $t$ and $t+dt$} \label{sandwich1}
\end{figure}
 We decompose $t^\mu$ into its  normal and tangential parts to $\Sigma_t$ (see Fig. \ref{sandwich1}) by defining the {\it lapse function} $N$ and the {\it shift vector} $N^\mu$ with respect to $t^\mu$ by
 \bes
 N&=&-t^\mu n_\mu, \nn \\
 N_\mu &=& q_{\mu \nu} t^\nu, \nn
 \ees
 where $n^\mu$ is the unit normal vector field to the hypersurfaces $\Sigma$ and $q_{\mu \nu}= g_{\mu \nu} +n_\mu n_\nu$ is a (3d Riemannian) {\it spatial metric} induced on each $\Sigma$ by the space-time metric $g_{\mu \nu}$. 
 \be
 t^\mu=N^\mu +Nn^\mu.
 \ee 
 We can interpret the vector field $t^\mu$ as the ``flux of time" across space-time. Moreover, it is then obvious that the lapse function $N$ and the shift vector $N^\mu$ are not considered dynamical since they merely prescribe how to ``move forward in time".  The metric $q_{\mu \nu}$ is spatial in the sense that $q_{\mu \nu}n^\mu=0$ and by convention, we now denote the induced spatial metric using only spatial coordinates $q_{ab}$ in the coordinates patch $\{x^a\}$ on the surface $\Sigma$. \\
The ADM formalism  characterizes the phase space of  general relativity when the degrees of freedom are encoded in the metric.
The object of interest is the Einstein-Hilbert action for  the metric $g_{\mu \nu}$ which propagates on the manifold $\mathcal{M}$
\be
S_{EH}=\f{1}{2\kappa} \int_{\mathcal{M}} d^4x \, \sqrt{|\det(g)|} \,R, 
\ee
where $R$ is the Ricci scalar associated with $g_{\mu \nu}$, and  $\kappa= 8 \pi G$ in units  $c=1$.

The ten components of the spacetime metric are then replaced by the six components of the induced Riemannian metric $q_{ab}$ of $\Sigma$, plus the three components of the shift vector $N^a$ and the lapse function $N$. Let us now illustrate this by defining the coordinates for two nearby hypersurfaces (see Fig. \ref{sandwich}) using the ADM variables $(q_{ab}, N, N_a)$ \cite{mtw}.
\begin{itemize}
\item The intrinsic metric of each surface $\Sigma$ is $q_{ab}=g_{ab}$. The metric of the  hypersurface $\Sigma_t$, 
\be
q_{ab}\,dx^adx^b=g_{ab}(t,x,y,z)\,dx^adx^b,
\ee
tells the square of the distance between two points in $\Sigma_t$. The metric on the upper hypersurface,  $\Sigma_{t+dt}$,  is similarly given by
\be
g_{ab}(t+dt,x,y,z)\,dx^adx^b.
\ee
This formalism thus suggests that the spatial metric on a three-dimensional hypersurface $\Sigma$ can be viewed as the dynamical variable in general relativity. Indeed, if we identify the hypersurfaces $\Sigma_t$, $\Sigma_{t+dt}$ by the diffeomorphism resulting from following integral curves of $t^\mu$, we may view the effect of ``moving forward in time" as that of changing the spatial metric on an {\it abstract} three-dimensional $\Sigma$ from $q_{ab}^{(t)}(x,y,z)=g_{ab}(t,x,y,z)$ to $q_{ab}^{(t+dt)}(x,y,z)=g_{ab}(t+dt,x,y,z)$. That is, the space-time $(\cM, g_{\mu \nu})$ may be viewed as representing the time development of $q_{ab}$ on a fixed three-dimensional manifold.
  \begin{figure}[ht]
\begin{center}
\includegraphics[width=7cm]{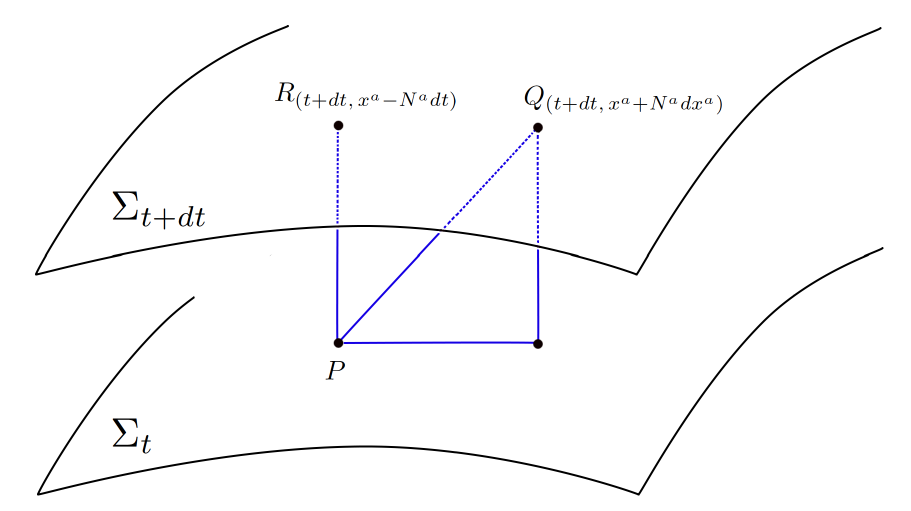}
\end{center}
\caption{Coordinates for two nearby hypersurfaces labelled by $t$ and $t+dt$} \label{sandwich}
\end{figure}
\item The shift vector $N^a=N^a(t,x^b)$ appears  when following the timelike normal vector from a point $P$ with co-ordinates $(t,x^a)$ to a point $R$ on $\Sigma_{t+dt}$, the upper surface with  coordinates:
\be
R=(t+dt, x^a-N^adt).
\ee
$N^a$ measures therefore the amount of ``shift" tangential to $\Sigma_t$ in the time flow vector field $t^\mu$. 
\item
The proper time elapsed from $P$ (lower surface) to $R$ (upper surface) can be defined  from the lapse function $N(t,x^a)$:
\be
(\textrm{lapse of proper time between lower and upper hypersurface})=N(t,x,y,z)dt.
\ee
That is, $N$ measures the rate of   proper time flow with respect to the unphysical coordinate time, $t$, as one moves normally to $\Sigma_t$.

\item In Fig. \ref{sandwich}, let $Q$ be a point on the final surface with coordinates $(t+dt, x^a+dx^a)$, then the squared interval $ds^2=g_{\mu \nu}dx^{\mu}dx^{\nu}$ for the infinitesimal vector $dx^{\mu}$ describing $PQ$ is deduced from the Lorentzian geometry in Fig. \ref{sandwich}: 
\bes \label{ADM}
ds^2&=&(\textrm{proper distance in base 3-geometry})^2-(\textrm{proper time from lower to upper 3-geometry})^2\nn\\
 \label{3-4metric}
&=& q_{ab}(dx^a+N^adt)(dx^b+N^bdt)-(Ndt)^2.
\ees
From this equation (\ref{ADM}), the four-metric can be constructed out of the three-metric and the lapse and shift functions:
\begin{equation} \label{4metric}
g_{\mu \nu}=
\left (
\begin{array}{cc}
g_{00} & g_{0b} \\
g_{a0} & g_{ab}
\end{array}
\right)=
\left (
\begin{array}{cc}
(N_aN^a-N^2) & N_b \\
N_a & q_{ab}
\end{array}
\right),
\end{equation}
where $N^a$ are the components of the shift in its contravariant form, and $N_a=q_{ab}N^b= g_{ab}N^b$ are the covariant components; they are calculated within the three-geometry with the three-metric. The inverted relation $N^b=q^{bc}N_c$ is obtained using the inverse three-metric $q^{bc}$ that has to be  distinguished from the inverse four-metric. One can verify that the inverse four-metric is
\begin{equation}\label{4inversemetric}
g^{\mu \nu}=
\left (
\begin{array}{cc}
g^{00} & g^{0b} \\
g^{a0} & g^{ab}
\end{array}
\right)=
\left (
\begin{array}{cc}
-1/N^2 & N^b/N^2 \\
N^a/N^2 & (h^{ab}-\frac{N^aN^b}{N^2})
\end{array}
\right),
\end{equation}
by expanding the relation $g_{\mu \nu}g^{\mu \lambda}=\delta_\nu^\lambda$.

\item The unit future-directed \emph{normal vector} $n^\mu$ to the hypersurface $\Sigma_t$ points along $PR$ in Fig. \ref{sandwich}. It corresponds to the one-form which has the value:
\begin{equation}
n_\mu dx^\mu=-Ndt. 
\end{equation}
Therefore,in covariant one-form representation, this unit timelike normal vector  has the components
\bes \label{unitn}
&& n_\mu=(-N, 0, 0, 0)\\
 \label{unitn2}
&& n^\mu=g^{\mu \nu}n_\nu=(\frac{1}{N}, -\frac{N^m}{N}).
\ees
\end{itemize}
In terms of these variables, after performing the standard Legendre transformation, the action of general relativity becomes, 
\be
S_{EH}[q_{ab}, \pi^{ab}, N^a, N]= \f{1}{2 \kappa} \int dt \int_{\Sigma} d^3x \, \left[\pi^{ab}\dot{q}_{ab} + 2 N_b\nabla^{(3)}_a (q^{-1/2}\pi^{zb}) +N(q^{1/2} [R^{(3)} -q^{-1}\pi_{cd} \pi^{cd} +\f12 q^{-1} \pi^2])\right]
\ee
where $\pi^{ab}=\f{\partial \mathcal{L}_{EH}}{\partial \dot{q}_{ab}}$ are the momenta canonically conjugated to the space metric $q_{ab}$, $\pi=\pi^{ab}q_{ab}$, $\nabla^{(3)}_a$ is the covariant derivative compatible with the metric $q_{ab}$, $q$ is the determinant of the space metric $q_{ab}$ and $R^{(3)}$ is the Ricci tensor of $q_{ab}$. 

\begin{figure}[ht]
\begin{center}
\includegraphics[width=7.5cm]{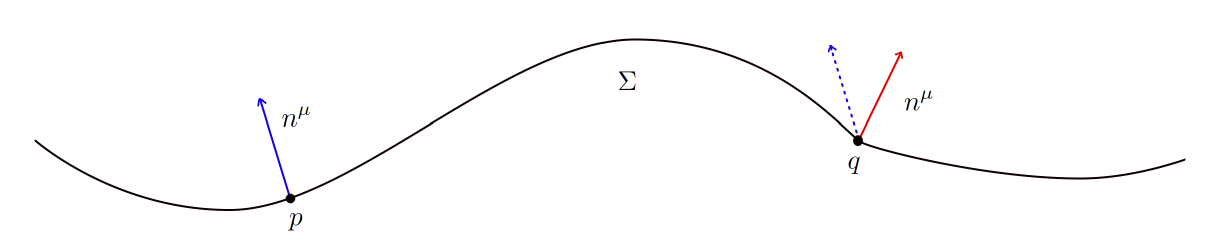}
\end{center}
\caption{This spacetime diagram illustrates the notion of the extrinsic curvaure of a hypersurface $\Sigma$. The dashed arrow at $q$ represents the parallel transport of the normal vector $n^\mu$ at $q$ along a geodesic connecting $p$ to $q$. The failure of this vector to coincide with $n^\mu$ at $q$ corresponds intuitively to the bending of $\Sigma$ in the space-time in which it is embedded. The extrinsic curvature $K_{ab}$ is a spatial entity; $K_{\mu \nu}$ the 4-dimensional entity is defined such that  $K_{ab}$ is the pull-back on $\Sigma$ of $K_{\mu \nu}$. Then, $K_{\mu \nu}\equiv q_\mu^\rho \nabla_\rho n_\nu$ shows that $K_{\mu \nu}$ directly measures this failure of the two vectors at $q$ to coincide for $p$ near $q$.} \label{rate}
\end{figure}

We now introduce the concepts of intrinsic and extrinsic curvatures which appear in the language associated to this $3+1$ space-time split. The intrinsic curvature gives the three-geometry of a spacelike hypersurface $\Sigma$. It may be defined and calculated by the same methods known for the calculation of four-dimensional curvature from the induced metric $q_{ab}$ on $\Sigma$. The extrinsic curvature, denoted $K_{ab}$,  of a three-dimensional hypersurface embedded in a four-geometry gives the rate of change of the three-metric in the normal direction (see Fig. \ref{rate}).

 The momenta $\pi_{ab}$ are related to the extrinsic curvature $K_{ab}$ of $\Sigma$ since we can check that 
\be
\pi^{ab}=q^{-1/2} (K^{ab}-Kq^{ab}),
\ee
with $K=K_{ab}q^{ab}$. 
Variations of the Einstein-Hilbert action with respect to the lapse function and the shift vector produce the following constraints.
\bes 
\textrm{The vector constraint: } &&-V^b(q_{ab}, \pi^{ab})\equiv 2\nabla^{(3)}_a(q^{-1/2}\pi^{ab})=0 , \label{Vconstraints} \\ 
\textrm{The scalar constraint: } && -S(q_{ab}, \pi^{ab})\equiv (q^{1/2} [ R^{(3)}-q^{-1} \pi_{cd}\pi^{cd}+1/2 q^{-1/2} \pi^2])=0. \label{Sconstraint}
\ees
The vector constraint $V^b$ and the scalar constraint are also respectively called the space-diffeomorphism constraint and the Hamiltonian constraint. Physical configurations (also called on-shell configurations) must satisfy these constraints that we can write under the more compact form $H^\mu=(S,V^b)$. $H^\mu$ are the  generators of the space-time diffeomorphism group Diff$\mathcal{M}$ on physical configurations. We can now  rewrite the action under the form.
\be \label{S_EH_hamilton}
S_{EH}[q_{ab}, \pi^{ab}, N^a, N]= \f{1}{2 \kappa} \int dt \int_{\Sigma} d^3x \, \left[\pi^{ab}\dot{q}_{ab}-N_bV^b(q_{ab}, \pi^{ab})-NS(q_{ab}, \pi^{ab})\right].
\ee
From this, we see that the Hamiltonian density of general relativity is 
$$
\mathcal{H}(q_{ab}, \pi^{ab}, N_a, N)= N_bV^b(q_{ab}, \pi^{ab})+NS(q_{ab}, \pi^{ab}).
$$ 
The Hamiltonian 
\be
{\bf H}=\f{1}{2 \kappa}  \int d^3x \, \left(N_bV^b(q_{ab}, \pi^{ab})+NS(q_{ab}, \pi^{ab})\right)
\ee
 vanishes on-shell since it is a linear combination of constraints. General relativity is an example of so-called constrained Hamiltonian system with no physical Hamiltonian  (for an introduction of this topic see \cite{book-dirac, teitelboim}). Classically, the constraints are equivalent to the  equations of motion.  The fact that the Hamiltonian is zero  is a consequence of the diffeomorphisms symmetry of the theory: the ``time" $t$ should not be regarded as a  physical quantity and there is no proper dynamics with respect to $t$. 

The formulation given by (\ref{S_EH_hamilton}) allows to study the phase space of general relativity parametrized by the pair $(q_{ab}, \pi^{ab})$ with the symplectic structure given by the canonical Poisson brackets:
\be
\{\pi^{ab}(t,x), q_{cd}(t,y) \}= 2 \kappa \delta^a_{(c}\delta^b_{d)}\delta(x-y), \quad \{\pi^{ab}(t,x), \pi^{cd}(t,y) \}=\{q_{ab}(t,x), q_{cd}(t,y) \}=0.
\ee
This defines the kinematical phase space. On this space, the constraints (\ref{Vconstraints}) and (\ref{Sconstraint}) will define a hypersurface --  the constraint surface -- where they are satisfied, i.e. the subspace of $(q_{ab}, \pi^{ab})$ such that $H^\mu(q, \, \pi)=0$. 
The constraints   $H^\mu$ have their Poisson brackets which vanish on this constraint surface and are therefore by definition \textit{first class constraints} \cite{book-dirac}.
There are six configuration variables $q_{ab}$ and four first class constraint equations given by (\ref{Vconstraints}) and (\ref{Sconstraint}) so that  we have  two physical degrees of freedom, which is the usual result.  Note that this counting of physical degrees of freedom is correct only because the constraints are first class. 

\begin{figure}[ht]
\centering
\subfigure[{}Generic situation (non relativistic picture): ${\bf H}=H_0+  constraints$]{\centering\includegraphics[width=8cm]{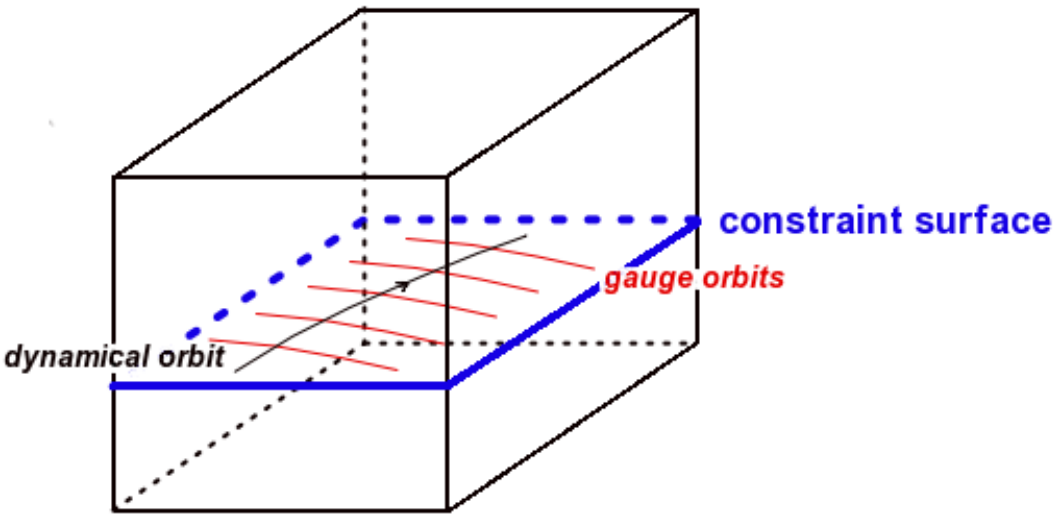}} 
\subfigure[{}Covariant theory (\eg general relativity): ${\bf H}= constraints$]{\centering\includegraphics[width=8cm]{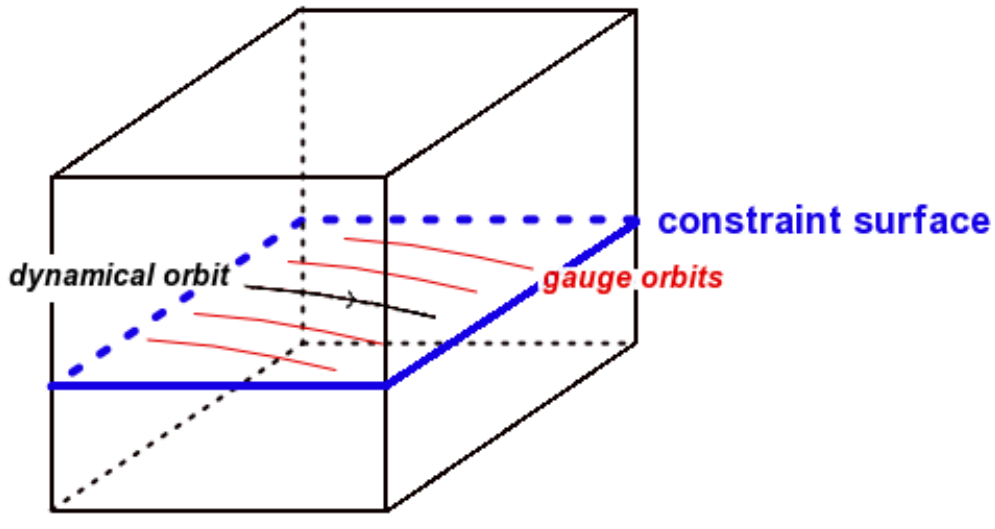}}
\caption{Illustration of dynamical orbits generated by ${\bf H}$ and gauge orbits generated by first class constraints.} \label{gaugeorbit}
\end{figure}

We recall that a first class constraint  generates a gauge transformation on the constraint surface \cite{book-dirac}. We have  illustrated in Fig.  \ref{gaugeorbit} the concepts of constraint surface and gauge orbits in two cases: the left-hand side  represents a generic situation where the Hamiltonian can be decomposed as ${\bf H}=H_0+ \,constraints$; the right-hand side figure illustrates the case of general relativity where the Hamiltonian is simply  given by ${\bf H}=\,constraints$.  We refer to the trajectories generated by the gauge transformations as \textit{gauge orbits}. Points along one gauge orbit correspond to the same physical configuration, only described in different coordinate systems.

The ADM formulation could not be used for a canonical quantization following Dirac's framework. Indeed, the constraints $H^\mu$ are non-polynomial functions of the variables $(q_{ab}, \pi^{ab})$ which makes therefore the theory too complicated to quantize. It was therefore  thought for many years that it was impossible to quantize gravity by other means that the perturbative quantum field   formalism, approach which was itself  plagued  by important issues such as  non-renormalizability or the fact that one only deals with linearized gravity in this context. This was until A. Ashtekar came and found a change of variables to simplify the shape of the constraints $H^\mu$,  that sparked the loop quantum gravity  revolution.

\section{Loop quantum gravity variables}
We will now make a very simple change of variables (see \cite{book-ashtekar, book-thomas, alej-spinfoam, simone-dona} for review articles).  A triad\footnote{The triad $e^i_a$ is the spatial part of the tetrad field $e^I=e^I_\mu dx^\mu$ which defines the space-time metric: $g_{\mu \nu}=e^I_\mu e^J_\nu \eta_{IJ}$.  $\eta_{IJ}$ is the flat metric  in the tangent Minkowski space,  $I, \cdots$ are internal indices living in the tangent Minkowski space. $i=1\cdots 3$ is an internal index  and the spatial component of $I$. $e^I_\mu$   thus represents the gravitational field and indeed, it can be viewed as the field that determines, at each point of space-time, the preferred frame in which motion is inertial. }  
$e^i_a$ is a set of three  1-forms defining a frame at each point in $\Sigma$ and such that the 3-metric $q_{ab}$ is given by
\be
q_{ab}=e^i_a e^j_b \delta_{ij}.
\ee
The crucial change of variables is defined using the densitized triad $E^a_i$
\be \label{densitizedtriad}
E^a_i\equiv \f12 \epsilon_{ijk}\epsilon^{abc}e^j_be^k_c,
\ee
 which  is related to the inverse metric $q^{ab}$ as follows
\be \label{qE}
qq^{ab}=E^a_iE^b_j\delta^{ij} \quad \textrm{ with } i, j =1,2,3.
\ee
Another relevant 1-form $K^i_a$ can be constructed by combining the extrinsic curvature  $K_{ab}$ and the densitized triad $E^a_i$
\be
K^i_a= \f{1}{\sqrt{\det(E)}} K_{ab}E^b_j\delta^{ij}.
\ee
We can then rewrite (\ref{S_EH_hamilton}) in terms of the new variables $(K^i_a, E^a_i)$. Indeed,  the canonical term is simply  
\be
\pi^{ab}\dot{q}_{ab}=-\pi_{ab}\dot{q}^{ab}=2E^a_i \dot{K}^i_a
\ee 
and the constraints (\ref{Vconstraints}) and (\ref{Sconstraint}) can also easily be written in terms of these variables $V^a(E^a_i, K^i_a)$ and $S(E^a_i, K^i_a)$. However, the six components of $q^{ab}$ are described from (\ref{qE}) using  nine $E^a_i$. The extra three degrees of freedom come from the fact that we have the freedom to perform internal $\SO(3)$ rotations. Indeed two different triads related by a local $\SO(3)$ transformation acting in the internal indices $i=1,2,3$ will define the same 3-metric $q_{ab}$. Since three degrees of freedom have been added to the configuration variables whereas the physical degrees of freedom have not changed, three new constraints have to be added. In fact, we missed the fact that $K_{(ab)}=K_{ab}$. which is equivalent to 
\be
G_{jk}(E^a_j, K^j_a):=K_{a[j}E^{a}_{k]}=0.
\ee
Therefore, the Einstein-Hilbert action on this extended phase space generated by $(K^i_a, E^a_i)$  becomes 
\be
S_{EH}[E^a_j, K_a^j, N^a, N, \lambda^{kl}]= \f{1}{ \kappa} \int dt \int_{\Sigma} d^3x \, \left[E^{a}_i\dot{K}^{i}_a-N_bV^b(E^a_j, K_a^j)-NS(E^a_j, K_a^j)-\lambda^{kl}G_{kl}(E^a_j, K_a^j)\right].
\ee
The symplectic structure is now given by
\be
\{E^a_j(t,x), K^i_b(t,y) \}= \kappa \delta^a_{b}\delta^i_{j}\delta(x-y), \quad \{E^a_j(t,x),E^b_i(t,y) \}=\{K^j_a(t,x), K_b^i(t,y) \}=0.
\ee
The vector and scalar constraints (\ref{Vconstraints}) and (\ref{Sconstraint}) have not been modified in the transition from the ADM phase space to the extended phase space. The addition of new degrees of freedom does not, by itself, simplify the constraints. 

The key-step in the simplification is  a  canonical transformation on the extended phase space which is performed to construct the Ashtekar-Barbero connection variables $(A^i_a, P_i^a)$ \cite{ABvariables1, ABvariables2, ABvariables3}.
\be \label{canontransfo}
(E^a_j, K^i_b) \; \longrightarrow \left\{\tabl{l}{A^i_a \equiv \Gamma^i_a +\gamma K^i_a \\
P^a_i\equiv \f{1}{\gamma} {E_i^a}
} \right. 
\ee
where $\gamma \in \R^\ast$  is a constant called Immirzi parameter \cite{immirziPara} and $\Gamma_a^i$ is a $\mathfrak{so}(3)$-spin connection which is characterized as the solution of Cartan's structure equations
\be
\partial_{[a}e^i_{b]}+\epsilon^i_{jk}\Gamma^j_{[a}e^k_{b]}=0.
\ee
A spin connection is defined as an extension of the spatial covariant derivative from tensors to generalized tensors with $\mathfrak{so}(3)$ indices. Then, $A^i$ is also a well-defined $\mathfrak{so}(3)$ connection. Indeed, $K_a^i$ is the conjugate momentum of the densitized triad $E^a_j$ which transforms in the vector representation of $\SO(3)$ under the redefinition of the triad $(q_{ab}=e^i_ae_b^j\delta_{ij})$. Consequently, $K^i_a=\f{1}{\sqrt{\det(E)}} K_{ab}e^b_j\delta^{ij}$ transforms also as a vector. This is why $A^i_a= \Gamma^i_a +\gamma K^i_a$ is also a $\mathfrak{so}(3)$ connection.

Let us emphasize that the constraint structure does not distinguish $\SO(3)$ from $\SU(2)$ as both groups have the same Lie algebra. Thus, from now on we choose to work with the more fundamental (universal covering) group $\SU(2)$. Moreover, from this point, we will rather use the variables $E_j^a$ instead of the variables $P^a_j$ as mostly used in the literature. 

Consider now the (rotational) constraint $G_{ij}$. It can be written  as
\be
G_i(E^a_j, K^j_a)=\epsilon_{ikl}K^k_aE^a_l,
\ee
then, using the new variables, it simplifies into
\be \label{Gauss}
G_i(E_j^a, A^j_a)=D_aE^a_i \equiv \partial_a E^a_i + \epsilon_{ij}^{\,\,\,k}A^j_aE^a_k=0,
\ee
\ie it is the covariant divergence of the densitized triad. Therefore, written with the Ashtekar-Barbero variables $(A^i_a, E_i^a)$, this constraint takes the structure of a Gauss law constraint for an $\SU(2)$ gauge theory. $G_i$ is called the Gauss constraint and generates gauge transformations. As we mentioned above, $E^b_i$ and $A^i_a$ transform respectively as an $\SU(2)$ vector and as an $\SU(2)$ connection under this transformation. The other constraints (\ref{Vconstraints}) and (\ref{Sconstraint}) which still form with the Gauss constraint a  first class algebra, become
\bes \label{VSconstraints}
V_a(E^a_j, A^j_a)&=& F^j_{ab}E^b_j-(1+\gamma^2)K_a^iG_i, \nn \\
S(E^a_j, A^j_a)&=&\f{E^a_iE^b_j}{\sqrt{\det(E)}} \left(\epsilon^{ij}_{\,\,k}F^k_{ab}-2(1+\gamma^2)K^i_{[a}K^j_{b]} \right) ,
\ees
where $F^i_{ab}=\partial_a A^i_b-\partial_bA^i_a+\epsilon^i_{\, jk}A^j_aA_b^k$ is the curvature of the connection $A^i_a$.
We can then rewrite the action 
\be
S_{EH}[E^a_j, A_a^j, N^i, N, \lambda^{k}]= \f{1}{ \kappa} \int dt \int_{\Sigma} d^3x \, \left[E^{a}_i\dot{A}^{i}_a-N_bV^b(E^a_j, A_a^j)-NS(E^a_j, A_a^j)-\lambda^{k}G_{k}(E^a_j, A_a^j)\right],
\ee
and the Poisson brakets of the Ashtekar-Barbero variables are given by the fundamental Poisson brackets:
\be \label{PoissonAB}
\{A^i_a(t,x), E^b_j(t,y) \}= \kappa \gamma \delta^b_{a}\delta^i_{j}\delta(x-y), \quad \{E^a_j(t,x),E^b_i(t,y) \}=\{A^j_a(t,x), A_b^i(t,y) \}=0.
\ee
The $\kappa \gamma$ factor arises in the first Poisson bracket because the conjugate momentum of the configuration  variable $A^i_a$ (obtained as the derivative of the Lagrangian with respect to the velocities) is actually given by $1/\kappa \times P^a_i$.  Moreover, the fact that the Immirzi parameter $\gamma$ can be arbitrary is important because the quantum theories obtained starting with different values of $\gamma$ will  lead to different physical predictions. Furthermore $\gamma$ enters the spectrum of geometrical observables such as areas and volumes (at the kinematical level) as well as the black hole entropy \cite{volume7, area2,  area3, blackhole1, blackhole2, blackhole3, blackhole4, blackhole5, blackhole6, blackhole7, blackhole8}.

\medskip

The next step is to smear this algebra. This is needed in order to proceed with the quantization. The densitized triad is a 2-form, thus we smear it on a two dimensional surface $S \subset \Sigma$ and define  its flux across $S$:
\be
E_i(S)\equiv \int_S n_a E^a_i d^2\sigma,
\ee
where $n_a= \f{\pp x^b}{\pp \sigma^1}\f{\pp x^c}{\pp \sigma^2}\epsilon_{abc}$ is the normal to the surface $S$ and $\sigma^1$, $\sigma^2$ are local coordinates on $S$. Moreover, $E^a_i$ encodes the full background independent Riemannian geometry of $\Sigma$ since the inverse metric $q^{ab}$ is related to $E^a_i$ as follows: $qq^{ab}=E^a_iE^b_j\delta^{ij}$.  \\
The connection $A^i_a$, which is a 1-form,  has also a simple geometrical interpretation: it provides a definition of parallel transport of $\SU(2)$ spinors on the space manifold $\Sigma$. Consider now a path $e \subset \Sigma$, we define the holonomy\footnote{Some useful properties of the holonomy are: \begin{itemize}
\item The holonomy of the composition of two paths $e_1,\, e_2$ is the product of the holonomies of each path: $h_{e_1 e_2}=h_{e_1}h_{e_2}$.
\item  Under a local gauge transformation, $g(x) \in \SU(2)$, the holonomy transforms as $h^g_e=g_{s(e)}h_eg_{t(e)}^{-1}$, where $s(e)$ and $t(e)$ are respectively the source and the target points of the path $e$.
\item Under the action of diffeomorphism, the holonomy  transforms as: $h_e[\phi^\ast A]=h_{\phi \circ e}[A]$.
\end{itemize}} 
of $A$ along $e$ by
\be \label{holonomy}
h_e[A]\equiv \mathcal{P} \exp\left(\int_e A  \right),
\ee 
where $\mathcal{P}$ denotes a path-order product.
\be
h_e= \sum_{n=0}^\infty\, \; \iiint\limits_{1>s_n>\cdots >s_1>0} A(e(s_1)) \cdots A(e(s_n))\, ds_1 \cdots ds_n,
\ee
where we parametrized the line with $s \in [0,1]$ and  $A=A^i_a\tau_i \f{dx^a(s)}{ds}$ with $\tau_i$ the $\su(2)$ generators and $x^a(s): [0,1]\rightarrow \Sigma \,$ a parametrization\footnote{For a given parametrization of the path $e$, $x^a(s): [0,1]\rightarrow \Sigma$ and for a given connection $A^i_a$, we integrate $A_a=A^i_a\tau_i$, a $\SU(2)$ element, along $e$  as a line integral $\int_e A\equiv \int_0^1 ds A^i_a(x(s))\f{dx^a(s)}{ds}\tau_i$.} of the path $e$.

The resulting smeared algebra of $h_e[A]$ and $E_i(S)$ is called the {\it holonomy-flux} algebra. It provides a regular version of the Poisson algebra (\ref{PoissonAB}), \ie no delta function appears anymore. This regularization step of   the Ashtekar-Barbero variables using paths and surfaces is the last step to prepare general relativity for the loop quantization.


Before giving a description of the implementation of Dirac's program in the case of gravity using the classical variables presented above,  we would like to comment on the fact that loop quantum gravity considers $\SU(2)$ as gauge group of general relativity instead of the non-compact Lorentz group $\SL(2,\C)$. In the next  section, we will introduce the classical framework which leads  to a  \textit{covariant}  loop quantum gravity theory, based on a $\SL(2,\C)$ connection.

The initial canonical formulation of loop quantum gravity used the (Ashtekar) variables $(A,E)$ as canonical variables \cite{book-ashtekar}: a self-dual complex connection $A= \Gamma \pm i K$ and its conjugate triad field $E$. The resulting theory is then invariant under the Lorentz group $\SL(2,\C)$ (seen as the complexified $\SU(2)$ group, and it is the covering group of the restricted Lorentz group $\SO(1,3)^+$) and under space-time diffeomorphism. However, a difficulty  comes from the reality constraints expressing that the imaginary part of the triad field $E$  vanishes and that the real part of the connection $A$ is actually a function of the $E$: $A^i_a +\overline{A}^i_a=\Gamma^i_a(E)$. \\
These constraints rendered the quantization complicated. The standard formulation of loop quantum gravity presented above avoids this reality constraint issue using the \textit{real} Ashtekar-Barbero connection $A= \Gamma +\gamma K$ where $\gamma \in \R$ and its real conjugate triad $E$. But then, the Ashtekar-Barbero connection $A$, on the spatial slice, is \textit{not} the pull-back of a space-time connection\footnote{Whereas in the case where $\gamma = \pm i$, $A_a$ is the pullback of $\omega^{+IJ}_\mu$ $(I,J=1, \cdots, 4)$ with $\omega^{+IJ}_\mu= \f12 (\omega^{IJ}_\mu -\f{i}{2} \epsilon^{IJ}_{\;\; KL}\, \omega_\mu^{KL})$ the self-dual part of a Lorentz connection $\omega^{IJ}_\mu$.} and from that point of view, the real connection $A$ cannot be considered as a genuige gauge field: $\SU(2)$ can not be viewed as the gauge group of gravity \cite{samuel}. 
In fact, deriving the theory from the original first order formalism, considering the   Palatini action  as starting point, 
\be \label{PalaAct}
S_{\textrm{Pal}}= \f{1}{2 \kappa} \epsilon_{IJKL}\int_{\mathcal{M}} e^I\wedge e^J \wedge F^{KL}(\omega)
\ee
we can see that the $\SU(2)$ gauge group appears because of a particular (partial) gauge fixing, the time gauge, which breaks the local Lorentz invariance down to a local $\SU(2)$ gauge invariance and allows to recover the Ashtekar-Barbero variables. In the previous action (\ref{PalaAct}), the tetrad field $e^I_\mu$  is defined as $e^I=e^I_\mu dx^\mu$ and $F(\omega)=d\omega+ \omega \wedge \omega$ is the curvature tensor of the connection $\omega$ with $\omega=\omega^{IJ}_\mu J_{IJ}dx^\mu$ a $\mathfrak{sl}(2,\C)$-valued 1-form, with  
$J_{IJ}\in\sl(2,\C)$. 
A tetrad $e^I_\mu(x), \; I=0,1,2,3$ is defined such that
\be \label{definertial}
g_{\mu \nu}(x)=e^I_\mu(x) e^J_\nu(x) \eta_{IJ}
\ee
and thus provide a local isomorphism between a general reference frame and an inertial one, characterized by the flat metric $\eta_{IJ}$. A local inertial frame is defined up to a Lorentz transformation: $e^I_\mu(x) \rightarrow \tilde{e}^I_\mu(x)= \Lambda_J^{I}(x) e^J_\mu(x)$. Notice that the definition (\ref{definertial}) is well invariant under such a transformation. Thus, the "internal" index $I$  carries a representation of the Lorentz group. Contracting vectors and tensors in spacetime with the tetrad, we get objects that transform under the Lorentz group. For example, we define the "internal time direction" $x^I$ from the unit normal vector field $n^\mu$ (\ref{unitn2}) to $\Sigma$ which appears in the 3+1 splitting of space-time $(\mathcal{M} \cong \R \times \Sigma)$ by:
\be \label{defX}
x^I=e^I_\mu n^\mu= \left(\f{e^0_0-e^0_a N^a}{N}, \f{e^i_0-e^i_a N^a}{N} \right)
\ee
We will see that this field is fundamental in the covariant framework presented in the following chapter \ref{CLQGclas}.
In (\ref{PalaAct}), the connection $\omega$ is considered as an independent variable 
and consequently, the action (\ref{PalaAct}) is invariant under local Lorentz transformations. When  varying the action with respect to $e$, we obtain the" torsion-free" equation
\be\label{torsion-free}
de^I+ \omega^I_{\; J} \wedge e^J=0,
\ee
On the other hand, varying with respect to $\omega$,  and considering $\omega$ as a solution $\omega(e)$ of \eqref{torsion-free}  one recovers  the solutions of the Einstein equations  (plus an additional sector with degenerate metrics in the case of a degenerate (i.e. non-invertible) tetrad),  further details can be found in \cite{simone-dona}.
 \\
For the Hamiltonian formulation of the Palatini action, we proceed as before, assuming a $3+1$ splitting of the space-time $(\mathcal{M} \cong \R \times \Sigma)$ and coordinates $(t,x)$. Using the definition of $x^I$ (\ref{defX}), we obtain that a tetrad for the ADM metric (\ref{ADM}) is given by
\be
e^I_0=Nx^I+N^a e^I_a, \quad q_{ab}=e^i_a e^j_b \delta_{ij}, \; i,j =1,2,3.
\ee
where the spatial part of the tetrad, $e^i_a$ is the triad.  At this stage, the structure is complicated, in particular because the constraint algebra is second class. Indeed, the fact to use the tetrad and the connection as independent fields implies that the conjugate variables are now functions of both $e^I_a$ and $\omega^{IJ}_a$ (and their time derivative) as opposed to be function of the metric $q_{ab}$ only. Furthermore, the canonically conjugated variables to the connection $\omega^{IJ}_a$ are not independent and satisfy the so-called \textit{simplicity constraints} which are second class constraints (\ie which Poisson brackets do not vanish weakly). This is why the "time gauge" $x^I=e^I_\mu n^\mu=\delta^I_0$ is used in order to simplify the discussion
\be
e^0_\mu=(N,0)\, \longrightarrow \; e^I_0=(N, N^a e^i_a).
\ee 
Working in the time gauge breaks the local Lorentz invariance but allows to define the densitized triad (\ref{densitizedtriad}) and the Ashtekar-Barbero connection\footnote{The link between the form $A^i_a=\f12 \epsilon^{i}_{\, jk} \omega_a^{\, jk}+ \gamma \omega_a^{\, 0i}$ of the Ashtekar-Barbero connection and the previous form $A^i_a= \Gamma^i_a +\gamma K^i_a$ is explicitly given in \cite{book-thomas}.} $A^i_a=\f12 \epsilon^{i}_{\, jk} \omega_a^{\, jk}+ \gamma \omega_a^{\, 0i}$ and to recover the first class constraint algebra with (\ref{Gauss}) and (\ref{VSconstraints}).  \\
In the following section, we present an alternative classical framework where no gauge is fixed. 

\section{Classical framework for a "covariant" loop quantum gravity theory} \label{CLQGclas}

We now review the classical framework of a Lorentz covariant approach to loop quantum gravity. The aim is to develop the canonical formalism for gravity with the full Lorentz group as a local symmetry.
This was first performed by Alexandrov \cite{sergei-immirzi} and was coined {\it covariant loop quantum gravity}. We will see that the canonical variables are in this approach a Lorentz connection $\omega$ and its conjugate triad $e$ valued in the Lorentz algebra.
We consider once again a first order formalism of general relativity with Palatini action (\ref{PalaAct}) where the connection $\omega$ is independent from the tetrad $e$. 
It is possible to take into account directly at this level the Immirzi parameter $\gamma$ which appear in the change of variables (\ref{canontransfo}).
Indeed, we can add  to the Palatini action a second term -- the Holst term -- which is compatible with all the symmetries and has mass dimension 4 \cite{holst}. This leads to the so-called Palantini-Holst action
\be \label{PalaHolst}
S[\omega, e]= \int_\mathcal{M}\left[ \f12 \epsilon_{IJKL}e^I\wedge e^J \wedge F^{KL}(\omega)-\f{1}{\gamma} e^I\wedge e^J \wedge F_{IJ}(\omega) \right].
\ee
As it can be checked directly, the equations of motion are not affected by the Holst term and we recover the  Einstein equations when the tetrad is not degenerated.
The Immirzi parameter $\gamma$ appearing in front of the Holst term has no effect on the equations of motion and thus does not matter at the classical level. As we mentioned previously, the difficulty in the canonical analysis comes from the second class nature of the constraints which appear. More precisely, the canonically conjugated variable $\Omega^a_{IJ}$ to the connection $\omega_a^{IJ}$ is $\Omega^a_{IJ}= \epsilon^{abc}\epsilon_{IJKL}e^K_be^L_c$. These variables are not independent and they satisfy the \textit{simplicity constraints}
\be
\forall \, a, \, b, \; \epsilon^{IJKL}\Omega^a_{IJ}\Omega^b_{KL}=0.
\ee
These constraints are the non trivial part of the canonical structure. In the ``time gauge" $x^I=e^I_\mu n^\mu=\delta^I_0$ we recover the Ashtekar-Barbero variables $(E^a_i, A^i_a)$, the first class constraint algebra with (\ref{Gauss}) and (\ref{VSconstraints}), and the simplicity constraints do not appear.
However, it is also possible to explicitly solve these constraints and by breaking the Lorentz covariance later,  we still recover the Ashtekar-Barbero formalism \cite{barros}.

Nevertheless, it is possible to write a canonical formulation of general relativity with the Immirzi parameter which preserves the full Lorentz gauge symmetry and to treat it in a covariant way. Such a formulation was constructed in \cite{sergei-immirzi}. 
The internal time direction $x^I$, defined by (\ref{defX}) or equivalently by 
\be \label{xnorma0}
x^I=e^{I\mu}n_\mu= -N \left( e^{00}, e^{i0} \right),
\ee
where we used  (\ref{unitn}) for the definition of $n_\mu$, can now be written under the general form
\be \label{xnorma}
x^I(\mathcal{X})=\f{1}{\sqrt{1-|\vec{\mathcal{X}}|^2}}{(1, \,\mathcal{X}^i)}.
\ee
Therefore, comparing the two previous equations \eqref{xnorma0} and \eqref{xnorma}, we deduce that $\mathcal{X}^i=\f{e^{i0}}{e^{00}}$. \\
We now start again with a space-time $\mathcal{M} \cong \R \times \Sigma$ where we distinguish the time direction from the three space dimensions. The lapse function $N$, the shift vector $N^a$, the triad $e^a_i$ and the new field $\mathcal{X}_i$ arise from the following decomposition of the tetrad
\bes
e^0&=&Ndt+\mathcal{X}_i \mathcal{E}_a^idx^a \nn \\
e^i&=&\mathcal{E}^i_aN^adt+ \mathcal{E}^i_adx^a.
\ees
where $\mathcal{E}^i_a$ is a triad.
The field $\mathcal{X}_i$ describes the deviation of the normal to the spacelike hypersurface $\{t=0\}$ from the time direction\footnote{The slice is respectively  spacelike,  lightlike or timelike when $|\vec{\mathcal{X}}|^2= \mathcal{X}^i\mathcal{X}_i$ is  respectively less than 1, equal to 1 or  bigger than 1. The presentation given here holds for a spacelike foliation but the timelike can also be treated in the same way \cite{CLQG3}.}.

$x$ defines a subgroup $H_x=\SU_x(2)$ of the gauge group $\SL(2,\C)$ which is the isotropy subgroup of $x^I$ with respect to the standard action of $\SL(2, \C)$ in $\R^4$. The field $\mathcal{X}$ is absent in the Ashtekar-Barbero formalism since the condition $\mathcal{X}=0$, which corresponds to the time gauge, is imposed from the very beginning. 

Let's call $X,Y, \cdots = 1.. 6$  the $\sl(2,\C)$-indices labeling antisymmetric couples $[I \,J]$. The first 3 components correspond to $(0,i)$ and the other 3 are obtained after contraction of the $(i,j)$ components with $\f12 \epsilon^{ijk}$. We define new  connection and triad variables $(\mathcal{A}^X, \mathcal{R}_X)$ valued in $\sl(2, \C)$ instead of the standard $\su(2)$ of Ashtekar-Barbero variables $(A^i, E_i)$. The Lorentz connection $\mathcal{A}^X_a$ is
\be
\mathcal{A}^X=\left( \f12 \omega^{0i}, \, \f12 \epsilon^i_{\, jk} \omega^{jk} \right).
\ee
It is just the space components of the spin-connection $\omega^{IJ}$. We define a ``rotational" triad 
and a ``boost" triad  
\be
\mathcal{R}^a_X=\left(-\epsilon_i^{\, jk} \mathcal{E}^i_a \mathcal{X}_k, \, \mathcal{E}_a^i \right), \quad \mathcal{B}^a_X=\left( \mathcal{E}_a^i,\, \epsilon_i^{\, jk} \mathcal{E}^i_a \mathcal{X}_k \right)= \left(\star \mathcal{R}^a \right)_X
\ee
where $\star$ is the Hodge operator on $\sl(2, \C)$ switching the boost and rotation part of the algebra. The action then reads
\be
S= \int dt \, d^3x\, \left( \left( \mathcal{B}^a_X - \f{1}{\gamma} \mathcal{R}^a_X \right) \partial_t \mathcal{A}_a^X +\lambda^X G_X + N^aV_a+ NS \right).
\ee
The phase space is therefore defined with the Poisson bracket,
\be
\left\{ \mathcal{A}^X_a(t,x), \left( \mathcal{B}^b_Y - \f{1}{\gamma} \mathcal{R}^b_Y \right)(t,y) \right\}=\delta^X_Y\delta^b_a \delta(x-y)
\ee
$\lambda^X$, $N^a$, $N$ are Lagrange multipliers enforcing the ten first class constraints 
\bes
G_X&=&D_{\mathcal{A}}\left( \mathcal{B}_X-\f{1}{\gamma} \mathcal{R}_X\right), \nn \\
V_a\, &=& -\left( \mathcal{B}^b_X - \f{1}{\gamma} \mathcal{R}^b_X \right) F_{ab}^X(\mathcal{A}), \nn \\
S \; &=& \f{1}{1+\f{1}{\gamma^2}} \left(  \mathcal{B} - \f{1}{\gamma} \mathcal{R} \right) \left(  \mathcal{B} - \f{1}{\gamma} \mathcal{R} \right) F(\mathcal{A}) 
\ees
$G_X$ generates the local Lorentz transformation,the analogue for the Gauss constraint (\ref{Gauss})  (there are effectively 6 constraints) and $V_a$ and $S$ generate space-time diffeomorphisms (4 constraints). The constraints have essentially the same form as the ones of the Ashtekar-Barbero formulation (\ref{Gauss}) and (\ref{VSconstraints}) with $E^a_i$, $A^i_a$ being replaced by $(\mathcal{B}^a_X-\f{1}{\gamma} \mathcal{R}_X^a)$ and $\mathcal{A}^X_a$, the structure constants of $\su(2)$ being replaced by the structure constants of $\sl(2,\C)$\footnote{ \ie in particular, $D_{\mathcal{A}}$ is defined as $D$ in (\ref{Gauss}) and $F({\mathcal{A}})$ as $F(A)$ replacing $A$ by ${\mathcal{A}}$ and $\epsilon^{i}_{jk}$ by the $\SL(2,\C)$ structure constants. The explicit definition of the $\sl(2,\C)$ structure constants $f^{XYZ}$ can be found in appendix of \cite{CLQG3, clqg21}.} and the last term in the Hamiltonian constraint $S$ involving the intrinsic curvature being dropped. 
However, in contrast to the  framework for loop quantum gravity, we also have to deal with second class constraints
\be \label{secondclass}
\phi^{ab}=\left( \star \mathcal{R}^a \right)^X \mathcal{R}^b_X,  \quad \psi^{ab} \equiv \mathcal{R}^a \mathcal{R}^b D_{\mathcal{A}} \mathcal{R}.
\ee
The constraints $\phi$ are the simplicity constraints. The constraints $\psi=0$ come from the Poisson bracket $\{S,\phi\}$ and is required in order that the constraint $\phi=0$ is preserved under gauge transformations (generated by $G, \, V_a, \, S$) and in particular under time evolution. $\psi$ correspond to the reality constraints of complex loop quantum gravity. 

The next step to deal with second class constraints is to define the Dirac brackets $\{f,g \}_D=\{f,g\}-\{f, \varphi_r\} \Delta^{-1}_{rs} \{ \varphi_s, g \}$ where the Dirac matrix $\Delta_{rs}=\{\varphi_r, \varphi_s \}$ is made of the Poisson brackets of the constraints $\varphi=(\phi, \psi)$. It is in fact at this step that computing the Dirac brackets of the smeared constraints $G, \, V_a, \, S$  we can show that the smeared $G$ constraints now generate $\SL(2, \C)$ gauge transformations and that the algebra of the first class constraints is not modified \cite{sergei-immirzi, clqg, clqg21}. \\
However, a new difficulty emerges from the fact that although the triad field $R$ is still commutative for the Dirac bracket, the properties of the connection $\mathcal{A}$ change drastically: it is not canonically conjugated to the triad and it does not commute with itself. The structure of the commutator which replaces the simple canonical commutation relation of loop quantum gravity (\ref{PoissonAB}) is now complicated.
To define covariant loop quantum gravity variables, the choice is to keep the triad $\mathcal{R}$ and to define a new connection $\tilde{\mathcal{A}}$. Requiring a good behavior of this Lorentz connection under Lorentz gauge transformations and space diffeomorphisms as well as requiring that it is conjugate to the triad\footnote{This is required in order that the states resulting from a loop quantization using the connection $\tilde{\mathcal{A}}$ diagonalize the area operators $A_{\mathcal{S}}= \int_{\mathcal{S}} d^2x \sqrt{n_an_b \mathcal{R}^a_X\mathcal{R}^{bX}}$ with $n_a$ the normal to the surface $\mathcal{S}$.} $\mathcal{R}$ allows to isolate a two-parameter family of possible connection variables: $\tilde{\mathcal{A}}(\alpha, \beta)$. Then requiring that the connection further behaves as a one-form under space-time diffeomophisms (i.e. which transforms as a 1-form under space-time diffeomophisms generated by the constraints $V_a$, $S$) allows to identify a unique covariant connection corresponding to $(\alpha, \beta)=(0,1)$  \cite{CLQG4, CLQG2} that we will simply denote $\tilde{\mathcal{A}}$ in the following parts. The problem with this connection is that it is non-commutative. Indeed, the bracket $\{ \tilde{\mathcal{A}}^X, \tilde{\mathcal{A}}^Y\}_D$ does not vanish and turns out to be complicated. This comes from the fact that the rotational part of $\tilde{\mathcal{A}}$\footnote{We can define $(P_\mathcal{R})^X_Y=\mathcal{R}_a^X\mathcal{R}^a_Y$, $(P_\mathcal{B})^X_Y=\mathcal{B}_a^X\mathcal{B}^a_Y$ which are respectively the projector on the Lie subalgebra of $H_x$ (rotations) and on its orthogonal complement (boosts). The rotational part of $\tilde{\mathcal{A}}$ is then given by $P_\mathcal{R}\tilde{\mathcal{A}}$ and its boost part by $P_\mathcal{B}\tilde{\mathcal{A}}$.} is not independent of the triad field $\mathcal{R}$ whereas the boost part of $\tilde{\mathcal{A}}$ is canonically conjugate to the boost triad $\mathcal{B}=\star \mathcal{R}$. The explicit relation between $\tilde{\mathcal{A}}$ and $\mathcal{R}$ which can be found in \cite{CLQG2, CLQG3} is reminiscent of the reality constraint   {$A^i_a + \bar{A}^i_a= \Gamma^i_a(E)$} of the complex loop quantum gravity formulation. On the other hand, both the rotation and the boost parts of $\tilde{\mathcal{A}}$ are commutative.
Therefore, the geometrical interpretation of $\tilde{\mathcal{A}}$  in quantum theory is not straightforward. However, we will see in the chapter \ref{chapCLQG} that it is possible to precisely define the Hilbert  and quantum states of space(-time) geometry at least at the kinematical level.


\chapter{Loop Quantum Gravity}
We now come back to the time gauge fixed 'triad-connection' formulation of general relativity  which was 
the key  to reformulate general relativity in a more amenable  way to Dirac's quantization  than the ADM formulation. The general aim is now to construct a Hilbert space of dynamical (physical) states which are annihilated  by the quantum version of the constraints  we derived earlier, using  the $\SU(2)$ Ashtekar-Barbero variables. 
To  recapitulate the classical Hamiltonian analysis we have recalled, we have formulated   general relativity as a $\SU(2)$ gauge theory, with Poisson brackets (\ref{PoissonAB}) and the three sets of constraints
\bes
G_i=0 &\qquad & \textrm{ Gauss law}, \nn \\
V_a=0 & \qquad & \textrm{ Spatial diffeomorphism}, \nn \\
S=0 & \qquad & \textrm{ Hamiltonian constraint}. \nn
\ees
Note however that  there is a key difference between general relativity and a gauge theory. Indeed, in a standard gauge theory like Yang-Mills,  after imposing the Gauss law, the dynamics  is determined by a physical Hamiltonian whereas  in general relativity  the whole dynamics content is in the four left constraints $(V_a, S)$.

The loop quantum gravity program   starts with the definition of the kinematical Hilbert space $\cH_{\textrm{kin}}$.
Once a well-behaved kinematical Hilbert space is constructed, we can follow Dirac's procedure and  we will have a well-posed problem of reduction by the constraints
\bes
\cH_{\textrm{kin}}\qquad \quad \stackrel{\hat{G}_i=0}{\longrightarrow}  & \cH^G_{\textrm{kin}} \quad \qquad \stackrel{\hat{V}_a=0}{\longrightarrow} \quad \qquad \cH_{\textrm{Diff}} \qquad & \stackrel{\hat{S}=0}{\longrightarrow}  \qquad \cH_{\textrm{phys}}. \\
\textrm{cylindrical functions } \qquad & \textrm{spin networks } \quad \textrm{ abstract spin networks} & \qquad  \qquad  \quad ? \nn
\ees

\section{The kinematical Hilbert space and cylindrical functions}\label{cylinfunc}

We regard the connection as the configuration variable. The kinematical Hilbert space consists of a suitable set of functionals $\Psi[A]$ of the connection  which have to be square integrable with respect to a suitable gauge invariant and diffeomorphism invariant measure $d\mu_{AL}[A]$. The kinematical inner product will then be given by:
\be
\la \Psi , \, \Phi  \ra= \mu_{AL}[\overline{\Psi} \Phi]= \int d\mu_{AL}[A]\,\overline{\Psi}[A]\, \Phi[A].
\ee
The difficulty at this stage comes from the fully dynamical feature of the metric. We do not have a background metric at disposal to define the integration measure. We need to define a measure on the space of connections without relying on any fixed background metric. The key to do this relies on the notion of cylindrical functions. We have already seen that a natural quantity associated with a connection consists in the holonomy along a path (\ref{holonomy}). We recall that geometrically the holonomy $h_e[A]$ is a functional of the connection that provides a rule for the parallel trandport of $\SU(2)$ spinors along the path $e$. If we think of it  as a functional of the path $e$ it is clear that it captures all the information about the field $A_a^i$. In addition, it has a very simple behavior under the transformations generated by six of the constraints.
\begin{itemize}
\item Under the  gauge transformation generated by the Gauss constraint  $G_i$, the holonomy transforms as
\be \label{gaugeHolo}
h^g_e=g_{s(e)}h_eg_{t(e)}^{-1}
\ee
where $s(e)$ and $t(e)$ are respectively the source and the target points of the path $e$.
\item  Under the  diffeomorphism action  generated by the vector constraint $V_a$, the holonomy transforms  given $\phi \in $ Diff($\Sigma$)
\be \label{spaceDiffeo}
h_e[\phi^\ast A]=h_{\phi \circ e}[A],
\ee
where $\phi^\ast A$ denotes the action of $\phi$ on $A$. Transforming the connection with a diffeomorphism is therefore equivalent to simply 'moving' the path with $\phi$.
\end{itemize}
For theses reasons, the holonomy is a natural choice of basic functional of the connection.

 We call \textit{cylindrical functions} the functionals that depend on the connection only through the holonomies $h_e[A]=\mathcal{P}\left( \int_e A\right)$ along some finite set of paths $e$. Given a graph $\Gamma \subset \Sigma$, defined as a collection of oriented paths $e \subset \Sigma$ meeting at most at their endpoints, we denote by $E$ the total number of edges it contains. A cylindrical function $\psi_{(\Gamma, \, f)}[A]$ is a functional of the connection $A$, labelled by a graph $\Gamma$ and a smooth function $f: \SU(2)^E \rightarrow \C$
\be
\psi_{(\Gamma, \, f)}[A]= f(h_{e_1}[A], \cdots, h_{e_E}[A]) 
\ee  
where $e_i$ with $i=1,\cdots, E$ are the edges of the corresponding graph $\Gamma$. Given a graph $\Gamma$, we denote  by $Cyl_\Gamma$ the space of cylindrical functions associated to $\Gamma$.

The next step is to equip $Cyl_\Gamma$ with a scalar product in order to turn it into an Hilbert space. 
The switch from the connection to the holonomy which is an $\SU(2)$ element, is here crucial since there exists a unique gauge-invariant and normalized measure $dg$ of $\SU(2)$, called the Haar measure\footnote{The Haar measure of $\SU(2)$ is defined by the following properties:
$$
\int_{\SU(2)}dg=1, \; \textrm{ and } dg=d(\alpha g)=d(g \alpha)=dg^{-1} \; \forall \alpha \in \SU(2).
$$}.  
Using $E$ copies of the Haar measure, we define on $Cyl_\Gamma$ the following scalar product
\be \label{loopScalarProd}
\la \psi_{(\Gamma, \, f)} | \psi_{(\Gamma, \, g)} \ra \equiv \int \left[\prod_{e \subset \Gamma} dh_e\right] \, \overline{f(h_{e_1}[A], \cdots h_{e_L}[A])} \,g(h_{e_1}[A], \cdots h_{e_L}[A]).
\ee
This turns $Cyl_\Gamma$ into an Hilbert space $\cH_\Gamma$ associated to a given graph $\Gamma$. Next, we need to consider $Cyl$ the algebra of the cylindrical functions of the connection $A$ which can be defined as
\be \label{cyl}
Cyl(A)= \bigcup_\gamma Cyl_\gamma
\ee
where $\bigcup_\gamma$ denotes the union on all graphs $\gamma$ in $\Sigma$.
We can then deduced from (\ref{loopScalarProd}) a scalar product between two cylindrical functions in $Cyl(A)$. Indeed,  any cylindrical function $\psi_{(\Gamma^\prime, \, f^\prime)}$ based on a graph $\Gamma^\prime$ can be viewed as a cylindrical function $\psi_{(\Gamma, f)}$ based on a larger graph $\Gamma$ containing $\Gamma^\prime$ by simply choosing $f$ to be independent from the links in $\Gamma$ but not in $\Gamma^\prime$. Moreover, any edge of a graph can be broken in two edges, separated by a (bivalent) node. As a consequence the scalar product for two cylindrical functions $\psi^{(1)}$, $\psi^{(2)}$ with graphs $\Gamma_1$ and $\Gamma_2$ is constructed as follows.
\begin{itemize}
\item If $\psi^{(1)}$, $\psi^{(2)}$ share the same graph,  the definition of the scalar product (\ref{loopScalarProd}) immediately applies.
\item  If they have different graphs, say $\Gamma_1$, $\Gamma_2$, they can be viewed as having the same graph  $\Gamma_3$ formed by the union of the two graphs  $\Gamma_3\equiv \Gamma_1 \cup \Gamma_2$ ($f_1$ and $f_2$ are  trivially extended on $\Gamma_3$) and we define the scalar product on $\Gamma_3$ as (\ref{loopScalarProd}).
\be \label{loopScalarProd2}
\la \psi_{(\Gamma_1, \, f_1)}^{(1)} | \psi^{(2)}_{(\Gamma_2, \, f_2)} \ra \equiv \la \psi_{(\Gamma_3, \, f_1)}^{(1)} | \psi^{(2)}_{(\Gamma_3, \, f_2)} \ra.
\ee
\end{itemize}
This last definition  can be interpreted  as a scalar product between cylindrical functionals of the connection:
\be
\int d\mu_{AL}[A]\; \overline{\psi^{(1)}_{(\Gamma_1, \, f_1)}}[A]\; \psi^{(2)}_{(\Gamma_2, \, f_2)}[A]\equiv \la \psi_{(\Gamma_1, \, f_1)}^{(1)} | \psi^{(2)}_{(\Gamma_2, \, f_2)} \ra 
\ee
with respect to $d\mu_{AL}$, the so-called Ashtekar-Lewandowski measure, which is an integration measure over the space of connections \cite{ALmeasure}. 
This key result, due to Ashtekar and Lewandowski, can be reformulated as follows.  The measure $d\mu_{AL}$ allows us to define an Hilbert space $\cH_{\textrm{kin}}$ 
as the Cauchy completion of the space of cylindrical functions $Cyl$ in the Ashtekar-Lewandowski measure $d\mu_{AL}$
\be
\cH_{\textrm{kin}}=\overline{Cyl(A)},
\ee
that is in addition to cylindrical functions we add to $\cH_{\textrm{kin}}$ the limits of all the Cauchy convergent sequences in the $\mu_{AL}$ norm.
\\
It is convenient to introduce an orthonormal basis for $\cH_{\textrm{kin}}$. This can be easily done thanks to the Peter-Weyl theorem. This theorem can be viewed as a generalization of Fourier theorem for functions on $\mathcal{S}^1$. It states that, given a function $f\in L_2[\SU(2)]$, it can be expressed as a sum over irreducible representations of $\SU(2)$, namely
\be \label{peterweyl}
f(g)=\sum_j d_j\,\hat{f}^j_{mn}D^{(j)}_{mn}(g), \quad j \in \N/2, \quad m =-j, \cdots, j, \quad d_j=2j+1,
\ee
where
\be \label{peterweylinv}
\hat{f}^{j}_{mn}= d_j\,\int_{\SU(2)} dg \, D^{(j)}_{mn}\, f(g),
\ee
with $dg$ is the Haar measure of $\SU(2)$ and the Wigner matrices $D^{(j)}_{mn}(g)$ give the spin-$j$ irreducible matrix representation of the group element $g$.  
The Peter-Weyl theorem can be applied to any cylindrical function $\psi_{(\Gamma, \, f)} \in \cH_\Gamma$, since $\cH_\Gamma$ is just a tensor product of $L_2(\SU(2), d\mu_{\textrm{Haar}})$
\bes
\psi_{(\Gamma, \, f)}[A]&=& f(h_{e_1}[A], \cdots, h_{e_E}[A])    \nn \\
&=& \sum_{j_e,\, m_e, \, n_e} \hat{f}^{j_1, \cdots, j_E}_{m_1,\cdots, m_E,\, n_1, \cdots, n_E}\, \tilde{D}_{m_1n_1}^{(j_1)}\left(h_{e_1}[A]\right) \cdots \tilde{D}_{m_E n_E}^{(j_E)}\left(h_{e_E}[A]\right),
\ees 
where we have introduced the normalized representation matrices $\tilde{D}^{(j)}_{mn}:= \sqrt{d_j} D^{(j)}_{mn}$. According to (\ref{peterweylinv}), $ \hat{f}^{j_1, \cdots, j_E}_{m_1,\cdots, m_E,\, n_1, \cdots, n_E}$ is given by the kinematical inner product of the cylindrical function $\psi_{(\Gamma, \, f)}$ with the tensor product of irreducible representations $ \tilde{D}_{m_1n_1}^{(j_1)} \cdots \tilde{D}_{m_E n_E}^{(j_E)} $
\be
 \hat{f}^{j_1, \cdots, j_E}_{m_1,\cdots, m_E,\, n_1, \cdots, n_E}=\la \tilde{D}_{m_1n_1}^{(j_1)} \cdots \tilde{D}_{m_E n_E}^{(j_E)} | \psi_{(\Gamma, \, f)} \ra,
\ee
where $\la \; | \; \ra$ is the kinematical inner product defined by (\ref{loopScalarProd}). Moreover, the Wigner matrix elements, matrix elements of the unitary irreducible representation of $\SU(2)$, form a complete set of orthogonal functions of $\SU(2)$
\be
\int dg \, D^{(j_1)}_{m_1 n_1}(g) \, D^{(j_2)}_{m_2 n_2}(g)=\f{1}{2j_1+1} {\delta_{j_1 j_2}\, \delta_{m_1 m_2} \,\delta_{n_1 n_2}}.
\ee
Thus, the product of components of 
irreducible representations $\prod_{i=1}^E D^{(j_i)}_{m_i n_i}[h_{e_i}]$ associated with the $E$ edges $e\subset \Gamma$  (for any graph $\Gamma$, for all values of the spin $j$, and $-j\leq m, n \leq j$) is a complete orthonormal basis of $\cH_{\textrm{kin}}$.

\section{Gauge-invariant Hilbert space and spin networks}

We are now interested in the solutions of the quantum Gauss constraint. They are characterized by the states in $\cH_{\textrm{kin}}$ which are $\SU(2)$ gauge invariant. They define a new Hilbert space $\cH_{\textrm{kin}}^G$ and an orthonormal basis in $\cH_{\textrm{kin}}^G$ is the so-called {\it spin network} basis which is a very important tool in the theory \cite{spinnet0, spinnet1, spinnet2, spinnet3, spinnet4}.

The simplest example of a $\SU(2)$ gauge invariant cylindrical function is given by the Wilson loop $W_\gamma [A]$. Given a closed loop $\gamma$, the Wilson loop is defined as the trace of the holonomy $h_\gamma[A]$ around the loop, namely
\be \label{Wilson}
W_\gamma [A]:=\tr [h_\gamma[A]].
\ee
The gauge invariance of a Wilson loop is straightforward from the behavior of the holonomy under $\SU(2)$ gauge transformation (\ref{gaugeHolo}) and the invariance of the trace. Using $\SU(2)$ unitary irreducible representation matrices of spin $j\in \N/2$ for $-j\leq m, n \leq j$, the cylindrical function $W_\gamma^j[A]$ is the simplest example of spin network function.
\be
W_\gamma^j[A]:=\sum_m \; D^j_{mm}(h_\gamma[A])
\ee

\begin{figure}[ht] 
\begin{center}
\includegraphics[width=9.5cm]{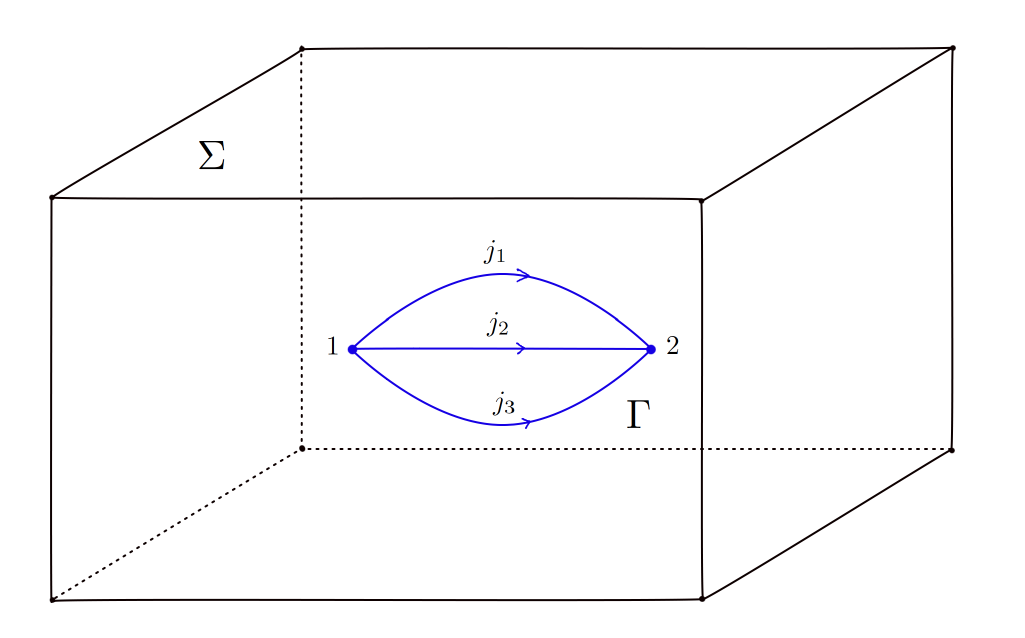}
\end{center}
\caption{An embedded theta graph } \label{theta}
\end{figure}

\begin{figure}[ht] 
\begin{center}
\includegraphics[width=7cm]{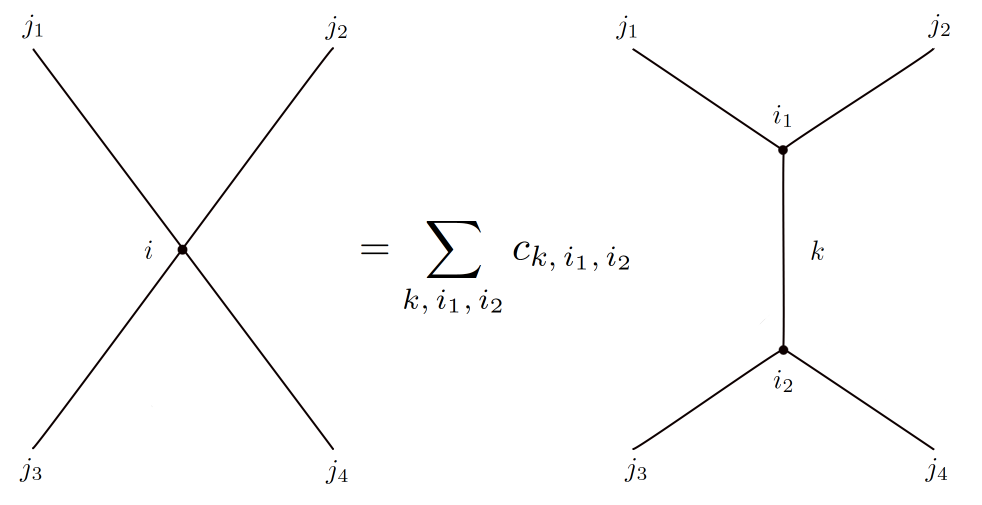}
\end{center}
\caption{Example of decomposition of an intertwiner: any intertwiner between four irreducible representations can be split into two intertwiners between three irreducible representations.} \label{decomposition}
\end{figure}

As already mentioned, spin network functions, which are $\SU(2)$ gauge invariant functionals of the connection, form a complete set of orthogonal solutions of the Gauss constraints. To construct a spin network, let us impose the gauge-invariance to a generic cylindrical function. 
This requires the spin network to be invariant under the gauge transformations that act on the nodes $n$ of the graph $\Gamma$:
\bes \label{gaugeInvariance}
\psi_{(\Gamma, \, f)}&=& f(h_{e_1}, \cdots, h_{e_E}) \nn \\
& = & f(g_{s_1}h_{e_1} g_{t_1}^{-1}, \cdots, g_{s_E}h_{e_E} g_{t_E}^{-1}).
\ees
This property can be easily implemented via group averaging. Given an arbitrary $F \in Cyl_\gamma$, 
\bes
\psi_{(\Gamma, \, f) }
& = & \int \prod_n dg_n \,F(g_{s_1}h_{e_1} g_{t_1}^{-1}, \cdots, g_{s_E}h_{e_E} g_{t_E}^{-1}) 
\ees
clearly satisfies (\ref{gaugeInvariance}). \\
 Let us first illustrate this procedure to construct  a more sophisticated example of spin network function than the Wilson loop. We consider a theta graph represented by the graph in Fig. \ref{theta}. 
We associate to this graph a generic cylindrical function that we decompose in Fourier modes as in (\ref{peterweyl}). Since the gauge transformations act only on the group elements, the gauge-invariant part is obtained by looking at the gauge-invariant part of the product of Wigner matrices, 
$$
\psi_{(\Gamma, \, f)}^{\textrm{inv}}[A]=\sum_{j_e, m_e, n_e} \hat{f}^{j_1,j_2, j_3}_{m_1,m_2, m_3, n_1, n_2, n_3}\left[D_{m_1n_1}^{(j_1)}(h_1[A])D_{m_2n_2}^{(j_2)}(h_2[A]) D_{m_3n_3}^{(j_3)}(h_3[A])\right]_{\textrm{inv}}.
$$
The invariant part of the basis is selected thanks to the group averaging
$$
\left[D_{m_1n_1}^{(j_1)}(h_1)D_{m_2n_2}^{(j_2)}(h_2) D_{m_3n_3}^{(j_3)}(h_3)\right]_{\textrm{inv}}=\int dg_1\, dg_2\, D_{m_1n_1}^{(j_1)}(g_1h_1g_2^{-1})D_{m_2n_2}^{(j_2)}(g_1h_2g_2^{-1}) D_{m_3n_3}^{(j_3)}(g_1h_3g_2^{-1}),
$$
where the projector on the gauge invariant space can be isolated
$$
\mathcal{P}_{m_1 m_2 m_3 \alpha_1 \alpha_2 \alpha_3}=\int dg_1\, D_{m_1\alpha_1}^{(j_1)}(g_1)D_{m_2\alpha_2}^{(j_2)}(g_1) D_{m_3\alpha_3}^{(j_3)}(g_1).
$$
We obtain therefore that 
$$
\left[D_{m_1n_1}^{(j_1)}(h_1)D_{m_2n_2}^{(j_2)}(h_2) D_{m_3n_3}^{(j_3)}(h_3)\right]_{\textrm{inv}}=\mathcal{P}_{m_1 m_2 m_3 \alpha_1 \alpha_2 \alpha_3} \ppp_{ \beta_1 \beta_2 \beta_3 n_1 n_2 n_3}\, D_{\alpha_1 \beta_1}^{(j_1)}(h_1)D_{\alpha_2 \beta_2}^{(j_2)}(h_2) D_{\alpha_3 \beta_3}^{(j_3)}(h_3).
$$
The projector $\mathcal{P}$ can be written in terms of  the Wigner's 3j-symbols (which are normalized Clebsch-Gordan coefficients\footnote{The explicit relation between the Wigner's 3j-symbol and the Clebsch-Gordan coefficient is given by
$$
\left( \tabl{ccc}{j_1 & j_2 & j_3 \\
m_1& m_2& m_3} \right)= \f{(-1)^{j_1-j_2-m_3}}{\sqrt{2j_3+1}} \la j_1 m_1 \, j_2 m_2 | j_3 -m_3\ra.
$$ })
 when the  triangular conditions $|j_2-j_3| \leq j_1 \leq j_2 + j_3 $ (also called  Clebsch-Gordan conditions)  hold,

\be \label{3j}
\int dg_1\, D_{m_1\alpha_1}^{(j_1)}(g_1)D_{m_2\alpha_2}^{(j_2)}(g_1) D_{m_3\alpha_3}^{(j_3)}(g_1)= \left( \tabl{ccc}{j_1 & j_2 & j_3 \\
m_1& m_2& m_3} \right) \overline{\left( \tabl{ccc}{j_1 & j_2 & j_3 \\
\alpha_1& \alpha_2& \alpha_3} \right)} .
\ee
With this notation,
$$
\left[D_{m_1n_1}^{(j_1)}(h_1)D_{m_2n_2}^{(j_2)}(h_2) D_{m_3n_3}^{(j_3)}(h_3)\right]_{\textrm{inv}}=\left( \tabl{ccc}{j_1 & j_2 & j_3 \\
m_1& m_2& m_3} \right)\overline{\left( \tabl{ccc}{j_1 & j_2 & j_3 \\
n_1& n_2& n_3} \right)} \prod_e D^{(j_e)}(h_e) \prod_n i_n ,
$$
where $i_n$ is here a short-hand notation for the 3j-symbols. More generally,   it will also denote  the invariant tensor in the space of $\otimes_{e \in n} j_e$ of all spins $j_e$ that enter in the node $n$. $i_n$ is called an {\it intertwiner}. Finally, we have
\be \label{thetaSpin}
\psi_{(\Gamma, \, f)}^{\textrm{inv}}= \sum_{j_e} \hat{f}^{j_1, j_2, j_3} \prod_e D^{(j_e)} \prod_n i_n ,
\ee
where $ \hat{f}^{j_1, j_2, j_3}=\sum_{m_e n_e} \hat{f}^{j_1,j_2, j_3}_{m_1,m_2, m_3, n_1, n_2, n_3} \left( \tabl{ccc}{j_1 & j_2 & j_3 \\
m_1& m_2& m_3} \right)$.
In this theta graph example, we have seen that  the projector on the gauge invariant space acts non trivially only at both the 3-valent nodes and it can be written as
\be \label{proj3val}
\mathcal{P}=i\, i^\ast,
\ee
with $i$ an intertwiner which is unique and given by Wigner's 3j-symbols in that   3-valent node case and $i^\ast$ its dual. This formula of the projector (\ref{proj3val}) can be generalized to a $N$-valent vertex:
\be \label{proj}
\mathcal{P} : \left\{ \tabl{rcl}{\displaystyle{\bigotimes_{e=1}^N} V^{(j_e)}=\displaystyle{ \bigoplus_i} V^{(j_i)} & \longrightarrow &\textrm{Inv} [\displaystyle{\bigotimes_e} V^{(j_e)}]= V^{(0)} \\
\displaystyle{\prod_e} D^{(j_e)}_{m_e n_e}(g)& \mapsto & \mathcal{P}_{m_1 \cdots m_N, n_1 \cdots n_N}\equiv \displaystyle{\int} dg \,  \displaystyle{\prod_e} D^{(j_e)}_{m_e n_e}(g)=\displaystyle{ \sum_{\alpha=1}^{\textrm{dim}V^{(0)}}}i^\alpha_{m_1\cdots m_N} \, i^{\alpha \ast}_{n_1 \cdots n_N},
}\right.
\ee 
with $\bigotimes_{e=1}^N V^{(j_e)}$ the tensor product of $\SU(2)$ irreducible representations. 
The integration in the definition (\ref{proj}) of the projector $\mathcal{P}$  selects the gauge invariant part of $\otimes_e V^{(j_e)}$, namely the singlet space $V^{(0)}$, if the latter exists. $i^{\alpha}_{m_1\cdots m_N}$ is an orthonormal set of invariant vectors (where $\alpha$ labels the basis elements and $\alpha \ast$ labels the dual basis elements), \ie  an orthonormal basis for $\textrm{Inv} [\bigotimes_e V^{(j_e)}]=V^{(0)}$. These invariants are the so-called \textit{intertwiners}. In fact, any intertwiner in the tensor product of an arbitrary number of irreducible representations can be expressed in terms of  basic intertwiners between three irreducible representations uniquely defined by the Wigner's 3j-symbols (see Fig. \ref{decomposition}). 

The result (\ref{thetaSpin}) can be generalized: any solution of the Gauss constraint can be written as a linear combination of products of representation matrices $D^{(j_e)}(h_e)$ contracted with intertwiners. 

\begin{figure}[ht] 
\begin{center}
\includegraphics[width=10cm]{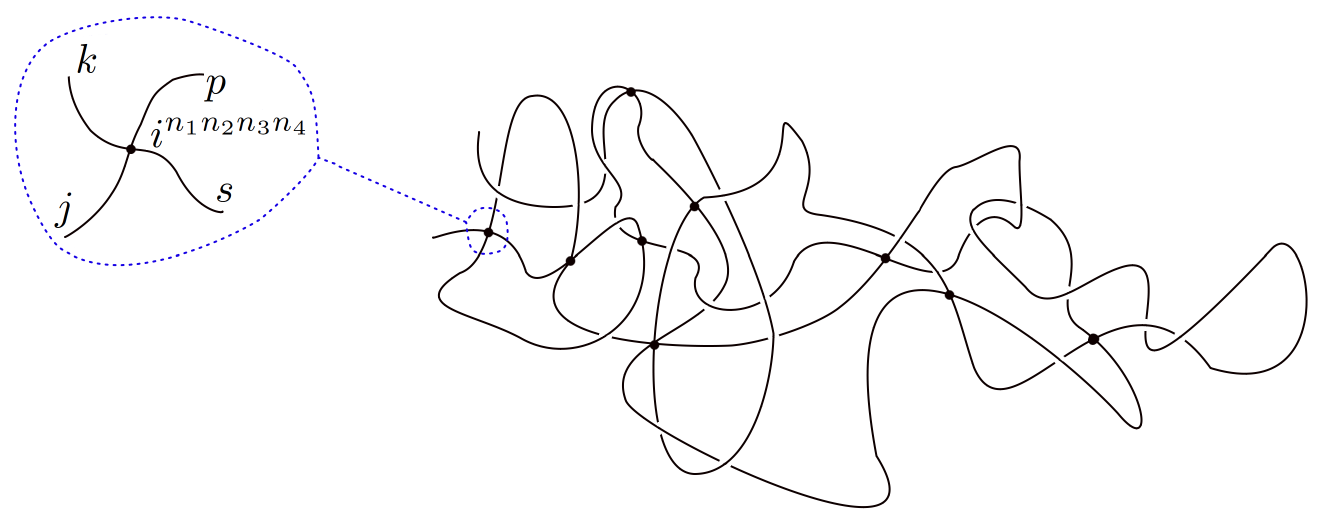}
\caption{Schematic representation of the construction of a spin network. To each node, we associated an invariant vector in the tensor product of irreducible representations converging at the node. For the node magnified on the left, we take $\imath^{n_1 n_2 n_3 n_4}\in j \otimes k\otimes p \otimes s $ and the relevant piece of spin network function is $D^j(h_{e_1}[A])_{m_1n_1}D^k(h_{e_2}[A])_{m_2n_2}D^p(h_{e_3}[A])_{m_3n_3}D^s(h_{e_4}[A])_{m_4n_4}\, \imath^{n_1 n_2 n_3 n_4}$.} \label{spinnet}
\end{center}
\end{figure}

To summarize, these states labeled with a graph $\Gamma$,  an irreducible representation $D^{(j)}(h)$ of spin $j$ of the holonomy $h$ along each link and an element $i_n$ of the intertwiner space $\textrm{Inv} [\bigotimes_e V^{(j_e)}]$ in each node, are called \emph{spin network states}, and are given by
\be
\psi_{\Gamma, \, j_e, \, i_n} [h_e]\equiv \displaystyle{\otimes_e} D^{(j_e)}(h_e) \displaystyle{\otimes_n}i_n
\ee
where the indices of representation matrices and intertwiners are left implicit in order to simplify the notation (see Fig. \ref{spinnet}).

 Spin network states form a complete basis of the Hilbert space of solutions of the quantum Gauss law $\cH^G_{\textrm{kin}}$ \cite{spinnet2}.

The geometrical interpretation of spin network states is provided by the properties of the area and volume operators.

\begin{figure}[ht] 
\begin{center}
\includegraphics[width=7cm]{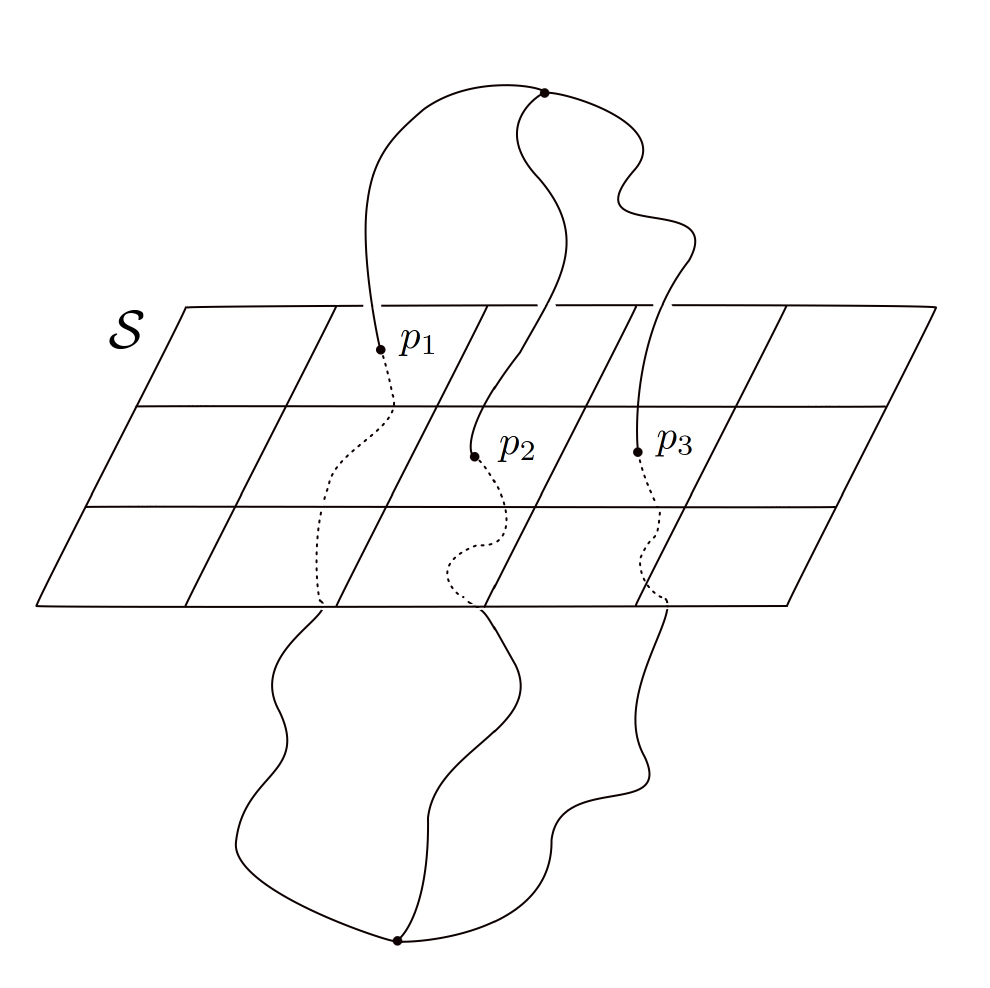}
\end{center}
\caption{A simple spin network intersecting a surface $\ss$. We have considered a partition of the surface $\ss$ sufficiently fine for which each puncture $p_k$ falls on a different small surface $\ss_k$ ($k \in \{1,2,3\}$). The punctures $p_k$ are the intersection points between the spin network $\psi_\Gamma$ and the surface $\ss$ (for further details, see \cite{gaul}).} \label{area}
\end{figure}

The  area operator $\hat{A}(\ss)$ \cite{volume7, area2, area3}  is a self-adjoint operator well-defined on $\cH^G_{\textrm{kin}}$, such that its classical  limit is the geometrical area of the two-dimensional surface $\ss \subset \Sigma$.  It is diagonalized by a basis of spin network states \cite{area3, area3}, more explicitely, its action on a spin network $ \psi_\Gamma$  is given by 
\be \label{areaOp}
\hat{A}(\ss)\, \psi_\Gamma= 
8\pi l_P^2 \gamma \sum_{p \in \ss \cup \Gamma}  \sqrt{ j_p(j_p+1)}\; \psi_\Gamma.
\ee
where $p$ are the intersection points between the spin network $\psi_\Gamma$ and the surface $\ss$ (see Fig. \ref{area}).

Notice that the spectrum of the area operator depends on the value of the Immirzi parameter $\gamma$. Moreover, the spectrum of the area operator is  discrete. It can be said that edges of spin networks carry a quanta of area. \\
We will not give any details concerning the volume operator  but the same kind of procedure can be applied to construct a quantum operator corresponding to the volume of a 3-hypersurface \cite{volume7, area2, volume1, volume6} and to find  that the eigenspectrum is again discrete. Two distinct mathematically well-defined volume operators have been proposed  in the literature \cite{volume7, area2, volume6}. Both of them act not trivially only at the nodes of a spin network state and the volume is given by the nodes (plus some additional labels to resolve degeneracy when needed) of the spin network inside the hypersurface \cite{volume7, area2, volume1, volume2, volume3, volume4, volume5}. \\
Therefore, all the information about the degrees of freedom of geometry (hence the gravitational field) is contained in the combinatorial aspects of the graph (what is connected to what) and in the discrete quantum numbers labeling area quanta (spin labels of edges) and volume quanta (linear combinations of intertwiners at nodes).

\section{Loop quantum gravity and dynamics}

\subsection{Solutions of the diffeomorphism constraint: abstract spin networks}
\begin{figure}[ht] 
\begin{center}
\includegraphics[width=14cm]{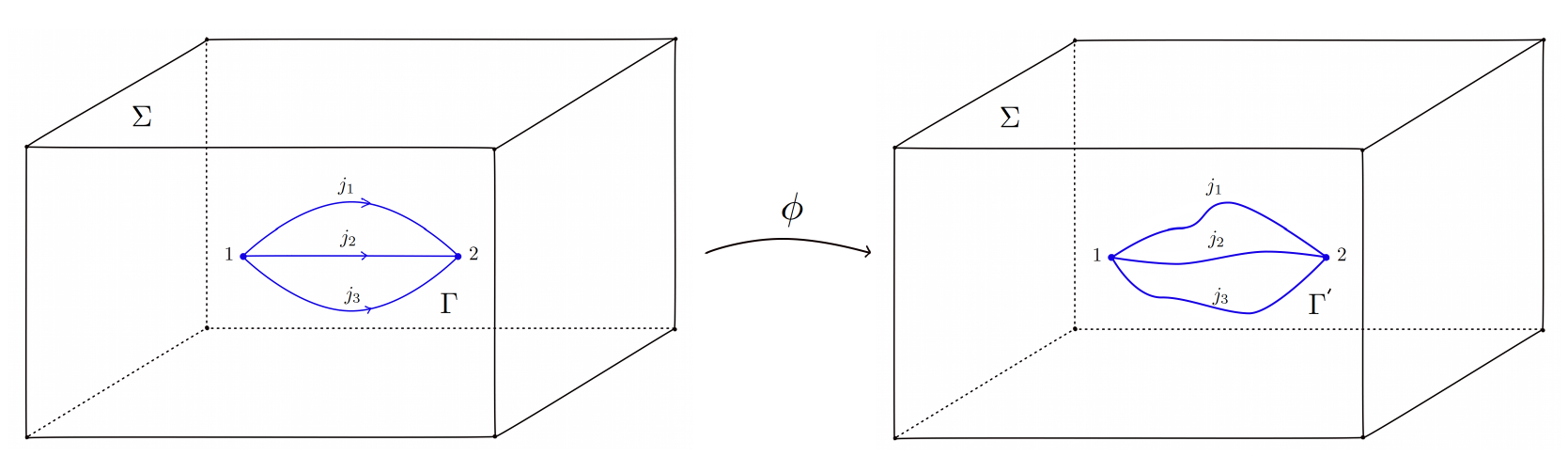}
\caption{$\phi \in \textrm{Diff}(\Sigma)$: $\phi \circ \Gamma = \Gamma^\prime$ therefore $\{\Gamma^\prime \} \approx \{\Gamma\}$ } \label{diffeo}
\end{center}
\end{figure}

From now on, we note $S$ the (embedded) spin network  given by  the triplet $(\Gamma,\, j_e,\, i_n)$. It is essentially  a colored graph as illustrated in Fig. \ref{theta}. It defines a quantum state $|S\ra$, represented in terms of the connection by a functional, a spin network state $\psi_{S}[A]$ such that $\psi_{S}[A]$ is in $\cH^G_{\textrm{kin}}$, i.e. $\hat{G}_i\,\psi_{S}[A]=0$. The next step in the Dirac program is to implement the spatial diffeomorphisms, namely to find gauge-invariant states such that $\hat{V}^a\psi_S[A]=0$.  We will not give the details of the construction of $\cH_{\textrm{Diff}}$, the Hilbert space of diffeomorphism invariant states. An analogous technique to the one used to obtain $\cH_{\textrm{kin}}^G$ can be applied. We will only emphasize the fact that  the orbits of the diffeomorphisms are not compact (contrary to the orbits of the Gauss constraint). Consequently, diffeomophism invariant states are not contained in the original $\cH_{\textrm{kin}}$. They have to be regarded as distributional states\footnote{ Diffeomorphism invariant states are therefore in the dual of the cylindrical functions $Cyl$. The Gelfand triple of interest is $Cyl \subset \cH_{\textrm{kin}} \subset Cyl^\ast$ and diffeomorphism invariant states have a well defined meaning as linear forms in $Cyl^\ast$. We denote $U(\phi)$ the operator representing the action of a diffeomorphism $\phi \in \textrm{Diff}(\Sigma)$ on $\psi_{\Gamma, \, f}[A]$. It is induced by the action of $\phi$ on the holonomy given by (\ref{spaceDiffeo}). This operator maps $Cyl_\Gamma$ to $Cyl_{\phi \circ\Gamma}$, that is $U(\phi)\psi_{\Gamma, \, f}[A]=\psi_{\phi \circ\Gamma, \, f}[A]$. Its action is well-defined and unitary, thanks to the fact that the Ashtekar-Lewandowski measure is diffeomorphism invariant. To explicitly obtain states which are invariant under $U$, one has to solve $U\psi=\psi$ for (distributional states) $\psi \in Cyl^\ast$. For details on the resolution, see \cite{ALmeasure, alej-spinfoam}.}. 
The diffeomorphism-invariant Hilbert space $\cH_{\textrm{Diff}}$ which can be considered as the space $\cH_{\textrm{kin}}/\textrm{Diff}(\Sigma)$  turns out to have a natural basis labeled by knots, or more precisely by ``$s$-knots" (also called \textit{ abstract spin network states}). A $s$-knots is an equivalence class $s$ of embedded spin networks $S$ under the action of diffeomorphisms Diff($\Sigma$), i.e. $S, \, S^\prime \in \, s$ if there exists a $\phi \in \textrm{Diff}(\Sigma)$ such that $S^\prime= \phi \cdot S$ (see Fig. \ref{diffeo}). 

 A $s$-knot is characterized by its ``abstract" graph (defined only by the adjacency relations between links and nodes), by the coloring, and by its knotting and linking properties, as in knot theory. To be more explicit, we still know how loops in the graph wind around holes in the manifold for example, or  how  edges intersect each other; but ``where" the graph sits inside the manifold, its location and its metric properties (e.g. length of edges,...) are no more well-defined concepts. Therefore, each $s$-knot defines  a gauge and diffeomorphism invariant state  $|s\ra$ of the gravitational field. It can be proven that the states $(1/\sqrt{is(s)}) |s\ra$, where $is(s)$ is the number of isomorphisms of the $s$-knot into itself, preserving the coloring and generated by a diffeomorphism of $\Sigma$, form an orthonormal basis  for $\cH_{\textrm{Diff}}$. Although  $s$-knots are labelled by discrete quantum numbers, the picture is not truly entirely discrete. It was pointed out in \cite{winston} that if the nodes of the $s$-knots have a valence higher than five, $s$-knots are labelled also by certain continuous moduli parameters (see \cite{winston} for details). These moduli do not seem to play any significant role in the theory (they are virtually undetectable by the hamiltonian operator that governs the dynamics and they do not seem to affect the physics of the theory) but they spoil the discretness of the picture. In fact they change the structure of  $\cH_{\textrm{Diff}}$, rather dratiscally, making it nonseparable. It has been shown that by  using the notion of ``extended" diffeomorphisms, the moduli disappear and the knot class become countable. The kinematical Hilbert space of loop quantum gravity, ``the extended $\cH_{\textrm{Diff}}$", becomes then separable. For further details see  \cite{winston}.
 
\begin{figure}[ht] 
\begin{center}
\includegraphics[width=8cm]{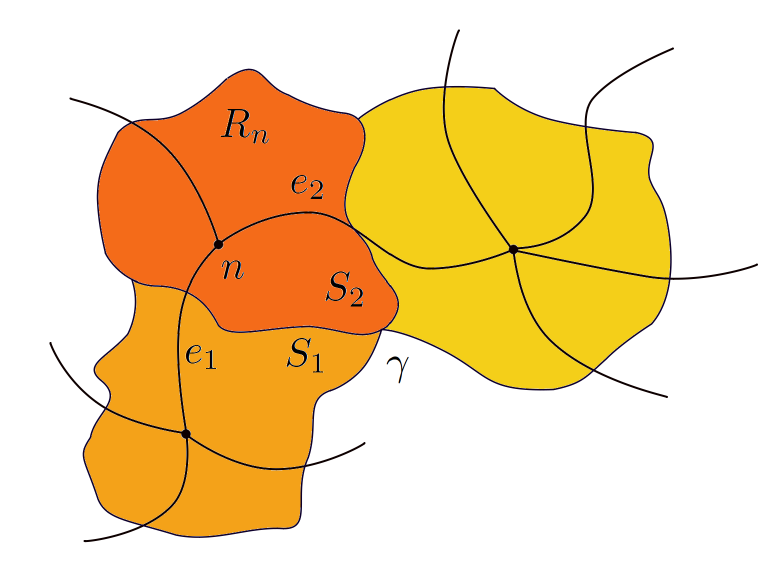}
\end{center}
\caption{A portion of a $s$-knot graph and the associated dual picture of quantum geometry. The region $R_n$ is dual to the node $n$. Two adjacent regions are shown. The surface $S_1$ and $S_2$ are dual to the links $e_1$ and $e_2$. They identify a curve $\gamma$ on the boundary of $R_n$.} \label{spacechunk}
\end{figure}

 A $s$-knot is a purely algebraic kinematical quantum state of the gravitational field: each link of the graph can be seen as carrying a quantum of area and the nodes carrying quanta of volume. A $s$-knot can therefore be  seen as an elementary quantum excitation of space formed by ``chunks" of space (the nodes) with quantized volume, separated by sheets of surface (corresponding to the links), with quantized area. The key point is that a $s$-knot does not live on a manifold. The quantized space does not reside ``somewhere" but it defines the ``where" by itself. It appears as the dual picture of the quantum geometry of a $s$-knot state (see Fig. \ref{spacechunk}). 
 
 This is the picture of quantum space-time that emerges from loop quantum gravity. 

\subsection{The Hamiltonian constraint}
The structure that gives the quantum states of the gravitational field at the kinematical level is therefore well understood. We still have to consider the scalar (Hamiltonian) constraint which encodes the dynamics. Indeed, the physical state of the theory should lie in the kernel of the quantum Hamiltonian constraint operator $\hat{S}$.  I will    give here neither  any details concerning the well posed definition of this constraint nor a strategy to compute  its kernel (for details see \cite{book-thomas, hamiltonian1, qsd, hamiltonian2}). I will only give the action of the Thiemann's Hamiltonian operator,  $\hat{S}$, on a spin network $|S\ra$ \cite{qsd}. 

It acts  only on the nodes of  the spin network. The result turns out to be given by
\be \label{Sop}
\hat{S}[N]|S\ra= \sum_{\textrm{nodes }n \textrm{ of }S} A_n N(x_n)\hat{D}_n|S\ra, \qquad  \begin{minipage}[h]{0.45\linewidth}
  \includegraphics[scale=0.4]{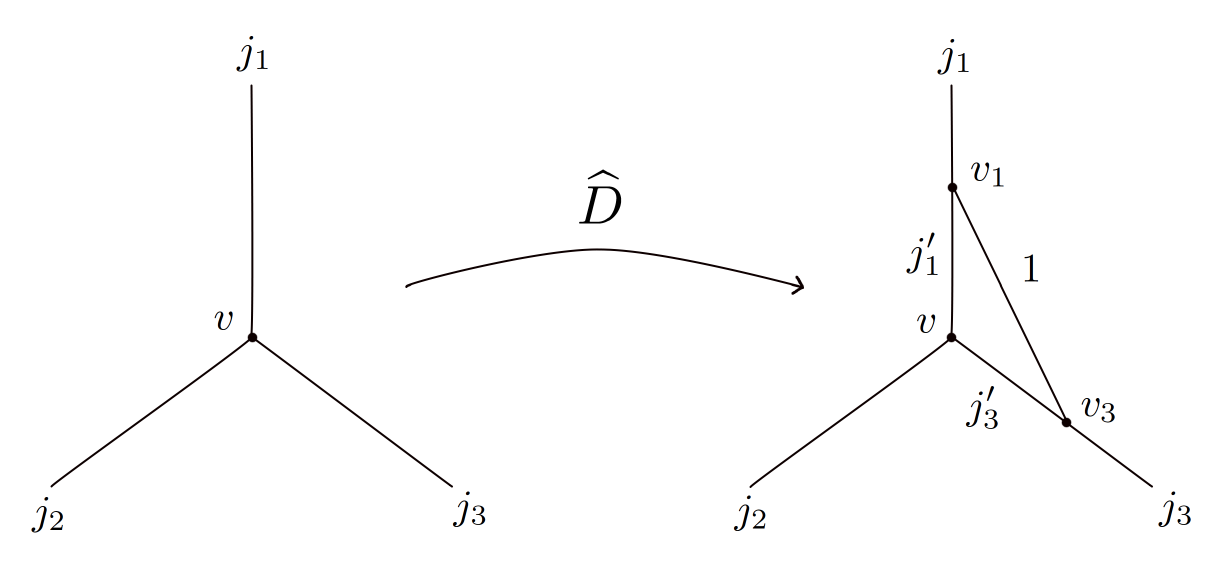} 
\end{minipage}
\ee
where $x_n$ refers to the point in which the node $n$ is located and we used $\hat{S}[N]$ the Hamiltonian constraint smeared with a scalar function $N(x)$ given by $\hat{S}[N]=\int d^3x N(x) \hat{S}(x)$. The action of  the operator $\hat{D}_n$ is illustrated in the figure on the left-hand side of equation (\ref{Sop}): an extra link with color one connecting two points $v_1$ and $v_3$ lying on distinct links adjacent to the node $v$ is created when acting on a single node. The color of the link between $v$ and $v_1$, as well as between $v$ and $v_3$ is altered and the state is multiplied by a coefficient $A_n$ (explicit expressions are computed in \cite{hamiltonian3}). The action of this Hamiltonian operator has been criticized to be too local \cite{HamiltonianLee} in the sense that the modifications performed by this Hamiltonian constraint operator at a given node of a spin network do not propagate over the whole graph but are confined to a neighborhood of the node.  Repeated actions of this Hamiltonian operator generate more and more new edges ever closer to the node, never intersecting each other and thus producing a fractal structure\footnote{Let us notice that there is no action of the Hamiltonian operator at the new created nodes and this is quite different from what happens in lattice gauge theory where no new links are created.}. Another criticism regarding this Hamiltonian operator is that it is highly ambiguous; several sources of ambiguities can be distinguished (factor ordering ambiguities, representation ambiguities, loop assignment ambiguities,...) \cite{book-thomas}.\\
To conclude this chapter, the kinematics of loop quantum gravity seems to be beautifully under control. One key result of loop quantum gravity, derived using kinematical states given by the spin network states, is that the space is discrete. However, the resolution of the dynamics is still an open question.
The effort to gain control over this issue 
has flourished into two main lines of research. The first one that we will not review here, is the idea of the Master Constraint \cite{master}.  One defines a unique constraint implementing simultaneously the vector (diffeomorphism) and scalar (Hamiltonian) constraints. The second one is the spin foam formalism \cite{carlo-look}. This framework will be introduced in details in the last chapter   since it is the basis of all my research results presented in the third and fourth parts. 
 However, before   reviewing the spin foam formalism, I will recall some important features concerning the kinematics of covariant loop quantum gravity, following the the classical set up which was presented in the section \ref{CLQGclas}.

\chapter{Covariant Loop Quantum Gravity} \label{chapCLQG}
Loop quantum gravity as presented above seems to be a promising approach for quantizing general relativity. It gives some interesting results like discrete quanta of area and volume. However, several problems appear (in addition to the implementation of the dynamics). First of all, it is based on the use of space triad and an $\SU(2)$ connection where $\SU(2)$ is the gauge group for the three dimensional space. Therefore, as we have previously pointed it out, this particular choice of variables looses the explicit covariance of the theory and a space-time geometrical interpretation. Moreover, there exists an additional puzzle: a free parameter in the theory, $\gamma$, the  Immirzi parameter. This parameter comes out of  the canonical transformation (\ref{canontransfo}) but creates a full one-parameter family of quantizations which are not unitarily equivalent \cite{Immirziambiguity}. Moreover, this unphysical parameter appears in the spectra of the area operator $\hat{A}(\ss)$.

One way to avoid this ambiguity is to consider the \textit{ covariant loop quantum gravity framework}.   One main difference with the standard loop quantum gravity is that the area operator has now   (at the kinematical level) a continuous spectrum which is now independent of the Immirzi parameter. However, it involves the variable $\mathcal{X}$, the time normal introduced in \ref{CLQGclas}. One of the main advantages of the formalism is that 
the covariance of the theory is kept; the main drawback of the approach is a non-compact gauge group and a non-commutative connection.

We are going now to  review the definition of the kinematical Hilbert space of covariant loop quantum gravity. We recall that the canonical variables are a Lorentz connection $\tilde{\mathcal{A}}$ , its conjugated triad $\mathcal{R}$ (a one-form valued in the Lorentz algebra) and the time normal $\chi \in \SL(2,\C)/\SU(2)$.  These variables have been defined in the section \ref{CLQGclas}.

\section{Cylindrical Functions and Gauge Invariance}
Let us now define the quantum states of space-(time) geometry. Let us consider an arbitrary oriented graph $\Gamma$ with $E$ edges and $V$ vertices. Since geometric observables (such as area) involve $\mathcal{X}$ and that $\mathcal{X}$ commutes with the connection $\tilde{\mathcal{A}}$ \cite{CLQG2, CLQG3}, the full configuration space is spanned by functionals  dependent on both $\tilde{\mathcal{A}}$ and $\mathcal{X}$. To define the Hilbert space structure, one considers generalized cylindrical functions which as the usual ones, are associated with graphs and whose dependence on the connection is supposed to be through the Lorentz group elements represented by holonomies $G_e\in \SL(2,\C)$. In addition, they also depend on the values of the field $\mathcal{X}$ at the vertices\footnote{Let emphasize that it is possible to chose any Lorentz connection as ``configuration" variable. However, we will have to be careful to define operators. For example, in the considered case, since the $\SL(2,\C)$ connection $\tilde{\mathcal{A}}$ does not commute with itself, the holonomy operators will not simply act by multiplication on the generalized quantum states -- the projected spin networks -- as it is the case for the action of loop quantum gravity holonomy operators on $\SU(2)$ spin networks.}. Note that $\mathcal{X}$ is naturally encoded in the unit vector $x^I=(1,\mathcal{X}^i)$ as defined in  (\ref{xnorma}). Since $x$ is normalized\footnote{We are using the signature $(-+++)$} $|x|^2=-1$,   it is an element of the (upper) hyperboloid $\hh_+= \{x^\mu\in \R^4, \, |x|^2=-1, \, x^0>0\}$, which is an homogenous space $\hh_+\sim \SL(2,\C)/SU(2)\sim\SO(3,1)/\SO(3)$. 
This implies that the generalized cylindrical functions $\vphi(G_e,x_v)$, the functionals of the Lorentz connection and the time-normal field, live on the homogenous space  $\SL(2,\C)^E\times \hh_+^V$. 

We further require that our functionals $\vphi(G_e,x_v)$ are invariant under the action of the Lorentz group  to solve the analogue of the Gauss law $G_X\vphi(G_e,x_v)=0$.
\be
\vphi(G_e,x_v)=\vphi(\Lambda_{s(e)}G_e\Lambda_{t(e)}^{-1},\Lambda_v\vartriangleright x_v),\quad
\forall \Lambda_v\in\SL(2,\C)^{\times V},
\ee
where $G_e\in\SL(2,\C)$,  $x_v\in \hh_+$. $s(e)$ and $t(e)$ are respectively the source and target vertices of the edge $e$.

The 4-vector $\Lambda\vartriangleright x$ is  obtained by acting  on $x$ by the $\SO(3,1)$ transformation corresponding to $\Lambda\in\SL(2,\C)$. The easiest way to write this action is to represent 4-vectors as 2$\times$2 Hermitian matrices:
\be
x=(x_0,x_1,x_2,x_3)\,\arr\,X=\mat{cc}{x_0+x_3 & x_1+ix_2 \\ x_1-ix_2 & x_0-x_3},
\ee
with $\tr X=2x_0$ and $\det X=-|x|^2$. Then $\SL(2,\C)$ group elements act by conjugation, $\Lambda\vartriangleright X \equiv\, \Lambda X \Lambda^{\dag}$. From there, we can act on the 4-vector $\om=(1,0,0,0)$, or equivalently on its corresponding matrix $\Om=\id$, to generate all elements in $\hh_+$:
\be
x= B\vartriangleright \om,\qquad
X=B\id B^{\dag}=B B^{\dag}, \quad B\in\SL(2,\C).
\ee
It is clear that this expression is invariant under the right $\SU(2)$ action $B\arr B h$ with $h\in\SU(2)$.
This actually shows the fact that $\hh_+$ is the coset $\SL(2,\C)/\SU(2)$.
From these various representations, we can equivalently see our functionals as depending on 4-vectors, 2$\times$2 Hermitian matrices or $\SL(2,\C)$ group elements (with an extra $\SU(2)$ invariance), i.e respectively $\vphi(G_e,x_v)$ or $\vphi(G_e,X_v)$ or $\vphi(G_e,B_v)$.

\medskip

A first important remark on these Lorentz invariant functions is that they are entirely determined by their section at $x_v=\om$ for all $v$. Indeed, let us define this section:
\be
\phi(G_e)\equiv\vphi(G_e,x_v=\om).
\ee
Effectively, these functions still satisfy a remaining $\SU(2)$-invariance, inherited from the full $\SL(2,\C)$-invariance:
\be
\phi(G_e)=\phi(h_{s(e)}G_eh_{t(e)}^{-1}),\quad\forall h_v\in\SU(2)^{\times V}.
\ee
And we can reconstruct the full functional from that particular section:
\be
\vphi(G_e,x_v)=\vphi(G_e,B_vB_v^{\dag})=\phi(B_{s(e)}^{-1}G_eB_{t(e)}).
\ee
The second remark is that if we integrate over the time-normals, then we recover the standard
$\SL(2,\C)$-invariant cylindrical functions, whose basis are $\SL(2,\C)$ spin networks.
More precisely, we define the group-averaged functional
\be \label{groupaverProj}
\vphi_g(G_e)=\int_{\hh_+^V} [dx_v]\, \vphi(G_e,x_v)=\int_{\SL(2,\C)^V} [d\Lambda_v] \, \vphi(G_e,\Lambda_v\vartriangleright\om),
\ee
where $[dx]$ is the translation-invariant measure on $\hh_+$ inherited from the Haar measure $[d\Lambda]$ on $\SL(2,\C)$. This new function satisfy a simple $\SL(2,C)$-invariance at the vertices:
\be
\vphi_g(G_e)=\vphi_g(\Lambda_{s(e)}G_e\Lambda_{t(e)}^{-1}),\quad
\forall \Lambda_v\in\SL(2,\C)^{\times V},
\ee

\medskip

We now have all the tools  to endow our space of cylindrical functions with a scalar product, following \cite{projected}.
\be
\la \vphi | \vphi'\ra
\,\equiv\,
\int [dG_e]\, \overline{\vphi}(G_e,x_v)\vphi'(G_e,x_v).
\ee
Due to the $\SL(2,\C)$ gauge invariance satisfied by the functionals, it is easy to see that this definition holds for any arbitrary choice of time-normals $x_v$ as long as both functionals are evaluated on the same set of $x_v$'s. Therefore, this scalar product can be entirely computed by setting all time-normals to the origin $\om$:
\be
\la \vphi | \vphi'\ra
\,=\,
\int [dG_e]\, \overline{\phi}(G_e)\phi'(G_e).
\ee
We call the corresponding $L^2$ space of functions as the Hilbert space of projected cylindrical functionals on the graph $\Gamma$, following the terminology introduced in \cite{projected}, and we will simply write $H$ for it (leaving implicit the dependence on the underlying graph $\Gamma$, since our whole analysis does not involve changing graph).

\section{The Basis of Projected Spin Networks}
The next step is to introduce the basis of the Hilbert space $H$ of Lorentz invariant functions $\vphi(G_e,x_v)$. To this purpose,  a few facts about the unitary representations of the Lorentz group $\SL(2,\C)$ are recalled in appendix \ref{appendix2}.

Following the original work \cite{projected}, we start with the section $\phi(G_e)=\vphi(G_e,\om)$, which fully determines the whole function $\vphi(G_e,x_v)$. We apply the Plancherel decomposition formula to $\phi(G_e)$ (see appendix \ref{appendix2} for details), thus attaching an irreducibe representation  $(n_e,\rho_e)$ and the corresponding matrix $D^{(n_e,\rho_e)}(G_e)$ to each edge $e$ of the graph. Then we glue these matrices at each vertex $v$ of the graph with vectors in the tensor product of the irreps attached to the incoming/outgoing edges. These tensors are not chosen entirely arbitrarily since the functions $\phi(G_e)$ are required to be $\SU(2)$-invariant at each vertex.

The final result of this procedure are the {\it projected spin networks}. A projected spin network on the graph $\Gamma$ is defined by the choice of a $\SL(2,\C)$ irrep $\cI_e=(n_e,\rho_e)$ for each edge, a choice of couple of $\SU(2)$ irrep $(j_e^s,j_e^t)$ attached to the source and target vertices of  each edge, and finally a $\SU(2)$-intertwiner (or equivalently $\SU(2)$-invariant tensor, or a singlet state in layman terminology) $i_v$ for each vertex $v$. The intertwiner $i_v$ lives in the tensor product of the $\SU(2)$ irreps coming in and going out the vertex $v$, or more precisely:
$$
i_v\,:\,
\bigotimes_{e|s(e)=v} V^{j_e^s}
\,\longrightarrow\,
\bigotimes_{e|t(e)=v} V^{j_e^t}.
$$
Then the functions is defined as:
\be \label{projectSpinNetdef}
\vphi_{\cI_e,j_e^{s,t},i_v}(G_e,x_v)
\,\equiv\,
\tr
\prod_e \la \cI_e,j_e^s,m_e^s|B_{s(e)}^{-1}G_e B_{t(e)}| \cI_e,j_e^t,m_e^t\ra
\,
\prod_v \la \otimes_{e|t(e)=v}\,\cI_e,j_e^t,m_e^t| i_v |\otimes_{e|s(e)=v}\cI_e,j_e^s,m_e^s\ra.
\ee
The trace is taken over the $\SU(2)$ representations i.e it amounts to summing over the basis labels $m_e^{s,t}$. We must require that the choice of spins $j_e^{s,t}$ be compatible with the choice of the $\SL(2,\C)$ irreps $\cI_e=(n_e,\rho_e)$, i.e that $j_e^{s,t}\ge n_e$, else the projected spin network functional would simply vanish.

\begin{figure}[h]
\begin{center}
\includegraphics[height=40mm]{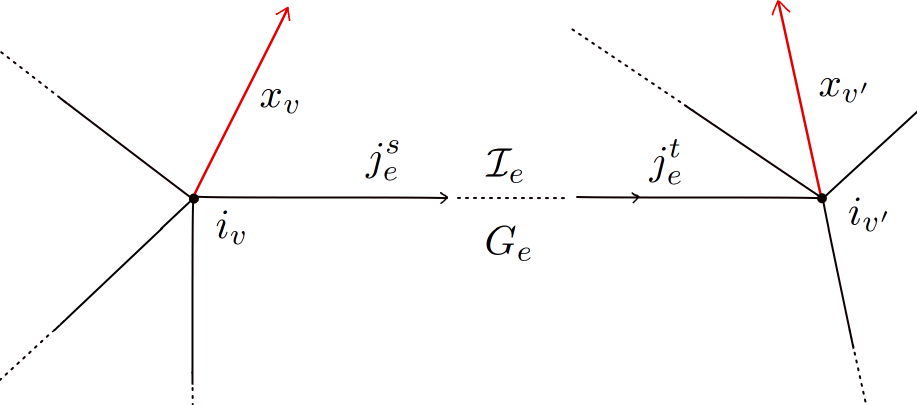}
\caption{An edge of a projected spin network.}
\end{center}
\end{figure}

First, to check that this function is well-defined, one must make sure that its definition is invariant under the right $\SU(2)$-action on the group elements $B_v$. It is actually the requirement of having $\SU(2)$-invariant intertwiners $i_v$ which ensures that the expression above is correctly invariant under the transformation $B_v\arr B_v h_v$ for all $h_v\in\SU(2)^{\times V}$.

Then, we would like to check that these projected spin networks are properly $\SL(2,\C)$-invariant. The Lorentz action at the vertices reads as:
$$
\left|
\begin{array}{l}
G_e \\
x_v \\
B_v
\end{array}
\right.
\,\longrightarrow\,
\left|
\begin{array}{l}
\Lambda_{s(e)}G_e\Lambda_{t(e)}^{-1} \\
\Lambda_v\vartriangleright x_v \\
\Lambda_v B_v
\end{array}
\right. \,.
$$
It is clear that the functions defined above are invariant under such transformations.

Finally, the Plancherel decomposition formula ensures that these projected spin network functionals cover the whole Hilbert space $H$ and provide us with an orthonormal basis. Indeed we can compute the scalar product between two such spin networks:
\be
\la \vphi_{\cI_e,j_e^{s,t},i_v}| \vphi_{\tilde{\cI}_e,\tilde{j}_e^{s,t},\tilde{i}_v}\ra
\,=\,
\int [dG_e]\,
\overline{\phi}_{\cI_e,j_e^{s,t},i_v}(G_e)
\phi_{\tilde{\cI}_e,\tilde{j}_e^{s,t},\tilde{i}_v}(G_e)
\,=\,
\prod_e\f{\delta_{n_e,\tilde{n}_e}\delta(\rho_e-\tilde{\rho}_e)}
{\mu(n_e,\rho_e)}
\,\delta_{j_e^{s,t},\tilde{j}_e^{s,t}}
\,\prod_v \la i_v|\tilde{i}_v\ra.
\label{scalarproj}
\ee
Thus a choice of orthonormal basis is given by a choice of an orthonormal basis of  $\SU(2)$ intertwiners, just as for the standard $\SU(2)$ spin networks of Loop Quantum Gravity.

Using these projected spin networks allow to project the Lorentz structures on specific fixed $\SU(2)$ representations. This allows to diagonalize the area operators. Considering a surface $\ss$ intersecting the graph $\Gamma$ only on one edge $e$, its area operator $A[\ss]$ will be diagonalized by the projected spin network basis:
\be
A[\ss] |\vphi_{\cI_e,j_e,i_v}\ra= l_P^2 \sqrt{j_e(j_e+1)-n_e^2+ \rho_e^2+1}\, |\vphi_{\cI_e,j_e,i_v}\ra
\ee
Interestingly, this spectrum contains a term coming from the Lorentz symmetry which makes it continuous and this spectrum does not depend on the Immirzi parameter which thus remains unphysical (at least at this kinematical level) as it was at the classical level. . 

Therefore, we now have described the kinematical structure of covariant loop quantum gravity. We still need to tackle the issue of defining the dynamics of the theory. On one hand, one can try to regularize and quantize \textit{\`a la Thiemann} the action of the  Hamiltonian constraint  on the covariant connection $\tilde{\mathcal{A}}$. On the other hand, one can turn to the spin foam formalism.

\chapter{Feynman's path integral approach: Spin Foams}

The concept of spin foam going from one spin network to another was introduced in order to address the problem of  dynamics of loop quantum gravity \cite{baez-spinfoam, baez, alej-spinfoam}. We have seen how spin networks describe the (quantum) geometry of space, we now introduce the spin foams which will  describe the (quantum) geometry of space-time. The definition of a spin foam is very analogous to a spin network, but everything is one dimension higher. 
The same algebraic and combinatorial structures as loop quantum gravity are used.
Just as a spin network is a graph with edges labeled by representations and vertices labeled by intertwiners, a spin foam is a 2-dimensional piecewise linear cell complex -- roughly a finite collection of polygons attached to each other along their edges -- with faces labeled by representations and edges labeled by intertwiners. 

The spin foam formalism attempts the construction of the path integral representation of the theory in order to provide an explicit tool to compute transition amplitudes in quantum gravity.

The solutions of the scalar constraint can be characterized by the definition of the generalized projection operator $P$ from the kinematical Hilbert space $\cH_{\textrm{Diff}}$ into the kernel of the scalar constraint $\cH_{\textrm{phys}}$. Formally, one can write $P$ as
\be
P=\int [dN] \, e^{i\hat{S}[N]}=\int [dN] \, e^{i\int N\, \hat{S}}.
\ee

$P$ can also be defined in a manifestly covariant manner \cite{SFDyn1, SFDyn2}, that amounts to giving  sense to the path integral of gravity 
\be \label{pathint}
P:=\int D[e] \, D[\omega] \, \mu[\omega, e] \, \exp[i S_{\textrm{GR}}(e, \omega)],
\ee
where we used the first order formulation of gravity given by (\ref{PalaAct}): $e$ is the tetrad field, $\omega$ is the space time connection and $\mu[\omega, e]$ denotes the appropriate measure. 

\begin{figure}[h]
\begin{center}
\includegraphics[height=70mm]{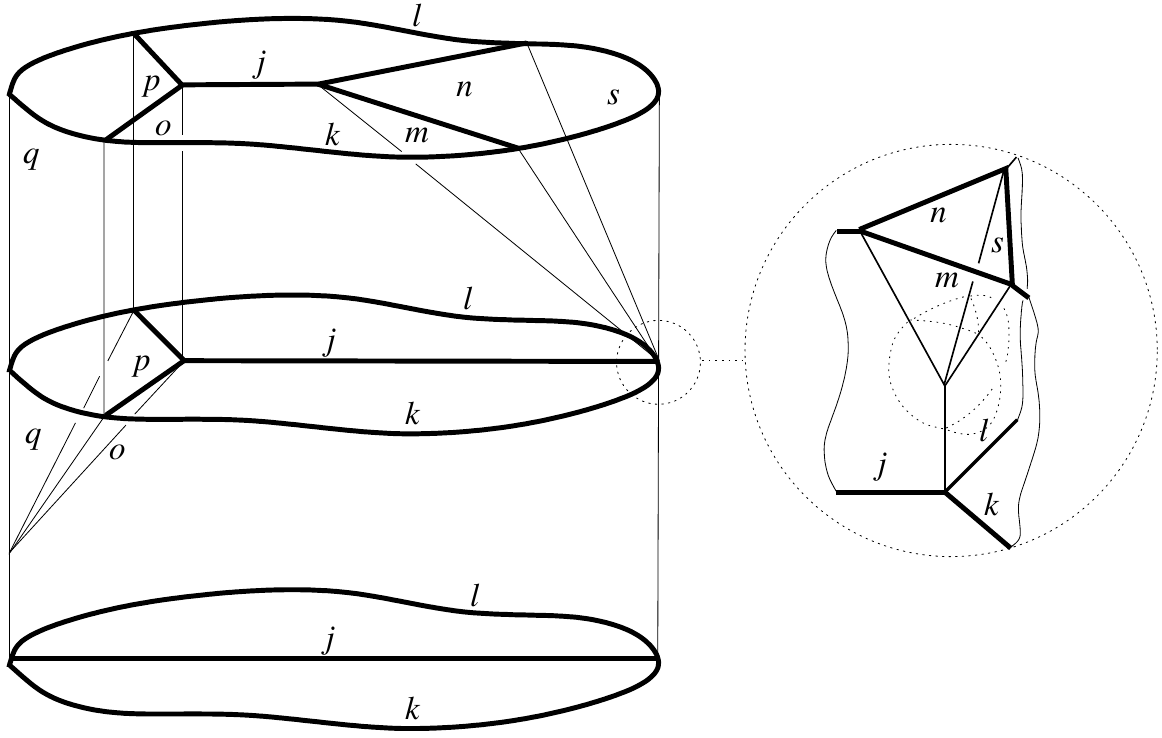}
\end{center}
\caption{A spin foam as a colored 2-complex representing the transition between three different spin network states. The vertex amplitude of the vertex magnified on the right is $A_v(j,k,l,m,n,s)$ \cite{alej-spinfoam}.} \label{SF}
\end{figure}

The matrix elements of $P$ define the physical scalar product  $\la\;, \; \ra_p$ providing the vector space of solutions $|\tilde{\psi}\ra$ of the quantum Einstein's equations $\hat{G}_i|\tilde{\psi}\ra=0, \; \hat{V}_a|\tilde{\psi}\ra=0, \; \hat{S}|\tilde{\psi}\ra=0$ with the Hilbert space structure that defines $\cH_{\textrm{phys}}$. 
\be
\la \psi, \psi^\prime \ra_p:=\la P\psi, \psi^\prime\ra,\quad \psi, \psi^\prime \in \cH_{\textrm{Diff}}.
\ee
 When these matrix are computed in the spin network basis, they can be expressed as a sum over amplitudes of ``spin network histories", \ie  spin foams. They are defined as colored 2-complexes. A 2-complex $C$ is a set of elements called ``vertices" $v$, ``edges" $e$ and ``faces" $f$, and a boundary relation among these, such that an edge is bounded by two vertices, and a face is bounded by a cyclic sequence of continuous  edges (edges sharing a vertex). A spin foam (see Fig. \ref{SF}) is a 2-complex plus a ``coloring" $c$, that is an assignment of an irreducible representation $j_f$ of $\SU(2)$ (or more generally of any given group $G$) to each face $f$ and of an intertwiner $i_e$ to each edge $e$. 
 
More generally, let us give the generic features of the spinfoam framework for an arbitrary gauge group $G$.
We have seen that the Hilbert space of quantum geometry states for loop quantum gravity $\cH_{\textrm{Diff}}$  is spanned by spin network states $|\psi \ra$ where  in $\psi=(\gamma, j_l, i_n)$: $\gamma$ is a graph, $j_l$ is a ``spin" labeling an irreducible representation of the gauge group $G$ associated to the link $l$ of the graph, and $i_n$ is associated to the node $n$ of $\gamma$ and labels  intertwiners. We consider a 4d space-time region $\mathcal{M}$  with a 3d boundary $\Sigma$. The spin network state $|\psi\ra$ defines the quantum state of geometry of the boundary $\Sigma$, then the spin foam amplitudes define the dynamical probability amplitude of that state and encode the whole quantum gravity dynamical content. More precisely, the standard ansatz for local spinfoam amplitudes can be presented in the following general form
\be \label{propaSF}
K[\psi]=\sum_{C|\partial C= \gamma_\psi} w(C) \sum_{j_f, i_e} \prod_f A_f(j_f) \prod_e A_e(j_f, i_e) \prod_vA_v(j_f, i_e),
\ee
where the sum is taken over two-complexes $C$ and their ``coloring"  $c\,\equiv\,\{j_f, i_e\}$. The 2-complexes are constraint to fit the graph $\gamma_\psi$ of the spin network $\psi$ at the boundary.
The representations $j_f$ are associated to the faces $f$ of $C$ and the intertwiners $i_e$ to its edges $e$. They are also constrained to fit the representation $j_l$ and intertwiners $i_n$ of the state $\psi$ living on the boundary.
 A spinfoam is then defined as a colored two-complex, namely a couple $(C, c)$.
Finally, the spinfoam amplitude is made of four types of factors. First, $w(C)$ is a statistical weight that depends only on the two-complex $C$ (similar to the symmetry factor for Feynman diagrams in standard quantum field theory). $A_f$ are weight factors  associated to the faces and $A_e$ are amplitudes associated to the edges of the 2-complex. These three types of factors can be interpreted as measure factors of the discrete path integral defined by $K[\psi]$. All the dynamical information is encoded in the vertex amplitude, $A_v(j_f, i_e)$, which is an amplitude associated with each vertex $v$ of the two-complex $C$ and depends on the spins $j_f$ and intertwiners $i_e$ living on the faces and edges around that vertex.

The amplitudes are usually assumed to be local, \ie they depend only on the coloring of adjacent simplicial elements. Thus, $A_f$ is a function of the representations located on the face $f$, $A_e$ is a function of the intertwiner assigned to $e$ and the representations on the faces containing $e$, whereas $A_v$  depends on the representations on the faces and on the intertwiners on the edges containing the vertex $v$.
Generally speaking, the choice of the vertex amplitude $A_v$ corresponds to the choice of a specific form of the hamiltonian operator in the canonical theory.
Since a spin network state defines a quantum state for the boundary geometry, spin foams are thus interpreted as representing the states of the bulk defining the quantum space-time interpolating between given boundary data. %
\\

Let us now focus on the special case where the boundary is made of two disconnected pieces an initial spin network $\psi$ and a final spin network $\psi^\prime$, which we can interpret as the initial boundary and the final boundary in order to make a clearer connection with the canonical framework, whose goal is to define transition amplitudes between initial and final states.
A spin foam $\mathcal{F}: \psi \rightarrow \psi^\prime$ defined by the 2-complex $C$ bordered by the supporting graphs of $\psi$ and  $\psi^\prime$, respectively $\gamma$ and $\gamma^\prime$, represents then  a transition from the spin network state $\psi=(\gamma, j_l, i_n)$ into $\psi^\prime=(\gamma^\prime, j_l^\prime, i_n^\prime)$. Nodes and links in the initial spin network $|\psi\ra$ evolve into 1-dimensional edges and faces. New links are created and spins are reassigned at vertexes. This defines a foam-like structure whose components inherit the spin representations from the initial spin network $\psi$ and are at the end compatible with the spin representations of the final spin network $\psi^\prime$ (see Fig. \ref{SF}). The propagator kernel $K[\psi,\psi^\prime]$ obtained by summing over all the spinfoams compatible with the boundary data  can  be then interpreted as the physical inner product between $|\psi\ra$ and $|\psi^\prime \ra$
\be \label{propaSFcompact}
\la \psi, \psi^\prime \ra_p=\la P\psi, \psi^\prime\ra=K[\psi,\psi^\prime]=\sum_{\mathcal{F}|_{\partial \mathcal{F}=\psi \cup \psi^\prime }} \mathcal{A}_\mathcal{F}[\psi,\psi^\prime],
\ee
where we have introduced  a more compact notation than in (\ref{propaSF}): the sum over the compatible spin foam $\mathcal{F}$ gathers the sum over the two-complexes $C$ compatible with the graphs $\gamma$ and $\gamma^\prime$ and the sums over the G-representations compatible with the representations $j_l$ and $j_l^\prime$ and over the intertwiners compatible with $i_n$ and $i_n^\prime$.  $\mathcal{A}_{\mathcal{F}}[\psi,\psi^\prime]$ is the spin foam amplitude associated to the spin foam $\mathcal{F}: \psi \rightarrow \psi^\prime$ interpolating between the initial and final states. It is given as above as a product of vertex, edge and face amplitudes: $A_v, \, A_e, \, A_f$. 
A spin foam thus represents a possible history of the gravitational field and can be interpreted as a set of transitions through different quantum states of space.
 

In quantum theory, the probability for any fundamental event is given by the absolute square of a complex amplitude; the amplitude for some event is given by adding together all the histories which include that event. In order to find the overall probability amplitude for a given process, one adds up, or integrates, the amplitude of each possible history over the space of all possible histories of the system in between the initial and final states. 
In the spinfoam formalism, the \emph{paths} summed over are the 4-geometries represented by the \emph{spin foam}  $\mathcal{F}=(C,c)$. The model is defined by the partition function
\be
Z=\sum_C w(C) \sum_{j_f, i_e} \prod_f A_f(j_f, i_e) \prod_e A_e(j_f, i_e) \prod_v A_v(j_f, i_e),
\ee
with again $A_f$, $A_e$ and $A_v$  the amplitude associated to faces, edges, vertices respectively and $w(C)$ a normalization factor for each 2-complex $C$. We will see in the next sections that this partition function can be derived directly as a regularization of the covariant path integral 
\be
Z=\int D[e] \, D[\omega] \, \mu[\omega, e] \, \exp[i S_{\textrm{GR}}(e, \omega)]
\ee
To summarize, quantum gravity formulated as a 'spin foam' consists of a rule for computing amplitudes from spin foam vertices, faces, and edges. 

Before going further, we have to emphasize that the sum over spinfoam $\mathcal{F}=(C,c)$ is not exactly well-defined. Usually, for fixed $C$, the sum over colorings $c$ is well-controlled. On the other hand, controlling the full sum over two-complexes is a much more subtle issue. It is nevertheless non-perturbatively defined as quantum field correlations  in the context of \textit{group field theory} (GFT) \cite{laurent-gft, daniele-gft}, which are non-local scalar field theories defined over some group manifolds. In this context, $\mathcal{A}_\mathcal{F}$ actually  depends on the GFT coupling constant $\lambda$. Indeed the statistical weight $w(C)$ for a two-complex $C$ is given by the symmetry factor of the two-complex (considered as a Feynman diagram for the GFT) times a factor $\lambda^V$ where $V$ is the number of vertices of the two-complex. The sum over 4-geometries in $K[\psi]$ can be written as a power series in $\lambda$.
\be
K[\psi]=\sum_{V=0}^\infty \lambda^{V} K_V[\psi],
\ee
where $K_V[\psi]=\sum_{\mathcal{F}^V|_{\partial \mathcal{F}^V=\psi }} \mathcal{A}^V_\mathcal{F}[\psi]$ is the sum over spinfoams $\mathcal{F}^V$ with $V$ vertices. 
Spin foams appear then naturally as higher-dimensional analogs of Feynman diagrams of the GFT. 


Spin foam models were  initially constructed as a history formalism for loop quantum gravity describing the evolution of spin network states. However since then, they have been developed  as a discretization of the path integral of general relativity. The more standard procedure is now to start with the path integral of general relativity  reformulated as a topological BF gauge theory plus constraints. 
In three dimensions, general relativity is simply given by the  BF theory whereas in 4 dimensions, general relativity can be viewed as a BF theory  with extra constraints. We will present this point of view in the following. \\
Since  BF theory is topological, it does not have any local degree of freedom and can be discretized and quantized exactly as a spin foam model without loosing any of its physical content. In the first following  section, we will illustrate the quantization of a BF theory by deriving the spin foam model for 3d Euclidean gravity, the so-called Ponzano-Regge model. \\
In four dimensions, we  attempt to impose consistently the constraints turning the BF theory into a geometrical theory and introducing local degrees of freedom directly at the quantum level in the spin foam framework. One approach is based on the MacDowell-Mansouri action, which writes general relativity as a BF theory for the gauge group $\SO(4,1)$ (or $\SO(5)$ in the Euclidean case) with a non-trivial potential in the $B$-field which breaks the symmetry down back to the Lorentz group \cite{artem}. Although this is a very interesting proposition, it has not yet led to a definite proposal for a spinfoam model. Therefore we will  focus on the reformulation using the Plebanski's action which is at the heart of the spinfoam models developed up to now. This will be presented in the second and third following sections.

\section{Three-dimensional gravity, the Ponzano-Regge model} \label{PRmodel}
As we have already mentioned it, in three space-time dimensions, gravity is a topological theory which can be exactly quantized as a spin foam model, the well-known Ponzano-Regge model \cite{PR}. We review here the construction of the model for a 3d Riemanian theory of gravity without cosmological constant (see \cite{louapre} for a review). We consider the general relativity action in first order formalism,  which   is simply a $BF$ action, on $\mathcal{M}$ a closed manifold. 
\be
S[B, \omega]= \int_\mathcal{M} \tr \left( B \wedge F[\omega] \right),
\ee
The triad $B$ is a $\su(2)$-valued 1-form, from which the metric is reconstructed as $g_{\mu \nu}=B^I_\mu B^J_\nu \eta_{IJ}$. The parallel transport on the manifold is given by the $\su(2)$-valued connection $\omega$ and its curvature $F[\omega]=d\omega+\omega\wedge \omega$ is an $\su(2)$-valued 2-form. $\tr$ is the trace over the Lie algebra $\su(2)$. 
The classical equations of motion impose that the connection is torsion-free and flat
\be
d_\omega B\equiv dB+[\omega, \, B]=0, \qquad F[\omega]=0,
\ee
where $[\cdot , \cdot ]$ is the $\su(2)$ Lie bracket. A simple counting of the degrees of freedom shows that there are no local degrees of freedom. This theory  has three types of symmetries: local Lorentz gauge symmetries, diffeomorphisms and translations \cite{PRrevisited1}. 

The formal path integral for 3d gravity is
\be
Z_{\mathcal{M}}=\int \mathcal{D} B \, \mathcal{D}\omega \,e^{i S[B, \omega]}.
\ee
Heuristically speaking, since  $BF$ theory is topological and does not have any local degree of freedom, we do not expect to lose any information by replacing the manifold $\mm$ by a simplicial manifold $\Delta_3$, with similar topology. 
More precisely, we consider a triangulation $\Delta_3$ built from gluing tetrahedra together along their respective triangles and edges. The different elements of the triangulation $\Delta_3$ are recalled in Table \ref{triang3d}. We also need the notion of  \textit{dual 2-complex} of $\Delta_3$, denoted by $\Delta_3^\ast$, which is a combinatorial object defined by a set of vertices $v^\ast \subset \Delta_3^\ast$ (dual to tetrahedra $t$ in $\Delta_3$), edges $e^\ast\subset \Delta^\ast_3$ (dual to triangles $f$ in $\Delta_3$) and faces $f^\ast \subset \Delta^\ast_3$ (dual to segments $e$ in $\Delta_3$).  

\begin{table}[h!]
    \centering
        \begin{tabular}{|ll|}
\hline
$\Delta_3$ & $\Delta_3^\ast$ \\
\hline
&\\
tetrahedron $\quad$ & {\it vertex} (4 edges, 6 faces) \\
triangle & {\it edge} (3 faces)\\
segment & {\it face} \\
point & 3d region \\
&\\
\hline
        \end{tabular}
\begin{minipage}[h]{0.5\linewidth}\vspace{0.2cm}
\centering\includegraphics[scale=0.4]{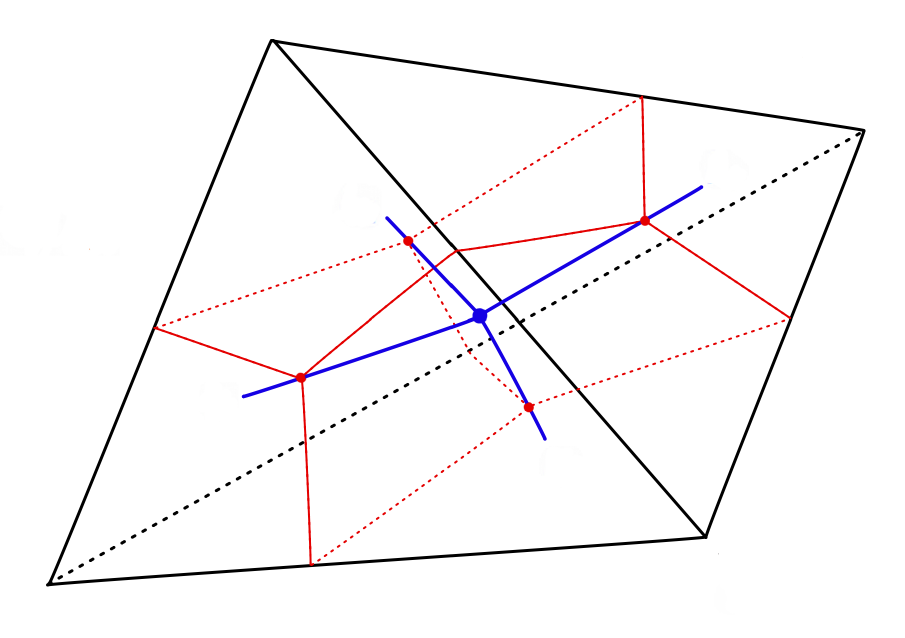} \vspace{0.1cm} 
\end{minipage}
 \caption{Relation between a triangulation $\Delta_3$ and its dual $\Delta_3^\ast$ in 3 dimensions. In italic, the two-complex. In parentheses: adjacent elements. The figure represents the dual two-complex (blue edges) dual to a tetrahedron (black segments): to each tetrahedron a dual vertex is associated $v^\ast \sim t$, to each triangle a dual edge is associated $e^\ast \sim f$ and to each segment a dual face is associated $f^\ast \sim e$.} \label{triang3d}
 \end{table}

The fields $B$ and $\omega$ are replaced by configurations which are distributional with support on subsimplicies of $\Delta_3$ and its topological dual $\Delta_3^\ast$ respectively. The triad $B$ is a 1-form and as such it is naturally integrated on one dimensional structures. We integrate it along the segments $e\in \Delta_3$ of the triangulation, and replace it by the collection of Lie-algebra elements $X_e$ obtained in this way. To discretize the connection field, one considers the dual 2-complex $\Delta_3^\ast$. One can naturally consider the holonomy of the connection along the edges $e^\ast$ of the dual two complex. This assigns a group element $g_{e^\ast}$ to each dual edge $e^\ast$. The discretized curvature is obtained as the holonomy of the connection around an edge $e$ of $\Delta_3$, which is defined as the ordered product of corresponding group elements living on the dual edges $e^\ast$ forming the dual face $f^\ast \sim e$. 

\begin{minipage}[h]{0.5\linewidth}\vspace{0.1cm}
\bes
B & \rightarrow & X_e=\int_eB \qquad \in \su(2), \nn\\
\omega & \rightarrow & g_{e^\ast}= \mathcal{P} e^{\int_{e^\ast} \omega} \qquad \in \SU(2),  \nn \\
F[\omega] & \rightarrow & U_e= \stackrel{\longrightarrow}{\prod_{e^\ast \subset e}} g^{\epsilon(e, e^\ast)}_{e^\ast} \qquad \in \SU(2). \nn 
\ees \vspace{0.2cm}
\end{minipage} 
 \begin{minipage}[h]{0.4\linewidth}
\centering\includegraphics[scale=0.56]{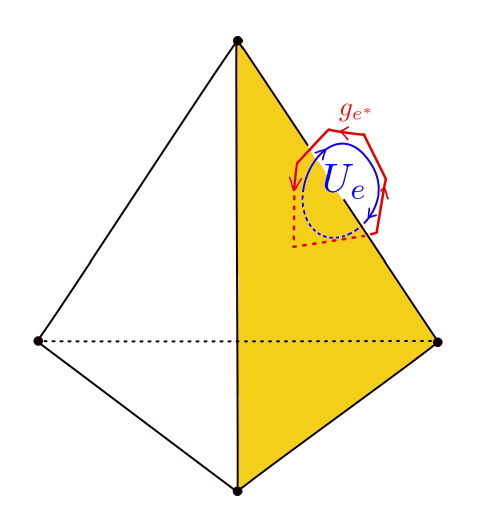} 
\end{minipage}

The flatness constraint $F[\omega]=0$ translates into the triviality of holonomies $U_e=\id$ (around closed loops, for trivial homotopy). After this discretization, the action on the triangulation  reads
\be
S[X_e, g_{e^\ast}]=\sum_{e \in \Delta_3} \tr(X_e U_e).
\ee
Actually the original path integral derivation of the Ponzano-Regge model introduced the Lie algebra element $Z_e=\log (U_e) \in \su(2)$ and considered the action $\sum_e \tr(B_e Z_e)$ \cite{laurent-kirill}. Nevertheless, one faces the issue of the non-continuity of the $\log$ map (and the choice of a particular branch and so on). Therefore, we prefer to work with the now more usual formulation presented in \cite{PRrevisited1} which uses directly the group element $U_e$.
The discrete path integral is then given by
\be \label{Z3d}
Z_{\Delta_3} = \left(\prod_{e\in \Delta_3}\int_{\su(2)} d^3X_e \right) \left(\prod_{e^\ast \in \Delta_3^\ast} \int_{\SU(2)} dg_{e^\ast}  \right)e^{iS[X_e, g_{e^\ast}]},
\ee
where the Lebesgue measure $d^3X_e$ on $\su(2) \sim \R^3$ and the Haar measure $dg_{e^\ast}$ on $\SU(2)$ are the natural choices for the discretized path integral measures. 
Integrating over the $X_e$ variables and using
\be \label{deltaSO3}
\int dX_e \exp\left(i \tr(X_e U_e) \right)= \delta(U_e)= \sum_j d_j \,\tr\left( D^j(U_e)\right), \quad d_j=2j+1
\ee
where we used the Peter-Weyl theorem in the last equality, one obtains
\be \label{Z3d1}
Z_{\Delta_3} = \sum_{\{j\}} \left(\prod_{e^\ast \in \Delta_3^\ast} \int_{\SU(2)} dg_{e^\ast}  \right) \prod_{e \in \Delta_3} d_{j_e}\, \tr\left[ D^{j_e}(U_e)\right].
\ee
Notice that the delta function $\delta(U_e)$ in (\ref{deltaSO3}) is the delta function on the group $\SO(3)$ and not $\SU(2)$, i.e. it is the distribution localizing the group element $U$ on the identity in $\SO(3)$ and it thus does not distinguish $\id$ and $-\id$ as $\SU(2)$ group elements ($\delta_{\SO(3)}(U)=\delta_{\SU(2)}(U) + \delta_{\SU(2)}(-U)$). Consequently, the sum of the Pancherel decomposition in (\ref{deltaSO3}) is over integer spin representations $j_e\in \N$. To recover the $\SU(2)$ Ponzano-Regge model with a sum over both integer and half-integer representations $j_e \in \N/2$, the amplitude in the path integral has to be slightly modified \cite{PRrevisited1, etera-jimmy} in order to kill the $\delta_{\SU(2)}(-U)$ term in the $\delta_{\SO(3)}(U)$ distribution. \\
We  consider that the sum over $j_e$ in (\ref{Z3d1}) is now a sum over $j_e \in \N/2$. It remains to integrate over the lattice connection $\{g_{e^\ast}\}$. Each edge $e^\ast \subset \Delta_3^\ast$ bounds 3 faces $f^\ast \subset \Delta_3^\ast$, therefore they will be 3 traces of the form $\tr\left[ D^{j_e}(\cdots g_{e^\ast}\cdots)\right]$ in (\ref{Z3d1}) containing $g_{e^\ast}$ in argument. In order to integrate over $g_{e^\ast}$, we can use the identity (\ref{3j}) recalled here
\bes \label{3dintert}
&&\int_{\SU(2)}dg \, D^{j_1}_{m_1 n_1}(g) D^{j_2}_{m_2 n_2}(g) D^{j_3}_{m_3 n_3}(g)= (-1)^{j_1+j_2+j_3}\left(\tabl{ccc}{j_1& j_2 & j_3 \\
m_1& m_2 &m_3} \right) \left(\tabl{ccc}{j_1& j_2 & j_3  \nn \\
n_1& n_2 &n_3} \right)\\
&&\qquad \qquad \qquad \qquad \quad \qquad\qquad\, \; \;\quad \qquad = i\, i^\ast  \\
&&\;  \includegraphics[scale=0.5]{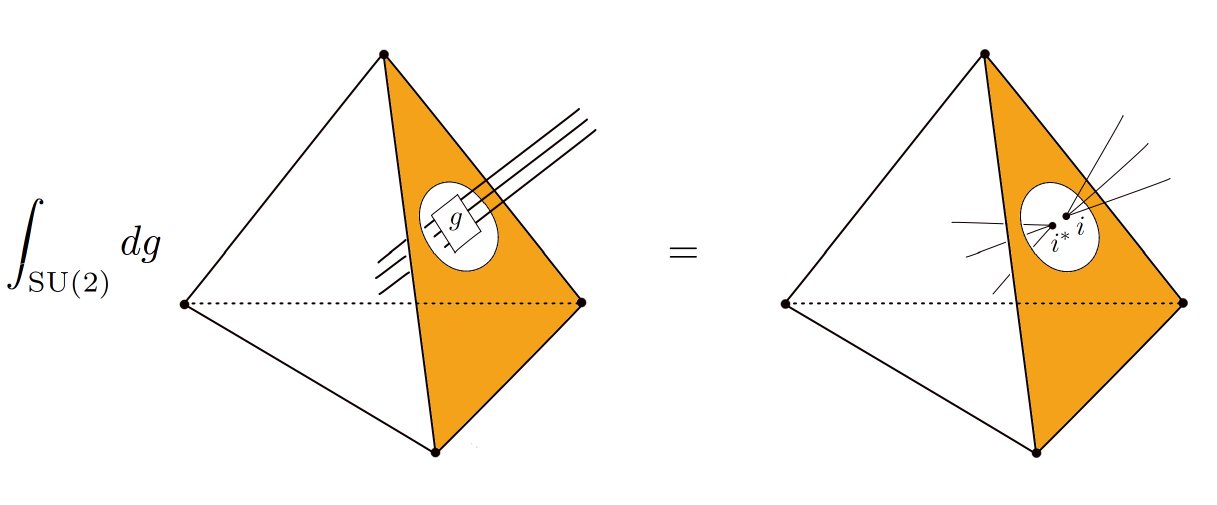}  \nn
\ees
where we can recognize the Wigner's 3j-symbol, which is the unique (up to normalization) three valent $\SU(2)$ intertwiner denoted by $i$. The figure below the equation (\ref{3dintert}) illustrates  that each 3j-symbol is in fact associated to a triangle $f\subset \Delta_3$ of a tetrahedron $t\subset \Delta_3$, \ie  in the  spin foam quantization procedure, a triangle becomes an intertwiner. The triangles of a given tetrahedron glued together form the boundary of the tetrahedron which is the key kinematical object. An intertwiner $i$ lives on the dual 1-skeleton of the triangle: the intertwiner vertex corresponds to the inside of the triangle while each leg crosses on edge of the triangle (it consequently inherits the same representation $j_e$). Then, the intertwiners associated to the four triangles of the tetrahedron are glued together into one spin network which defines the quantum state of the boundary of the tetrahedron.

\begin{figure}[ht]
\begin{center}
\includegraphics[scale=0.6]{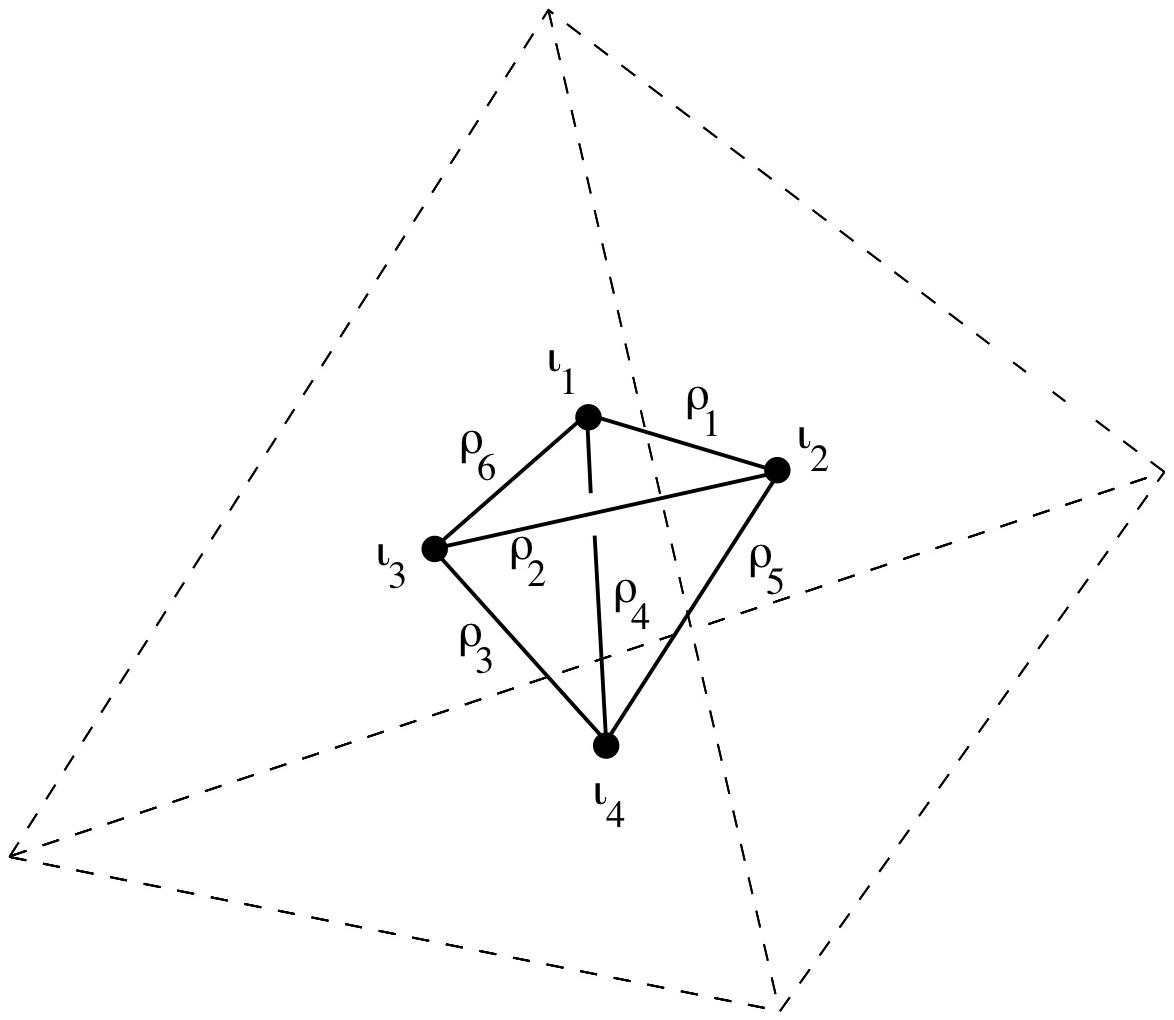} \label{tetra3d3} 
\caption{Tetrahedron}\label{tetra3d3}
\end{center}
\end{figure}

 In the case of a tetrahedron, the spin network dual to this original tetrahedron is also a tetrahedron.
By combining the four normalized Clebsh-Gordan coefficients $\imath_1, \cdots \imath_4$ corresponding to these four triangles, we then get a well-known object in recoupling theory of $\SU(2)$ given by the six variables $j_1, \cdots, j_6$, irreducible representations of the group, associated to the segments of the tetrahedron (or equivalently to the edges of the dual tetrahedral spin network (see Fig. \ref{tetra3d3} )).
\bes
\left\{ \tabl{ccc}{j_1& j_2 & j_3 \\
j_4 & j_5 & j_6}\right\} &= &\sum_{m_1, \cdots, m_6} (-1)^{j_4+j_5+j_6+m_4+m_5+m_6} \left(\tabl{ccc}{j_1& j_2 & j_3 \\
m_1& m_2 &m_3} \right) \left(\tabl{ccc}{j_3& j_5 & j_4 \\
m_3& -m_5 &m_4} \right) \nn \\
&&\qquad \quad \left(\tabl{ccc}{j_4& j_2 & j_6 \\
-m_4& m_2 &m_6} \right)\left(\tabl{ccc}{j_6& j_5 & j_1 \\
-m_6& m_5 &m_1} \right) ,\nn
\ees
where $\left\{ \tabl{ccc}{j_1& j_2 & j_3 \\
j_4 & j_5 & j_6}\right\} $ is the so-called $\{6j\}$-symbol. Finally, the Ponzano-Regge amplitude for a given colored triangulation is simply given by the product of the $\{6j\}$-symbol
\be \label{Z3df}
Z_{\Delta_3} = \sum_{\{j_e\}} \prod_{e} d_{j_e} \prod_t \left\{ \tabl{ccc}{j^{(t)}_1& j^{(t)}_2 & j^{(t)}_3 \\
j^{(t)}_4 & j^{(t)}_5 & j^{(t)}_6}\right\}_N,
\ee
where the index $N$ means that it is the normalized $\{6j\}$ symbol which appears\footnote{$\left\{ \tabl{ccc}{j_1& j_2 & j_3 \\
j_4 & j_5 & j_6}\right\}_N= (-1)^{j_1+j_2+j_3+j_4+j_5+j_6}\left\{ \tabl{ccc}{j_1& j_2 & j_3 \\
j_4 & j_5 & j_6}\right\}. $}. 
Moreover, let us recall that the starting point (\ref{Z3d}) of the construction presented above gives the Ponzano-Regge model for $\SO(3)$ and not $\SU(2)$. In (\ref{Z3d1}), the sum over half-integer representations $j_e\in \N/2$ instead of the sum over integer representations $j_e\in \N$ was introduced by hand. In this last formula (\ref{Z3df}) a sign factor only visible for the half-integer representations is still missing \cite{ileana} and the Ponzano-Regge model is actually given by
\be
Z_{\Delta_3} = \sum_{\{j_e\}} \prod_{e} (-1)^{2j} d_{j_e} \prod_t \left\{ \tabl{ccc}{j^{(t)}_1& j^{(t)}_2 & j^{(t)}_3 \\
j^{(t)}_4 & j^{(t)}_5 & j^{(t)}_6}\right\}_N.
\ee
Historically, this model is the first spin foam model ever built. It was proposed by Ponzano and Regge in 1968 \cite{PR}. The simple quantization ansatz was that the edge lengths are quantized as $l_e=(j_e+1/2)l_p$ (with $l_P$ the Planck length) and that the $\{6j\}$-symbol, which is thus the basic building block of their model, is associated to each ``quantized" tetrahedron. 

\begin{minipage}[h]{0.2\linewidth}\vspace{0.01cm}
$$\qquad \qquad {\Large \left\{ \tabl{ccc}{j_1& j_2 & j_3 \\
j_4 & j_5 & j_6}\right\} } \qquad \qquad$$
\end{minipage}
\begin{minipage}[h]{0.8\linewidth}\vspace{0.4cm}
\;\;\;\;\;
\end{minipage}
\begin{minipage}[h]{0.2\linewidth}\vspace{0.4cm}
$ \qquad \qquad {\Huge \longleftrightarrow }$
\end{minipage}
\begin{minipage}[h]{0.2\linewidth}\vspace{0.45cm}
\centering\includegraphics[scale=0.5]{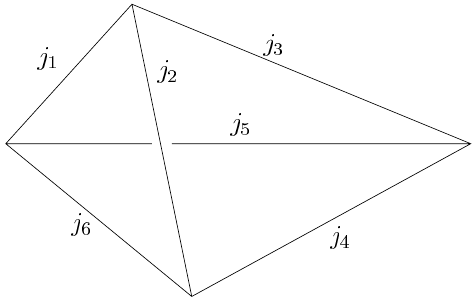} 
\vspace{0.02cm}
\end{minipage}

This ansatz was justified by the asymptotic of the $\{6j\}$ symbol.
\be
\left\{ \tabl{ccc}{j_1& j_2 & j_3 \\
j_4 & j_5 & j_6}\right\} \sim_{j \rightarrow \infty} \f{1}{\sqrt{12\pi V}} \cos \left( S_R + \f{\pi}{4}\right),
\ee
where $V$ is the volume of the tetrahedon and $S_R= \sum_e \f{d_{j_e}}{2}\, \theta_e$ is the Regge action for discrete gravity, with the sum being over the edges of the tetrahedron, the $\theta_e$ being the dihedral angle of the edge $e$, \ie the angle between the two outward normals of the two faces incident on that edge. The large spin limit is a semi-classical limit. Indeed, since the $\SU(2)$ irreducible representations are interpreted as lengths 
\be
l_e=(j_e+\f12)l_p=(j_e + \f12)G \hbar,
\ee
the large spin limit for a fixed length $l_e$ is equivalent to make $\hbar$ (or $l_p$) approaching 0.
This justifies that the Ponzano-Regge model is a quantum  gravity model. More details concerning the $\{6j\}$-symbol and the recoupling theory of $\SU(2)$ can be found in appendix \ref{6j} and a detailed study of some of its properties and its asymptotical behavior  has been done in  \cite{article1} and \cite{article2}. The results which are relevant to extracting the semi-classical information from the spin foam formalism will be presented in  Chapters \ref{groupInt6j} and \ref{recursion6j}.

Let us now consider the 4 dimensional case and see how we can write general relativity as a BF theory plus some additional constraints.

\section{Reformulating 4d gravity as a BF theory: Plebanski's action}
The Plebanski action (without cosmological constant)\footnote{The cosmological constant can be added to the Plebanski action which thus becomes 
$$
S_{\textrm{Plebanski}}[B, \omega, \lambda]= \int_\cM B^{IJ} \wedge F_{IJ}[\omega] -\f{\Lambda}{4} \epsilon_{IJKL} B^{IJ}\wedge B^{KL}- \f12 \lambda_{IJKL} B^{KL}\wedge B^{IJ}
$$.} reformulates  general relativity as a constrained BF gauge theory for the Lorentz group, $\SO(3,1)$ (or $\SO(4)$ in the Euclidean case)
\be \label{PlebanAct}
S_{\textrm{Plebanski}}[B, \omega, \lambda]= \int_\cM B^{IJ} \wedge F_{IJ}[\omega] - \f12 \lambda_{IJKL} B^{KL}\wedge B^{IJ},
\ee
where we recall that $\cM$ is the space-time manifold, $I,J$ are Lorentz indices  running from 0 to 3, $\omega$ is a  $\mathfrak{so}(3,1)$-valued 1-form (or $\mathfrak{so}(4)$-valued in the Euclidean theory), $\omega= \omega^{IJ}_\mu J_{IJ}dx^\mu$, $J_{IJ}$ are the Lorentz generators and $F[\omega]$ is its strength tensor. The $B$ field is a $\mathfrak{so}(3,1)$ valued 2-form, $B=B^{IJ}_{\mu \nu} \,J_{IJ}\,dx^\mu \wedge dx^\nu$.
The constraints $\mathcal{C}_{IJKL}[B]\equiv B^{IJ}\wedge B^{KL}$ are enforced by the Lagrange multipliers $\lambda_{IJKL}$ which satisfy $\lambda_{IJKL}=-\lambda_{JIKL}=-\lambda_{IJLK}=\lambda_{KLIJ}$ and $\lambda_{IJKL}\epsilon^{IJKL}=0$. The $B$ field is thus constrained 
in a such way that we recover general relativity in its first order formalism as given in (\ref{PalaAct}). 
 Indeed, the associated equations of motion are
\bes
\f{\delta S}{\delta \omega} & \longrightarrow & \mathcal{D}B=dB+ [ \omega, \, B]=0 \\
\f{\delta S}{\delta B} & \longrightarrow & F^{IJ}(\omega)= \lambda^{IJKL}B_{KL} \\
\f{\delta S}{\delta \lambda} & \longrightarrow & B^{IJ} \wedge B^{KL}=e\, \epsilon^{IJKL}
\ees 
 where $e=\f{1}{4!} \epsilon_{IJKL} B^{IJ}\wedge B^{KL}$. The relation with gravity arises because of this last constraint. When $\tilde{e}=\f{1}{4!} \epsilon_{IJKL} \epsilon^{\mu \nu \rho \delta }B^{IJ}_{\mu \nu} B^{KL}_{\rho \delta}\neq 0$, the last constraint is equivalent to 
 \be \label{constraintB}
 \epsilon_{IJKL}B^{IJ}_{\mu \nu}B^{KL}_{\rho \delta}=\epsilon_{\mu \nu \rho \delta}\tilde{e},
 \ee
which can be decomposed in three parts
\bes  \label{simplicityB}
a) \quad \epsilon_{IJKL} B^{KL}_{\mu \nu} B^{IJ}_{\mu \nu} & =& 0 \nn \\
b) \quad \epsilon_{IJKL} B^{KL}_{\mu \nu} B^{IJ}_{\mu \rho} & =& 0  \\
c) \quad \epsilon_{IJKL} B^{KL}_{\mu \nu} B^{IJ}_{\rho \delta} & =& \pm \tilde{e} \nn
\ees
where the indices $\mu, \, \nu, \, \rho, \, \delta$ are all different and the sign in the last equation is determined by the sign of their permutation. It is usually this system which is referred as the simplicity constraints. $\epsilon_{IJKL}B^{IJ}_{\mu \nu}B^{KL\mu \nu}=0$ is equivalent to say that $B_{\mu \nu}$ is a simple bivector (\ie of the form $B=u\wedge v$). Moreover,  the constraint $B^{IJ} \wedge B^{KL}=e\, \epsilon^{IJKL}$ (with $e \neq 0$) is satisfied if and only if there exists a real tetrad field $e^I=e^I_\mu \, dx^\mu$ so that one of the following equations holds:
 \bes
 (\star) && B^{IJ}= \pm e^I \wedge e^J  \label{Bconstr1} \\
 (s) && B^{IJ} = \pm \epsilon^{IJ}_{\; \; KL} e^K \wedge e^L. \label{Bconstr2}
 \ees
The action (\ref{PalaAct}) is obtained if we restrict the field $B$ to be always in the sector (s) (with the plus sign), and substitute the expression for $B$ in terms of the tetrad field into (\ref{PlebanAct}). We denote by $(\star B)^{IJ}\equiv \f12 \epsilon^{IJ}_{\; \; KL}B^{KL}$ the Hodge duality operation. Therefore, the Plebanski model is not exactly pure gravity. The restriction on the $B$ field is always possible classically, so the two theories do not differ at the classical level, but they are different at the quantum level, since in the quantum theory one cannot avoid interference between different sectors. Indeed, this interference is due to the existence in the Plebanski action of a $\Z_2 \times \Z_2$ symmetry:
 \be\label{sym-plebanski}
 B \rightarrow -B, \quad B \rightarrow \star B.
 \ee

Before considering the spin foam level, let us emphasize that the simplicity constraints $(\star B) \cdot  B=0$ correspond to the simplicity constraints (\ref{secondclass}) of the canonical analysis. 
$$ (\star B) \cdot  B=0 \longleftrightarrow\phi=(\star \mathcal{R}) \cdot \mathcal{R}=0.$$
  They are thus second class constraints. Furthermore,    there is no analogue here of the other second class constraints we have given in (\ref{secondclass}), \ie  $\psi= \mathcal{R}\mathcal{R} D_{\mathcal{A}}\mathcal{R}=0$  since the spin foam formalism is fully covariant. Indeed, the $\psi$ are secondary constraints, coming from the Poisson bracket $\{S, \phi\}$. In the canonical approach, $\phi=0$ is only imposed on the initial hypersurface and $\psi=0$ is then needed to ensure $\phi=0$ under the Hamiltonian evolution (given by the Hamiltonian constraint) whereas in the spin foam formalism approach $\phi=0$ is directly imposed on all  space-time structures. That is that we have projected  on $\phi=0$ at all stages of the evolution (\ie on all hypersurfaces) and therefore we do not need the secondary constraints $\psi$.

Another important point which appears in the canonical approach for quantum gravity is the Immirzi parameter. A way to introduce it in the spin foam formalism is to consider a generalized BF-type action for gravity instead of the Plebanski action (\ref{PlebanAct})  as starting point of the quantization procedure. Indeed, it is possible to recover the Palatini-Holst action (\ref{PalaHolst}) for general relativity with Immirzi parameter considering a more general constraint on the $\lambda$ field \cite{etera-daniele},
\be
a\, \lambda_{IJ}^{\;\;\, IJ}+b\, \lambda_{IJKL}\,\epsilon^{IJKL}=0.
\ee
This constraint $H\equiv a \, \lambda_{IJ}^{\;\;\, IJ}+b\, \lambda_{IJKL}\,\epsilon^{IJKL}$ can be taken into account in a generalized BF-type action for gravity given by 
\be \label{BFgeneral}
S[B, \omega, \lambda, \mu]= \int_\cM B^{IJ} \wedge F_{IJ}[\omega] - \f12 \lambda_{IJKL} B^{KL}\wedge B^{IJ} +\mu \, H,
\ee
where the condition $H=0$ is enforced by the Lagrange multiplier $\mu$ while $\lambda_{IJKL}$ still enforces the constraints on the $B$-field. The equations of motion for $\omega$ and $B$ are the same as those coming from the Plebanski action, but the constraints on the $B$-field are now:
\bes 
&&B^{IJ}\wedge B^{KL}= \f16( B^{MN}\wedge B_{MN}) \eta^{[I|K|}\eta^{J]L}+\f{\epsilon}{12}(B^{MN}\wedge (\star B)_{MN} )\epsilon^{IJKL},   \label{ImmiBconstr1}\\
&&2\, a\, B^{IJ}\wedge B_{IJ} -\epsilon\, b\, B^{IJ}\wedge (\star B)_{IJ}=0, \label{ImmiBconstr2}
\ees
where $\eta^{IJ}$ is the flat metric and  the constant $\epsilon$ is $1$ in the Euclidean case and $-1$ in the Lorentzian case. 

Let us note that we are back to the Plebanski action in the case $a=0$.  By  choosing $a=4 \gamma$ and $b=-(\epsilon+\gamma^2)$ with $\gamma$ a non-zero parameter,  the solution of \eqref{ImmiBconstr1} and \eqref{ImmiBconstr2}, for non-degenerate $B$ ($B^{IJ}\wedge (\star B)_{IJ} \neq 0$), is then
 \bes
 (\star) && B^{IJ}= \pm \f{1}{\gamma}\left(\star(e^I\wedge e^J)-\epsilon\, \gamma \, e^I \wedge e^J)\right)\label{pos1}\\
 (s) && B^{IJ} = \pm  \star(e^I\wedge e^J)-\f{1}{\gamma} e^I \wedge e^J. \label{pos2}
 \ees
The same symmetry  \eqref{sym-plebanski} as  in the Plebanski case ,   brings us  four sectors.  Both  positive sectors \eqref{pos1} and \eqref{pos2}  yield to general relativity with different Immirzi parameters. Indeed, we get the Palatini-Holst action (\ref{PalaHolst}) inserting any of the two positive solutions \eqref{pos1}, \eqref{pos2}, into (\ref{BFgeneral}). In the $(s)$ sector,  the   Immirzi parameter is as usual $\gamma$ whereas  in the $(\star)$ sector the Immirzi parameter is instead $\epsilon/ \gamma$.   In the following, we will  work with a \textit{finite} (non-zero) Immirzi parameter, unless specified otherwise.

\section{The spin foam framework}
The starting point of the spin foam framework is  a BF-like action 
\be \label{actionBFplus}
S= \int_{\mathcal{M}}[\tr(B\wedge F) + \Phi(B)].
\ee
A general prescription for the definition of constrained BF theories on the lattice has been studied in \cite{laurent-kirill}. 
The three dimensional case,   with $\Phi(B\equiv 0)$,    has been discussed in the previous Section \ref{PRmodel}. In the following we shall study the  four dimensional case where  $\Phi(B)$ is a quadratic function which also depends on some Lagrange multipliers ($\Phi(B)=- \f12 \lambda_{IJKL} B^{KL}\wedge B^{IJ}+\mu \, H$). We will also discuss in details the implementation of the simplicity constraints encoded in $\Phi(B)$. 

\medskip

The strategy usually followed in four dimensions can be sketched in three main stages
\begin{enumerate}
\item discretize the classical theory by putting it on a cellular decomposition;
\item quantize the topological BF part of the discretized theory;
\item  impose the simplicity constraints at the quantum level.
\end{enumerate}
In this sense, the classical relation between BF theory and gravity is directly translated to the quantum level.

As we have seen in the three dimensional case, the first step is the discretization of the manifold $\cM$. We start with a cellular decomposition  $\Delta_4$ of the 4d space-time manifold $\cM$ built from  gluing 4-cells  together.  

The 4-cells are  4-simplices, the 3-cells are tetrahedra and the other elements of $\Delta_4$ are recalled in  Table \ref{triang4d}. Since  BF theory is topological, it does not have any local degree of freedom. It can be discretized on the cellular decomposition of $\cM$ and quantized exactly as a spinfoam model without losing any of its physical content. 
Then one works directly at the quantum level in the spinfoam framework and attempts to impose consistently the constraints turning the BF theory into a geometrical theory and introducing local degrees of freedom. We will see that there exists ambiguities in the way constraints can be imposed. Up to now, several theories (such as the  EPRL-FK models \cite{EPR0, EPR, FK, LS, EPRL}) have been proposed and any of these theory can at this moment be considered as a quantum theory of gravity. I will give a short review focusing only on the models relevant for the  next parts. 

Before going on to the specific quantization, let us remark that it is quite possible that one should construct in a different way, following a different strategy, the relevant spinfoam to encode 4d quantum gravity.   Two recent results point to   this direction. Firstly, B. Dittrich and J. Ryan have recently proposed an interesting new procedure\cite{sf_jimmy}  -- that we will not explain here --  which seems to give even at the discretized level a different model than the EPRL-FK models constructed using the usual startegy and  which we shall describe in the next section.  Secondly, this strategy we have recalled above  to construct spinfoam models in 4d  -- firstly  quantize, secondly  implement the constraints --  has been  criticized in details recently  in \cite{Critic-Alexandrov-Roche}. An alternative method would thus be to discretize the simplicity constraints and include them at the classical level in a discrete BF action. I will not give more details on this procedure used in \cite{sf_valentin} which however seems cleaner but which leads to spin foam amplitudes much more complicated and not always local.

%
\begin{table}[h!]
    \centering
        \begin{tabular}{|ll|}
\hline
$\Delta_4$ & $\Delta_4^\ast$ \\
\hline
&\\
4-simplex & {\it vertex} (5 edges, 10 faces) \\
tetrahedon $\quad$ & {\it edge} (4 faces)\\
triangle & {\it face} \\
segment & 3d region \\
point & 4d region \\
&\\
\hline
        \end{tabular} \begin{minipage}[h]{0.5\linewidth}\vspace{0.35cm}
\centering\includegraphics[scale=0.2]{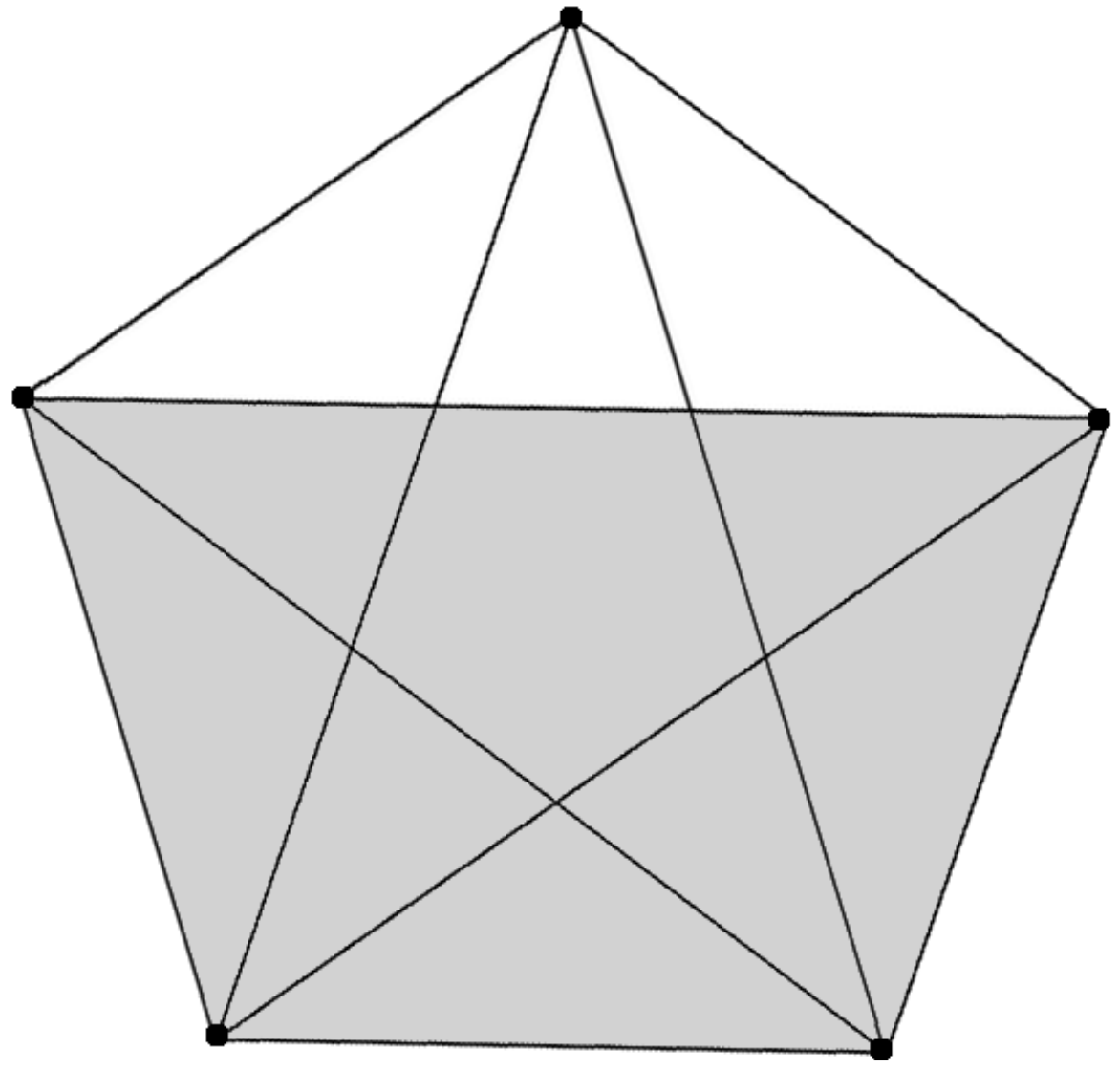} 
\vspace{0.02cm}
\end{minipage}
    \caption{Relation between a triangulation $\Delta_4$ and its dual $\Delta_4^\ast$ in 4 dimensions. 
    In italic, the two-complex. In parentheses: adjacent elements. 
    The figure represents a 4-simplex. In grey, a tetrahedron of the 4-simplex.}
    \label{triang4d}
\end{table} 

\subsection{Spin foam quantization of the 4d BF theory} \label{SFBF}
\paragraph{The Euclidean case.}  
In this case, we use   $\Spin(4)$ as  gauge group. The action of the BF theory is given by the first term of the action (\ref{actionBFplus}) 
\be
S_{BF}[B, \omega]= \int_{\mathcal{M}} B^{IJ} \wedge F_{IJ}[\omega]= \int_{\cM} \tr \left( B \wedge F[\omega]\right)
\ee
where  $B^{IJ}_{\mu \nu}$ is a $\spin(4)$ Lie algebra valued 2-form, $\omega^{IJ}_\mu$ is a connection on a $\Spin(4)$ principle bundle over $\cM$. The theory has no local excitations and its properties are  analogous to the case of three dimensional gravity studied in Section \ref{PRmodel}.
For simplicity, we will work with  the discretization of $\cM$ given by a triangulation $\Delta_4$ (instead of a general cellular decomposition). 

The next step is to discretize the fields,  the $B$-field and the Lorentz gauge connection $\omega$.  The connection curvature $F[\omega]$ is associated to the dual surfaces. Indeed, if we introduce a dual link variable $g_{e^\ast}=e^{i \omega_{e^\ast}}$ for each dual link $e^\ast$, through the holonomy of the $\mathfrak{spin}(4)$-connection $\omega$ along the link, then consequently, the product of dual link variables along the boundary $\partial f^\ast$ of a dual plaquette $f^\ast$ leads to a curvature located at the center of the dual plaquette: $\prod_{e^\ast \in \partial f^\ast}g_{e^\ast}\equiv 
U_{f^\ast}$.  The $B$-field is a 2-form and is naturally discretized on the faces $\Delta$ of the cellular decomposition
\be
B^{IJ}_\Delta=\int_\Delta B_{\mu \nu}^{IJ}dx^\mu \wedge dx^\nu.
\ee
The faces $\Delta$ of the cellular decomposition in the case of a triangulation $\Delta_4$ are triangles and in 4 dimensions triangles $\in \Delta_4$ are dual to faces $f^\ast \in \Delta_4^\ast$. This one-to-one correspondence allows us to denote the discrete $B$ by either a triangle $B_\Delta$ or a face $B_{f^\ast}$ subindex respectively. $B_{f^\ast}$ can be interpreted as the smearing of the continuous 2-form $B$ on triangles in $\Delta_4$. Using these discretized variables, the path integral becomes
\be
Z^{\textrm{BF}}_{\Delta_4}= \int \prod_{e^\ast \in \Delta^\ast_4} dg_{e^\ast}\, \prod_{f^\ast \in \Delta^\ast_4} dB_{f^\ast}\, e^{iB_{f^\ast}U_{f^\ast}},
\ee
which is analogous to (\ref{Z3d}) in the 3 dimensional case. The result of the $B$ integration is given by
\be
Z^{\textrm{BF}}_{\Delta_4}= \int \prod_{e^\ast \in \Delta^\ast_4} dg_{e^\ast}\, \prod_{f^\ast \in \Delta^\ast_4} \delta (U_{f^\ast}),
\ee
where $\delta$ is the delta distribution defined on $L^2(\Spin(4))$. This interpretation of this result is simple: we have to sum over the flat connection. We can use  the Peter-Weyl theorem as in (\ref{deltaSO3}), to obtain
\be \label{BFpartitionPart}
Z^{\textrm{BF}}_{\Delta_4}= \sum_{\cI_1 \cdots \cI_P}\int \prod_{e^\ast \in \Delta^\ast_4} dg_{e^\ast}\, \prod_{f^\ast \in \Delta^\ast_4} d_{\cI_{f^\ast}}\tr \left[ D^{\cI_{f^\ast}}(g^1_{e^\ast} \cdots g_{e^\ast}^n)\right]
\ee
where $\cI_1 \cdots \cI_P$ are the quantum numbers associated to the $P$ faces of $\Delta_4^\ast$. $\cI$ denotes an arbitrary unitary irreducible representation appearing in the Plancherel measure of the group $\Spin(4)$ and $d_{\cI}$ is its dimension.
There is an integration for each $g_{e^\ast}$. In $\Delta^\ast_4$, each $e^\ast \in \Delta^\ast_4$ bounds precisely four different faces; therefore, the $g_{e^\ast}$ in (\ref{BFpartitionPart}) appears in four different traces.We can then introduce the intertwinners as in (\ref{proj}) using 
$$
\int dg \prod_{e=1}^N D^{j_e}_{m_e n_e}= \sum_i  i_{m_1\cdots m_N} i^\ast_{n_1 \cdots n_N},
$$  
to obtain that the partition function is equal to
\be\label{BF-4d}
Z^{\textrm{BF}}_{\Delta_4}= \sum_{\{\cI_{f^\ast}\}, \{i_{e^\ast}\}} \prod_{f^\ast} d_{\cI_{f^\ast}} \prod_{v^\ast} \begin{minipage}[h]{0.1\linewidth}\vspace{0.1cm}
\centering\includegraphics[scale=0.25]{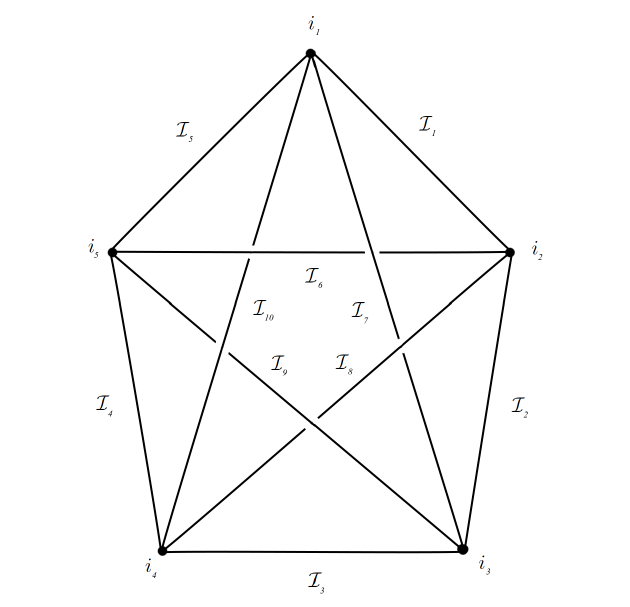} 
\vspace{0.01cm}
\end{minipage}
\ee
The 4-simplex in the formula is dual to the 4-simplex in the original triangulation that contains $v^\ast \subset \Delta_4^\ast$.  Its edges are labelled by the representations labeling the ten dual faces incidents to $v^\ast$ and its vertices are labelled by the intertwiners labeling the dual edges incidents to $v^\ast$. We can rewrite this formula for the partition function by splitting each 4-valent vertex into two trivalent vertices using the relation described in Fig. \ref{decomposition}. The resulting equation involes a trivalent spin network with 15 edges: $(\cI_{f^\ast}, \cI_{{e^\ast}})$ where $\cI_{{e^\ast}}$ is the spin coming from the splitting of the 4-valent intertwiner $i_{e^\ast}$. This trivalent spin network is called a $\{15\}$\textit{-symbol} since it depends on 15 spins. 

Let us notice that $\Spin(4) \sim \SU_L(2) \times \SU_R(2)$. An irreducible representation $\cI_{f^\ast}$ of $\Spin(4)$ is thus labeled by a couple $(j^L_{f^\ast}, j^R_{f^\ast})$. Consequently,  an intertwiner $i_{e^\ast}= i_{j^L_{e^\ast}}\otimes   i_{j^R_{e^\ast}}$ where $(j^L_{e^\ast}, j^R_{e^\ast})\equiv \cI_{{e^\ast}}$ and  the different weights 
\be \label{factorize}
 A_v^{\Spin(4)}(\mathcal{I}_{f^\ast}, \mathcal{I}_{e^\ast})\equiv\begin{minipage}[h]{0.09\linewidth}\vspace{0.1cm}
\centering\includegraphics[scale=0.15]{4simplexBis} 
\vspace{0.01cm}
\end{minipage}= \{15j\}_{SU_L(2)} \{15j\}_{SU_R(2)} = A_v^{\SU(2)}(j^L_{f^\ast}, j^L_{e^\ast})A_v^{\SU(2)}(j^R_{f^\ast}, j^R_{e^\ast}).
\ee
As in the 3-dimensional case,  due to  the absence of local degrees of freedom in the topological theory, the partition function is  invariant under changes of triangulation keeping the topology fixed (\ie invariant under the Pachner moves).

\paragraph{The Lorentzian case.} 
The gauge group is now $\SL(2, \C)$.  The resulting partition function constructed in this case  is formally the same as in (\ref{BF-4d}).  The principal series of unitary, irreducible representations of $\SL(2, \C)$ are labelled by two parameters $(n, \rho)$, with $n$ a half-integer and $\rho$ a real number. To describe explicitly the partition function in this Lorentzian case,  we consider the partition function obtained in   (\ref{BF-4d}). We associate the couple $2j^{L}+1= n+ i \rho$, $2j^R+1=n-i \rho$ to each unitary principal representation $(n, \rho)$ and use the factorized formula from the Euclidean case (\ref{factorize}).



\subsection{Some ambiguities in the 4d gravity spin foam quantization procedure} \label{ambiguitiesSF}

The idea is now to provide a definition of the path integral of gravity formally written as
\be
 \int  \mathcal{D}B \, \mathcal{D}\omega \; \delta[B \rightarrow \star(e\wedge e)]\; \exp \left[ i \int \tr [ B \wedge F] \right]
\ee
where the measure $ \mathcal{D}B \, \mathcal{D}\omega \, \delta[B \rightarrow \star(e\wedge e)]$ restricts the sum of the BF theory path integral  
$$Z^{\textrm{BF}}= \int  \mathcal{D}B \, \mathcal{D}\omega \, \exp \left[ i \int \tr [ B \wedge F] \right]$$ 
to those configurations of the topological theory satisfying the constraints $B= \star (e \wedge e)$ for some tetrad $e$. The constraints $B= \star (e \wedge e)$ are going to be directly implemented on the spin foam configurations of $Z^{\textrm{BF}}_{\Delta_4}$ by the appropriate restriction on the allowed spin labels and intertwiners.

In the spinfoam quantization process a number of ambiguities appear. The most severe ones are related to the discretization process, while another appears due to Immirzi parameter dependence in the action.

The discretization ambiguities arise in the different steps of the spinfoam quantization. The first important one, and most obvious, is the choice of cellular decomposition of the manifold. There is a very large number of decompositions available -- a triangulation, a square lattice or an arbitrary irregular cellular decomposition -- and we have to pick up one.
One proposal to solve this issue is the use of GFT which provides automatically the sum over the relevant complexes \cite{laurent-gft, daniele-gft}. The most common approach to this issue, which we will adopt here,  consists in simply postponing this issue and  working either with a triangulation to allow to write explicitly the spin foam amplitudes or when possible with an arbitrary cellular decomposition in order to remain as general as possible.

Once such a manifold discretization as been chosen, another key ambiguity appears since there is no unique prescription to discretize the curvature $F$. We will not discuss this point here. We will mostly focus on the simplicity constraints which only concern the discretized $B$ field, $B_\Delta$, and which are not affected by the definition of the discretized curvature.  We will only specify for each approach how the Lorentz gauge connection is discretized and if some constraints are consequently added.

The third discretization ambiguity appears when discretizing the dynamical content of the theory, \ie the constraints. We will address this issue at length in Part \ref{part2}.

Finally, we recall that the  classical Palatini-Holst action contains some ambiguity  in the Immirzi parameter choice. This parameter is important to have the spin connection as a well-defined configuration variable \cite{book-ashtekar}. To relate the spinfoam formalism with the loop quantum gravity approach, it is therefore important to take this parameter into account in the spinfoam quantization. However, just as in loop quantum gravity, it then generates some ambiguities in the formalism: we have a free parameter dependent theory.
We consider the action  (\ref{BFgeneral}) as the starting point of the procedure.  We have seen that the continuum constraints on the $B$-fields are then modified compared to the Plebanski constraints (\ref{constraintB}). For clarity we denote   {$B^{IJ}$ the bivector  $\star (e^I \wedge e^J)$ and $\Sigma^{IJ}$ the  bivector $\ast (e^I \wedge e^J) - \f{1}{\gamma} e^I \wedge e^J=B^{IJ}- \f{1}{\gamma} \star B^{IJ}$, which  depends on the Immirzi parameter $\gamma$,  and is solution to the constraints (\ref{ImmiBconstr1}) and (\ref{ImmiBconstr2}). }


Two different procedures of quantization are usually followed. 
\begin{itemize}
\item The first approach consists in taking  directly into account the Immirzi parameter at the discretized level. We focus on the variables $\Sigma^{IJ}$ constrained by (\ref{ImmiBconstr1}) and (\ref{ImmiBconstr2}). Then, noticing that the constraints (\ref{ImmiBconstr1}) and (\ref{ImmiBconstr2}) (with $a=4\gamma$, $b=-(\epsilon +\gamma^2)$) can be recast in an equivalent form
\be \label{Immisimpli}
\left( \epsilon_{IJKL} +\f{4 \gamma}{\epsilon + \gamma^2} \eta_{[I|K|} \eta_{J]L}\right) \Sigma^{KL}_{cd} \Sigma^{IJ}_{ab}= e \epsilon_{abcd} \left(1-\epsilon \f{4 \gamma^2}{(\epsilon+ \gamma^2)^2} \right),
\ee 
with $e= \f{1}{4!}\epsilon_{IJKL} B^{IJ} \wedge \Sigma^{KL}$, we consider the discretized version of this new constraint. The quantization step is simply to replace $\Sigma^{IJ}_\Delta$  with the canonical generators $J^{IJ}_\Delta$ of $\mathfrak{spin}(4)$. Finally, writing the simplicity constraints for the generators $J^{IJ}_\Delta$ translates into a condition on the Casimirs of $\mathfrak{spin}(4)$ or $\mathfrak{sl}(2,\C)$. 
\item The second approach consists in  considering as discretized variable the $B$-field constrained to be a simple bivector. Contrary to the previous case,  the Immirzi parameter will be taken into account only at the quantized level. This is the  procedure that we are going to detail in the following.
\end{itemize}
Let us first review the discretized setting.  The simplicity constraints of the 2-form, $\epsilon_{IJKL}B^{IJ}_{\mu \nu} B^{KL}_{\rho \delta}= e \epsilon_{\mu \nu \rho \delta}$ ensure that $B$ comes from a tetrad field $e$. We want to translate this constraint to the discrete setting. We recall that the $B$-field is a 2-form and is naturally discretized on the faces $\Delta$ of the cellular decomposition
\be
B^{IJ}_\Delta=\int_\Delta B_{\mu \nu}^{IJ}dx^\mu \wedge dx^\nu.
\ee
 Let us  look in details at the simplicity constraints within a 4-cell. For any two faces $\Delta, \, \tilde{\Delta}$, we have:
\be
\epsilon_{IJKL} \,B^{IJ}_\Delta \, B^{KL}_{\tilde{\Delta} } =\int_{\Delta, \tilde{\Delta}} e\;  d^2\sigma  \wedge  d^2\sigma^\prime = V(\Delta,\tilde{\Delta}),
\ee 
where $V(\Delta,\tilde{ \Delta})$ is the 4-volume spanned by the two faces. Like for the continuum constraints appearing in the Plebanski action (\ref{simplicityB}), we can again distinguish three different cases.
\begin{enumerate}
\item The faces are the same $\Delta=\tilde{\Delta}$, then $\epsilon_{IJKL}B_\Delta^{IJ} B_\Delta^{KL}=0, \; \forall \, \Delta$, \ie  the associated bivector $B_\Delta$ must be simple. In the case of a triangulation $\Delta_4$, within a 4-simplex, this constraint imposes to each bivector associated to a triangle to be simple.
 $$\includegraphics[scale=0.3]{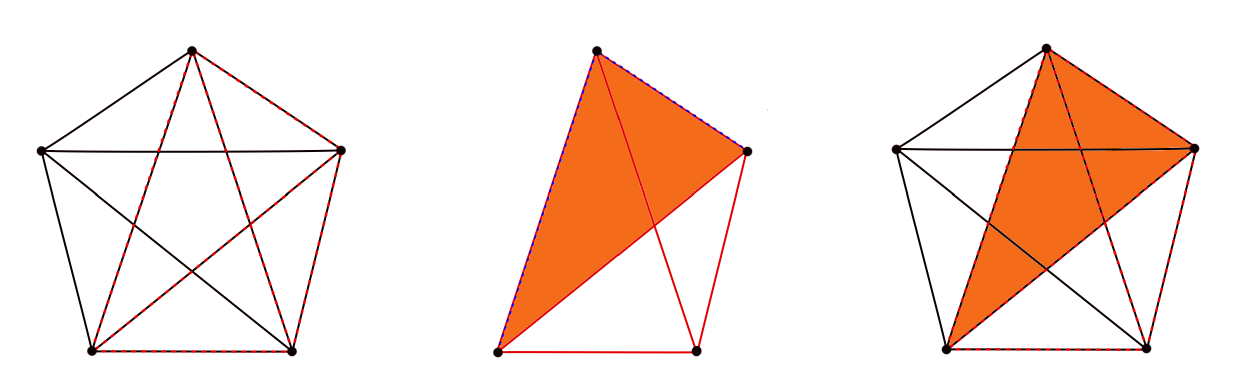}$$
\item  $\Delta$ and $\tilde{\Delta}$ belong to the same 3-cell,  then $\epsilon_{IJKL}B_\Delta^{IJ} B_{\tilde{\Delta}}^{KL}=0 ,\; \forall \, \Delta \neq \tilde{\Delta}$, therefore the sum  $B_\Delta + B_{\tilde{\Delta}}$ must be simple. In the case of a triangulation $\Delta_4$, this constraint acts on two bivectors associated to two different triangles belonging to the same tetrahedron, \ie sharing a common edge and imposes to the sum   to be simple.
 $$\includegraphics[scale=0.3]{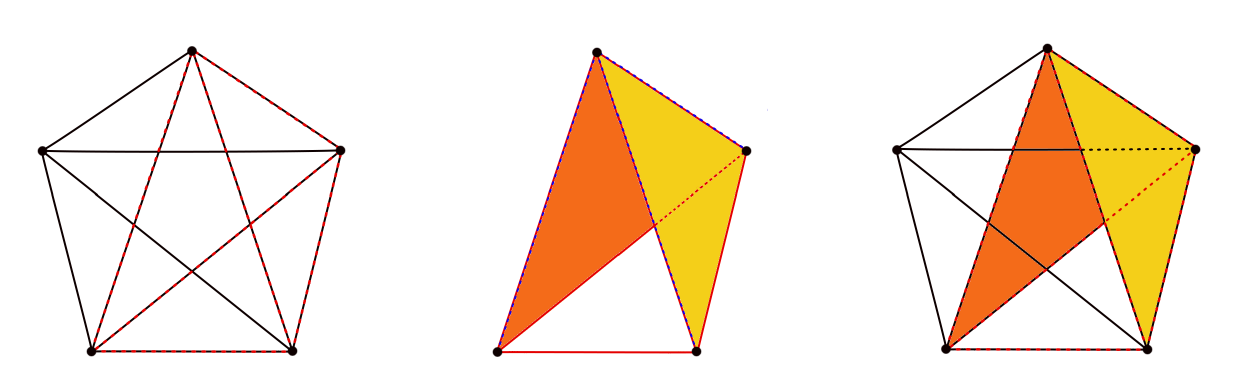}$$
\item  $\Delta$ and $\tilde{\Delta}$ do not belong to the same 3-cell, then  $\epsilon_{IJKL}B_\Delta^{IJ} B_{\tilde{\Delta}}^{KL}= V(\Delta,\tilde{\Delta})$. In the case of a triangulation $\Delta_4$, this constraint acts on two bivectors associated to two different triangles which only share a common vertex and consequently do not belong to the same tetrahedron. It imposes that the quantity $\epsilon_{IJKL}B_\Delta^{IJ} B_{\tilde{\Delta}}^{KL}$ is equal to the 4-simplex volume (up to a factor).
 $$\includegraphics[scale=0.3]{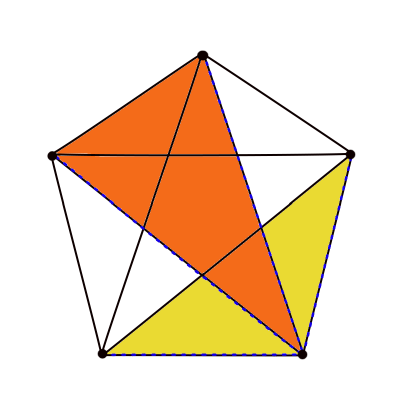}$$
\end{enumerate}
Moreover, the discrete $B$-fields are constraints to satisfy a closure constraint for each 3-cell
\be\label{closureConstraint}
\sum_{\Delta\in 3-\textrm{cell}}B^{IJ}_\Delta =0.
\ee
This is the discrete equivalent of the Gauss law ensuring the $G=\Spin(4)$ or $\SL(2, \C)$ gauge invariance. \\

Let us now discuss the simplicity constraints. We are going to see that the constraints 1. are straightforward to deal with while discussing the other two.   As we pointed out  in the previous section, the simplicity constraints 2. $(\star B_\Delta) \cdot  B_{\tilde{\Delta}}=0$ correspond to the continuum simplicity constraint (\ref{secondclass}), \ie  $\phi=(\star \mathcal{R}) \cdot \mathcal{R}=0$ of the canonical analysis. Hence they are   primary constraints. 

We  also recalled  that the other second class constraints (\ref{secondclass}), \ie $\psi= \mathcal{R}\mathcal{R} D_{\mathcal{A}}\mathcal{R}=0$, the secondary constraints are not needed here since the spin foam formalism is fully covariant. 

At this discrete level, this can be justified by noticing that the simplicity constraints 3. $(\star B_\Delta) \cdot  B_{\tilde{\Delta}}=V$ give the secondary constraints. Indeed, assuming that the constraints 2. $(\star B_\Delta) \cdot  B_{\tilde{\Delta}}=0$ hold on the initial hypersuface (e.g. one a 3-cell of the 4-cell) and that the simplicity constraints 3. are satisfied, then using the closure constraints \eqref{closure}, it is possible to show that $(\star B_\Delta) \cdot  B_{\tilde{\Delta}}=0$ is also true on the final hypersurface (e.g. all remaining 3-cells). 
Therefore the simplicity constraints 3. ensure that the spatial constraints 2. are satisfied under time evolution and are the secondary constraints  analogue. Furthermore, still using the closure constraints \eqref{closure}, it is possible to show that when the constraints 2. hold for all 3-cells of the 4-cell, the constraints 3. are true. That is why we only need to solve the constraints 2., the primary constraints, in the spin foam formalism. Moreover, these constraints will usually be solved separately  on each 3-cells of the cellular decomposition. 

Let us therefore focus on one 3-cell, which we  denote  $v$, for the  vertex  the dual element of the 3-cell of interest.  The bivectors $B^{IJ}_\Delta$ associated to each face $\Delta$ on the (3d) polyhedron boundary define the geometry of the polyhedron. We distinguish two sectors: the normal bivector to the face (the normal to a plane embedded in a 4d manifold is indeed a bivector) is either $B^{IJ}_\Delta$, which we call the standard sector $(s)$, or it is given by its Hodge dual $(\star B)^{IJ}_\Delta\,\equiv\f12 \epsilon^{IJ}{}_{KL}B^{KL}_\Delta$, which we call the dual sector $(\star)$. In both cases, the norm of the bivector $|B^{IJ}_\Delta|$ gives the area of the face $\Delta$ and the bivectors satisfy the closure constraint (\ref{closureConstraint}).
%
These two sectors correspond, at the discrete level, to the two sectors of solutions (\ref{Bconstr1}) and (\ref{Bconstr2}) of the simplicity constraints of the Plebanski action. A discussion on how these two sectors correspond to gravity or not has been given  in the previous section.
We have seen that the usual discrete {\it quadratic simplicity constraints} (the discretized version of the Plebanski constraints (\ref{simplicityB}))
\be  \label{quadraSimplicity}
\forall \Delta,\tDelta\in\pp v,\quad
\epsilon_{IJKL}\,B^{IJ}_\Delta B^{KL}_\tDelta=0
\ee
do not distinguish the two sectors $(s)$ and $(\star)$ since they are invariant under taking the Hodge dual $B\,\arr\, \star B$. This means that we have to recognize the two sectors by hand when solving these simplicity constraints. 

However, it is possible to rewrite the quadratic simplicity constraints in such a way that they distinguish the two different sectors  \cite{EPR}.
Indeed, geometrically, the simplicity constraints come from the fact that all the faces of a given 3-cell $v$  lay in the same hypersurface. Let us call $x_I$ the 4-vector normal to that hypersurface. This leads to very simple constraints on the bivectors $B^{IJ}_\Delta$. We will call them the {\it linear simplicity constraints}\footnote{The quadratic constraints (\ref{Immisimpli}) can also be replaced, at the discrete level, by some linear simplicity constraints where in order to take into account the Immirzi parameter, we consider a linear combination of these two sectors, ($s$) and ($\star$), thus leading to a linear simplicity constraint of the type $x_I\cdot (\Sigma^{IJ}_\Delta+\gamma\epsilon^{IJ}{}_{KL}\Sigma^{KL}_\Delta)=0$. We have chosen not to follow this procedure because the geometrical interpretation of the $B_\Delta$ variable is not as obvious as in the presented case.}
\be \label{linConst}
(s)\quad x_I\cdot B^{IJ}_\Delta = 0,\forall \Delta\in\pp v,\qquad\qquad
(\star)\quad \epsilon^{IJ}{}_{KL}x_J\cdot B^{KL}_\Delta = 0,\forall \Delta\in\pp v.
\ee
We will see in the next part that these linear simplicity constraints are at the root of the construction of the  recent EPRL-FK model and other new spinfoam models \cite{EPR,EPRL,FK,LS,sergei}. A clear link between the quadratic and the linear simplicity constraints has been given in \cite{EPR}. 
These linear simplicity constraints involve explicitly the time normal $x_I$, which is an extra field. We have  seen that this field variable appears also in the definition of the projected spin network (see chapter \ref{chapCLQG}). This is therefore the first evidence of the possible relationship between the projected spin networks and the spinfoam models.

\medskip

Let us now go on with the last step of the spin foam quantization procedure which is the quantization itself. The key kinematical ingredient is the boundary of the 4-cell.  More precisely, the boundary of a given 4-cell is made of  3-cells glued together.  We have seen that we can solve the simplicity constraints within a 3-cell. We thus consider the discretized $B$-field, the bivectors $B_\Delta$ associated to each face of a given boundary 3-cell. The bivectors are then translated as elements of $\spin(4)^\ast$ or $\sl(2, C)^\ast$. 
There exists an ambiguity at this level. Indeed, there exists a family of isomorphisms between bivectors and Lie algebra elements.
\be
B \leftrightarrow  \tilde{J}^{IJ}= \alpha J^{IJ} + \beta \star J^{IJ}.
\ee
We will actually discuss this ambiguity in the next part \ref{part2}. The rest of the quantization procedure for the 3-cell is very simple: we associate an irreducible representation of our gauge group $\Spin(4)$ or $\SL(2, \C)$ to each face $\Delta$ and we quantize the bivectors $B^{IJ}_\Delta$ as the Lie algebra generators $J^{IJ}_\Delta$ (or more generally as $\alpha J_\Delta^{IJ} + \beta \star J_\Delta^{IJ}$  acting in that representation. The states of the 3-cell $v$ live then in the tensor product of the representations of all its faces. Moreover, taking into account the closure constraint,
$$
\sum_{\Delta\in\pp v}B^{IJ}_\Delta =0
\quad\longrightarrow\quad
\sum_{\Delta\in\pp v}J^{IJ}_\Delta =0,
$$
we require that the states associated to the 3-cell must be invariant under the global $\Spin(4)$ or $\SL(2, \C)$ action which acts simultaneously on all faces. Thus, the Hilbert space of quantum states of the 3-cell is the space of $\Spin(4)$  or $\SL(2, \C)$ intertwiners between the representations attached to its faces. We still need to implement the simplicity constraints. It is one of the key issue of the spin foam program to implement them at the quantum level in the regularized path integral. For example when using   the quadratic simplicity constraints, we have to implement the constraints
\be
\forall \Delta,\tDelta\in\pp v,\quad
\epsilon_{IJKL}\,\tilde{J}^{IJ}_\Delta \tilde{J}^{KL}_\tDelta
\,=\,0,
\ee
which  translate into  conditions on the Casimir operators on the intertwiners states. The next chapter is devoted to this issue of imposing consistently the simplicity constraints at this quantum level.

\part{Spin foam models for 4d gravity} \label{part2}

In the previous Part, we ended the presentation of the spin foam quantization procedure just before the last step which is the implementation of the simplicity constraints at the quantized level. At this penultimate step of the procedure,  the Hilbert space of quantum states of a 3-cell is the space of $\Spin(4)$ or $\SL(2,\C)$ intertwiners between the representations attached to its faces. The last step consists in imposing the simplicity constraints on these intertwiners. This last step is still a topic of discussion, no final consensus has been reached on the "correct" way to proceed.
   \\
The two projects presented in this part (Chapter \ref{ProjectChap} and Chapter \ref{UNChap}) aimed at understanding better these simplicity constraints and in particular focused on the two specific questions:
\begin{itemize}
\item What is a consistent implementation of the simplicity constraints at the quantum level? 
\item What are the  consequences  of their implementation on the boundary Hilbert space of the 3-cell (or more generally of the whole boundary space)?
\end{itemize}

The main difficulty comes from the fact that the algebra of these simplicity constraints -- which translate into conditions on the Casimir operators -- does not close  at the quantized level. This reflects  that they correspond to second class constraints.  

We will distinguish the diagonal simplicity constraints that we will denote $\hat{\cC}_\Delta$ or $\cC_i$ where $i$ refers to the intertwiner's leg dual to the face $\Delta$ and the cross simplicity constraints that we will denote $\hat{\cC}_{\Delta \tilde{\Delta}}$ or $\cC_{ij}$ with $\Delta \neq \tilde{\Delta}$ (or $i\neq j$). We recall that there is a one-to-one correspondence between a 3-cell and its dual vertex $v$ and between a face $\Delta$ of the 3-cell and its dual link, the leg $i$ of the vertex, which  explains that we can equivalently use the face subindex $\Delta, \tilde{\Delta}, \cdots$ or the leg subindex $i,j, \cdots$. 

In the Euclidean case, the quantum version of the quadratic simplicity constraints (\ref{quadraSimplicity}) is simply given by
\be \label{quadraSimpliIII}
\cC_{\Delta \tilde{\Delta}}\equiv \vJ^L_{\Delta}\cdot\vJ^L_{\tilde{\Delta}} \, 
-
\,\vJ^R_{\Delta}\cdot\vJ^R_{\tilde{\Delta}}=0, 
\ee
where $\vJ^L$, $\vJ^R$ are the $\su(2)$ generators of the two commuting $\su(2)$ algebra of  the $\spin(4)$ algebra decomposition $\spin(4)= \su_L(2) \oplus \su_R(2)$.   
 Looking at the cross simplicity constraints, it is fairly easy to check that the Lie algebra of the quadratic simplicity constraints does not close
\be \label{commutatorSimplicity}
\left[
\,\vJ^L_{\Delta_1}\cdot\vJ^L_{\Delta_2} \,
-
\,\vJ^R_{\Delta_1}\cdot\vJ^R_{\Delta_2} \,,
\,\vJ^L_{\Delta_1}\cdot\vJ^L_{\Delta_3} \,
-
\,\vJ^R_{\Delta_1}\cdot\vJ^R_{\Delta_3} \,
\right]
\,=\,
\vJ^L_{\Delta_1}\cdot\left(\vJ^L_{\Delta_2}\w \vJ^L_{\Delta_3}\right)
+
\vJ^R_{\Delta_1}\cdot\left(\vJ^R_{\Delta_2}\w \vJ^R_{\Delta_3}\right).
\ee
In this way, we generate higher and higher order constraints by computing further commutators. 


Naturally, the first attempt was to impose these constraints strongly to obtain intertwiner states $|\psi\ra$ vanishing under the simplicity constraints
\be
\cC_{\Delta \tilde{\Delta}}\,|\psi\ra=0 \quad \forall \Delta, \, \tilde{\Delta}.
\ee
The resolution to this system of equations led to the first spin foam model for 4 dimensional gravity, the Barrett-Crane spin foam model, which was actually initially built as a geometrical quantization of individual 4-simplices \cite{BC,BC2}. Both Euclidean and Lorentzian versions of this model -- which remained the leading proposal during 10 years -- will be detailed in the next Chapter. It was shown (at least in the 4-valent case corresponding to a tetrahedron) in \cite{unique} that these equations have a unique solution once the representations associated to each  $\Delta$ are specified. 
This unique or "frozen" intertwiner property leads to several problems in the interpretation of the Barrett-Crane model, especially when looking at its relation with the canonical loop quantum gravity framework and when studying the graviton propagator derived in the asymptotical semi-classical regime of the model. The uniqueness of the intertwiner state can be traced back to the fact that the Lie algebra of the quadratic simplicity constraints does not close. Indeed, since the commutators between two different cross simplicity constraints (\ref{commutatorSimplicity}) generate  higher order constraints, when we impose strongly the quadratic simplicity constraints on the intertwiner state $|\psi\ra$, we are actually also imposing all these higher order constraints. Then it is not surprising to find a unique solution, although it could actually be considered surprising  to find at least one solution. 

To remedy this issue, it was proposed to solve the crossed simplicity constraints weakly, either by some coherent state techniques \cite{coh1} or by some Gupta-Bleuler-like method using the linear simplicity constraints \cite{EPR0,EPR,EPRL}. 

In the coherent state approach, one seeks semi-classical states such that the simplicity constraint is solved in average \cite{coh1,FK},
\be
\la\psi|\cC_{\Delta \tilde{\Delta}}|\psi\ra
\,=\,
0,
\ee
while minimizing the uncertainty of these operators. 

In the Gupta-Bleuler-like approach\footnote{We will review in the section \ref{EPRGupta} that in fact in this procedure (followed in \cite{EPR0, EPR, EPRL}) the simplicity constraints are imposed prior to the imposition of the closure constraint (\ref{closureConstraint}), and that consequently the Hilbert space of the 3-cell on which the simplicity constraints will be imposed is not an intertwiner space yet.}, one looks for a Hilbert space $\cH_s$ such that the matrix elements of the simplicity constraints on this  Hilbert space all vanish \cite{EPR0,LS}
\be \label{guptalike}
\forall \phi,\psi\in \cH_s,\quad\,
\la\phi|\cC_{\Delta \tilde{\Delta}}|\psi\ra
\,=\,
0.
\ee
These two approaches 
were shown to lead to the same spinfoam amplitudes \cite{FK,LS} apart from some subtle cases \cite{engle}. \\
Before detailing these different ways of  implementing the simplicity constraints, let us emphasize that a true Gupta-Bleuler method would involve decomposing the constraints $\cC_{\Delta \tilde{\Delta}}$ under a sum of holomorphic and anti-holomorphic factors (or equivalently of creation and annihilation operators), \ie $\cC_{\Delta \tilde{\Delta}}=\sum_\alpha (\cF^\alpha_{\Delta \tilde{\Delta}})^\dagger \cF^\alpha_{\Delta \tilde{\Delta}}$. Then, from this decomposition, the next step would be to extract algebraic constraints $\cF^\alpha$ which are  strongly solved in the sense: $\cF^\alpha_{\Delta \tilde{\Delta}} |\psi\ra=0$. And finally, (\ref{guptalike}) would appear as the consequence of the resolution of the algebraic constraints $ \cF^\alpha_{\Delta \tilde{\Delta}}$. Indeed, $\forall \, |\psi\ra$, $|\phi \ra \in \cH_s$ such that $ \cF^\alpha_{\Delta \tilde{\Delta}}|\psi \ra=0$ and $ \cF^\alpha_{\Delta \tilde{\Delta}}|\phi \ra=0$, then $\la\phi|(\cF^\alpha_{\Delta \tilde{\Delta}})^\dagger \cF^\alpha_{\Delta \tilde{\Delta}}|\psi\ra=0$ and $\la\phi|\cC_{\Delta \tilde{\Delta}}|\psi\ra=0$. \\
In the next two Chapters of this part, the different approaches, \cite{BC,BC2}, \cite{coh1,FK}, \cite{EPR0,LS}  are presented. In Chapter \ref{UNChap}, we will present the right Gupta-Bleuler method to implement the second class simplicity constraints at the quantum level \cite{article3}.


\chapter{Simplicity constraints and the Barrett-Crane model} \label{BCmodel}
\section{The Euclidean Barrett-Crane model}
We are now going to review the construction of the Barrett-Crane model of 4d Euclidean quantum gravity without Immirzi parameter \cite{BC}. The manifold $\cM$ is replaced by a cellular decomposition.  Explicit results will be mostly given in the simplest case of a triangulation $\Delta_4$ (see Table \ref{triang4d}) since it is  the framework for the results  exposed in the next part, Chapter \ref{4dphysical}, regarding the physical boundary state for the quantum 4-simplex. However, as often as possible the definitions will be given for the general case of a cellular decomposition.

Going from the continuum theory to the Barrett-Crane model is achieved in three main steps. We recall that the first step is the discretization of the two-form into bivectors associated to each face to the cellular decomposition. Bivectors are then translated as elements of $\mathfrak{so}(4)^\star$. 
The last step is the quantization itself, using techniques from geometric quantization. It  gives representation labels to the faces of the cellular decomposition and gives the Barrett-Crane model (to some normalisation factors). It is this last step that we are  going to detail now.

Since $\Spin(4)\sim\SU_L(2)\times\SU_R(2)$ ,
the irreducible representations of $\Spin(4)$  are labeled by a couple of half-integers $(j^L,j^R)$. Hence, we attach a pair of spin to every face $\Delta$ and the intertwiner space for the 3-cell is the tensor product of the space of $\SU_L(2)$ intertwiners between the spins $j^L_\Delta$ and the space of $\SU_R(2)$ intertwiners between the spins $j^R_\Delta$
\be
H_{j^L_\Delta,j^R_\Delta}\,\equiv\,
\textrm{Inv}\left[\bigotimes_\Delta V^{j^L_\Delta}\right]\,\otimes\,
\textrm{Inv}\left[\bigotimes_\Delta V^{j^R_\Delta}\right].
\ee

Finally, we now come to the crucial step of implementing the simplicity constraints (\ref{quadraSimplicity}) at the level of the spin foam model of BF theory. In the Barrett-Crane model, the quantum version of these constraints is given by
\bes
\widehat{\cC}_\Delta=  \epsilon_{IJKL}J^{IJ}_\Delta J^{KL}_\Delta=0 & \quad \forall \Delta &\qquad \textrm{ diagonal simplicity constraints,} \nn \\
\widehat{\cC}_{\Delta\tilde{\Delta}}=  \epsilon_{IJKL}J^{IJ}_\Delta J^{KL}_{\tilde{\Delta}}=0 &\quad \forall \Delta \neq \tilde{\Delta} \in \partial v & \qquad \textrm{ cross simplicity constraints.}
\ees
That is the bivectors $B_\Delta^{IJ}$ of the discretized constraints (\ref{quadraSimplicity}) have been replaced by generators $J^{IJ}$ of the $\spin(4)$ Lie algebra in the representation $\mathcal{I}_\Delta=(j_\Delta^L, j^R_\Delta)$.  In the expression of the diagonal simplicity constraints, $\f14 \star J \cdot J$ is the pseudo-scalar quadratic Casimirs of $\spin(4)$ and it is equal to  the difference of the quadratic Casimir  of $\su_L(2)$  and $\su_R(2)$. This gives a restriction on the representations $\cI_\Delta=(j^L_\Delta, j^R_\Delta)$  of $\Spin(4)$
\be
2j^L_\Delta (j^L_\Delta+1)-2j^R_\Delta (j^R_\Delta+1)=0 \quad \forall \Delta,
\ee
leading to $j^L=j^R$. The irreducible representations satisfying this condition are called \textit{simple representations}. The strong imposition of the diagonal simplicity constraints implies that only simple representations are associated to faces $\Delta$ of the 3-cell \cite{baez-spinfoam, quantumtetra}. \\
The cross simplicity constraint is already imposed  at the level of a 3-cell and induces a restriction on possible intertwiners. In the simple case where the cellular decomposition is a triangulation $\Delta_4$, a 3-cell is  a tetrahedron and taking into account that the intertwiners couple simple representations, it is easy to conclude that the restriction means that, given a decomposition of any two $\cI_\Delta$'s, the intertwiner has support only on simple intermediate representations. Indeed, we have four simple representations $(j_\alpha, j_\alpha)$ for the four triangles $\Delta_{\alpha=1 \cdots 4}$ on the tetrahedron boundary.
\begin{figure}[h]
\begin{center}
\includegraphics[height=20mm]{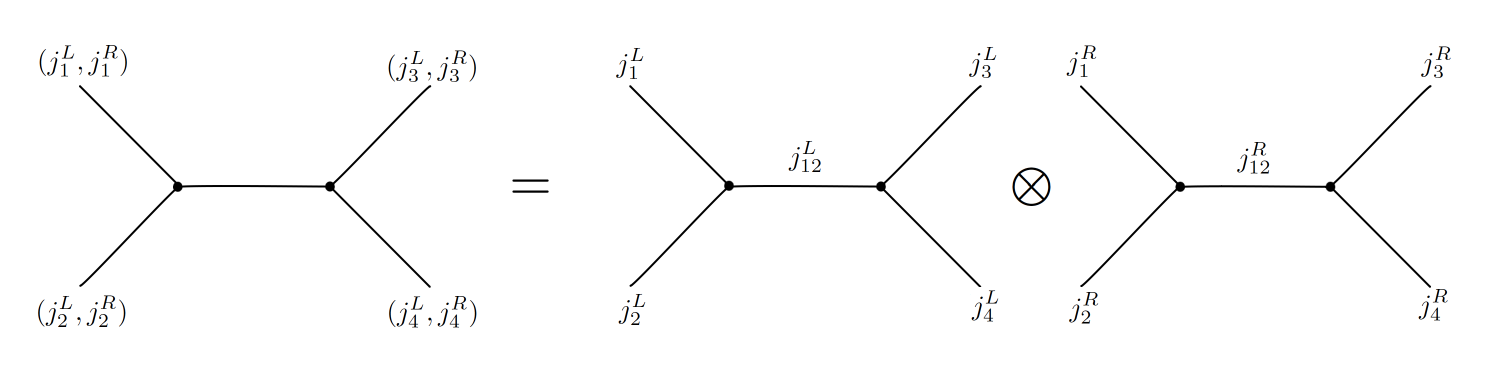}
\end{center}
\caption{The label $(j_{12}^L, j_{12}^R)$ stands  for the representation $(J_1+ J_2)$. } \label{recoupling}
\end{figure} 
 In $H_{j_\alpha}\,\equiv\,
\textrm{Inv}\left[\bigotimes^4_{\alpha=1} V^{j_\alpha}\right]\,\otimes\,
\textrm{Inv}\left[\bigotimes_{\alpha=1}^4 V^{j_\alpha}\right]$ with $j_\alpha\equiv j_\alpha^L= j_\alpha^R$, we have to impose the three independent  cross simplicity constraints $\cC_{\alpha\beta}= \epsilon J_\alpha J_\beta=0$ for all couples of triangles $(\Delta_\alpha, \Delta_\beta)$. These constraints mean that  the sum $(J_\alpha+J_\beta)$ is required to remain simple. Using the standard recoupling basis of intertwiner as in Fig. \ref{recoupling}, we denote the representation of $J_\alpha + J_\beta$, $(j_{\alpha \beta}^+, j_{\alpha, \beta}^-)$. Strongly imposing the simplicity conditions $\cC_{1,2}$ forces the recoupled representation to be simple, $j_{12}^L=j_{12}^R$. Further imposing $\cC_{1,3}$ and $\cC_{1,4}$ then leads to a single intertwiner \cite{unique}: the Barrett-Crane intertwiner, $i_{BC}$. 
One can now implement these restrictions at the level of the partition function. For this it is sufficient to take the spin foam representation of the BF partition function (\ref{BF-4d}) and  restrict the sum over $j_\Delta$ to only simple representations and to remove the sum over intertwiners substituting for them $i_{BC}$. We recall that the volume simplicity constraint (quantized version of the third equation in \ref{simplicityB}) can be ignored because it is a consequence of the other simplicity constraints (\ref{quadraSimplicity}) supplemented by the closure (\ref{closureConstraint}) and that the latter is already satisfied since we have started from the state space of BF theory.

Once we have determined the state space, the spin foam amplitude for the Euclidean Barrett-Crane model (valid also for an arbitrary cellular decomposition) is explicitely given by the evaluation of a spin network $\psi$ with group $\Spin(4)$ 
such that each node $n$ of the graph $\gamma$ of $\psi$ is associated to a $\Spin(4)$ group element $G_n$ and each link $l$ is labelled by a simple irreps of $\Spin(4)$ $\cI_l=(j_l,j_l)$. Using the homomorphism $\Spin(4)=\SU(2)_L\times\SU(2)_R$, $\Spin(4)$ group elements decompose as the product of two left and right rotations $G=g_Lg_R$. The evaluation then reads \cite{Barrett-spinnet}
\be
\mathcal{A}[\psi] \equiv \int_{\Spin(4)} \prod_n [dG_n] \prod_i
\mathcal{K}_{\mathcal{I}_i} (G^{-1}_{s(i)}G_{t(i)}),
\ee
where we use the Haar measure $dG=dg_L dg_R$.
%
%
The kernel $\mathcal{K}_{\mathcal{I}}(G)$ is the matrix element of $G$ on the $\SU(2)$-invariant vector $|\cI, 0\ra$ in the $\cI$ representation with $\cI=(j, j)$ a simple representation of $\Spin(4)$. Here $\SU(2)$ is the diagonal rotation group, corresponding to the subgroup of 3d rotations.
Expressing the invariant vector in term of left/right components
\be
|\cI, 0\ra = \f{1}{\sqrt{d_j}} \sum_m (-1)^{j-m} |j, m\ra_L |j, -m\ra_R, 
\ee
where $d_j=2j+1$ is the dimension of the $\SU(2)$ representation of spin $j$. It is straightforward to show that the kernel $\mathcal{K}_{\mathcal{I}}$ is simply given by  the $\SU(2)$ character $\chi_j(g)$, defined defined as the trace of the group element in the $j$-representation of $\SU(2)$.
Parameterizing the $\SU(2)$ group elements as
\be
g(\theta, \hat{n})= \cos\theta \,\id + i \sin\theta \,\hat{n} \cdot \vec{\sigma}, \quad \theta \in [0, \pi],
\ee
the characters depend entirely on the class angle $\theta$ (half the rotation angle) and are expressed as
\be
\chi_j(g)=\f{\sin d_j \theta}{\sin \theta}.
\ee
Using the properties of invariance of the Haar measure, it can then be proved that the relativistic spin network evaluation is actually a 3d object involving only integrals over $\SU(2)$ \cite{anal1}
\be
\mathcal{A}= \int_{\SU(2)} \prod_n dg_n \prod_i \f{1}{d_{j_i}} \chi_{j_i}(g^{-1}_{s(i)} g_{t(i)}).
\label{evaluation}
\ee

The Barrett-Crane model was originally derived for  specific two-complexes $C$, which are the dual 2-skeleton of 4-dimensional triangulations $\Delta_4$ (see Table \ref{triang4d}).
%
%
Starting with the triangulation  $\Delta_4$, the dual two-skeleton $C$ is constructed  by associating to each simplex a point in its interior. An edge of the skeleton connects two points, which corresponds to two 4-simplices having a common tetrahedron. Then, a triangle of the triangulation $\Delta_4$ correspond to a face of the two-complex $C$ formed by the edges of the dual tetrahedra having this triangle in common. This duality allows to derive the spin foam amplitudes from a quantization of the geometry of a 4-simplex.
%
%
Indeed, the amplitude for a single 4-simplex defines the vertex amplitude of the spinfoam model. Arbitrary 2-complexes are constructed by gluing 4-simplices together and the corrresponding spinfoam amplitudes are given by the product of these vertex amplitudes corresponding to each 4-simplex.

\begin{figure}[ht]
\begin{center}
\includegraphics[width=5cm]{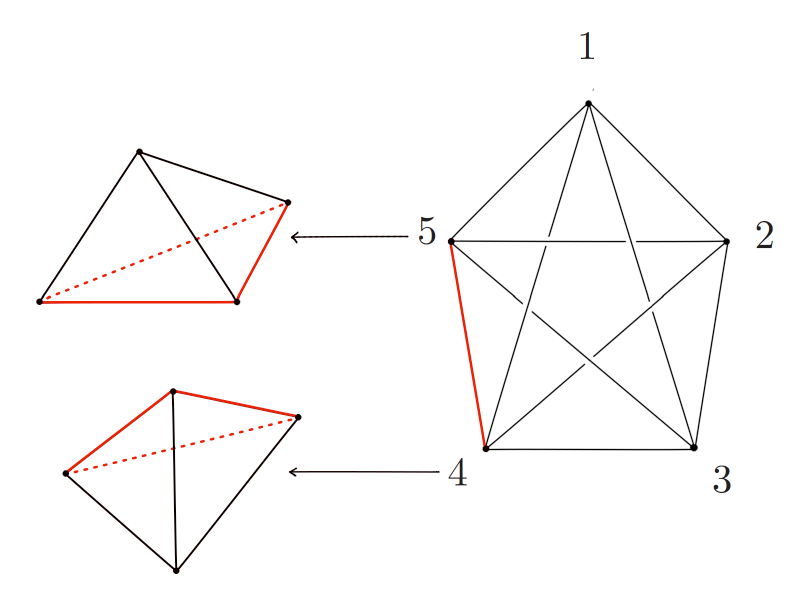}
\end{center}
\caption{The 4-simplex (or pentahedral) boundary spin network. We label the nodes $a= 1, . . . 5$. In the dual picture, they are in correspondence with tetrahedra of the boundary triangulation. Two of them are represented. The ten links $ab$, on the other hand, are dual to triangles. Consider for instance the link 45: this is dual to the triangle shared by the tetrahedra 4 and 5. The dihedral angle $\theta_{45}$ between the tetrahedra 4 and 5 is associated with the link 45.} \label{4simplexPlusTetra}
\end{figure}

Considering a single 4-simplex, the vertex amplitude is constructed as the evaluation of its boundary spin network. The boundary graph is constructed as the dual of the boundary of the 4-simplex: each tetrahedron is mapped on a node of the graph and each triangle shared by two tetrahedra is mapped to a link between these nodes. This gives a graph $s$ with 5 nodes $n \in \{1, \cdots, 5\}$ and 10 links between them, as represented in Fig. \ref{4simplexPlusTetra}.

%
%
The vertex amplitude $\mathcal{A}[s]$ is the given -- up to normalization factors --  by the evaluation of the boundary spin network $s$ and defines the Barrett-Crane $\{10j\}$-symbol,  following (\ref{evaluation}).
\be \label{10j}
\{10j\}\equiv  \int_{\SU(2)} \prod_{n=1}^{5} dg_n \prod_{i=1}^{10} \chi_{j_i}(g^{-1}_{s(i)} g_{t(i)}),
\ee
The normalization ambiguity leads to an ambiguity in the definition of the Barrett-Crane model. It corresponds to arbitrary gluing factors between 4-simplices, or equivalently to ambiguity in defining the spinfoam edge amplitude ${\cal A}_e$. We will come back to this issue in the last part, Chapter \ref{4dphysical}.  \\
Moreover, the $\{10j\}$ symbol will be  the central object of the discussion about the physical boundary state for the quantum 4-simplex in the context of the Barrett-Crane model presented in the Chapter \ref{4dphysical}.

This symbol admits a geometrical interpretation associated to a 4-simplex which we explain now. We introduce  the notation $(a\, b)$ for $i$ linking the nodes $a$ and $b$.  The indices $a, b=1\cdots 5$ label the five nodes of the pentahedral; in the triangulation picture, there are in correspondence with tetrahedra of the boundary  4-simplex and links $i\equiv (a\, b)$ are dual to triangles (see Fig. \ref{4simplexPlusTetra}). Thus, in the triangulation picture,  while the five group elements $g_a \, \in$ $\SU(2)$ are associated to the 5 tetrahedra of the 4-simplex, the ten representations $j_i \equiv j_{ab}$ for $a\ne b$ can be seen as attached to the triangles of the 4-simplex .
The key fact is that it can be written as an integral over ten class angles $\theta_{ab} \in [0, \pi]$ of the group elements $g^{-1}_{a}g_{b}$:  $\theta_{ab}$ is the dihedral angle and we use the convention $\theta_{aa}=0$. In the triangulation picture, it is the angle between the outward normals to the tetrahedron $a$ and the tetrahedron $b$ of the 4-simplex. The $\{10j\}$ symbol then reads
\be
\{10j\}=\int d\mu[\theta_{ab}] \prod_{a<b} \chi_{j_{ab}}(\theta_{ab}),
\ee
where the measure on the ten class angles $d\mu[\theta_{ab}]$ takes into account that the group elements $g^{-1}_{a}g_b$ are obviously not independent. The measure is in fact simply given by a constraint \cite{6jsaddle2}
\be
d\mu[\theta_{ab}] = \prod_{a<b} d\theta_{ab} \sin \theta_{ab} \delta \left(\det G_{ab} \right),
\ee
where $G$ is the Gram matrix, a symmetric $5\times 5$ matrix defined as $G_{ab} = \cos \theta_{ab}$. The constraint $\delta(G)$ contains all the geometric information and allows to relate the $\{10j\}$-symbol to the Regge amplitude (which describes discretized gravity) for a Euclidean 4-simplex in the large $j$ asymptotics \cite{6jsaddle2, anal1}. 

The independent variables are the areas of the triangles of the 4-simplex representing the triangulation or equivalently the dimension of the representation $j_{ab}$, $d_{j_{ab}}=2j_{ab}+1$, since $A_{ab}=l_P^2d_{j_{ab}}$. The geometry of the 4-simplex is fixed by giving the ten areas $A_{ab}=d_{j_{ab}}$ and the dihedral angles can be considered as functions $\theta_{ab}(d_{j_{cd}})$ of the areas. Actually, we scale the area $d_{j_{ab}}$ instead of the spins $j_{ab}$, then the asymptotic behaviour on a single 4-simplex at the leading order\footnote{To scale $d_{j_{ab}}$ instead of $j_{ab}$ has no effect on the leading order. 
} is given by \cite{6jsaddle2}
\be \label{asympt}
\{10j\}\, \sim \, P(d_{j_{ab}}) \cos \left( \f{1}{l_P^2}S_R[d_{j_{ab}}]\right) + D(d_{j_{ab}}),
\ee
where $S_R[d_{j_{ab}}]= \sum_{a<b}d_{j_{ab}}\theta_{ab}$. $S_R$ is the action derived by Regge as a discretization of general relativity \cite{regge}. The function $P(d_{j_{ab}})$ is a slowly varying factor, that grows as $\zeta^{-9/2}$ when scaling all triangle areas $d_{j_{ab}}$ by $\zeta$.
$D(d_{j_{ab}})$ is a contribution coming from degenerate configurations of the 4-simplex. This is a non-oscillating term which has no interpretation in term of geometrical 4-simplices \cite{asymptSpinNet}. It was found to scale like $1/\zeta^2$ and thus dominate the large spin limit of the $\tj$-symbol. Moreover, it was shown in a recent paper \cite{numerics2} that even after having removing the dominating contribution of order $1/\zeta^2$, there are additional non-oscillating contributions still hiding the Regge term. This result indicates that the connection between this Barrett-Crane model and general relativity is not so simple.

\section{The Lorentzian Barrett-Crane model} \label{BCLorentz}
Let us now introduce the Lorentzian Barrett-Crane model  \cite{BC2, bcl_carlo}. 
In the Lorentzian case, we need to start with the $\SL(2,\C)$ BF theory  where $\SL(2, \C)$, regarded as a real Lie group,  is the spin cover  of the Lorentz group. We recall that the principal series of unitary, irreducible representations of $\SL(2, \C)$ are labelled by two parameters $(n, \rho)$, with $n$ a half-integer and $\rho$ a real number. 
As for the Euclidean case, the diagonal simplicity constraints impose a restriction on representations $\cI_\Delta=(n_\Delta, \rho_\Delta)$,
\be
2 n_\Delta \rho_\Delta= 0 \quad \forall \Delta,
\ee 
leading to\footnote{The solutions $\rho_\Delta=0$ have been disregarded in the initial model \cite{BC2}, but incorporated later in \cite{bcl2_carlo}.} $n_\Delta =0 \; \forall \Delta$. The cross simplicity constraints are also imposed strongly at the level of a tetrahedron and induce a restriction on possible intertwiners. The  analysis done in the Euclidean case is valid and the Lorentzian model is based on a vertex amplitude given by a $\{10j \}$ symbol, labelled by ten numbers $\rho_i$, one per edge of the 4-simplex boundary spin network. Let us introduce again the notation $(ab)$ for the edge $i$ linking the nodes $a$ and $b$ ($a, b =1, \cdots, 5$). The vertex amplitude is then explicitly given by
\be
\{10j\}(\rho_{ab})\equiv \int_{\mathcal{H}_+^{\times 4}} \prod_{a=1}^4 dx_a \prod_{ab} K_{\rho_{ab}} (x_a, x_b).
\ee
$\mathcal{H}_+=\{x_\mu\in \R^4, \, x^2=-1, \, x_0>0\}$ is the unit upper hyperboloid in Minkowski space and the propagator $K_{\rho_{ab}}(x_a, x_b)$ is defined by 
\be \label{10jlorentz}
K_\rho(x,y)\equiv \f{\sin(\rho r(x,y))}{\rho \sinh (r(x,y))},
\ee
where $r(x,y)$ is the hyperbolic distance between $x$ and $y$.  To integrate  over $\mathcal{H}_+^{\times 4}$ instead of $\mathcal{H}_+^{\times 5}$ allows to regularize the amplitude which is otherwise divergent \cite{BC2}.

\medskip

Let us now introduce the Immirzi parameter in this model. Indeed we will see in chapter \ref{ProjectChap} that a canonical basis of quantum states of geometry for the Lorentzian Barrett-Crane spinfoam model for quantum gravity can be given by projected spin networks (introduced in Chapter  \ref{chapCLQG}). In  Chapter \ref{ProjectChap}, we will define in particular a map from these projected spin networks to the standard SU(2) spin networks of loop quantum gravity. We will show that this map is not one-to-one and that the corresponding ambiguity is parameterized by the Immirzi parameter \cite{article4}. 
\\
In both the Euclidean and  Lorentzian Barrett-Crane models, the Immirzi parameter $\gamma$ can be taken into account by modifying the correspondence relation between the bivector $B$ and the element of the Lie algebra 
\be \label{correspondence}
B \leftrightarrow  \tilde{J}^{IJ}= \alpha J^{IJ} + \beta \star J^{IJ},
\ee
with $\alpha$ and $\beta$ functions of $\gamma$ \cite{etera-daniele}. It has been shown in \cite{etera-daniele} that the Barrett-Crane spin foam amplitude remains valid for all constructions taking as correspondence relation any relation of the form (\ref{correspondence}). Therefore, the fact to take into account the Immirzi parameter does not modify the final Barrett-Crane spinfoam model defined by (\ref{10j}) in the Euclidean case and by (\ref{10jlorentz}) in the Lorentzian case.
\chapter{Simplicity constraints and the EPRL-FK models} \label{EPRLFK}
It seems that the Barrett-Crane presented above is not able to capture the dynamics of general relativity. In the Euclidean case, the vertex amplitude given by a $\{10j\}$ symbol does not reproduce the semi-classical Regge action. This important property was identified in \cite{6jsaddle3} where it was shown that the Barrett-Crane model is not able to reproduce the structure of the graviton propagator. This result is linked to the fact that each 4-simplex contribution in the partition function is independent because of the uniqueness of the Barrett-Crane intertwiner. Hence, such unique intertwiner seems to be inadequate to carry information from one 4-simplex to another. Considered from the point of view of loop quantum gravity, this intertwiner also has the drawback that it seems to freeze too many degrees of freedom of the 3d space geometry.  Since the uniqueness of the Barrett-Crane intertwiner is a consequence of the imposition of the simplicity constraints, it became clear that one should modify the way these constraints are imposed. It was then proposed to solve the crossed simplicity constraints \textit{weakly}, either by using some coherent state techniques \cite{coh1} or by using a Gupta-Bleuler-like method involving the linear simplicity constraints \cite{EPR0,EPR,EPRL}. The "weak" sense means in this context that one only requires that $\la\phi|\cC_{\Delta \tilde{\Delta}}|\psi\ra
\,=\,
0$ for any allowed boundary spin network states. These two approaches were shown to lead to the same spinfoam amplitudes \cite{FK,LS} apart from some subtle cases \cite{engle}. The weak imposition is justified by noting that after identification of the bivectors $B_\Delta$ with the generators $J$ of the gauge group (or a linear combination of the generators and of its Hodge dual due to the ambiguity in the correspondence step, see Section \ref{ambiguitiesSF}), the simplicity constraints become non-commutative and imposing them strongly leads to inconsistencies. 

The EPRL-FK models rely on a linear reformulation of the simplicity constraints (\ref{linConst}) which distinguishes the two sectors ($\star$ given in  (\ref{Bconstr1})) and ($s$  given in  (\ref{Bconstr2})).
\be \label{linearConstraints}
(s)\quad x_I\cdot B^{IJ}_\Delta = 0,\, \, \forall \Delta\in\pp v,\qquad\qquad
(\star)\quad \epsilon^{IJ}{}_{KL}x_J\cdot B^{KL}_\Delta = 0,\,\,\forall \Delta\in\pp v,
\ee
 which replace the off simplicity constraints, \ie the quadratic simplicity constraints (\ref{quadraSimplicity}) for the case where the two faces $\Delta$, $\Delta^\prime$ are involved. The diagonal simplicity constraints remain under their quadratic form
 \be \label{diagquadra}
\epsilon_{IJKL}B^{IJ}_{\Delta}B^{KL}_{\Delta}=0.
\ee

\section{A quantization \`a la Gupta-Bleuler} \label{EPRGupta}
We will review  the derivation of the EPRL model both in the Euclidean case and the Lorentzian case with a finite Immirzi parameter. In these models, the Immirzi parameter is incorporated in the Plebanski formulation by modifying the BF part of the action. We will see that this will imply a modification of the correspondence rule used in the Barrett-Crane model above. We will not simply have $B \rightarrow J$ with $J$ the generators of $\mathfrak{g}=\spin(4)$ in the Euclidean case and $\mathfrak{g}=\sl(2,\C)$ in the Lorentzian case but a relation of the type (\ref{correspondence}) where the parameters $\alpha$, $\beta$ can be determined studying the symplectic structure of the boundary space \cite{EPR, EPRL}.
\\
We  review here the construction of \cite{EPRL} done in the case of a triangulation $\Delta_4$ and for a group $G=\Spin(4), \, \SL(2,\C)$. We will specialize the analysis in the Euclidian and Lorentzian cases to identify their specific features after introducing the general procedure to  quantize the model \`a la Gupta-Bleuler.  The model has  been then generalized to the case of an arbitrary cellular decomposition in \cite{eprl_jerzy}. Furthermore,  we will present the model defined for a cellular decomposition in the next section \ref{SUcoh} where we will build the model using coherent states.
\\
In order to translate the simplicity constraints (\ref{linearConstraints}) at the quantum level, we first need to fix the ambiguity in the correspondence relation,
\be \label{correspondence0}
B \leftrightarrow  \tilde{J}^{IJ}= \alpha J^{IJ} + \beta \star J^{IJ},
\ee
that is we need to determine $\alpha$, $\beta$ and we shall proceed using  the symplectic structure of the boundary space. Let us first precise the boundary variables \cite{EPR}. 
As in the BF theory or in the Barrett-Crane model, the $B$-fields are discretized on the triangles $\Delta$ of the triangulation $\Delta_4$. We are now going to explain the way the curvature is discretized. Geometry is assumed to be flat on each 4-simplex, therefore a cartesian coordinate patch  can be chosen in order to cover the whole 4-simplex. However, the triangulation is usually not flat and this patch cannot be extend to cover several 4-simplexes $\{v^1_{f^\ast}, v^2_{f^\ast}, \cdots v^n_{f^\ast} \}$ which surround the dual face $f^\ast \in \Delta_4^\ast$ . The curvature is thus concentrated on the dual faces $f^\ast$ 
in the dual 2-complex of $\Delta_4$ and is coded in the holonomy around the link of each $f^\ast$. 

The variables to describe this geometry are chosen as follows. We denote by 
\be
e(t)=e^I_\mu(t) v_Idx^\mu,
\ee
an orthonormal basis for the tetrahedron $t$ where $v_I$ is a basis in $\R^4$ chosen once and for all and 
\be
e(v)=e^I_\mu(v) v_I dx^\mu,
\ee 
an orthonormal basis describing the geometry in the 4-simplex $v$. We denote by $t_a$, $a=1, \cdots, 5$, the five tetrahedra that bound the single 4-simplex $v$, then there exists five matrices  $V_{vt_a} \in G$ such that each $e(t_a)$ is related to $e(v)$ by 
\be \label{ev}
e(v)^I_\mu= (V_{vt_a})^I_{\; J} \, e(t_a)^J_\mu,
\ee
in the common coordinate patch. We can also define the elements $U_{t_a t_b}\in G$ which describe the parallel transport between two tetrahedra $t_a$, $t_b$ belonging to the 4-simplex $v$.
\be \label{et}
e(t_a)=U_v(t_a t_b) \,e(t_b) \quad \textrm{ with } \quad U_v(t_a t_b)= (V_{vt_a})^{-1}V_{vt_b}=V_{t_a v} V_{vt_b},
\ee
the last equality  reflecting the fact that the 4-simplices are supposed to be flat. If we consider now two tetrahedra $t$ and $t^\prime$  sharing a face $\Delta$ but not necessarily at the same 4-simplex, we can also define the parallel transport between $t$ and $t^\prime$ by  $U_\Delta(t,t^\prime)\in G$  such that 
\be
U_\Delta(t,t^\prime)\equiv V_{tv_1} V_{v_1 t_1}V_{t_1v_2} \cdots V_{v_n t^\prime},
\ee
where the product is around the link  of $\Delta$ in the clock-wise direction from $t^\prime$ to $t$.
We can then write for each tetrahedron $t$
\be
B_\Delta(t)^{IJ}\equiv \int_\Delta \star \left( e(t)^I \wedge e(t)^J \right).
\ee
Two bivectors $B_\Delta(t)$, $B_\Delta(t^\prime)$ associated to the same triangle $\Delta$ represent the triangle $\Delta$ in two different internal frames associated to two distinct tetrahedra $t$ and $t^\prime$. These two bivectors are different but related by
\be
U_\Delta(t,t^\prime)\, B_\Delta(t^\prime)= B_\Delta(t)\, U_\Delta(t,t^\prime).
\ee
For a given triangle $\Delta$, we denote $t_1 \cdots t_n$  the set of the tetrahedra in the link around $\Delta$ and $v_{12}$, $v_{23}, \cdots, v_{n1}$ the corresponding set of simplicies in this link, where $t_2$ bounds $v_{12}$ and $v_{23}$ and so on cyclically. We call the link around the face $\Delta$, the cyclic sequence of 4-simplices separated by the tetrahedra that meet at the  triangle $\Delta$. The curvature associated to $\Delta$ is then defined by
\be
U_\Delta(t_1)=U_{v_{12}}(t_1,t_2)\cdots U_{v_{n_1}}(t_n,t_1)=V_{t_1v_{12}} V_{v_{12}t_2} \cdots V_{t_nv_{n1}} V_{v_{n1}t_1} 
\ee
that is by the product of the  matrices obtained turning along the link of the triangle $\Delta$, beginning with the tetrahedron $t_1$. The $V$ matrices are associated to the tetrahedra or equivalently to the dual edges $e^\ast \in \Delta_4^\ast$, $U_\Delta(t)\equiv U_\Delta(t,t)$ is the holonomy around the triangle $\Delta$ and the $V$ matrices are the analogous of the group elements $g_{e^\ast}$, defined in the discrete BF theory (see Section \ref{SFBF}),  which also appear in the derivation of the Barrett-Crane model (see Chapter \ref{BCmodel}).
\\
The classical discrete action is \cite{EPRL, coh_roberto}
\be \label{clasDiscritAct}
S=-\sum_{\Delta \in \textrm{int}\Delta_4} \tr\left[ B_\Delta(t)\, U_\Delta(t) + \f{1}{\gamma} \star B_\Delta(t)\, U_\Delta(t)\right] - \sum_{\Delta \in \partial\Delta_4} \tr\left[B_\Delta(t) \, U_\Delta(t, t^\prime) + \f{1}{\gamma} \star B_\Delta(t) \, U_\Delta(t,t^\prime)\right].
\ee
This action with the simplicity constraints (\ref{linearConstraints}) and closure constraint (\ref{closure}) defines a discretization of general relativity \cite{EPR,EPRL}. The boundary variables are in this formulation $B_\Delta(t) \in \mathfrak{g}$ and $U_\Delta(t, t^\prime) \in G$. The conjugated variable\footnote{Note that this not the conjugated variables in the 3+1 canonical sense. They are conjugated in the general sense where 
$J$ and $U$ are conjugated  means that the action can be written as $S=\int U J$ \cite{quantumtetra}.} to $U_\Delta(t, t^\prime)$ is
\be \label{BJ}
J_\Delta(t)= B_\Delta(t) +\f{1}{\gamma} \star B_\Delta(t) \quad \Rightarrow \quad B_\Delta(t) = \f{\gamma^2}{\gamma^2-s} \left(J_\Delta(t) -\f{1}{\gamma} \star J_\Delta(t)\right),
\ee
where $s=1$ for $G=\Spin(4)$ and $s=-1$ for $G=\SL(2,\C)$ and where  we have assumed $\gamma$ finite and  $\gamma\neq 0,1$. Thus a group element $U_\Delta\in G$ and as conjugated variable  element $J_\Delta \in\mathfrak{g} $ are associated to each boundary triangle $\Delta$ in the boundary  $ \Delta_\partial$ of the triangulation  $\Delta_4$.
$$ G\times \mathfrak{g}\ni (U_\Delta,J_\Delta)\longrightarrow \Delta\subset\Delta_\partial. $$
It has been shown in \cite{EPR} that this choice of variables $(U_\Delta$, $J_\Delta)$ defines the same boundary phase space as for a  lattice Yang-Mills theory with gauge group $G$. The explicit expression of the symplectic structure can be found in \cite{EPR}. 

Before quantizing the theory, let us express the constraints for the $J$ field. Indeed, the constraints (\ref{linearConstraints}) and (\ref{diagquadra}) as well as the closure constraint (\ref{closureConstraint}) on the $B$ field translate into constraints on the $J$ field. 
\\
The closure for the $B$ is equivalent to the closure for the $J$ and this constraint will be imposed automatically by the dynamics. The simplicity constraints (\ref{diagquadra}) and  (\ref{linearConstraints}) become respectively
\bes
\cC_{\Delta \Delta} &\equiv& \star J_\Delta \cdot J_\Delta \left( 1+ \f{s}{\gamma^2}\right) - s\f{2}{\gamma} J_\Delta \cdot J_\Delta =0 , \label{Jdiag}\\
\cC_\Delta^J& \equiv  &x_I \left((\star J)^{IJ} - \f{s}{\gamma} J^{IJ}_\Delta \right) =0 \label{Jcross},
\ees
where  in the first equation, the dot stands for the scalar product in the algebra. To solve the second, we fix the gauge to $x_I= \delta^0_I$
\be \label{Jcross1}
\cC_\Delta^j=L_\Delta^j-\f{s}{\gamma} K^j_\Delta =0,
\ee
where
\be
L_\Delta^j\equiv \f12 \epsilon^j_{\; kl} J^{kl}_\Delta \quad \textrm{ and } \quad  K^j_\Delta \equiv J^{0j}_\Delta
\ee
are respectively the generators of the $\SU(2)$ subgroup that leave $x_I$ invariant  and the generators of the corresponding boosts. The choice of this gauge corresponds in the Lorentzian case to the restriction of spacelike tetrahedra. The general case is recovered thanks to gauge invariance. 

\medskip

Let us now focus on the quantization step. 
We first need to write the Hilbert space associated with the boundary state space. For a given boundary, a 3-surface denoted by $\Sigma$, the associated Hilbert space is 
$$
\cH_\Sigma=L^2\left( G^L, d\mu_{\textrm{Haar}}\right),
$$ 
where $L$ is the number of links of the graph of the dual 2-complex $\Delta^\ast_\partial$ of the boundary triangulation $\Delta_\partial$ (see Table \ref{triang3d} for the precise relation between a three-dimensional triangulation and its dual two-complex) and $d\mu_\textrm{Haar}$  is the Haar measure on the group $G$. In fact, we can write $\cH_\Sigma$ under the form
\be
\cH_\Sigma= \bigoplus_{\cI_\Delta} \bigotimes_t \cH_t,
\ee
where $\cH_t$ is the associated Hilbert space to the tetrahedron $t$.
\\
Let us focus on a single tetrahedron $t$ of the boundary triangulation $\Delta_\partial$ and on its  associated Hilbert space $\cH_t$. We associate an irreducible representation of $G$, $\cI_\Delta$, to each triangle $\Delta$ of the tetrahedron. We denote $\Delta_{\alpha=1\cdots 4}$ the four triangles associated to $t$, then
\be
\cH_t \equiv \bigotimes_{\alpha=1}^4\cH_{\cI_\alpha},
\ee
where $\cH_\cI$ is the carrier space of the representation $\cI$.
It is the bivector conjugated to the curvature,  $J_\Delta$ in our case, that is quantized being replaced by the generator of the symmetry algebra $\mathfrak{g}$ in this representation\footnote{This is justified in \cite{EPR, EPRL} by the symplectic structure of the boundary phase space which reproduces the one of a  lattice Yang-Mills theory with gauge group $G$.}. We denote $\hat{J}$ the $\mathfrak{g}$ generators. The closure constraint on the $J$-fields associated to the faces of the tetrahedron translates into a closure constraint on the $\mathfrak{g}$-generators $\hat{J}_\Delta$. However, contrary to the Barrett-Crane model (see section \ref{BCmodel}) in which the simplicity constraints were imposed on the $G$-intertwiner space -- \ie the closure constraint was imposed before the simplicity constraints --  the procedure followed in \cite{EPR0, EPR, EPRL} consists in imposing firstly the simplicity constraints and in imposing the closure constraint secondly. 
%

\medskip

%
Let us precise  the procedure followed to impose the different constraints in this context. 
The quadratic diagonal simplicity constraints (\ref{Jdiag}) imposed strongly  translates into an equation on the Casimirs $\tilde{C}_G$ and  $C_G$.
\be \label{diagCasi}
\left(1+\f{s}{\gamma^2} \right)\, \tilde{C}_G(\cI_\Delta) -\f{2s}{\gamma}\, C_G(\cI_\Delta)=0,
\ee
where 
\be \label{Casi1}
\tilde{C}_G(\cI_\Delta)=\f12 \epsilon^{IJKL}\hat{J}_{IJ}^{\cI_\Delta}\hat{J}_{KL}^{\cI_\Delta}=4sL^jK_j,
\ee
 is the pseudo-scalar Casimir of the group $G$,  and 
 \be \label{Casi2}
 C_G(\cI_\Delta)=(\hat{J}^{\cI_\Delta})^{IJ}\hat{J}_{IJ}^{\cI_\Delta}=2(L^2+sK^2)
 \ee
 is the scalar Casimir of $G$. We have respectively  $s=1, -1$ for $G=\Spin(4), \SL(2,\C)$.  This strong constraint will restrict the allowed $G$-representations to $\gamma$-simple representations. The definition of a $\gamma$-simple representation for the group $G$ will be explicitly given in the two next sections where we will distinguish the Euclidean case $G=\Spin(4)$ and the Lorentzian case $G=\SL(2,\C)$.
\\
Regarding the gauge-fixed cross simplicity constraints (\ref{Jcross1}), we look for a Hilbert space $\cH_s$, a subspace of $\cH_t$, such that the matrix elements of the cross simplicity constraints (\ref{Jcross1}) on this smaller Hilbert space all vanish 
\be \label{Hspace}
\forall \phi,\psi\in \cH_s,\quad\,
\la\phi|\cC_{\Delta}^j|\psi\ra
\,=\,
0.
\ee
The strategy followed in \cite{coh_roberto, EPRL} is inspired by \cite{master, master1, master2, master3}. The set of constraints (\ref{Jcross1}) is replaced by the single "master" constraint 
\be \label{mast}
\cM_\Delta\equiv \sum_j \left( \cC_\Delta^j\right)^2= \sum_j \left( L^j-\f{s}{\gamma}K^j \right)^2=0.
\ee
This new constraint can be written in terms of the Casimir operators $\tilde{C}_G$ and $C_G$ respectively given by (\ref{Casi1}) and (\ref{Casi2})
\be \label{mast1}
\cM_\Delta= L^2 \left( 1 -\f{s}{\gamma} \right) + \f{s}{2\gamma^2}  \tilde{C}_G - \f{1}{2 \gamma} C_G=0,
\ee
which combined with (\ref{diagCasi}) simplifies into
\be \label{masterCasi}
C_G- 4 \gamma L^2=0.
\ee
This new equation will select one subspace of the Hilbert space associated to the tetrahedron. This subspace corresponds to $\cH_s$ -- this is explicitly shown in \cite{eprl_ding1} for the Euclidean case and in \cite{eprl_ding} for the Lorentzian case.
\\
Finally, once the simplicity has been implemented, the closure constraint (\ref{closureConstraint}) is imposed on $\cH_s$ turning it into an intertwiner space. In terms of the $\SU(2)$ rotation and boost generators, $L$ and $K$, this constraint (\ref{closureConstraint}) becomes
\bes  
\sum_{\Delta \in t} L^i_\Delta=0  \quad \forall i \label{closureL} ,\\
\sum_{\Delta \in t} K^i_\Delta=0  \quad \forall i \label{closureK}.
\ees
In fact, these closure constraints are imposed only weakly turning $\cH_s$ into a $\SU(2)$ intertwiner space, denoted by
\be \label{Kspace}
\mathcal{K}_s \equiv \textrm{Inv}_{\SU(2)} \left[\cH_s\right].
\ee
More precisely, by definition of $\mathcal{K}_s$, the left-hand-side of (\ref{closureL}) which is the generator of global $\SU(2)$ transformations vanishes strongly on $\mathcal{K}_s$ whereas the left-hand-side of (\ref{closureK}) is weakly proportional  to $\sum_{\Delta \in t} L^i_\Delta$ due to the weak relation (\ref{Hspace}) and therefore vanishes weakly. 
\\
The total physical boundary space $\cH_{\textrm{ph}}$ of the theory is finally obtained as the span of spin networks in $L^2[G^L/G^V, d\mu_{\textrm{Haar}}]$ -- $V$ is the number of vertices of the graph of the dual 2-complex $\Delta_\partial^\ast$ of the boundary triangulation $\Delta_\partial$ -- with $\gamma$-simple representations on edges and with intertwiners in the spaces $\mathcal{K}_s$ at each vertex. Thus, by definition, the boundary states correspond to projected spin networks $|\vphi_{\cI_e,j_e^{s,t},i_v}\ra$ defined in (\ref{projectSpinNetdef}) with $\cI_e\equiv \cI_\Delta$  $\gamma$-simple representations of $G$ for any triangulation\footnote{We recall that edges $e \in \Delta_\partial^\ast$ in the dual picture are dual to triangles $\Delta \in \Delta_\partial$ and this one-to-one correspondence allows us to denote the discrete variables by either a triangle $\Delta$ or an edge $e$ subindex respectively.} $\Delta$ and $j_e^s=j_e^t\equiv j_\Delta$ a $\SU(2)$ irreducible representations such that $(\vec{L}_\Delta)^2=j_\Delta(j_\Delta+1)$. The embedding of the irreducible representations $j_\Delta$ into the $G$ irreducible representations $\cI_\Delta$  given by the relation (\ref{mast1}) is explicitly detailed in the two next sections, respectively for $G=\Spin(4)$ and $G=\SL(2,\C)$. Moreover, $i_v$ is the $\SU(2)$ intertwiner chosen in $\mathcal{K}_s$ associated to each vertex $v$. We will see in the Chapter \ref{ProjectChap} that the projected spin networks are in addition very easily related to the $\SU(2)$ spin networks of loop quantum gravity. 
\\
Finally, to recover $G$ spin networks the last step is a group averaging (as defined in (\ref{groupaverProj})). 
\\
The EPRL-FK vertex amplitude can then be defined using the BF amplitude (\ref{factorize}). The explicit expressions, which depends on the considered group $G$, are given in the two following sections: let us identify now the different aspects of the model according to the choice of group  $G=\Spin(4)$ or $G=\SL(2,\C)$.

\subsection{The Euclidean case with a finite Immirzi parameter}
We recall that unitary representations of $\Spin(4)$ are labelled by a couple of half integers $\cI=(k^L,k^R)$. \begin{itemize}
\item The diagonal simplicity constraints (\ref{diagCasi})  impose then the following restriction on the representations associated to each face $\Delta$
\be \label{relation1}
k^L_\Delta= \left| \f{\gamma +1}{ \gamma -1}\right| k^R_\Delta.
\ee
Let us notice that in order to obtain this solution -- more precisely in oder that (\ref{diagCasi}) for the $\Spin(4)$ Casimir does have solutions -- the authors of \cite{EPRL} had to choose a different ordering than the usual one used to define  Casimirs operators\footnote{With the usual ordering, we have 
$$
\tilde{C}_{\Spin(4)}= 4k^L(k^L+1)+4k^R(k^R+1) \quad \textrm{ and } C_{\Spin(4)}= 4k^L(k^L+1)-4k^R(k^R+1).
$$.} of $\Spin(4)$.
\item Let us restrict to the case $\gamma >0$ which implies $k^L>k^R$. Inserting  (\ref{relation1}) into the "master" constraint (\ref{masterCasi}) constraints the quantum number $j$ associated to the $\SU(2)$ Casimir $L^2$ (again up to an ordering ambiguity) to
\be \label{relation2}
j^2=\left(\f{2k^L}{1+ \gamma} \right)^2=\left(\f{2k^R}{1- \gamma} \right)^2 \qquad \Rightarrow \qquad  j= \left\{ \tabl{lc}{k^L+k^R \qquad & 0<\gamma<1 \\
k^L-k^R \qquad & \gamma>1} \right. .
\ee
This constraint selects a subspace $\cH_s$ of $\cH_t=\bigotimes_{\alpha=1}^4\cH_{\cI_\alpha}$. Indeed, for each triangle $\Delta_{\alpha=1 \cdots 4}$, the Clebsch-Gordan decomposition of $\cH_{\cI_\alpha}$ gives
\be
\cH_{\cI_\alpha}=\cH_{k^L_\alpha \otimes k^R_\alpha}= \bigoplus_{j_\alpha=|k_\alpha^L-k^R_\alpha|}^{k^L_\alpha+k^R_\alpha} \cH_{j_\alpha},
\ee
and the constraint (\ref{relation2}) selects in $\cH_{\cI_{\alpha}}$ the "extremum" subspace leading to 
\be
\cH_s = \left\{ \tabl{lc}{\displaystyle{\bigotimes_{\alpha=1}^{4}} \cH_{k^L_\alpha+k^R_\alpha}, &\qquad \textrm{ for } \gamma <1,\\
\displaystyle{\bigotimes_{\alpha=1}^{4}} \cH_{k^L_\alpha-k_\alpha^R}, &\qquad \textrm{ for } \gamma >1.
}\right.
\ee
The simplicity constraints (\ref{Jcross}), when (\ref{relation1}) is holding, is then satisfied weakly in $\cH_s$ \cite{eprl_ding1}. That is, the action of the constraints on the states in $\cH_s$ (\ref{Hspace}) results in states orthogonal to $\cH_s$.
\item The last step is to solve weakly the closure constraints given by (\ref{closureL}) and (\ref{closureK}) in the space $\cH_s$ in order to get an intertwiner space $\mathcal{K}_s$. We get that
\be
\mathcal{K}_s=\textrm{Inv}_{\SU(2)} \left[ \cH_s\right].
\ee
\item Then, to get a $\Spin(4)$ spin network a group averaging on $\Spin(4)$ is performed. We can now define the vertex amplitude. We will not give any details on its construction which can be obtained starting from the $\Spin(4)$ BF vertex amplitude (\ref{factorize}), see \cite{EPR, EPRL}.
\be
A_v^{\Spin(4) \textrm{ EPRL-FK}}(j_{f^\ast}, i_{v^\ast})= \sum_{i^L_{e^\ast}, i^R_{e^\ast}} 15j\left(\f{(1+\gamma)}{2} j_{f^\ast}, i^L_{v^\ast}\right)15j\left(\f{|1-\gamma|}{2} j_{f^\ast}, i^R_{v^\ast}\right) \displaystyle{\bigotimes_{v^\ast}}f^{i_{v^\ast}}_{i^L_{v^\ast}i^R_{v^\ast}}
\ee
where the $15j$ are the standard $\SU(2)$ Wigner symbols and $f^{i_{v^\ast}}_{i^L_{v^\ast}i^R_{v^\ast}}$ are the fusion coefficient obtained contracting $\SU(2)$ intertwiners $i_{v^\ast}$ and $\Spin(4)$ intertwiners $(i^L_{v^\ast}, i^R_{v^\ast})$ (for further details see \cite{EPRL}). 
\end{itemize}

\subsection{The Lorentzian, with a finite Immirzi parameter}
We recall that the principal series of unitary representations of $\SL(2,\C)$ are labelled by two parameters $\cI=(n, \rho)$ with $n$ a half-integer and $\rho$ a real number.  \begin{itemize}
\item The diagonal simplicity constraints (\ref{diagCasi})  impose  the following restriction on the representations associated to each face $\Delta$
\be \label{relation1L}
\rho=\gamma (n+1).
\ee
Let us notice that this solution first proposed in \cite{eprl_ding} differs from the usual solution $\rho=\gamma n$ found in the literature but it is this ansatz we use in the Chapter \ref{ProjectChap}. Moreover, there exists a second branch of solutions to these constaints (\ref{diagCasi}) given by $\rho=-n/\gamma$ but it is eliminated by the master constraint (\ref{masterCasi}).
\item Inserting this relation (\ref{relation1L}) into the "master" constraint (\ref{masterCasi}) constraints the quantum number $j$ associated to the $\SU(2)$ Casimir $L^2$ to
\be \label{relation2L}
j=n.
\ee
This constraint selects a subspace $\cH_s$ of $\cH_t=\bigotimes_{\alpha=1}^4\cH_{\cI_\alpha}$. Indeed, for each triangle $\Delta_{\alpha=1 \cdots 4}$,  $\cH_{\cI_\alpha}$ splits into the irreducible representations $\cH_{j_\alpha}$ of the $\SU(2)$ subgroup as
\be
\cH_{\cI_\alpha}=\cH_{(n_\alpha,\rho_\alpha)}= \bigoplus_{j_\alpha=n_\alpha}^{\infty} \cH_{j_\alpha},
\ee
with $j_\alpha$ increasing in steps of $1$ and the constraint (\ref{relation2L}) selects in $\cH_{\cI_{\alpha}}$ the "minimal" subspace leading to 
\be
\cH_s = \displaystyle{\bigotimes_{\alpha=1}^{4}} \cH_{n_\alpha}.
\ee
The simplicity constraints (\ref{Jcross}), when (\ref{relation1L}) is holding, is then satisfied weakly in $\cH_s$ \cite{eprl_ding}. That is, the action of the constraints on the states in $\cH_s$ (\ref{Hspace}) results in states orthogonal to $\cH_s$.
\item The last step is to solve weakly the closure constraints given by (\ref{closureL}) and (\ref{closureK}) in the space $\cH_s$ in order to get an intertwiner space $\mathcal{K}_s$. We get that
\be
\mathcal{K}_s=\textrm{Inv}_{\SU(2)} \left[ \cH_s\right].
\ee
\item As in the Euclidean case, the last step is a group averaging over $\SL(2,\C)$ in order to obtain $\SL(2, \C)$ spin networks and the vertex amplitude is given by \cite{EPRL, eprl_ding}
\be
A^{\SL(2,\C) \textrm{ EPRL-FK}}_v(j_{f^\ast}, i_{v^\ast})=\sum_{n_{e^\ast}} \int d\rho_{e^\ast}(\rho_{e^\ast}^2+n_{e^\ast}^2)
15j_{\SL(2,\C)}(j_{f^\ast}, \gamma(j_{f^\ast}+1); (n_{e^\ast}, \rho_{e^\ast}))  \left(\prod_{e^ast} f^{i_{e^\ast}}_{n_{e^\ast}, \rho_{e^\ast}}(j_{f^\ast}) \right) 
\ee
where we now use the $15j$ of $\SL(2,\C)$ and $ f^{i_{e^\ast}}_{n_{e^\ast}, \rho_{e^\ast}}$ are fusion coefficients obtained contracting $\SU(2)$ intertwiners $i_{v^\ast}$ and $\SL(2,\C)$ intertwiners $(n_{e^\ast}, \rho_{e^\ast})$.
\end{itemize}
This concludes our overview of the way introduced in \cite{EPR0, EPR, EPRL} to deal with the simplicity constraints.
\section{Using coherent states} \label{SUcoh}
We will now introduce the coherent intertwiners used to solve weakly the cross simplicity constraints  in the EPRL-FK models. The problem of defining "coherent states" for loop quantum gravity has raised an increasing interest over the last few years. 
Indeed, although the fact that spin network states, the building block of loop quantum gravity, provide a basis of the kinematical Hilbert space and diagonalize some geometric operators, they lack a low-energy interpretation. We need to bridge  the Planck scale quantum geometry  to a smooth and classical three dimensional geometry, which  is why the construction of coherent states is very important. These semi-classical states should be the analogue of wave packets or coherent states that approximate classical configurations in ordinary quantum theory.
We recall that the phase space of general relativity can be described by the triad and $\SU(2)$ connection $(E_i^a, A_a^i)$ used in loop gravity. Consequently, a coherent state for loop gravity should be a superposition of spin networks peaked on $(E_i^a, A_a^i)$, with small fluctuations. This approach has been developed by T. Thiemann and collaborators \cite{hamiltonian1, complexifier, thiemann7, thiemann8, volume8}. \\
In the context of spin foam models, the approach to find semi-classical quantum states that approximate a given classical geometry is different. In the spin foam framework, a classical phase space point is described in terms of quantities referring to discrete geometries, \eg areas and dihedral angles, as opposed to holonomies and fluxes. The precise connection between the loop quantum variables associated to a given graph $\Gamma$ and labels describing the spin foam discrete geometry has been done using twisted geometries \cite{twisted, twisted1}. Twisted geometries are quantities describing the intrinsic and extrinsic geometry of the cellular decomposition dual to the graph $\Gamma$.

We now define the $\SU(2)$ coherent states introduced by E. Livine and S. Speziale in \cite{coh1}. These coherent intertwiners, labelled by $N$ unit vectors in $\R^3$ satisfying a closure condition -- in the case of a $N$ valent  vertex --, have very interesting semi-classical properties and allow to peak intertwiners on specific classical convex polyhedra \cite{coh2, coh3, UN2}.
\subsection{The $\SU(2)$ coherent intertwiners}\label{SU2coh}
The $\SU(2)$ coherent states  are derived by acting with $\SU(2)$ rotations on the highest weight vectors $|j,j\ra$
\be
\forall g\in\SU(2),\quad
|j,g\ra \,\equiv\, g\,|j,j\ra .
\ee
These states are coherent states \`a la Perelomov and transform consistently under the $SU(2)$ action
\be
h\,|j,g\ra\,=\,
|j,hg\ra.
\ee
They also satisfy a very simple tensorial property
\be
|j,g\ra\otimes|\tj,g\ra
\,=\,
g\,(|j,j\ra\otimes |\tj,\tj\ra)
\,=\,
g\,|j+\tj,j+\tj\ra
\,=\,
|j+\tj,g\ra.
\ee
It is easy to compute their expectation values
\be
\la j,g|\vJ|j,g\ra
\,=\,
j\,\hat{n},
\qquad
\hat{n}=g\,\hat{z},
\ee
where the unit vector $\hat{n}\in\cS^2$ is obtained by rotating the $z$ axis by the  $\SU(2)$ group element $g$. They are actually coherent states on the 2-sphere $\cS^2\sim \SU(2)/\U(1)$ since $\hat{n}$ only depends on $g$ up to a $\U(1)$ phase. Actually, the standard definition of the $\SU(2)$ coherent states involves a choice of section and we usually choose the unique group element $g(\hat{n})$ for a given unit vector $\hat{n}$ such that the rotation axis lays in the $(xy)$ plane. Then the coherent state is defined as $|j,\hat{n}\ra \,\equiv\, g(\hat{n})\,|j,j\ra=|j,g(\hat{n})\ra$.
Finally, these states minimize the uncertainty relation
\be
\la j,g|\vJ^2|j,g\ra-\la j,g|\vJ|j,g\ra^2
\,=\,
j(j+1)-j^2
\,=\,
j.
\ee

Then $N$-valent coherent intertwiners are defined following \cite{coh1} by tensoring $N$ such $\SU(2)$ coherent states and group averaging this tensor product in order to get an intertwiner. More precisely, we choose $N$ representations of $\SU(2)$ labeled by the spins $j_1,..,j_N$ and $N$ unit 3-vectors $\hat{n}_1,..,\hat{n}_N$, then we define
\be
||j_i,\hat{n}_i\ra
\,\equiv\,
\int_{\SU(2)} dg\, g\triangleright \bigotimes_{i=1}^N
|j_i,\hat{n}_i\ra
\,=\,
\int_{\SU(2)} dg\,\bigotimes_{i=1}^N
gg(\hat{n}_i)\,|j_i,j_i\ra.
\ee

\subsection{Intertwiner states as weak solutions of the simplicity constraints}
Now coming back to the $\Spin(4)\sim\SU_L(2)\times\SU_R(2)$  and the simplicity constraints,
we use simple $\Spin(4)$ representations with $j^L_i=j^R_i$ and double the labels of the coherent states introducing unit 3-vectors $\hat{n}^{L,R}_1,..,\hat{n}^{L,R}_N$. Considering the tensor product of $\SU(2)$ coherent states $|j,\hat{n}^L\ra\otimes|j,\hat{n}^R\ra$, the expectation values of the $\spin(4)$ generators $\vJ_i^{L,R}$ are $j\,\hat{n}^{L,R}$. These two 3-vectors with equal norm define a single bivector $B\in\w^2\R^4$ being its self-dual and anti-self dual components. Then considering $N$ such coherent states $|j_i,\hat{n}_i^L\ra\otimes|j_i,\hat{n}_i^R\ra$, their expectation values give $N$ bivectors $B_i$. We now impose the simplicity constraints on these classical bivectors.

\medskip

First looking  at the sector $(s)$ (defined by (\ref{Bconstr1}), the fact that the bivectors $B_i$ all share a same   time normal $x$ translates to the existence of a $\SU(2)$ transformation mapping simultaneously all the self-dual part  (left) onto the anti-self-dual part (right). Moreover, this $\SU(2)$ group element defines uniquely the normal 4-vector,
\be
\forall i,\, x\cdot B_i=0
\qquad\Leftrightarrow\qquad
\exists g\in\SU(2),\quad \forall i,\,\hat{n}_i^L=g\,\hat{n}_i^R
,\quad
x=(g,\id)\triangleright \omega,
\ee
where $\omega=(1,0,0,0)$ is the  reference unit time vector and the $\Spin(4)$ group element $(g,\id)$ is defined through its left/right factors.
Imposing this constraint on the labels of the coherent states, we define a $\Spin(4)$ coherent intertwiner by group averaging. This coherent intertwiner is labeled by the representation label, plus the unit 3-vectors $\hat{n}_i^R$, plus the group element $g$ which gives the time normal and the rotation which  defines the components $\hat{n}_i^L$ from the $\hat{n}_i^R$
\bes
||j_i,\hat{n}_i,g\ra_s
&=&
\int_{\SU_L(2)\times\SU_R(2)} dg_Ldg_R\,
\left[g_L\triangleright \bigotimes_{i=1}^N g|j_i,\hat{n}_i\ra\right]
\otimes \left[g_R\triangleright\bigotimes_{i=1}^N |j_i,\hat{n}_i\ra\right]\nn\\
&=&
\left[\int_{\SU(2)} dg_L\,g_L\triangleright \bigotimes_{i=1}^N |j_i,\hat{n}_i\ra\right]
\otimes \left[\int_{\SU(2)} dg_R\,g_R\triangleright\bigotimes_{i=1}^N |j_i,\hat{n}_i\ra\right].
\ees
Two things are obvious from this expression.
\begin{itemize}
\item The $\Spin(4)$ group averaging erases the group element $g$ and thus all the data about the precise time normal $x$. In particular, we can drop the label $g$ and call these states simply $||j_i,\hat{n}_i\ra_s$.
\item This states $||j_i,\hat{n}_i\ra_s$ are the tensor product of two identical $\SU(2)$ intertwiners for the left and right parts. In particular, they obvious satisfy the quadratic simplicity constraints
    \be
    \forall i,j,\quad
    \la\vJ^L_i\cdot\vJ^L_j\ra
    \,=\,
    \la\vJ^R_i\cdot\vJ^R_j\ra.
    \ee
    Moreover, they minimize the uncertainty by definition.
\end{itemize}
We can go one step further by re-writing these states,
\be
||j_i,\hat{n}_i\ra_s
\,=\,
\int_{\Spin(4)}dG\,G\triangleright \bigotimes_{i=1}^N |j_i,\hat{n}_i\ra_L\otimes|j_i,\hat{n}_i\ra_R
\,=\,
\int_{\Spin(4)}dG\,G\triangleright \bigotimes_{i=1}^N |2j_i,\hat{n}_i\ra,
\ee
where we used the tensorial property of the $\SU(2)$ coherent states $|j,\hat{n}\ra\otimes|\tj,\hat{n}\ra=|j+\tj,\hat{n}\ra$. This shows that the coherent states are exactly the EPR states \cite{EPR0,EPR} which form a Hilbert space $H_s[j_\Delta]$ solving weakly the simplicity constraints \cite{FK,LS}.

\medskip

Second, we consider the dual sector $(\star)$ (defined by (\ref{Bconstr2})) and we follow the same procedure. The only difference is a sign in solving the corresponding linear simplicity constraints.
\be
\forall i,\, \eps x B_i=0
\qquad\Leftrightarrow\qquad
\exists g\in\SU(2),\quad \forall i,\,\hat{n}_i^L=-g\,\hat{n}_i^R.
\ee
This leads to similar coherent states \cite{FK,LS}.
\bes
||j_i,\hat{n}_i\ra_\star
&=&
\left[\int_{\SU(2)} dg_L\,g_L\triangleright \bigotimes_{i=1}^N |j_i,-\hat{n}_i\ra\right]
\otimes \left[\int_{\SU(2)} dg_R\,g_R\triangleright\bigotimes_{i=1}^N |j_i,\hat{n}_i\ra\right],\\
&=&
\left[\int_{\SU(2)} dg_L\,g_L\triangleright \bigotimes_{i=1}^N \overline{|j_i,\hat{n}_i\ra}\right]
\otimes \left[\int_{\SU(2)} dg_R\,g_R\triangleright\bigotimes_{i=1}^N |j_i,\hat{n}_i\ra\right].
\ees
Once again, it is obvious to check that these states satisfy the quadratic simplicity constraints in expectation value. Also, the information about the time normal is completely erased by the group averaging. Finally, the key difference with the previous sector $(s)$ is that these intertwiner states generate the whole intertwiner space and do not form a subspace. This ansatz looks more like a fuzzy version of the Barrett-Crane intertwiner.

\medskip

This concludes our quick overview of the standard way to deal with the crossed simplicity constraints using the $\SU(2)$ coherent intertwiners.

\chapter{Simplicity constraints and $\SU(2)$ spin networks} \label{ProjectChap}

\section{Back and forth between projected and $\SU(2)$ spin networks}
We focus in this Chapter on the Lorentzian case, \ie $G=\SL(2,\C)$.  Let us consider an arbitrary oriented graph  $\Gamma$ with $E$ edges and $V$ vertices. We denote again $H=L^2_{\SU(2) \textrm{ inv.}} [dG^L]$ where $dG$ is the $\SL(2,\C)$ Haar measure, the Hilbert space of projected cylindrical functionals on the graph $\Gamma$ introduced in Section \ref{chapCLQG}. We have left implicit the underlying graph $\Gamma$ since the whole following analysis is done for the fixed graph $\Gamma$.
The projected spin networks
\be
\vphi_{\cI_e,j_e^{s,t},i_v}(G_e,x_v)
\,\equiv\,
\tr
\prod_e \la \cI_e,j_e^s,m_e^s|B_{s(e)}^{-1}G_e B_{t(e)}| \cI_e,j_e^t,m_e^t\ra
\,
\prod_v \la \otimes_{e|t(e)=v}\,\cI_e,j_e^t,m_e^t| i_v |\otimes_{e|s(e)=v}\cI_e,j_e^s,m_e^s\ra,
\ee
form an orthonormal basis of $H$ and are the natural boundary states for Spin Foam models (as explained in Section \ref{EPRGupta} for  the EPRL-FK spin foam models and in \cite{projected} for the Barrett-Crane model). In the previous formula, we recall that   $\cI=(n, \rho)$ are $\SL(2,\C)$ irreducible representations, $j$ are $\SU(2)$ irreducible representations and $i$ are $\SU(2)$ intertwiners.  Our goal is to compare the projected spin networks with the $\SU(2)$ spin network basis of loop quantum gravity. As we have seen, the projected spin networks are Lorentz-invariant functionals of the $\SL(2,\C)$ connection and of the time-normal field. Nevertheless, as soon as we fix the value of the time-normal field (at the vertices of the graph used to construct the spin network), they are only required to satisfy an effective $\SU(2)$ invariance and thus they are built using $\SU(2)$-intertwiners and not $\SL(2,\C)$-intertwiners. Since $\SU(2)$ spin networks are also built from $\SU(2)$-intertwiners, this hints towards a direct path between the two sets of states. From this perspective, projected spin networks seems to be extensions of $\SU(2)$ spin networks, allowing to evaluate them on the whole Lorentz group $\SL(2,\C)$ and not only on the $\SU(2)$ subgroup.

\subsection{Projecting down to $\SU(2)$ spin networks}

Let us start by reminding the definition of $\SU(2)$ cylindrical functions on the graph $\Gamma$ introduced in Section \ref{cylinfunc}. They are functions of $E$ group elements in $\SU(2)$ living on the edges of the graph and satisfying a $\SU(2)$ invariance at every vertex
\be
\psi(g_e)=\psi(h_{s(e)}g_e h_{t(e)}^{-1}),\quad\forall h_v\in\SU(2)^{\times V}.
\ee
The natural scalar product on this space of functions is
\be
\la\psi|\psi'\ra_{\SU(2)}
\,=\,
\int_{\SU(2)}[dg_e]\,
\overline{\psi}(g_e)\psi'(g_e),
\ee
where $dg$ is the Haar measure on the $\SU(2)$ Lie group. Let us call $\Hs$ the $L^2$ space of such $\SU(2)$ invariant cylindrical functions. Then this Hilbert space $\Hs$ is spanned by the usual spin network states defined in Section \ref{spinnet}. A spin network is labeled by a set of spins $j_e$ for each edge and $\SU(2)$-intertwiners $i_v$ for every vertex. Then we define
\be
\psi_{j_e,i_v}(g_e)
\,\equiv\,
\tr
\prod_e \la j_e,m_e^s|g_e| j_e,m_e^t\ra
\,
\prod_v \la \otimes_{e|t(e)=v}j_e,m_e^t| i_v |\otimes_{e|s(e)=v}j_e,m_e^s\ra,
\ee
which simply amounts to contracting the Wigner matrices $D^j_{m^sm^t}(g)=\la j,m^s|g| j,m^t\ra$ along every edge $e$ with the intertwiners sitting at the vertices. We point out that this definition is almost the same as the one of projected spin networks: the difference is that we evaluate projected spin networks on the whole $\SL(2,\C)$ group and this requires the choice of an extra  $\SL(2,\C)$ irrep $\cI_e$ for each edge of the graph.

The scalar product between two such $\SU(2)$ spin networks is easily computed
\be
\la\psi_{j_e,i_v}|\psi_{\tj_e,\tilde{i}_v}\ra_{\SU(2)}
\,=\,
\prod_e\f{\delta_{j_e,\tilde{j}_e}}{d_{j_e}}
\,
\,\prod_v \la i_v|\tilde{i}_v\ra,
\label{scalarsu}
\ee
where we remind that $d_j=(2j+1)$ is the dimension of the $\SU(2)$-irrep of spin $j$.

\medskip

Since the projected cylindrical functions and the $\SU(2)$ cylindrical functions share the same $\SU(2)$ invariance, it is natural to introduce the following projection
\be
\begin{array}{llcl}
\cM:\,&H&\rightarrow&H_S \\
&\vphi(G_e,x_v)&\mapsto&
\psi(g_e)=\vphi(g_e,\om)=\phi(g_e),
\end{array}
\ee
which is simply the restriction of the projected cylindrical function to the $\SU(2)$ subgroup.
Considering the invariance property of the function $\vphi$ and its section $\phi$ at $x_v=\om,\,\forall v$, the map $\cM$ is well-defined and the resulting function $\psi$ is correctly $\SU(2)$-invariant as wanted.

It is straightforward to compute the image of the projected spin network by the map $\cM$. First, considering the case of functions with $j_e^s\ne j_e^t$, the corresponding $\SU(2)$ function vanishes:
\be
\forall j_e^s\ne j_e^t,\quad \cM\vphi_{\cI_e,j_e^{s,t},i_v}\,=\,0,
\ee
since a $\SU(2)$ group element could never trigger a transition between two different $\SU(2)$ irreps (by definition). On the other hand, now assuming that the two spins are equal for all edges so that we can drop the index $s,t$, $j_e^s=j_e^t=j_e$, then the image of the corresponding projected spin network is as expected simply a $\SU(2)$ spin network:
\be
\forall j_e^s=j_e^t=j_e,\quad
\cM\vphi_{\cI_e,j_e,i_v}\,=\,\psi_{j_e,i_v},
\ee
as long as the spin $j_e$ is compatible with the $\SL(2,\C)$ irrep, i.e $j_e\ge n_e$ (or more exactly $j_e\in n_e+\N$).

\medskip

In the next sections, we investigate the inverse map(s) to $\cM$, that is how to lift $\SU(2)$ cylindrical functions to functions on the whole Lorentz group $\SL(2,\C)$. Understanding in details how this lifting is achieved is crucial to the construction of the EPR-FK class of spin foam models and their interpretation as an ansatz for the dynamics of Loop Quantum Gravity.

In the following, we will focus on projected spin networks satisfying the ``matching" constraints $j_e^s=j_e^t$. We call $\Hp$ the Hilbert spanned by these ``proper" projected spin network functionals (whose evaluation on the $\SU(2)$ subgroup does not trivially vanish). As seen from the last equation above, inverting the map $\cM$ would more or less simply amount to choosing a $\SL(2,\C)$ irrep $\cI_e$ into which to embed the $\SU(2)$ irrep $j_e$. We analyze this in details below.

\subsection{Lifting back Spin Networks}
\label{lifting}

Starting with a $\SU(2)$ cylindrical function $\psi(g_e)$ invariant under the $\SU(2)$ action at every vertex, the goal is to construct a Lorentz invariant extension for it. Following the insight of the previous section, the simplest way to proceed would be to decompose the function $\psi$ in $\SU(2)$ irrep $j_e$ and to choose a $\SL(2,\C)$ irrep for every spin. At the level of the groups, these operations are done through convolutions with $\SU(2)$ and $\SL(2,\C)$ characters.

More precisely, starting with $\psi(g_e)$, we construct the following projected cylindrical function:
\be
\vphi(G_e,B_v)
\,\equiv\,
\sum_{\{j_e\}}\Delta_{j_e}\int_{\SU(2)} [dh_edk_e]\,
\psi(k_e)\,\chi^{j_e}(h_ek_e)\,\Theta^{\cI_e}(B_{s(e)}^{-1}G_eB_{t(e)}h_e).
\label{lift}
\ee
$\Delta_{j}$ is a weight depending on the spin $j$ that we will uniquely fix below by requiring that $\cM\vphi=\psi$ or more explicitly $\vphi(g_e,\id)=\psi(g_e)$. The label $\cI_e$ is an arbitrary function of the spin $j_e$ and it does not need to be the same for all the edges $e$ of the graph. The only constraint is that the $\SU(2)$-irrep $j_e$ needs to be in the $\SL(2,\C)$-irrep $\cI_e$, i.e we require that $n_e\le j_e$ always (more exactly, $j_e\in n_e+\N$).

First, we check that the constructed function is invariant under $\SU(2)$ shifts $B_v\arr B_vh_v$. This is true thanks to the $\SU(2)$ invariance of the original function $\psi$.
Then, we easily see that this function is invariant under Lorentz transformations acting simultaneously on both $G_e$ and $B_v$.
Finally, we would like to ensure that $\vphi$ is a proper lifting of $\psi$, i.e that $\cM\vphi=\psi$. To check this, we compute straightforwardly the value of $\vphi$ for $G_e=g_e\in\SU(2)$ and $B_v=\id$:
\be
\vphi(g_e,\id)
\,\equiv\,
\sum_{\{j_e\}}\Delta_{j_e}\int_{\SU(2)} [dh_edk_e]\,
\psi(k_e)\,\chi^{j_e}(h_ek_e)\,\Theta^{\cI_e}(g_eh_e).
\ee
As we reviewed earlier, we can express the $\SL(2,\C)$-character in term of the $\SU(2)$ characters when evaluated on $\SU(2)$ group elements:
$$
\Theta^{(n_e,\rho_e)}(g_eh_e)
\,=\,
\sum_{l_e\in n_e+\N}\chi^{l_e}(g_eh_e).
$$
We can then proceed to the integration over $h_e$ using the known convolution formula\footnotemark{ } for $\SU(2)$-characters :
\footnotetext{
The convolution formula for $\SU(2)$-characters is:
$$
\int_{\SU(2)} dh\,\chi^{j}(hk)\chi^{l}(gh)=\,\f{\delta_{j,l}}{d_j}\,\chi^j(gk^{-1}),
$$
where $d_j=(2j+1)$ is the dimension of the $\SU(2)$-irrep of spin $j$. This follows from the orthonormality of matrix elements with respect to the Haar measure. When $g=k$ in particular, we recover the usual orthonormalization condition for characters $\int \chi^j\chi^l=\delta_{j,l}$.
}
$$
\vphi(g_e,\id)
\,=\,
\int_{\SU(2)} [dk_e]\,
\psi(k_e)\,\prod_e\sum_{j_e}\f{\Delta_{j_e}}{d_{j_e}}\,\chi^{j_e}(g_ek_e^{-1})
\,=\,
\psi(g_e),
$$
as long as we fix the weights $\Delta_j\,\equiv\, d_j^2=(2j+1)^2$ in order to recover the $\delta$-distribution, $\sum_j d_j\chi^j(gk^{-1})=\delta(gk^{-1})$.

\medskip

Finally, we have checked that our formula \Ref{lift} correctly defines a lift of $\SU(2)$ cylindrical functions to Lorentz-invariant projected cylindrical functions and properly inverses the projection map $\cM$. The parameters of this lifting are a choice of $\cI_e$ irrep label for each spin $j_e$ on each edge $e$. There have been two typical choices for this parameter in the spin foam literature reviewed in Sections \ref{BCLorentz} and \ref{EPRGupta}:
\begin{itemize}

\item {\bf The Barrett-Crane ansatz}: $n_e=0$ for all spins $j_e$ on all edges \\

This restricts to irreps of the type $(0,\rho)$  used in the (Lorentzian) Barrett-Crane model \cite{BC2,bcl_carlo}. Let us emphasize that the label of the $\SL(2,\C)$ $n_e$ is not the spin $j_e$, which can still vary freely. If we further fix $j_e=0$, then we recover the simple spin networks usually used as boundary states of the Barrett-Crane model. Nevertheless, our analysis here suggests that we should {\it not} proceed to such a restriction and we would have a Hilbert space of projected spin networks for the BC model which would be isomorphic to the space of $\SU(2)$ spin networks. This interpretation of the BC model in term of projected spin networks and time-normals was already pushed forward in \cite{projected,clqg1,morecoupling}. In particular, in \cite{morecoupling}, it was speculated that spins $j_e\ne 0$ would correspond to particle insertions in the Barrett-Crane model, but we will not pursue in this direction.

\item {\bf The EPRL-FK ansatz}: $n_e=j_e$ for all spins $j_e$ on all edges \\

This is the condition to build $\SL(2,\C)$ coherent states  used in the construction of the Lorentzian spinfoam models of the EPRL-FK type \cite{EPRL,FK,roberto}.
We will study this case in details in the next section, and see how the Immirzi parameter enters our definition of the inverse lift.

\end{itemize}

\section{Simplicity Constraints and the Immirzi Parameter}

\subsection{Weak Constraints}

Following the approach used for constructing the EPRL-FK spinfoam models, we look at weak constraints that are satisfied by the projected spin network states \cite{EPR,EPRL,FK,LS}.
More precisely, we compare the matrix elements of the $\SU(2)$ rotation generators $\vJ$ and of the boost generators $\vK$ \cite{EPR,EPRL,eprl_ding}. The simplicity constraints amounts to requiring that the matrix elements of these two operators are the same up to a global factor, which would be identified as the Immirzi parameter.

We start with $\SU(2)$ spin network states $\psi$ and $\tpsi$, which we lift to projected spin networks $\vphi$ and $\tvphi$ using the same mapping i.e the same choice of $\SL(2,\C)$ irreps.
Then considering a fixed edge $e$, let us start by looking at the matrix elements of the left action of the boost generators $\vK_e$ on these projected spin networks:
\bes
\la \vphi | \vK_e^{(L)} | \tvphi\ra
&\equiv&
\int[dG_e]\,
\overline{\phi}(G_e)\vK_e\vartriangleright_L\tphi(G_e)\\
&=&
\int[dG_e][dh_ed\th_edk_ed\tk_e]\,
\overline{\psi}(k_e)\tphi(\tk_e)\,
\prod_e\sum_{j_e} d_{j_e}^2d_{\tj_e}^2\chi^{j_e}(h_ek_e)\chi^{\tj_e}(\th_e\tk_e)
\overline{\Theta^{(n_e,\rho_e)}}(G_eh_e)
\Theta^{(n_e,\rho_e)}(\vK_e G_e\th_e). \nn
\ees
The integral over the $\SL(2,\C)$ group elements $G_e$ can be done using the orthonormality of the $\SL(2,\C)$ matrix elements with respect to the Haar measure and give $\Theta^{(n_e,\rho_e)}(\vK_e h_e^{-1}\th_e)$ up to a measure factor depending solely on $(n,\rho)$.
Let us have a closer look at this term:
$$
\Theta^{(n_e,\rho_e)}(\vK_e h_e^{-1}\th_e)
\,=\,
\sum_{l_e,m_e} \la (n_e,\rho_e) l_e m_e| \vK_e h_e^{-1}\th_e |(n_e,\rho_e) l_e m_e\ra.
$$
First, the group variable $h_e^{-1}\th_e$ is in the $\SU(2)$ subgroup and therefore doesn't change the spin $l_e$. Thus only the matrix elements of the boost generators $\vK_e$ in the $\SU(2)$-irrep of spin $l_e$ matter. Next, due to the integration over $h_e$ and the insertion of the character $\chi^{j_e}(h_ek_e)$, only the component $l_e=j_e$ enters the calculation of the expectation value above. Similarly, the integration over $\th_e$ and the insertion of the character $\chi^{\tj_e}(\th_e\tk_e)$ forces $l_e=\tj_e=j_e$. Finally, we refer to the explicit action of the boost and rotation generators in a $(n,\rho)$-irrep given in \Ref{actionJ} and \Ref{actionK},
\be
\forall\, l,m,m',\quad
\la (n,\rho) l,m |\vK|(n,\rho) l,m' \ra
\,=\,
\beta_j^{(n,\rho)}\la (n,\rho) l,m |\vJ|(n,\rho) l,m' \ra,
\ee
where the coefficient $\beta_j$ is given in \Ref{actionbeta}. This was already noticed in \cite{clqg1,EPRL,eprl_ding}. We would like to use this fact in order to relate the values of the expectation values $\la \vphi | \vK_e^{(L)} | \tvphi\ra$ and $\la \vphi | \vJ_e^{(L)} | \tvphi\ra$. The obvious issue is that $\beta_{j_e}^{(n_e,\rho_e)}$ depends on $j_e$ and the precise choice of embedding $(n_e,\rho_e)$ chosen for each value of $j_e$.

\medskip

Considering the Barrett-Crane ansatz $n_e=0$ for all values of $j_e$, we get the trivial value of the proportionality coefficients, $\beta_{j_e}^{(0,\rho_e)}=0$. This leads to the identity:
\be
\textrm{{\bf Barrett-Crane ansatz} }n_e=0
\quad\Rightarrow\quad
\la \vphi | \vK_e^{(L)} | \tvphi\ra =0.
\ee
We do not consider this ansatz particularly useful, but at least worth mentioning considering the attention that the Barrett-Crane model has received over the past decade.

\medskip

The case of the EPRL-FK ansatz is much more interesting. We choose the maximal value for the label of the $\SL(2,\C)$ irrep, $n_e=j_e$. Then we would like to fix the value of the coefficients $\beta_{j_e}$ to a fixed value $\beta_e$ which does not depend on the value of the spin $j_e$ but only on the considered edge $e$. This leads to a unique solution for $\rho_e$ as a function of the spin $j_e$:
\be
n_e(j_e)=j_e,\quad
\rho_e(j_e)=\beta_e(j_e+1),\quad
\Rightarrow\quad
\beta_{j_e}^{(n_e,\rho_e)}=\f{n_e\rho_e}{j_e(j_e+1)}\,=\,\beta_e.
\ee
This leads to the final equality:
\be \label{EPRansatz}
\textrm{{\bf EPRL-FK ansatz} } (n_e,\rho_e)=(j_e,\beta_e(j_e+1))
\quad\Rightarrow\quad
\la \vphi | \vK_e^{(L)} | \tvphi\ra
\,=\,
\beta_e\,\la \vphi | \vJ_e^{(L)} | \tvphi\ra.
\ee
The same equality holds if considering  the right action of the boost and rotation generators.
This is exactly the (linear) simplicity constraints that are imposed in the EPRL-FK spinfoam model with Immirzi parameter $\beta_e$.
Let us underline that we do not need to choose the same proportionality coefficient $\beta_e$ for all edges $e$.

This is what is usually assumed in the EPRL-FK spinfoam model. However, in our framework, we are free to choose a different value $\beta_e$ for each edge of the graph, i.e a different value of the Immirzi parameter along the edges of the projected spin networks. This makes it more like an Immirzi field than an Immirzi parameter.

Finally, we introduce the precise lift inverting the projection map $\cM$ in the EPRL-FK ansatz. This lift is parameterized by a choice of coefficients $\{\beta_e\}\in\R^E$ for all edges of the graph. Then we define:
\be
\begin{array}{llcl}
\cL_{\{\beta_e\}}:
\,&H_S&\rightarrow&H_p \\
&\psi(g_e)&\mapsto&
\displaystyle{\vphi(G_e,B_v)=
\int_{\SU(2)} [dh_edk_e]\,
\psi(k_e)\,\sum_{j_e}d_{j_e}^2\,
\chi^{j_e}(h_ek_e)\,\Theta^{(j_e,\beta_e(j_e+1))}(B_{s(e)}^{-1}G_eB_{t(e)}h_e).}
\end{array}
\ee
As already shown in section \ref{lifting}, this provides us with a proper inverse for the map $\cM$:
\be
\forall \{\beta_e\},\quad
\forall \psi\in H,\quad
\cM\cL_{\{\beta_e\}}\psi\,=\,\psi.
\ee

\medskip

We can even go further by noticing by all possible values for $(n_e,\rho_e)\in\N/2\times\R$ are reached as $j_e$ and $\beta_e$ vary respectively in $\N/2$ and $\R$. Indeed, we can inverse the relations given above to get:
\be
j_e=n_e,\quad
\beta_e=\f{\rho_e}{j_e+1}.
\ee
This means that we can use the maps $\cL_{\{\beta_e\}}$ to obtain a full foliation of the Hilbert of (proper) projected spin network:
\be
H_p\,=\,
\bigoplus_{\{\beta_e\}\in\R^E}\,\cL_{\{\beta_e\}}H.
\ee
In words, this means that choosing arbitrary values of the Immirzi parameter $\beta_e$ for each edge of the graph, we will cover the whole space of proper projected spin networks by applying the lifting map $\cL_{\{\beta_e\}}$ to the standard $\SU(2)$ spin networks. We underline that we are restricted to {\it proper} projected spin networks since we always require that $j_e^s=j_e^t$ on all edges of the graph.

From the point of view of Loop Quantum Gravity's dynamics, we believe that the dynamical LQG operators would act on the Hilbert space $H$ of  standard $\SU(2)$ spin networks. This hints towards considering each subspace $\cL_{\{\beta_e\}}H$ of projected spin networks as {\it super-selection sectors} for the dynamics. A spinfoam model would then work in a given $\cL_{\{\beta_e\}}H$ subspace with all the parameters $\beta_e$ fixed, and would not mix these different sectors. Since spinfoam models are usually built for arbitrary graphs $\Gamma$, the simplest restriction would be to require that the Immirzi parameter be fixed and the same for all edges on all graphs, i.e $\beta_e=\beta,\,\forall e,\Gamma$. This can be obtained by imposing in addition the closure constraints (\ref{closureL}) and (\ref{closureK}) on the projected spin network $|\vphi\ra$, $|\tilde{\vphi}\ra$. (\ref{closureL}) holds strongly by defintion of a projected spin network and (\ref{closureK}) can only be satisfied weakly because of (\ref{EPRansatz}) as it is the case on the EPRL-FK boundary Hilbert space. This additional condition imposes that
\be
\beta_e= \beta \quad \forall e.
\ee Then we recover the boundary states for the usual (Lorentzian) EPRL-FK spinfoam models with fixed Immirzi parameter.

Nevertheless, our framework leaves us the freedom of attributing a different value of the Immirzi parameter for each edge of the graph. Let us speculate on the possibility that the Immirzi parameter provides us with a (length/area) scale which we would vary when coarse-graining or renormalizing LQG's transition amplitudes and dynamics. Then our framework for boundary states would allow to coarse-grain various regions of space independently.

\subsection{Strong Constraints}

From the perspective of the construction of spinfoam models, the weak constraints can be translated to strong constraints in the spirit of ``master constraints". The logic is to replace the weak constraints  $\la \vphi|\vK_e-\beta_e\vJ_e|\tvphi\ra=0$ by strong constraints using the $\SU(2)$ and $\SL(2,\C)$ Casimir operators \cite{EPR,EPRL}.

Considering the EPRL-FK ansatz, $n(j)=j$ and $\rho(j)=\beta\,(j+1)$, we can easily express the values of the $\SL(2,\C)$ Casimir operators in term of the $\SU(2)$ Casimir operator:
\be
\begin{array}{lcl}
C_2= &\vJ\cdot\vK & = 2n\rho=2\beta j(j+1)=2\beta \vJ^2 \\
C_1= &\vK^2 -\vJ^2 & = \rho^2-n^2+1=(\beta^2-1)j(j+1)+(\beta^2+1)(j+1)=
(\beta^2-1)\vJ^2+(\beta^2+1)(\sqrt{\vJ^2+\f14}+\f12).
\end{array}
\ee
The expression of the second quadratic Casimir looks much simpler and it is straightforward to check that the explicit definition that the projected spin networks $\vphi=\cL_{\{\beta_e\}}\psi$ indeed satisfy strong (simplicity) constraints:
\be
\forall \vphi=\cL_{\{\beta_e\}}\psi,\quad
\left(\vJ_e\cdot\vK_e\,-2\beta_e\vJ_e^2\right)\,\phi=0.
\ee
Here, it does not matter whether we consider the left or right action of the boost and rotation operators as long as we take them all as acting on the same side of the group variable $G_e$. Moreover, we wrote the constraint as acting on the section $\phi(G_e)=\vphi(G_e,\om)$. This constraint can be rotated by the suitable Lorentz transformations to apply it on the whole function $\vphi(G_e,B_v)$.

As long as we require by hand that $n_e=j_e$, this strong constraint is sufficient to impose that $\rho_e=\beta_e\,(j_e+1)$. However, in order to impose $n_e=j_e$ through an operator constraint as well, we need to impose the other constraint involving the first Casimir operator. The drawback is that this constraint involve a rather ugly ``quantum correction" term in $\sqrt{\vJ^2}$ operator, which is nevertheless necessary if we want an exact constraint at the quantum level.

\subsection{Comparing $\SU(2)$ and $\SL(2,\C)$ Scalar Products}

Since we have constructed a map between $\SU(2)$ spin networks and projected spin networks, it is natural to wonder if these lifts are unitary and preserve the scalar products. It is straightforward to see that this is a priori not the case. Indeed, considering two projected cylindrical functions, $\vphi$ and $\tvphi$, and their projections $\psi=\cM\vphi,\tphi=\cM\tvphi$, the scalar products are best expressed in term of the sections $\phi,\tphi$:
\bes
\la \vphi |\tvphi\ra &=&\int_{\SL(2,\C)} \overline{\phi}(G_e)\tphi(G_e),\\
\la \psi |\tpsi\ra_{\SU(2)} &=&\int_{\SU(2)} \overline{\phi}(g_e)\tphi(g_e). \nn
\ees
These two evaluations are a priori very different. This can be seen also from the scalar product between the basis states \Ref{scalarproj} and \Ref{scalarsu}:
$$
\la \vphi_{\cI_e,j_e^{s,t},i_v}| \vphi_{\tilde{\cI}_e,\tilde{j}_e^{s,t},\tilde{i}_v}\ra
\,=\,
\prod_e\f{\delta_{n_e,\tilde{n}_e}\delta(\rho_e-\trho_e)}
{\mu(n_e,\rho_e)}
\,\delta_{j_e^{s,t},\tilde{j}_e^{s,t}}
\,\prod_v \la i_v|\tilde{i}_v\ra,
\qquad
\la\psi_{j_e,i_v}|\psi_{\tj_e,\tilde{i}_v}\ra_{\SU(2)}
\,=\,
\prod_e\f{\delta_{j_e,\tilde{j}_e}}{d_{j_e}}
\,
\,\prod_v \la i_v|\tilde{i}_v\ra,
$$
which differ in their measure and normalization. The key difference is due to the extra $\delta$-functions due to the $\SL(2,\C)$-irrep label, more specifically $\delta(\rho_e-\trho_e)$ which potentially could lead to divergences.

To illustrate this, we start with two $\SU(2)$ cylindrical functions $\psi,\tpsi$ and respectively apply the generalized lifts $\cL_{\{\beta_e\}}$ and $\cL_{\{\tbeta_e\}}$ (the following analysis remains the same for two lifts $\cL_\beta$ and $\cL_{\tilde{\beta}}$). Then a straightforward calculation leads to:
\bes
\la\cL_{\{\beta_e\}}\psi|\cL_{\{\tbeta_e\}}\tpsi\ra
&=&
\int_{\SU(2)} [dk_e d\tk_e]\,
\overline{\psi}(k_e)\tpsi(\tk_e)\,
\prod_e \sum_{j_e} \f{\Delta_{j_e}^2}{d_{j_e}^2}\f{\delta(\rho_e-\trho_e)}{(\rho_e^2+j_e^2)}\chi^{j_e}(k_e^{-1}\tk_e) \nn\\
&=&
\prod_e\delta(\beta_e-\tbeta_e)\,
\int_{\SU(2)} [dk_e d\tk_e]\,
\overline{\psi}(k_e)\tpsi(\tk_e)\,
\prod_e \sum_{j_e} \f{\Delta_{j_e}^2}{d_{j_e}^2(j_e+1)(\beta_e^2(j_e+1)^2+j_e^2)}\chi^{j_e}(k_e^{-1}\tk_e).
\ees
Assuming the standard definition $\Delta_{j_e}=d_{j_e}^2$ ensuring that the lifts $\cL_{\{\beta_e\}}$ correctly invert the projection map $\cM$, then it is clear that the two scalar products do not match.
Then the natural question is which scalar product (between the Lorentz scalar product and the $\SU(2)$ scalar product) should we use on our kinematical Hilbert space of boundary states? This question should ultimately not matter so much since the final physical scalar should a priori be neither of them. Nevertheless, it is a crucial issue when building spinfoam amplitudes.

\medskip

An alternative would be to give up the requirement that a lift should be the inverse of the projection map $\cM$, i.e give up the idea that the restriction of the projected cylindrical function to the $\SU(2)$ subgroup be equal to the original $\SU(2)$ cylindrical function. Then we can modify the definition of the weight $\Delta_{j_e}$ and choose the new renormalized value, which now depends on the value of the Immirzi parameter $\beta_e$:
\be
\Delta_{j_e}^{\beta_e}\,\equiv\,
d_{j_e}^2\,\sqrt{(j_e+1)(\beta_e^2(j_e+1)^2+j_e^2)}.
\ee
This would define modified lifting maps, which would still send $\SU(2)$ cylindrical functions onto projected cylindrical functions, but that would conserve scalar products. Indeed, explicitly defining the new maps,
\be
\begin{array}{llcl}
L_{\{\beta_e\}}:
\,&H_S&\rightarrow&H_p \\
&\psi(g_e)&\mapsto&
\displaystyle{\vphi(G_e,B_v)=
\int_{\SU(2)} [dh_edk_e]\,
\psi(k_e)\,\sum_{j_e}\Delta_{j_e}^{\beta_e}\,
\chi^{j_e}(h_ek_e)\,\Theta^{(j_e,\beta_e(j_e+1))}(B_{s(e)}^{-1}G_eB_{t(e)}h_e).}
\end{array}\,,
\ee
using the new definition of the weight $\Delta_{j_e}^{\beta_e}$ given above, we will have the exact equality:
\be
\la L_{\{\beta_e\}}\psi|L_{\{\tbeta_e\}}\tpsi\ra
\,=\,
\la \psi|\tpsi\ra_{\SU(2)}\,\prod_e \delta(\beta_e-\tbeta_e).
\ee
Let us insist on the fact that this lifting map will still send the basis of $\SU(2)$ spin networks on projected spin network states satisfying the EPRL-FK ansatz, but with a different normalization that the lifting maps $\cL_{\{\beta_e\}}$ inverting $\cM$.

Finally, the natural issue is which lifting maps should we use to send LQG's $\SU(2)$ cylindrical functions onto the projected cylindrical functions of spinfoam models: should we enforce the matching condition that the restriction of projected cylindrical function to the $\SU(2)$ subgroup be equal to  the $\SU(2)$ cylindrical function or should we simply require the matching of the two scalar products and the unitarity of the lifting?

\chapter{Simplicity constraints and the $\U(N)$ framework} \label{UNChap}
Let us now tackle another open problem in the spin foam quantization procedure which is  how to decide the "strength" with which the simplicity constraints turning BF theory into 4d gravity are imposed. In short, the first try has been to impose  all the simplicity constraints strongly, leading to the Barrett-Crane model (see chapter \ref{BCmodel}) which seems to be an over-constrained model.  A more recent proposal has been to  impose the cross simplicity constraints weakly whereas the diagonal simplicity constraints are still impose strongly, which had led to the EPRL-FK models (see chapter \ref{EPRLFK}).  In this chapter, we present results published in \cite{article3} obtained using the $\U(N)$ framework initially developed for $\SU(2)$ intertwiners \cite{UN1, UN2, UN3}. One major interest of this new point of view is that within this $\U(N)$ framework one can achieve a precise control on the "strength" with which we decide to impose the simplicity constraints.

In the next section, we start by giving a short review on the $\U(N)$ framework. Then, using these $\U(N)$ tools, we recall an alternative definition of  the $\SU(2)$ coherent states used in the section \ref{SUcoh} as well as  the definition of the $\U(N)$ coherent states and their link with the usual $\SU(2)$ coherent states. The end of the second section is devoted to technical results summarizing a first part of work regarding the analysis of the action of geometrical observables   on some $\U(N)$ coherent states, elements of the space of $\SU(2)$ intertwiners.

The second part of this  work consisted in recasting the simplicity constraints for 4d gravity within the $\U(N)$ framework.
The first idea has been to follow the line of the EPRL-FK model, that is to impose weakly the cross simplicity constraints but to keep strong diagonal simplicity constraints. Different closed algebra of simplicity constraints used to solve weakly the cross simplicity constraints are identified. Details as well as the analysis of their advantages and disadvantages can be found in the third section of this chapter.  

The final proposition is exposed in the last section of this chapter. It emphasizes  the fact that all simplicity constraints should be treated on the same footing and  give a solution in terms of $\U(N)$ coherent states in this case. Moreover, this result takes into account the Immirzi parameter $\gamma$. The different results  are therefore  relevant for the definition of a spin foam model for 4 dimensional Euclidean gravity with an arbitrary value of the Immirzi parameter.  
The investigation of this approach  has been done only for the Euclidean case, i.e. constrained BF theory with gauge group $\Spin(4)$  and the corresponding intertwiners.  
 
 \medskip 
 
 We recall that in the case of Euclidean 4d gravity, the Hilbert space of quantum states of a 3-cell of the boundary cellular decomposition,  before having implemented the simplicity constraints, is the space of $\Spin(4)$  intertwiners between the representations attached to its faces or equivalently to the legs of the dual vertex of this 3-cell.
Since $\Spin(4)\sim\SU_L(2)\times\SU_R(2)$ is the direct product of its two $\SU(2)$ subgroups,
the irreducible representations of $\Spin(4)$  are labeled by a couple of half-integers $(j^L,j^R)$. So  a pair of spin to every leg $i$ is attached and the intertwiner space for the vertex is the tensor product of the space of $\SU_L(2)$ intertwiners between the spins $j^L_i$ and the space of $\SU_R(2)$ intertwiners between the spins $j^R_i$
\be
H_{j^L_i,j^R_i}\,\equiv\,
\textrm{Inv}\left[\bigotimes_\Delta V^{j^L_i}\right]\,\otimes\,
\textrm{Inv}\left[\bigotimes_\Delta V^{j^R_i}\right].
\ee
We still need to implement the simplicity constraints
\begin{itemize}
\item The {\it diagonal simplicity constraints} obtained when the legs are the same $i=j$~,
\be \label{diagI}
\forall i,\quad
(\vJ^L_i)^2
\,=\,
\rho^2(\vJ^R_i)^2,
\ee 
with $\rho=\f{\gamma+1}{|\gamma-1|}$ the proportionality coefficient between the left  and the right parts of the scalar products and $\gamma$ $(\gamma >0, \gamma\neq 1)$ the Immirzi parameter.
\item The {\it crossed simplicity constraints} in the case that the two legs are different $i\neq j$,
\be \label{crosseddiagI}
\forall i\ne j,\quad
(\vJ^L_i\cdot\vJ^L_j )
\,=\,
\rho^2(\vJ^R_i\cdot\vJ^R_j ).
\ee
\end{itemize}

\section{$\SU(2)$ intertwiners and the $\U(N)$ framework}
As recalled above, the discrete simplicity constraints are usually formulated in term of the scalar product of $\SU(2)$ generators.  
These scalar product operators, $\vJ_i\cdot\vJ_j$, are the standard invariant operators on the space of $\SU(2)$ intertwiner. They act on pairs of legs $(i,j)$ of a $\SU(2)$ intertwiner.  An important issue is that the algebra of the scalar product operators does not close. More precisely, the commutator of two scalar product operators gives a operator of order 3 in the $\vJ$'s:
\be
[ \vec{J}_i\cdot \vec{J}_j\,, \, \vec{J}_i\cdot \vec{J}_k]=\,i\vec{J}_i\cdot(\vec{J}_j\wedge \vec{J}_k),
\ee
and so on generating higher and higher order operators. This leads directly to problems when one tries to define coherent intertwiner states minimizing the corresponding uncertainty relations or when one attempts to solve constraints such as the simplicity constraints.  The approach followed in \cite{UN1} was to use Schwinger's representation of the $\su(2)$ algebra in term of harmonic oscillators to identify a new family of invariant operators, whose Lie algebra closes and which would still generate the full algebra of invariant operators acting on the intertwiner space. This leads to the $\U(N)$ framework for $\SU(2)$ intertwiners \cite{UN1,UN2,UN3}, which actually goes beyond this initial objective and offers a whole new perspective on the intertwiner space. It shows that the intertwiner space carries a natural representation of the $\U(N)$ unitary group and allows to build semi-classical coherent states transforming consistently under the $\U(N)$ action. It also uncovers a deep relation between the intertwiner space and the Grassmannian spaces, which could prove very useful to understand the geometry of the intertwiner space and its (semi-)classical interpretation. We review this formalism below.

\medskip

Let us consider the Hilbert spaces of intertwiners between $N$
irreducible $\SU(2)$-representations of spin $j_1,..,j_N$~:
\be
\cH_{j_1,..,j_N}\,\equiv\, \textrm{Inv}[V^{j_1}\otimes..\otimes V^{j_N}].
\ee
We further introduce the space of intertwiners with $N$ legs and fixed total area $J=\sum_i j_i$~:
\be
\cH_N^{(J)}\,\equiv\,\bigoplus_{\sum_i j_i=J}\cH_{j_1,..,j_N},
\ee
and the full Hilbert space of $N$-valent intertwiners:
\be
\cH_N\,\equiv\,\bigoplus_{\{j_i\}}\cH_{j_1,..,j_N}\,=\, \bigoplus_{J\in\N}\cH_N^{(J)}.
\ee
The key result of the $\U(N)$ formalism is that there is a natural action of $\U(N)$ on the intertwiner space $\cH_N$ \cite{UN1}. More precisely the intertwiner spaces $\cH_N^{(J)}$
carry irreducible representations of $\U(N)$ \cite{UN2}. Finally the full space $\cH_N$ can be endowed with a Fock space structure with creation and annihilation operators compatible with
the $\U(N)$ action \cite{UN3}.

\begin{figure}[h]
\begin{center}
\includegraphics[height=50mm]{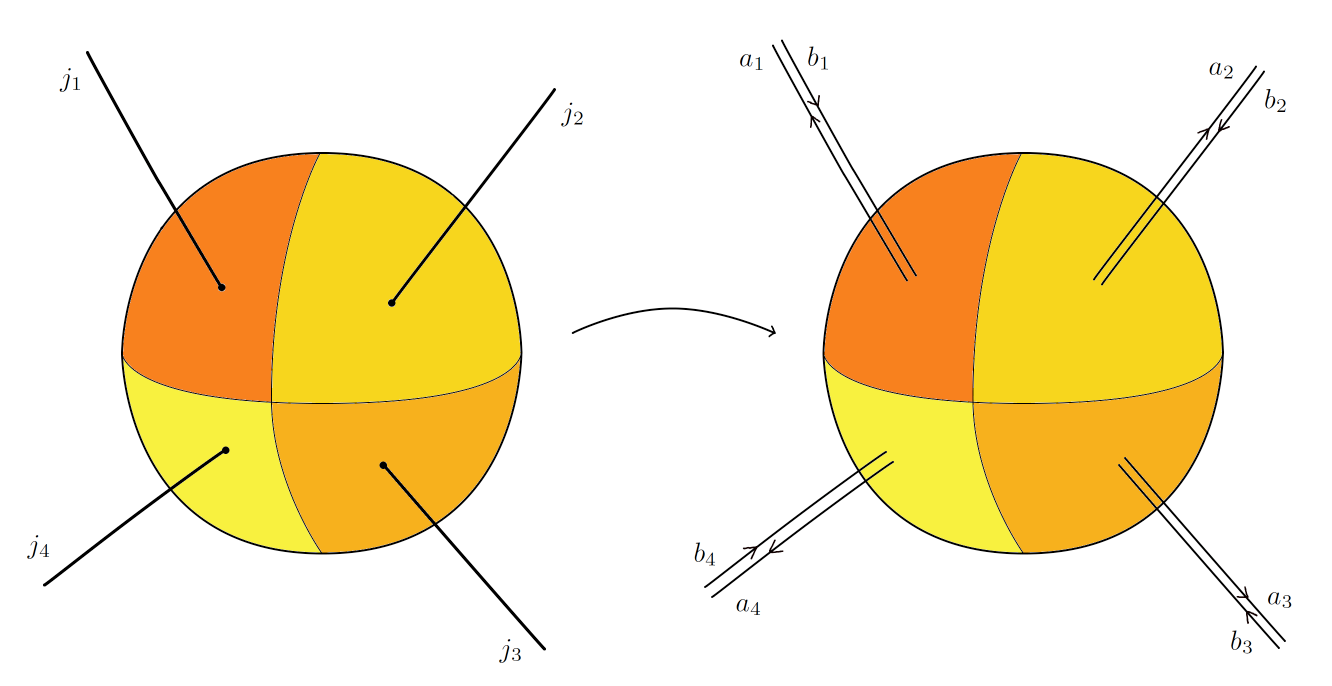}
\end{center}
\caption{Four legs intertwiner  and harmonic oscillators.}
\end{figure}

This $\U(N)$ formalism is based on the Schwinger representation of the $\su(2)$ Lie algebra in term of harmonic oscillators. Let us introduce $2N$ oscillators with creation operators $a_i,b_i$ with
$i$ running from 1 to $N$:
$$
[a_i,a\dag_j]=[b_i,b\dag_j]=\delta_{ij}\,,\qquad [a_i,b_j]=0.
$$
The generators of the $\SU(2)$ transformations acting on each leg of
the intertwiner are realized as quadratic operators in term of the
oscillators:
\be
J^z_i=\f12(a\dag_i a_i-b\dag_ib_i),\qquad
J^+_i=a\dag_i b_i,\qquad
J^-_i=a_i b\dag_i,\qquad
E_i=(a\dag_i a_i+b\dag_ib_i).
\ee
The $J_i$'s satisfy the standard commutation algebra while the total
energy $E_i$  is a Casimir operator:
\be
[J^z_i,J^\pm_i]=\pm J^\pm_i,\qquad
[J^+_i,J^-_i]=2J^z_i,\qquad
[E_i,\vec{J}_i]=0.
\ee
The correspondence with the standard $|j,m\ra$ basis of $\su(2)$
representations is simple:
\be
|n_a,n_b\ra_{HO}=|\f12(n_a+n_b),\f12(n_a-n_b)\ra\,,\qquad
|j,m\ra=|j+m,j-m\ra_{HO}
\ee
where $m$ is the eigenvalue of $J^z$ defined as
the half-difference of the energies between the two oscillators,
while the total energy $E_i$ gives twice the spin, $2j_i$, living on
the $i$-th leg of the intertwiner.

Intertwiner states are by definition invariant under the global
$\SU(2)$ action, generated by:
\be
J^z=\sum_{i=1}^N J^z_i,\qquad
J^\pm=\sum_i J^\pm_i.
\ee
Then operators acting on the intertwiner space need to commute with
these operators too. The simplest family of invariant operators was
identified in \cite{UN1} and are quadratic operators acting on
couples of legs:
\be
E_{ij}=a\dag_ia_j+b\dag_ib_j, \qquad
E_{ij}\dag=E_{ji}.
\ee
The main result is that these operators are invariant under
global $\SU(2)$ transformations and form a $\u(N)$ algebra:
\be
[\vec{J},E_{ij}]=0,\qquad
[E_{ij},E_{kl}]\,=\,
\delta_{jk}E_{il}-\delta_{il}E_{kj}.
\ee
The diagonal operators $E_i\equiv E_{ii}$ form the Cartan
sub-algebra of $\u(N)$, while the off-diagonal operators $E_{ij}$
with $i\ne j$ are the raising and lowering operators. As said
earlier, the generators $E_i$ give twice the spin $2j_i$ while the
$\U(1)$ Casimir $E=\sum_i E_i$ will give twice the total area,
$2J\equiv \sum_i 2j_i$. Then all operators $E_{ij}$ commute with the
$\U(1)$ Casimir, thus leaving the total area $J$ invariant:
\be
[E_{ij},E]=0.
\ee
The usual $\SU(2)$ Casimir operators have a simple expression in
term of these $\u(N)$ generators:
\be
(\vec{J}_i)^2=\f{1}{2}E_i\left(\f{E_i}{2}+1\right),\qquad
\forall i\ne j,\,\,
(\vec{J}_i\cdot \vec{J}_j)
\,=\,
\f12E_{ij} E_{ji} -\f14E_iE_j-\f12 E_i.
\label{scalaropE}
\ee
Let us point out that case $i=j$ of $(\vec{J}_i\cdot
\vec{J}_j)$ does not give back exactly the formula for
$(\vec{J}_i)^2$ due to the ordering of the oscillator operators. The
two formula agree on the leading order quadratic in $E_i$ but
disagree on the correction linear in $E_i$.

The next point is that explicit definition of the $E_{ij}$'s in term of
harmonic oscillators leads to quadratic constraints on these
operators as shown in \cite{UN2}~:
\be
\forall i,\quad
\sum_j E_{ij}E_{ji}=E_{i} \left(\f E2+N-2\right),
\ee
where we have assumed that the global $\SU(2)$ generators $\vec{J}$
vanish.
These quadratic constraints on the $E_{ij}$ operators lead to non-trivial restrictions on the representations of $\u(N)$ obtained from this construction. To solve them, the method
used in \cite{UN2} is to apply them to a highest weight vector.
This allows to identify the representations corresponding to the intertwiner spaces $\cH_N^{(J)}$ at fixed total area $J=\sum_i j_i$. The highest weight vector $v_N^{(J)}$ diagonalizes the generators of the Cartan sub-algebra $E_i$ and vanishes under the action of the raising operators $E_{ij}\,v=0$ for all $i<j$. The $N$ eigenvalues depend very simply on the area $J$~:
\be
E_1\,v_N^{(J)}=E_2\,v_N^{(J)}=J\,v_N^{(J)},\qquad
E_k\,v_N^{(J)}=0,\,\forall k\ge 3.
\ee
This highest weight vector corresponds to a bivalent intertwiner between two legs  both carrying the spin $\f J2$.
One can compute the corresponding value of the quadratic $\U(N)$ Casimir using the previous quadratic constraints:
\be
\sum_{i,j} E_{ij}E_{ji}=E\left(\f E2+N-2\right)= 2J(J+N-2),
\ee
and the dimension of $\cH_N^{(J)}$ in term of binomial coefficients using the standard formula for $\U(N)$ representations:
\be
D_{N,J}\,\equiv\,
\dim \cH^{(J)}_N
\,=\,
\f{1}{J+1}\bin{J}{N+J-1}\bin{J}{N+J-2}
\,=\,
\frac{(N+J-1)!(N+J-2)!}{J!(J+1)! (N-1)!(N-2)!}\,.
\label{dimNJ}
\ee

\medskip

Next, we introduce annihilation and creation operators to move
between the spaces $\cH_N^{(J)}$ with different areas $J$ within the
bigger Hilbert space of all intertwiners with $N$ legs \cite{UN3}.
We define the new operators:
\be
F_{ij}=(a_i b_j - a_j b_i),\qquad
F_{ji}=-F_{ij}.
\ee
These are still invariant under global $\SU(2)$ transformations, but
they do not commute anymore with the total area operator $E$. They nevertheless form a closed algebra together with the operators $E_{ij}$:
\bea
{[}E_{ij},E_{kl}]&=&
\delta_{jk}E_{il}-\delta_{il}E_{kj}\nn\\
{[}E_{ij},F_{kl}] &=& \delta_{il}F_{jk}-\delta_{ik}F_{jl},\qquad
{[}E_{ij},F_{kl}\dag] = \delta_{jk}F_{il}\dag-\delta_{jl}F_{ik}\dag, \\
{[} F_{ij},F\dag_{kl}] &=& \delta_{ik}E_{lj}-\delta_{il}E_{kj} -\delta_{jk}E_{li}+\delta_{jl}E_{ki}
+2(\delta_{ik}\delta_{jl}-\delta_{il}\delta_{jk}), \nn\\
{[} F_{ij},F_{kl}] &=& 0,\qquad {[} F_{ij}\dag,F_{kl}\dag] =0.\nn
\eea
The annihilation operators $F_{ij}$ allow to go from $\cH_N^{(J)}$ to
$\cH_N^{(J-1)}$ while the creation operators $F\dag_{ij}$ raise the
area and go from $\cH_N^{(J)}$ to $\cH_N^{(J+1)}$.
We can re-express the scalar product operators in term of this new set of operators:
\bes
\label{scalaropF}
\vec{J}_i\cdot \vec{J}_j&=& -\f12 F_{ij}^\dagger F_{ij} +\f14 E_iE_j,  \\
&=& -\f12 F_{ij}F_{ij}^\dagger +\f14 (E_i+2)(E_j+2). \nn
\ees
This formula is explicitly symmetric in the edge labels $i \leftrightarrow j$ contrary to the previous formula \Ref{scalaropE} in terms of the $E_{ij}$-operators.
Finally, as shown in \cite{UN3} and as we review below, these operators can be used to construct coherent states transforming consistently under $\U(N)$ transformations. These $\U(N)$ coherent states will turn out very convenient in order to re-express and solve the simplicity constraints.

\section{Coherent states and the $\U(N)$ framework}
\subsection{Revisiting the $\SU(2)$ Coherent Intertwiners}

%
%
%

To define coherent intertwiner states, we attach a spinor $z_i$ to each leg of the intertwiner:
$$
z_i=\mat{c}{z^0_i\\z^1_i}.
$$
Basically, the first component $z^0_i$ is the label of the coherent state for the oscillator $a_i$ while the second component $z^1_i$ corresponds to the oscillator $b_i$.
Let us first clear up the geometrical meaning of the spinors $z_i$. Considering a spinor $z$, the matrix $|z\ra\la z|$ is a $2\times 2$ matrix which can be decomposed on the Pauli matrices $\sigma_a$ (taken Hermitian and normalized so that $(\sigma_a)^2=\id$). This defines a 3-vector $\vec{V}(z)$:
\be \label{vecV}
|z\ra \la z| = \f12 \left( {\la z|z\ra}\id  + \vec{V}(z)\cdot\vec{\sigma}\right),
\ee
where the norm of the vector is $|\vec{V}(z)| = \la z|z\ra= |z^0|^2+|z^1|^2$ and its components are given explicitly as~\footnotemark:
\be
V^z=|z^0|^2-|z^1|^2,\qquad
V^x=2\,\Re\,(\bar{z}^0z^1),\qquad
V^y=2\,\Im\,(\bar{z}^0z^1).
\ee
\footnotetext{
Remember the convention for the $\pm$ generators:
$$
\sigma_\pm=\sigma_x\pm i\sigma_y,\quad
\sigma_x=\f12(\sigma_++\sigma_-), \quad
\sigma_y=-i\f12(\sigma_+-\sigma_-).
$$
}
The spinor $z$ is entirely determined by the corresponding 3-vector $\vec{V}(z)$ up to a global phase. Following \cite{UN3}, we also introduce the map $\varsigma$ between spinors:
\be
\varsigma\begin{pmatrix}z^0\\ z^1\end{pmatrix}
\,=\,
\begin{pmatrix}-\bar{z}^1\\\bar{z}^0 \end{pmatrix},
\qquad \varsigma^{2}=-1.
\ee
This is an anti-unitary map, $\la \varsigma z| \varsigma w\ra= \la w| z\ra=\overline{\la z| w\ra}$, and we will write the related state as
$$
|z]\equiv \varsigma  | z\ra.
$$
This map $\varsigma$ maps the 3-vector $\vec{V}(z)$ onto its opposite:
\be
|z][  z| = \f12 \left({\la z|z\ra}\id - \vec{V}(z)\cdot\vec{\sigma}\right).
\ee
Finally coming back to the intertwiner with $N$ legs, we have $N$ spinors and their corresponding 3-vectors $\vV(z_i)$. Typically, we can require that the $N$ spinors satisfy a closure constraint, $\sum_i \vec{V}(z_i)=0$. This can be written in term of $2\times 2$ matrices:
\be
\sum_i |z_i\ra \la z_i|=A(z)\id,
\qquad\textrm{with}\quad
A(z)\equiv\f12\sum_i \la z_i|z_i\ra=\f12\sum_i|\vec{V}(z_i)|.
\ee
It translates into quadratic constraints on the spinors:
\be
\sum_i z^0_i\,\bar{z}^1_i=0,\quad
\sum_i \left|z^0_i\right|^2=\sum_i \left|z^1_i\right|^2=A(z),
\label{closure}
\ee
which means that the two components of the spinors, $z^0_i$ and $z^1_i$, are orthogonal $N$-vectors of equal norm.

\medskip

Then we can define coherent intertwiner states \cite{coh1,coh2,coh3}. First, for a given leg, we define the  $\SU(2)$ coherent states labeled by the spin $j_i$ and the spinor $z_i$:
\be
|j_i,z_i\ra\,\equiv\,
\f{(z^0_ia\dag_i+z^1_ib\dag_i)^{2j_i}}{\sqrt{(2j_i)!}}\,|0\ra.
\ee
This vector clearly lives in the irreducible $\SU(2)$-representation of spin $j_i$ since it's an eigenvector of the energy $E_i$ with value $2j_i$. To show that it transforms coherently under $\SU(2)$, we compute the  $\SU(2)$ action. Dropping the index $i$, $\SU(2)$ rotations are parameterized by an angle $\theta$ and a unit 3-vector $\hat{v}\in\cS_2$:
\be
g(\theta,\hat{v})
\,\equiv\,
e^{i\theta \hat{v}\cdot\vec{J}}
\,=\,
e^{i\theta (v_zJ_z+\f{\bv}2J_++\f{\bar{\bv}}2J_-)},
\qquad
|\vec{v}|^2=v_z^2+|\bv|^2=1,\qquad
v_z=\cos\phi,\quad \bv=e^{i\psi}\sin\phi.
\ee
It is represented by a $2\times 2$ matrix in the fundamental spin-$\f12$ representation:
\be
g(\theta,\hat{v})
\,=\,
e^{i\theta \hat{v}\cdot\f{\vec{\sigma}}2}
\,=\,
\mat{cc}{\cos\f\theta2+i\cos\phi\sin\f\theta2 & i e^{i\psi}\sin\phi\sin\f\theta2 \\
i e^{-i\psi}\sin\phi\sin\f\theta2 & \cos\f\theta2-i\cos\phi\sin\f\theta2}
\,\in\SU(2).
\ee
To compute the action of $\SU(2)$, we first compute the following commutator:
\be
\left[\vec{v}\cdot \vJ, (z^0a\dag+z^1b\dag)\right]=
\left((\vec{v}\cdot\f{\vec{\sigma}}2)\, z\right)^0a\dag+\left((\vec{v}\cdot\f{\vec{\sigma}}2)\, z\right)^1b\dag,
\ee
which gets easily exponentiated:
\be
e^{i\theta \hat{v}\cdot\vec{J}}\,(z^0a\dag+z^1b\dag)\,e^{-i\theta \hat{v}\cdot\vec{J}}
=e^{[i\theta \hat{v}\cdot\vec{J},\cdot]}\,(z^0a\dag+z^1b\dag)
\,=\,
(\tz^0a\dag+\tz^1b\dag),
\qquad\textrm{with}\quad
\tz=e^{i\theta \hat{v}\cdot\f{\vec{\sigma}}2}\,z=g(\theta,\hat{v})\,z.
\ee
This shows that the states introduced above are proper $\SU(2)$ coherent states:
\be
g(\theta,\hat{v})\,|j,z\ra
\,=\,
|j,\,g(\theta,\hat{v})\,z\ra
\ee
This means that these are the standard $\SU(2)$ coherent states \`a la Perelomov. Indeed, one can always set $\tz^1$ to 0, or reversely get any arbitrary state from the initial state without any $b$-excitation. Such an initial state actually corresponds to the highest weight vector $|j,j\ra$ of the $\SU(2)$-representation of spin $j$. More precisely, we act on that highest weight vector with a $\SU(2)$ transformation parameterized by $\alpha$ and $\beta$ satisfying $|\alpha|^2+|\beta|^2=1$:
\be
|j,j\ra=|2j,0\ra_{HO}=\f{(a\dag)^{2j}}{\sqrt{(2j)!}}\,|0\ra\,\arr\,
\mat{cc}{\alpha &\beta \\ -\bar{\beta}&\bar{\alpha}}\,|j,j\ra
\,=\,
|j,\,\mat{c}{\alpha \\ -\bar{\beta}}\ra,
\ee
\be
|j,z\ra=\,\la z|z\ra^j
\mat{cc}{\f{z^0}{\sqrt{\la z|z\ra}} &\f{-\bar{z}^1}{\sqrt{\la z|z\ra}}  \\ \f{z^1}{\sqrt{\la z|z\ra}} &\f{\bar{z}^0}{\sqrt{\la z|z\ra}} }\,|j,j\ra.
\ee
%
We also give the scalar product between two such $\SU(2)$ coherent states:
\be
\la j,w|j,z\ra=\la w|z\ra^{2j},
\ee
and the expectation values of the $\su(2)$ generators:
\be
\la J_z \ra\,\equiv\,
\f{\la j,z|J_z|j,z\ra}{\la j,z|j,z\ra}
\,=\,
j\,\f{|z^0|^2-|z^1|^2}{|z^0|^2+|z^1|^2},
\quad
\la J_+ \ra\,=\,
2j\,\f{\bar{z}^0z^1}{|z^0|^2+|z^1|^2},
\quad\Rightarrow\quad
\la \vJ \ra\,=\,
\,=\,
j\,\f{\vV}{|\vV|},
\ee
as expected.
Finally, expanding these states explicitly on the standard basis for harmonic oscillators,
$$
|j,z\ra\,=\,\sum_{n=0}^{2j}\sqrt{\bin{n}{2j}}\,(z^0)^n(z^1)^{2j-n}\,|n,2j-n\ra_{HO},
$$
and following the usual calculation done with oscillator coherent states (as shown in appendix \ref{cohHO}), we can decompose the identity on the Hilbert space $V^j$ in term of these $\SU(2)$ coherent states:
\be
\id_j \,=\,
\sum_{n=0}^{2j} |n,2j-n\ra_{HO}\,{}_{HO}\la n,2j-n|
\,=\,
\f{1}{(2j)!}\int [d^2 z^0d^2z^1]\,\f{e^{-\la z|z\ra}}{\pi^2}\,|j,z\ra\la j,z|.
\ee
One can check that taking the trace of this expression and using the formula for the scalar product between coherent states give back as expected $\tr \,\id_j=(2j+1)$. Let us emphasize a last point that the projector $|j,z\ra\la j,z|$ does not depend on the overall phase of the spinor $z$ but only on the corresponding 3-vector $\vV(z)$.

Coherent intertwiners are then defined following \cite{coh1} by group averaging the tensor product of $\SU(2)$ coherent states attached to every leg of the intertwiner:
\be
||\{j_i,z_i\}\ra
\,\equiv\,\int_{\SU{2}}dg\, g\rhd\bigotimes_{i=1}^N|j_i,z_i\ra.
\ee
These are the standard coherent intertwiners used in the construction of the EPRL-FK spinfoam models and their boundary states \cite{FK,LS}. Following the logic of \cite{coh1}, we can write the identity on the intertwiner space $\cH_{j_1,..,j_N}$ in term of these coherent intertwiners:
\be
\label{idcohint}
\id_{\cH_{j_1,..,j_N}}
\,=\,
\int \prod_i \f{e^{-\la z_i|z_i\ra}\,d^4z_i}{(2j_i)!\pi^2}\,
||\{j_i,z_i\}\ra\la\{j_i,z_i\}||,
\ee
where we used the fact that the spinor norm $\la z|z\ra$ is invariant under the $\SU(2)$ action. Finally, the main point shown in \cite{coh1} is that this integral is dominated by intertwiners satisfying the closure constraint $\sum_i j_i \vV(z_i)/|\vV(z_i)|=0$ while the norm of intertwiners not satisfying this closure constraint is exponentially suppressed. It is also possible to write a decomposition of the identity on the intertwiner space restricting the integral to intertwiners satisfying exactly the closure constraint by modifying slightly the measure \cite{coh2,coh3}. This is achieved through considering and gauging out the $\SL(2,\C)$ action on spinors complexifying the previous $\SU(2)$ action.

\medskip

\subsection{The $\U(N)$ coherent states}
We are now ready to define the $\U(N)$ coherent states. Their definition is slightly more complicated. Following \cite{UN3}, we define the following antisymmetric matrix $Z_{ij}$, which is holomorphic in the spinors $z_i$  and anti-symmetric in $i\leftrightarrow j$~:
\be
Z_{ij}\,\equiv\,
[z_i|z_j\ra
\,=\,
(z^0_iz^1_j-z^0_jz^1_i),
\ee
and the corresponding creation operator:
\be
F\dag_Z
\,\equiv\,
\f12\sum Z_{ij} F\dag_{ij}
=\f12\sum (z^0_iz^1_j-z^0_jz^1_i)\, F\dag_{ij},
\ee
which is also holomorphic in $z$.
A crucial property of this matrix $Z$ is the Pl\"ucker relation $Z_{ik} Z_{jl}-Z_{il} Z_{jk}= Z_{ij} Z_{kl}$ which holds for any indices $i,j,k,l$.
The $\U(N)$ coherent states are then labeled by the total area $J$ and the $N$ spinors $z_i$:
\be
|J,\{z_i\}\ra\,\equiv\,
\f{1}{\sqrt{J!(J+1)!}}\,(F\dag_Z)^J\,|0\ra.
\ee
This state clearly lives in $\cH_N^{(J)}$ since it is an intertwiner (invariant under global $\SU(2)$ transformation) and is an eigenvector of the total area operator $E$ with value $2J$.
Notice that the behavior under rescaling of this coherent state is very simple:
\be
z_i\arr \lambda z_i,\quad
Z_{ij}\arr \lambda^2 Z_{ij},\qquad
|J,\{\lambda z_i\}\ra =\lambda^{2J}\,|J,\{z_i\}\ra.
\ee

Now we assume that the spinors $z_i$ satisfy exactly the closure condition $\sum_i \vV(z_i)=0$ introduced earlier in \Ref{closure}. We can compute the norm of these states:
\be
\la J,\{z_i\}|J,\{z_i\}\ra=(A(z))^{2J}
\,=\,
\left(\f12\sum_i \la z_i|z_i\ra\right)^{2J}
\,=\,
\left(\f12\sum_i |\vec{V}(z_i)|\right)^{2J}.
\ee
Then, as shown in \cite{UN3}, these states are coherent under the action of $\U(N)$:
\be
\forall u\in\U(N),\quad
\hat{u}\,|J,\{z_i\}\ra
\,=\,
|J,\{(uz)_i\}\ra,
\ee
where $\hat{u}$ is the operator representing the unitary transformation $u$, that is for an arbitrary anti-Hermitian matrix $\alpha$~:
\be
u=e^\alpha \quad\rightarrow\quad
\hat{u}\equiv e^{E_\alpha}\equiv e^{\sum_{ij} \alpha_{ij}E_{ij}}.
\ee
The $\U(N)$ action on the $N$ spinors  is the natural one:
\be
z_i\,\rightarrow\, (uz)_i=\sum_j u_{ij}z_j,\qquad
z\arr uz,\quad Z\arr uZu^t.
\ee
This $\U(N)$-action is proved by computing explicitly the action of $\hat{u}$ on the $F\dag$-operators \cite{UN3}:
\be
[E_\alpha,F\dag_Z]=F\dag_{\alpha Z + Z\alpha^t}\quad\Rightarrow\quad
e^{E_\alpha}F\dag_Ze^{-E_\alpha}=F\dag_{e^\alpha Z e^{\alpha^t}}.
\ee
Here is a summary of the properties  of these $\U(N)$ coherent states already proved in \cite{UN3}:
\begin{itemize}
\item They transform simply under $\U(N)$-transformations:
$u\,|J,\{z_i\}\ra\,=\,|J,\{(u\,z)_i\}\ra$. This key property actually holds also if the spinors do not satisfy the closure condition.

\item They are coherent states {\it \`a la} Perelomov  and are obtained by the action of $\U(N)$ on highest weight states. These highest weight vectors correspond to bivalent intertwiners such as the state defined by the spinors $z_1=(z^0,z^1)$, $z_2=\varsigma z_1=(-\bar{z}^1,\bar{z}^0)$ and $z_k=0$ for ${k\ge 3}$. This only holds if one assumes that the spinors satisfy the closure constraint. Indeed, $\U(N)$ transformations conserve the closure condition and the spinors defining the bivalent intertwiner satisfy it.

\item For large areas $J$, they are semi-classical states peaked around the expectation values for the $\u(N)$ generators and the scalar product operators:
    \be
    \la E_{ij}\ra= 2J\,\f{\la z_i|z_j\ra}{\sum_k \la z_k|z_k\ra}=\f{J}{A(z)}\la z_i | z_j\ra,
    \ee
    \be \label{expectJJ}
    \forall i\ne j,\quad 4\,\la \vJ_i\cdot\vJ_j\ra\,=\, \f{J^2}{A(z)^2}\vV(z_i)\cdot\vV(z_j)
    +\f{J}{2\,A(z)^2} \left(\vV(z_i)\cdot\vV(z_j)-3|\vV(z_i)|\,|\vV(z_j)|\right).
    \ee

\item The scalar product between two coherent states is easily computed:
$$
\la J,\{z_{i}\}|J,\{w_{i}\}\ra = \mathrm{det}\left(\sum_{i}|z_{i}\ra\la w_{i}|\right)^{J}
\,=\,
\left(\f12\tr \,Z\dag W\right)^J.
$$

\item They minimize the uncertainties on the $E_{ij}$ operators. The interested reader can find the various uncertainties computed in \cite{UN3}. The simplest is the $\U(N)$-invariant uncertainty:

\be
\Delta
\,\equiv\,
\sum_{ij}  \la E_{ij}E_{ji}\ra- \la E_{ij}\ra \la E_{ji}\ra
\,=\,
2J\,(J+N-2)-2J^2
\,=\,
2J\,(N-2).
\ee


\item They are related to the coherent  intertwiners discussed above:
    \be
    \f{1}{\sqrt{J!(J+1)!}}|J,\{z_i\}\ra
    \,=\,
    \sum_{\sum j_i=J}
    \frac{1}{\sqrt{(2j_{1})!\cdots (2j_{N})!}} \,||\{j_i,z_i\}\ra.
    \label{sumcoh}
    \ee

\item The coherent states $|J,\{z_i\}\ra$ at fixed $J$ provide an over-complete basis on the space $\cH_N^{(J)}$. This gives a new decomposition of the identity on that space $\id_N^{(J)} = \int [d\mu(z_i)]\,|J, \{z_{i}\}\ra\la J,\{z_{i}\}|$ where $[d\mu(z_i)]$ is a $\U(N)$-invariant measure (on $\C\mathbb{P}_{2N-1}$). For more details, the interested reader can refer to \cite{UN3}.

\end{itemize}

\subsection{Relaxing the Closure Conditions}

Discussing the $\U(N)$ coherent states in the previous section, we have assumed that the spinor labels satisfy the closure condition \Ref{closure} that requires that $\sum_i \vV(z_i)=0$ or equivalently that $\sum_i |z_i\ra\la z_i| \propto \id$, or even equivalently that the two components of the spinors $z^0_i$ and $z^1_i$ are orthonormal $N$-vectors. It has been shown in \cite{UN3} how to relax this closure condition using the $\SL(2,\C)$ invariance of the coherent states. Let us review this procedure.

We consider the $\GL(2,\C)$ action acting simultaneously on all spinors $z_i$. It has a simple rescaling action on the $Z_{ij}$ matrix, which means that the $\U(N)$ coherent states also get simply rescaled:
\be
\forall \Lambda\in\GL(2,\C),\quad z_i\arr \Lambda z_i,\qquad
Z_{ij}\arr \det\Lambda\, Z_{ij},\qquad
|J,\{z_i\}\ra\arr (\det\Lambda)^J\,|J,\{z_i\}\ra\,.
\ee
Thus two coherent states labeled by spinors related through a $\GL(2,\C)$ action define the same quantum state, up to normalization. In particular, if the transformation $\Lambda$ lies in $\SL(2,\C)$ then the coherent state is exactly the same. The moot point is that $\GL(2,\C)$ transformations allow to go in and out of the closure constraint. Indeed, following \cite{UN3}, given an arbitrary set of $N$ spinors, we consider the matrix:
\be
X(z)\,\equiv\,\sum_i |z_i\ra\la z_i|.
\ee
Since $X(z)$ is obvious a positive Hermitian matrix, there exists a matrix $\Lambda\in\SL(2,\C)$ which takes its square-root, $X=\sqrt{\det X}\,\Lambda\Lambda\dag$. This matrix is unique up to $\SU(2)$ transformations. It allows to define a new set of spinors $\tz_i\equiv \Lambda^{-1}\,z_i$ which induce the same coherent state but also satisfy the closure condition:
\be
\tX=\sum_i |\tz_i\ra\la \tz_i|
\,=\,
\Lambda^{-1}\,X\,(\Lambda\dag)^{-1}
\,=\,
\sqrt{\det X}\,\id,\qquad
\det\tX=\det X.
\ee
This is the exact same $\SL(2,\C)$ action used in \cite{coh2,coh3} to take the standard coherent intertwiners in and out of the closure constraint. Let us point out that the $\SL(2,\C)$ action is simply the complexified $\SU(2)$-action still generated by the operators $J^{z,\pm}$ quadratic in the harmonic oscillators. In the $\U(N)$ framework, this simply mean that we can drop the closure condition on the spinor label, when defining $\U(N)$ coherent states and integrating over spinor labels, e.g. in the decomposition of the identity.
Moreover, the coherent states $|J,\{z_i\}\ra$ still transform covariantly under $\U(N)$ whether they satisfy the closure condition or not, and their norm is easily computed:
\be
\la J,\{z_i\}|J,\{z_i\}\ra\,=\, (\det X)^J.
\ee
Since the projectors $|z_i\ra\la z_i|$ are easily expressed in term of the classical 3-vectors $\vec{V}(z_i)$, we give similar expressions for the matrix $X$ and its determinant:
\be
X=\f12\left(
\sum_i |\vec{V}(z_i)|\,\id
+\sum_i \vec{V}(z_i)\cdot\vec{\sigma}
\right)
\quad\Rightarrow\quad
\det X
\,=\,
\f14\left(
\left(\sum_i |\vec{V}(z_i)|\right)^2 -\left|\sum_i \vec{V}(z_i)\right|^2
\right),
\ee
so that $(\det\,X)^J=A(z)^{2J}$ as before when the closure condition $\sum_i \vec{V}(z_i)=0$ is satisfied. Let us underline that $\det\,X\ge0$ can be interpreted as a measure of how far from the closure condition we are: the larger the total 3-vector $\sum_i \vec{V}(z_i)$ is, the smaller $\det\,X$ gets.

Finally, we can write a decomposition of the identity on the intertwiner space $\cH_N^{(J)}$ as an integral over $\C^{2N}$:
\be
\label{iduNcoh}
\id_{\cH_N^{(J)}}
\,=\,
D_{N,J}\,\int_{\C^{2N}}
\prod_i \f{e^{-\la z_i|z_i\ra}\,d^4z_i}{\pi}\,
\f{|J,\{z_i\}\ra\la J,\{z_i\}|}{\left(\det X(z)\right)^J}.
\ee
This is to be compared with the decomposition of the identity on $\cH_{j_1,..,j_N}$ in term of coherent intertwiners \Ref{idcohint}. To check this identity, it is enough to check that this integral commutes with the $\U(N)$-action and that its trace is equal to the dimension $D_{N,J}$ of the Hilbert space $\cH_N^{(J)}$. As explained in more details in \cite{UN3}, we can gauge-fix this integral by the $\GL(2,\C)$-action and restrict it to an integral over the Grassmanian space ${\rm Gr}_{2,N}=\C^{2N}/\GL(2,\C)=\U(N)/\U(N-2)\times \U(2)$. The $\SL(2,\C)$-action allows to gauge-fix to spinors satisfying the closure condition; then rescaling the state allows to fix the matrix $X(z)=\id$ and the total area $A(z)=1$ thus to restrict the integral to coherent states of unit norm.

\subsection{The $F$-action on Coherent Intertwiners}


In order to discuss the simplicity constraints in the $\U(N)$ framework, we need the explicit action of the operators $E_{ij},F_{ij},F\dag_{ij}$ on the coherent states. This technical result has been a first part of my work. Let us start by looking closer at the $F$ annihilation operators. We first compute the action of $F_{ij}$ on coherent intertwiners:
\be
F_{ij}\,||\{j_k,z_k\}\ra
\,=\,
\int_{\SU(2)} dg\,g\vartriangleright
F_{ij}\,\otimes_k\f{(z^0_ka\dag_k+z^1_kb\dag_k)^{2j_k}}{\sqrt{(2j_k)!}}\,|0\ra,
\ee
since the operator $F_{ij}$ is invariant under global $\SU(2)$ transformations and thus commutes with the action of group elements $g\in\SU(2)$. Making $F_{ij}=(a_ib_j-a_jb_i)$ commute through the creation operators, we obtain after a straightforward calculation:
\be
F_{ij}\,||\{j_k,z_k\}\ra\,=\,
Z_{ij}\,\sqrt{(2j_i)(2j_j)}\,
||\{j_i-\f12,..,j_j-\f12,j_k,z_k\}\ra,
\ee
where we remind the reader that $Z_{ij}=(z^0_iz^1_j-z^1_iz^0_j)$.
Then using the formula \Ref{sumcoh} of $\U(N)$ coherent states in term of coherent intertwiners, we easily get:
\be
F_{ij}\,|J,\{z_k\}\ra
\,=\,
F_{ij}\,\sum_{\sum j_k=J}
\frac{\sqrt{J!(J+1)!}}{\sqrt{(2j_k)!}} \,||\{j_k,z_k\}\ra
\,=\,
Z_{ij}\sqrt{J(J+1)}\,|J-1,\{z_k\}\ra.
\ee
This fits with the $F$-action on $\U(N)$ coherent states derived in \cite{UN3}. Moreover we can use these relations to diagonalize the $F_{ij}$ operators. We introduce the vectors $|\beta,\{z_k\}\ra$ for $\beta\in\C$~:
\be
|\beta,\{z_k\}\ra\,\equiv\,
\sum_{J\in\N}\f{\beta^{2J}}{\sqrt{J!(J+1)!}}\,|J,\{z_k\}\ra
\quad\Rightarrow\quad
F_{ij}\,|\beta,\{z_k\}\ra\,=\,\beta^2 Z_{ij}\,|\beta,\{z_k\}\ra.
\ee
These new intertwiners $|\beta,\{z_k\}\ra$  diagonalize all $F_{ij}$ operators simultaneously. This is normal since the $F_{ij}$'s all commute with each other. We can also give other convenient expressions for these vectors in term of creation operators acting on the vacuum:
\bes
|\beta,\{z_k\}\ra
&=&
\sum_J \f{(\beta^2F\dag_Z)^J}{J!(J+1)!}\,|0\ra\\
&=&
\int dg\, g\vartriangleright
\otimes_k e^{\beta(z^0_ka\dag_k+z^1_kb\dag_k)}\,|0\ra,
\ees
which makes a clear link between these vectors and coherent states for the harmonic oscillator. Finally, we can compute the norm of these states, which is easily expressed as a Bessel function:
\be
\la\beta,\{z_k\}|\beta,\{z_k\}\ra
=\sum_J \f{(|\beta|^2)^{2J}}{J!(J+1)!} \,\la J,\{z_k\}|J,\{z_k\}\ra
=\sum_J \f{(|\beta|^2A(z))^{2J}}{J!(J+1)!}
=\f{I_1(2|\beta|^2A(z))}{|\beta|^2A(z)},
\ee
where we assumed the closure condition on the spinors so that the norm of the $\U(N)$ coherent state is given directly by $A(z)^{2J}$ (else we should in general replace $A(z)$ by the determinant $\sqrt{\det\,X(z)}$).
Here $I_1$ is the first modified Bessel function of the first kind. This clears up the action of the $F$-operators. Below, we further investigate the actions of the $E$ and $F\dag$ operators on the $\U(N)$ coherent states.

\subsection{Operator Algebra on Coherent Intertwiners}

We already have the action of the annihilation operators $F_{ij}$ on the $\U(N)$ coherent states. Now we need to complete the algebra to derive the action of the operators $F\dag_{ij}$ and $E_{ij}$. To this purpose, we use the standard action as differential operators of the creation and annihilation operators for the harmonic oscillator (see in appendix for some details):
$$
a_i \arr z^0_i,\qquad a\dag_i \arr \f{\pp}{\pp z^0_i},\qquad
b_i \arr z^1_i,\qquad b\dag_i \arr \f{\pp}{\pp z^1_i}.
$$
Using this on the definition of the operators $E$ and $F$, we guess the following action of these operators on the $\U(N)$ coherent states:
\bes
F_{ij} |J,\{z_k\}\ra&=& \sqrt{(J+1)J}Z_{ij} |J-1,\{z_k\}\ra,\\
F^\dagger_{ij} |J,\{z_k\}\ra&=&\f{1}{\sqrt{(J+2)(J+1)}} \Delta^z_{ij}|J+1, \{z_k\}\ra, \\
E_{ij}|J,\{z_k\}\ra&=&\delta^z_{ij}|J,\{z_k\}\ra,
\ees
where $Z_{ij}=(z^0_iz^1_j-z^1_iz^0_j)$ as before and where we have defined the following differential operators with respect to the spinor $z_i$:
\bes
\Delta^z_{ij}&= &\f{\pp^2}{\pp z_i^{0}\pp z_j^{1}}-\f{\pp^2}{\pp z_i^{1}\pp z_j^{0}},\\
\delta^z_{ij}&=&z_j^{0}\f{\pp}{\pp z_i^{0}}+z_j^{1}\f{\pp}{\pp z_i^{1}}.
\ees
The $J$-factors in front of the actions of $F$ and $F\dag$ come from the normalization factor $\sqrt{J!(J+1)!}$ of the coherent states.

The multiplication action of $F$ on the $\U(N)$ coherent states can be derived by using the commutation relation between the creation and annihilation operators:
\bes \label{Fact_1}
&&\left[ F_{ij}, F_Z^\dagger \right]= \mathcal{E}^Z_{ij} +2Z_{ij},\\
&&\left[\mathcal{E}^Z_{ij}, F^\dagger_Z \right]=2Z_{ij}F_Z^\dagger.
\ees
where we have defined the auxiliary operator $\mathcal{E}^Z_{ij}= \sum_m \left(Z_{im}E_{mj}-Z_{jm}E_{mi} \right)$. To show the second commutator, we have used that the antisymmetric matrix $Z$  satisfies that $Z_{ik} Z_{jl}-Z_{il} Z_{jk}= Z_{ij} Z_{kl}$. This allows the straightforward calculation:
\bes \label{Fact_2}
F_{ij}\left(F_Z^\dagger \right)^J|0\ra
&=&
\sum_{k=0}^{J-1} \left(F_Z^\dagger \right)^{J-1-k} \mathcal{E}^Z_{ij} \left( F^\dagger_Z\right)^k|0\ra +2J Z_{ij} \left(F_Z^\dagger \right)^{J-1}|0\ra, \nn\\
&=&
\left(\sum_{k=0}^{J-1} 2k +2J\right) Z_{ij} \left(F_Z^\dagger \right)^{J-1}|0\ra, \nn\\
&=&
J(J+1) Z_{ij}\left(F_Z^\dagger \right)^{J-1} |0\ra,
\ees
which gives the expected result. Moreover, we recover the same action for the $F_{ij}$ operators that we had already computed in the previous section using that the $\U(N)$ coherent states are superpositions of coherent intertwiners.

As for the $F\dag$-action, it is straightforward to compute the action of the differential operator $\Delta^z_{ij}$ on the coherent state taking into account that:
\be
\Delta^z_{ij}\,(Z_{kl})=2(\delta_{ik}\delta_{jl}-\delta_{il}\delta_{jk}),\qquad
\Delta^z_{ij}\,(F\dag_Z)^J
\,=\,
\Delta^z_{ij}\,\left(\f12\sum_{kl}Z_{kl}F\dag_{kl}\right)^J
\,=\,
J(J+1)F\dag_{ij}\,(F\dag_Z)^{J-1}.
\ee
This leads to the following action for the creation operator $F\dag_{ij}$
\be
\Delta^z_{ij} |J+1, \{z_k\} \ra= \sqrt{(J+1)(J+2)}F\dag_{ij}\,|J, \{z_k\}\ra,
\ee
since $F\dag_{ij}$ commutes with $F\dag_Z$ because they only involve oscillator creation operators $a\dag$ and $b\dag$.
At this stage, we can also check that the differential $F\dag$-action is indeed the adjoint of the multiplicative  $F$-action on the $\U(N)$ coherent state. That is straightforward to show. First, considering the matrix element $\la K, \{w_k\}| F_{ij}|J,\{z_k\}\ra$, it doesn't vanish iff $K=(J-1)$. Then, on the one hand,  we can compute:
\be
\la J-1, \{z_k\}| F_{ij}|J,\{w_k\}\ra
=\sqrt{J(J+1)}\, W_{ij} \,\la J-1,\{z_k\}|J-1, \{w_k\}\ra
= \sqrt{J(J+1)}\, W_{ij} \,\left(\f12\tr\,Z\dag W\right)^{J-1}.
\ee
On the other hand, we have:
\be
\la J,\{w_k\} |F\dag_{ij}|J-1, \{z_k\}\ra
=\f{1}{\sqrt{J(J+1)}}\,\Delta^z_{ij}\,\la J,\{w_k\} |F\dag_{ij}|J, \{z_k\}\ra
=\f{1}{\sqrt{J(J+1)}}\,\Delta^z_{ij}\,\left(\f12\tr\,W\dag Z\right)^{J}.
\ee
To evaluate this expression, we calculate explicitly the action of the differential operator on the $J$-power of the trace:
\be
\Delta^z_{ij}\,\left(\tr\,W\dag Z\right)^{J}
\,=\,
2J(J+1)\,\overline{W}_{ij}\left(\tr\,W\dag Z\right)^{J-1}.
\ee
This allows to conclude that we have as expected:
\be
\overline{\la J-1, \{z_k\}| F_{ij}|J,\{w_k\}\ra}=\la J,\{w_k\} |F\dag_{ij}|J-1, \{z_k\}\ra.
\ee

Finally, let us now compute the action of the $E$-operators on the U(N) coherent states. First we compute the commutator $\left[ E_{ij}, F_Z^\dagger\right] = \sum_{m} Z_{jm} F^\dagger_{im}$, which easily gets generalized to arbitrary power of the creation operator:
\be
[E_{ij},(F_Z^\dagger)^J]
\,=\,
\sum_{k=0}^{J-1} (F_Z^\dagger)^{J-1-k}
\left[ E_{ij}, F_Z^\dagger\right]
(F_Z^\dagger)^k
\,=\,
J\,\left(\sum_{m} Z_{jm} F^\dagger_{im}\right)\,
(F_Z^\dagger)^{J-1},
\ee
since all $F\dag$ commute with each other. This proves directly that the $E$-action on $\U(N)$ coherent states is simply related to the $F\dag$-action:
\be
E_{ij} (F_Z^\dagger)^{J}\,|0\ra
\,=\,
J\,\left(\sum_m Z_{jm} F\dag_{im}\right)\, (F_Z^\dagger)^{J-1}\,|0\ra.
\ee
Then we can easily compute the action of the differential operator:
\bes
\delta^z_{ij} \left(F_Z^\dagger \right)^J&=& \left(z_j^{0}\f{\pp}{\pp z_i^{0}}+z_j^{1}\f{\pp}{\pp z_i^{1}}\right) \left(\f12\sum_{kl} Z_{kl}F\dag_{kl} \right)^J \nn \\
&=& J\sum_{m}\left(Z_{jm} F^\dagger_{im}\right) \left(F_Z^\dagger \right)^{J-1}
\ees
This allows us to deduce the actions of the $E$-operators and of the differential operators $\delta^z_{ij}$ match on the $\U(N)$ coherent states:
\be
E_{ij}|J,\{z_k\}\ra=\delta^z_{ij}|J,\{z_k\}\ra.
\ee
It is possible to check directly that these differential operators actually generate the correct $\U(N)$ action on the spinors. Let us for instance consider the infinitesimal unitary transformation $u=\exp(\eps(E_{ij}-E_{ji}))$ where $i,j$ are arbitrary but fixed indices. It acts at first order on the spinors as:
$$
(u\,z)_k\sim z_k + \eps\delta_{ik}z_j - \eps\delta_{jk}z_i.
$$
It is very easy to check that this fits with the action of the previous differential operator:
$$
\eps(\delta^z_{ij}-\delta^z_{ji})\,z_k
\,=\,
 i \eps\delta_{ik}z_j - \eps\delta_{jk}z_i
\,\sim\,
(u\,z)_k- z_k.
$$
Following the same steps with the unitary transformations $u=\exp(i\eps(E_{ij}+E_{ji}))$ allows to prove completely that the differential operators $\delta^z_{ij}$ generate as expected the $\U(N)$ action on our coherent states.

Finally, it is also possible to check that the commutation relation between the differential operators corresponding to $E$, $F$ and $F\dag$ satisfy the correct commutation relations (see in appendix).

\section{The $\U(N)$ setting for $\Spin(4)$ intertwiner and a Gupta-Bleuler quantization} 

We have reviewed the whole $\U(N)$ formalism for $\SU(2)$ intertwiners and we gave the explicit action of the operators $E_{ij},F_{ij},F\dag_{ij}$ on the $\U(N)$ coherent states.  I now present the main part of my work concerning the issue of the way the simplicity constraints can be implement on a $\Spin(4)$ intertwiner within the $\U(N)$ framework. The results obtained show that there is a lot of freedom in the way the simplicity constraints can be implemented.\\
Since the Lie algebra $\spin(4)=\su_L(2)\oplus\su_R(2)$ simply splits into two copies of the $\su(2)$ algebra, it is straightforward to adapt the whole $\U(N)$ framework to $\spin(4)$. We double all the operators, introduce harmonic oscillators $a_i^L,b_i^L$ and $a_i^R,b_i^R$ and build two sets of $\u(N)$ operators $E^L_{ij},F^L_{ij},F^{L\dagger}_{ij}$ and $E^R_{ij},F^R_{ij},F^{R\dagger}_{ij}$.
These two $\u(N)$ sectors don't speak to each other and are a priori decoupled. It is the simplicity constraints that will couple them. \\
Several ways of imposing the constraints are explored and their advantages and disadvantages are analyzed. The idea  presented in this section is to replace the simplicity constraint algebra which does not close by a new equivalent constraint algebra which does close defined using the $\u(N)$ operators  keeping a strong imposition of the diagonal simplicity constraint and  imposing weakly the cross simplicity constraints as it is done in the EPRL-FK model. In this section, we only treat  the case without Immirzi parameter ($\rho=1$ in (\ref{diagI}) and (\ref{crosseddiagI})). The issue of the extension to the case with a finite Immirzi parameter ($\rho \neq 1$ in (\ref{diagI}) and (\ref{crosseddiagI})) is discussed in the next part of this section as well as in the last section of this chapter.

Let us start with the diagonal simplicity constraints: $\forall i,\quad
(\vJ^L_i)^2
\,=\,
(\vJ^R_i)^2$. Imposed strongly, they constraint  the $\spin(4)$ living on the legs of the intertwiners to be simple. This means that the spins in the left and right sectors  match: $j_i^L=j_i^R$. This translates into very simple constraints in the $\u(N)$ framework:
\be
\cC_i\,\equiv\,
E^L_{i}-E^R_{i}.
\ee
This diagonal simplicity  definitively couples the two sectors. This constraint is actually the same than the matching conditions for spin networks on the 2-vertex graph and the whole construction will be very similar \cite{2vertex}. Every (constraint) operator that we will now introduce to solve the simplicity constraints will have to commute (at least weakly) with these diagonal simplicity constraints $\cC_i$.

Now moving to the crossed simplicity constraints, $
\forall i\ne j,\quad
(\vJ^L_i\cdot\vJ^L_j )
\,=\,
(\vJ^R_i\cdot\vJ^R_j )$
, they refer to couples of legs of the intertwiners and to their scalar product. They amount to impose strongly, weakly or semi-classically, the equality between the scalar products of the left and right sectors, $\vJ_i^L\cdot\vJ_j^L=\vJ_i^R\cdot\vJ_j^R$. Dropping the $L/R$ index, we remind the expression for the scalar product operator in term of $\u(N)$ operators for $i\ne j$:
\bes \label{scalarProdOp}
\vJ_i\cdot \vJ_j
&=&
\f12E_{ij} E_{ji} -\f14E_iE_j-\f12 E_i,\\
&=&
\f12E_{ji} E_{ij} -\f14E_iE_j-\f12 E_j,\nn\\
&=& -\f12 F_{ij}^\dagger F_{ij} +\f14 E_iE_j,\nn\\
&=& -\f12 F_{ij}F_{ij}^\dagger +\f14 (E_i+2)(E_j+2).\nn
\ees
This expression clearly suggests two things. First, we could replace the $\vJ_i^L\cdot\vJ_j^L=\vJ_i^R\cdot\vJ_j^R$ constraints by some constraints of the type $E^L=E^R$ or $F^L=F^R$. We will explore these various possibilities below. Second, we then expect that the equality $\vJ_i^L\cdot\vJ_j^L=\vJ_i^R\cdot\vJ_j^R$ will only hold semi-classically at first order and will usually have corrections linear in the $j_i$'s (terms in $E_i$ and $E_j$).


\subsection{The Closed Algebra of Simplicity Constraints}

%
%
%
%
%
%

One big issue about the standard crossed simplicity constraints $\vJ_i^L\cdot\vJ_j^L-\vJ_i^R\cdot\vJ_j^R=0$ for all couples of legs $i\ne j$ is that their algebra doesn't close. The $\U(N)$ framework was precisely introduced to close the algebra of scalar product operators and provide an alternative algebra for invariant observables on the space of intertwiners. Indeed, considering the operators $E_{ij}$ instead of $\vJ_i\cdot \vJ_j$ allowed to have a closed algebra of invariant observables and to build coherent intertwiner states \'a la Perelomov that transforms nicely under the operators of that algebra.
This naturally suggests to replace the simplicity constraints $\vJ_i^L\cdot\vJ_j^L-\vJ_i^R\cdot\vJ_j^R=0$ by a simpler constraint expressed in term of the operators $E_{ij}^{L,R}$.
We propose to consider a new set of constraints, that we name the $\u(N)$ simplicity constraints:
\be
\cC_{ij}\,\equiv\,
E^L_{ij}-E^R_{ji}=E^L_{ij}-(E^R_{ij})\dag,
\qquad\qquad
\cC\dag_{ij}=\cC_{ji}.
\ee

The two important facts about these new proposed contraints are:
\begin{itemize}

\item They naturally include the diagonal simplicity constraints:
$$
\cC_{ii}=\cC_i=E^L_{i}-E^R_{i}.
$$

\item They form a closed $\u(N)$ algebra:
\be
[\cC_{ij},\cC_{kl}]
\,=\,
\delta_{jk}\cC_{il}-\delta_{il}\cC_{kj}.
\ee

\end{itemize}

Moreover, let us $\cH_\cC$ be the Hilbert space  of states satisfying these $\u(N)$ simplicity constraints:
\be
\cH_\cC
\,\equiv\,
\{|\psi\ra\,\textrm{ such that }\, \cC_{ij}\,|\psi\ra=0,\,\forall i,j\}.
\ee
Then this solves weakly the crossed simplicity constraints at leading order (i.e for large spins). Indeed, for all solution states $\phi,\psi\in\cH_\cC$, we have for $i\ne j$:
\bes
\la\phi|\vJ_i^L\cdot\vJ_j^L|\psi\ra
&=&
\la\phi|\f12E_{ij}^L E_{ji}^L -\f14E_i^LE_j^L-\f12 E_i^L|\psi\ra
\,=\,
\la\phi|\f12E_{ji}^R E_{ij}^R -\f14E_i^RE_j^R-\f12 E_i^R|\psi\ra,\nn\\
&=&
\la\phi|\vJ_i^R\cdot\vJ_j^R|\psi\ra\,+\,\la\phi|\f12(E_j^R-E_i^R)|\psi\ra.
\ees
Therefore, the crossed simplicity constraints are solved approximatively at first order. Indeed the expectation values $\la\vJ_i^L\cdot\vJ_j^L\ra$ are of order $\cO(j^2)$ while the correction term is of order $(j_j-j_i)\sim\cO(j)$.
This is not a very big obstacle since we only expect the simplicity constraints to be satisfied semi-classically in the large spin regime. Let us still point out that the diagonal simplicity constraints are still strongly and exactly enforced on all invariant states in $\cH_\cC$.

\medskip

As we said above, the operators $\cC_{ij}$ generate a $\u(N)$ Lie algebra: they actually generate the $\U(N)$ action $(u,\bar{u})$ on the coupled $L,R$ system such that the $\U(N)$ transformation acting on the right sector is the complex conjugate  of the transformation acting on the left sector.
Indeed, a finite transformation generated by the constraints $\cC_{ij}$ will read, for a anti-Hermitian matrix $\alpha_{ij}=-\bar{\alpha}_{ji}$:
$$
U
\,\equiv\,
e^{\sum_{ij}\alpha_{ij}\cC_{ij}}
\,=\,
e^{\sum_{ij}\alpha_{ij}E_{ij}^L}\,e^{-\sum_{ij}\alpha_{ij}E_{ji}^R}
\,=\,
e^{\sum_{ij}\alpha_{ij}E_{ij}^L}\,{e^{\sum_{ij}\overline{\alpha_{ij}}E_{ij}^R}}.
$$
%
%
Thus, states which are solution to the $\cC_{ij}$-constraints are $\U(N)$-invariant and the Hilbert space $\cH_\cC$ can be obtained by performing a $\U(N)$ group averaging on the space of intertwiners $\bigoplus_J \cH_N^{(J),L}\otimes \cH_N^{(J),R}$.
An over-complete basis of solutions can be obtained by group averaging the $\U(N)$ coherent states $|J,\{z^L_k \} \ra \otimes |J, \{ z^R_k \} \ra$.
However, we can give a more precise description of the $\U(N)$-invariant space. Indeed, since the spaces $\cH_N^{(J)}$ are irreducible $\U(N)$-representations, the Schur's lemma implies that there exists a unique invariant vector in the tensor product $\cH_N^{(J), L}\otimes \cH_N^{(J),R}$. Calling $|J\ra$  this unique state solution to the $\u(N)$-constraints for every total spin $J$, we have a complete basis of our solution space:
\be
\cH_\cC =\bigoplus_J \C\,|J\ra.
\ee
This construction is exactly the same than the definition of isotropic states in the 2-vertex loop quantum gravity model constructed in \cite{2vertex} using the $\U(N)$ formalism. Thus following that approach, we won't perform the $\U(N)$-group averaging to construct our $\U(N)$-invariant states but we will use using the following symmetric operator~:
\be
f^\dagger \equiv \sum_{kl} F_{kl}^{L \dagger} F_{kl}^{R \dagger}.
\ee
Indeed, this operator commute with all generators $\cC_{ij}$:
\be
\left[\cC_{ij}, f^{ \dagger}  \right] = \sum_{kl} \left(\left[ E_{ij}^L, F^{L \dagger}_{kl}\right] F^{R \dagger}_{kl} - F^{L \dagger}_{kl} \left[ E^R_{ji}, F^{R \dagger}_{kl} \right] \right) =0,
\ee
therefore,  the operator $f^{ \dagger}$ is $\U(N)$-invariant.
%
%
Since the vacuum state is also $\U(N)$-invariant, we can define the invariant states by applying this creation operator $f\dag$ to the vacuum state $|0\ra$~:
\be
|J\ra \equiv \left(f^\dagger \right)^J |0\ra
\ee
is obviously $\U(N)$-invariant. We also check that $|J\ra \in \cH_N^{(J),L}\otimes \cH_N^{(J),R}$.
Indeed, a straightforward calculation of the action of the total spin operator  $E \equiv \sum_i E^{L}_i=\sum_i E^{R}_i$ (the left and right total spin operators are obviously equal on the invariant space $\cH_\cC$) gives~:
\be
E |J\ra=2J |J\ra.
\ee
Finally, following the computations done in \cite{2vertex}, we also give the norm of these invariant vectors:
\be
\la J | J \ra= 2^{2J} J! (J+1)! \f{(N+J-1)!(N+J-2)!}{(N-1)! (N-2)!}=2^{2J}(J!(J+1)!)^2 D_{N,J}
\ee
where $D_{N,J}$ is the dimension of the intertwiner space $\cH_N^{(J)}$ given by (\ref{dimNJ}). The details of this calculation can be found in the appendix.

The fact that we get a single state for each total spin means that we are imposing constraints which are too strong. In the next parts, we will try too impose less constraints using the $E$ operators then different constraints in terms of the $F$ and $F^\dagger$ operators in order to get a bigger set of solutions to the simplicity constraints. Finally, we will see in the last section how we can use the $\U(N)$ coherent states in order to solve weakly all the simplicity constraints.

%
%




\subsection{Highest weight vectors for the $\u(N)$-simplicity constraints}

As we said in the previous section, it seems that the $\u(N)$-simplicity constraints are too strong. Following the idea that we might have imposed too many constraints, we propose  a new restricted set of $\u(N)$ constraints and consider only the raising operators of our $\u(N)$ algebra. This is also consistent with the line of thoughts that such a procedure usually leads to the construction of proper coherent states with the expected semi-classical properties.
Thus we try with the following new set of constraints:
\be \label{Erelax}
\{ \cC_{i<j}\equiv \cC_{ij} \textrm{ for } i< j \textrm{ and } \cC^\sigma_i=\cC_i-\sigma_i  \}
\ee
where we have chosen to require that only the raising operators\footnotemark{ }vanish
$\cC_{ij}\,|\psi\ra\,=0$ for $i<j$. We have also relaxed the diagonal simplicity constraints:  $\cC_i\,|\psi\ra\,=\sigma_i\,|\psi\ra$ where the parameters $\sigma_i\in\Z/2$ are arbitrary but fixed. In general, we will require that  $|\sigma_i|<< j_i$, so that the diagonal simplicity constraint are still satisfied at leading order.
\footnotetext{
This new set of  constraints \ref{Erelax} still forms a closed algebra. Indeed, let  be $i\leq j$ and $k\leq l$ then:
\be
\left[\cC_{ij}, \cC_{kl} \right]= \delta_{jk} \cC_{il} -\delta_{il} \cC_{kj}
\ee
where $i \leq j, \, k \leq l$ and $j=k$ imply $i \leq l$ or $k \leq l, \, i \leq j$ and $i=l$ imply $k\leq l$. Therefore, $\cC_{il}$ and $\cC_{kj}$ are also raising operators.}

Even we do not impose the full $\u(N)$ simplicity constraints, the cross simplicity constraints are still weakly satisfied. Indeed, let us define the Hilbert space $\cH_{\cC^<_\sigma}$ of states which satisfy the restricted set of constraints (\ref{Erelax}).  For all states $\phi, \, \psi \in \, \cH_{\cC^<_\sigma}$, we have:
\be
\forall i<j, \qquad  \la\phi|\vJ_i^L\cdot\vJ_j^L|\psi\ra
\,=\,
\la\phi|\vJ_i^R\cdot\vJ_j^R|\psi\ra + \la\phi|\f12 (E_i^R-E_j^R)-\f14 (\sigma_i E_j^R +\sigma_j E_i^R + \sigma_i \sigma_j)-\f12 \sigma_j |\psi\ra
\ee
Therefore, the weak cross simplicity constraints are still satisfied approximatively at leading order: the matrix elements $\la \phi |\vJ_i^R \cdot \vJ_j^R | \psi \ra$ are of order $\cO(j^2)$ while  the correction terms are of order $\cO(j)$.

The meaning of the Hilbert space $\cH_{\cC^<_\sigma}$ is straightforward in term of the theory of representations of the $\u(N)$ Lie algebra: it is the space of highest weight vectors. More precisely, let us consider the full space of $\spin(4)$ intertwiners defined as the tensor product of the uncoupled intertwiner spaces for $\su(2)_L$ and $\su(2)_R$. It is given by the direct sum over possible total area labels $J_L,J_R$ of the corresponding irreducible $\u(N)$ representations:
\be
\cH^{\spin(4)}_N
\,=\,
\bigoplus_{J_L,J_R}
\cH_N^{J_L}\otimes \cH_N^{J_R}.
\ee
Now our constraint algebra generates the diagonal $\u(N)$ action  which acts simultaneously on both the left and right sectors. Then we decompose the tensor products $\cH_N^{J_L}\otimes \cH_N^{J_R}$ into irreducible representations of this diagonal $\U(N)$ action and the vectors that are annihilated by the raising operators $\cC_{i<j}$ are the highest weight vectors  of these irreducible representations. The parameters $\sigma_i$ are the eigenvalues of the diagonal $\u(N)$ generators, they are the values of the highest weight and select the relevant representations.

For instance, the most natural case, $\sigma_i =0,\,\forall i$ , corresponds to $\U(N)$-invariant representations and we recover the space $\cH_\cC$ considered in the previous section. Then for a generic choice of $\sigma_i$, we do not necessarily require that $J_L=J_R$ as before, but this condition is slightly shifted to $J_L=J_R +\sum_i\sigma_i$.
The next step would be to decompose the product tensor of the two $\U(N)$ representations $\cH_N^{J_L}\otimes \cH_N^{J_R}$ into $\U(N)$ irreducible representations and then to extract the highest weight vector of this decomposition which correspond to our choice of $\sigma_i$'s. This can be done using the Gelfand-Zetlin basis and the Gelfand patterns \cite{tensorUN}. I have not investigated further in this direction yet.


\subsection{Using $F^L-F^R$ Constraints} \label{Fconstraints}

%
%
%

Another possibility to identify new simplicity constraints within the $\U(N)$ framework is to use the $F_{ij}$-operators instead of the $E_{ij}$ operators. Moreover, introducing simplicity constraints defined in terms of the $F_{ij}^{L,R}$ would be more in the spirit of the Gupta-Bleuler procedure since the $F$'s are indeed the annihilation operators.
Following this intuition, we define $F$-constraints:
\be
f_{ij}
\,\equiv\,
F_{ij}^{L}-F_{ij}^{R}.
\ee
First, these constraints all commute with each other, $[f_{ij},f_{kl}]=0$.
Moreover, these constraints are straightforward  to solve since we know how to diagonalize explicitly and simultaneously the operators $F_{ij}$ using the superposition of coherent states $|\beta,\{z_i\}\ra$.

Furthermore, solving these constraints seem to allow to solve weakly  the exact original quadratic simplicity constraints (without correction terms). Indeed, for all states $\phi,\psi$ in the kernel of $f_{ij}$ for all $i\ne j$, we have
\bes
\la\phi|\vJ_i^L\cdot \vJ_j^L|\psi\ra
&=&
\la\phi|-\f12 F^L_{ij}{}^\dagger F^L_{ij} +\f14 E^L_iE^L_j|\psi\ra \nn\\
&=&
\la\phi|\vJ_i^R\cdot \vJ_j^R |\psi\ra
\,+\,
\la\phi|\f14 (E^L_iE^L_j-E^R_iE^R_j)|\psi\ra.
\ees
If we also assume that the diagonal simplicity constraints hold, i.e that the operators $E^L_i-E^R_i$ vanish on both states $\psi,\phi$, then the second term vanishes and it all works out. Unfortunately, the $F$-constraints do not form a closed algebra with the diagonal constraints $\cC_i$:
\be
[\cC_i,f_{kl}]
\,=\,
[E^L_i, F^L_{kl}]+[E^R_i, F^R_{kl}]
\,=\,
\delta_{il}(F^L_{ik}+F^R_{ik})-\delta_{ik}(F^L_{il}+F^R_{il}).
\ee
Thus, if we require both $\cC_i=0$ and $F_{kl}^{L}-F_{kl}^{R}=0$, then we automatically also require $F_{kl}^{L}+F_{kl}^{R}=0$, which means that we are actually imposing the much stronger constraints $F_{kl}^{L}=F_{kl}^{R}=0$. These constraints are obviously only satisfied by the vacuum state $|0\ra$. Thus the $f_{kl}$ constraints are not consistent with the diagonal simplicity constraints. However, we will see in the last section that if we drop the requirement of imposing strongly the diagonal simplicity constraints then these $f$ constraints appear to be the right constraints to consider: they allow to impose all the (diagonal and crossed) simplicity constraints weakly.

\subsection{Using $F^L-(F^R)\dag$ Constraints}

%
%
%
%
%

We now consider  ``holomorphic" constraints defined in terms of the $F_{ij}$ and $F_{ij}^\dagger$ operators by:
\be
\cF_{ij}\,\equiv\,
F_{ij}^{L}-F_{ij}^{R}{}\dag,\qquad
\cF_{ij}=-\cF_{ji}.
\ee

These new operators commute with each other:
\be
[\cF_{ij},\cF_{kl}]=0
\ee
and the commutator of these new constraints with the $\u(N)$ generators $\cC_{ij}$ is now given by:
\bes
[\cC_{ij},\cF_{kl}]
&=& [E_{ij}^{L}-E_{ji}^{R}\,,\,F_{kl}^{L}-F_{kl}^{R}{}\dag] \nn\\
&=& \delta_{il}\cF_{jk}-\delta_{ik}\cF_{jl}.
\ees
This shows two things. First, if we take $i=j$, the $\u(N)$ generators are the diagonal simplicity constraints. This means that the holomorphic constraints are compatible with the diagonal simplicity constraints and together they form a closed Lie algebra: we can impose $\cC_{i}=0$ on the space of solutions to $\cF=0$ without obvious obstacle. Second, let us call $\cH_\cF$ the Hilbert space of states $\psi$ satisfying $\cF_{ij}\,|\psi\ra=0$ for all indices $i,j$. Then the previous commutator also means that there is a natural $\U(N)$ action on this solution space $\cH_\cF$ generated by the operators $\cC_{ij}$. In particular, once we identify a single solution to the holomorphic constraints $\cF$ then this induces a whole family of solutions obtained by acting with $\U(N)$ transformations on that initial solution.

We introduce the Hilbert space $\mathcal{H}_{\cF}^0$ of states satisfying the holomorphic constraints and the diagonal simplicity constraints $\cC_i$. Then, for all solution states $\psi$, $\phi \, \in \cH_{\cF}^0$, the expectation values of the left and right scalar product operators are equal up to a correction of order $\cO(j)$:
\bes
\la\phi|\vJ_i^L\cdot \vJ_j^L|\psi\ra
&=&
\la\phi|-\f12 F^L_{ij}{}^\dagger F^L_{ij} +\f14 E^L_iE^L_j|\psi\ra \nn\\
&=&
\la\phi|-\f12 F^R_{ij} F^R_{ij}{}^\dagger +\f14 E^R_iE^R_j|\psi\ra \nn\\
&=&
\la\phi|\vJ_i^R\cdot \vJ_j^R |\psi\ra
\,-\,
\la\phi| 1+\f12(E_i+E_j) |\psi\ra.
\ees
To identify solution states in $\mathcal{H}_{\cF}^0$, we start by the simplest case, which is to construct $\U(N)$-invariant states solution of this new set of constraints.  We recall that while the $E_{ij}$-operators leave invariant the total sum of spins $E^{L,R}$, the $F_{ij}^{L,R}$ operators decrease by $-1$ respectively the left and right total areas $E^{L,R}$ and the $F_{ij}^{L,R}{}^\dagger$ operators increase them by $+1$. That is why we consider a linear combination of $\U(N)$-invariant states  for different areas $J$; we use the $U(N)$-invariant basis $|J\ra$. It is straightforward to compute that \footnote{The computation is similar to the computation of the multiplication action of $F$ on the $\U(N)$ coherent states, done from (\ref{Fact_1}) to (\ref{Fact_2}), replacing $Z_{kl}$ by $2 F_{kl}^{R \dagger}$: $F^{R\dagger }_{kl}$ is also antisymmetric in $k \leftrightarrow l$ and satisfies the Pl\"ucker relation ($F^{R\dagger }_{ik}F^{R\dagger }_{jl}-F^{R\dagger }_{il}F^{R\dagger }_{jk}=F^{R\dagger }_{ij}F^{R\dagger }_{kl}$). Therefore, we just recall the main steps:
$$
[F_{ij}^L, f^\dagger] = 2\underbrace{\sum_k F_{ik}^{R \dagger} E_{kj}-F_{jk}^{R \dagger} E_{ki}}_{= \mathcal{E}_{ij}^{L,F^{R\dagger}}} + 4 F_{ij}^{R \dagger}, \qquad\qquad
[\mathcal{E}_{ij}^{L,F^{R\dagger}}, f^\dagger] =4F^{R \dagger}_{ij} f^\dagger,
$$
therefore we get:
$$
F^{L}_{ij} |J \ra=2J(J+1) F_{ij}^{R \dagger} |J-1 \ra.
$$
}:
\be
F^{L}_{ij} |J \ra=2J(J+1) F_{ij}^{R \dagger} |J-1 \ra.
\ee
Then if we define the states:
\be \label{alpha}
|\alpha\ra= \sum_J \f{\alpha^J}{2^J J! (J+1)!} |J\ra= \sum_J \f{\alpha^J}{2^J J! (J+1)!} (f^\dagger)^J|0\ra \quad \textrm{ with } \alpha \in \C
\ee
they satisfy:
\be
F_{ij}^L |\alpha\ra =\alpha \, F_{ij}^{R \dagger} |\alpha\ra \quad \forall \, i , j \quad \textrm{ i.e. } \cF_{ij} |\alpha \ra =0 \quad \forall \, i,j.
\ee
Thus for $\alpha=1$, they are solution of the $\cF$-constraints: $F_{ij}^L |\alpha=1\ra = F_{ij}^{R \dagger} |\alpha=1\ra$. Let us notice that these new states $|\alpha\ra$ for the coupled $L/R$ system are very similar to the coherent states $|\beta,\{z_k\}\ra$ diagonalizing the $F_{ij}$ operators acting on a single (left or right) sector. It's actually the exact same expression if we replace the spinor parameters $Z_{ij}$ by the creation operators $F\dag_{ij}$ of the other sector: instead of imposing by hand the values of the expectation values using the spinor labels, the behavior of the left sector is entirely dictated by the right sector, and vice-versa. As underlined in \cite{2vertex} in the context of loop quantum gravity on the 2-vertex graph, these states $|J\ra$ and $|\alpha\ra$ maximally entangle the left and right sectors.

Therefore, we have determined the unique $\U(N)$-invariant state solution to the $\cF$ constraints.
The natural question is whether there exist  other solutions to these $\cF$-constraints, which would necessarily be non-$\U(N)$-invariant. At this point, we have not been able to identify such solutions and we would like to conjecture that they do not exist. We however postpone the precise analysis of such conjecture to future investigation. Nevertheless, we would like to point out that a promising line of tackling this issue would be to work in the coherent intertwiner basis and use the expression of the operators $E,F,F\dag$ as differential operators on the spinor labels.

\subsection{Including the Immirzi Parameter?}\label{immparam?}





The next step is to extend our construction to the Euclidean case with a finite Immirzi parameter $\gamma$ ($\gamma>0, \, \gamma\neq 1$). At the discrete level, there is no equality between the left and the right parts of the scalar products anymore imposed by the simplicity constraints but the relation between the left and the right parts becomes a proportionality relation given by  (\ref{diagI}) and (\ref{crosseddiagI}).  We recall that the proportionality coefficient $\rho$ is simply related to the Immirzi parameter $\gamma$ by: $\rho\equiv \f{\gamma+1}{|\gamma-1|}$ \cite{FK, EPRL}.
Once again, we would like to use the $\U(N)$ formalism to solve these constraints. We therefore focus on a $\Spin(4)$ intertwiner with $N$ legs labelled by $i\, \in \{1, \cdots, N \}$. The issue remains that the crossed simplicity constraints $\vJ_i^L \cdot \vJ_j^L- \rho^2 \vJ_i^R \cdot \vJ_j^R=0$ for all couples of legs $i \neq j$ do not form a closed algebra. Following the same idea as previously we would like to replace the simplicity constraints by simpler constraints expressed in term of the operators $E^{L,R}_{ij}$ or $F^{L,R}_{ij}$ and $(F_{ij}^{L,R})^\dagger$ which form a closed algebra. We tried all possible combinations of $E$ and $F$ constraints and the only way to get a closed algebra including all the simplicity constraints is to consider constraints of the form:
\bes
\cC_i&= &E_{i}^L-E_i^R =0, \quad \forall \, i, \quad \textrm{ for the diagonal simplicity constraints.} \nn \\
\cF_{ij}^{\rho}&\equiv& F_{ij}^L - \rho (F_{ij}^R)^{\dagger}=0 \quad \forall \, i, j \quad \textrm{ for the cross simplicity constraints}
\ees
Then,
\be
[ \cF_{ij}^\rho, \cF_{kl}^\rho]=0 \quad \textrm{ and } [\cC_i, \cF^\rho_{kl}]= \delta_{il}\cF^\rho_{ik}-\delta_{ik}\cF^\rho_{il}.
\ee
We can again define a Hilbert space $\cH^\rho$ of states satisfying these constraints: $\cC_i |\psi \ra=0 \; \forall i, \, \cF_{ij}^{\rho} |\psi \ra=0 \;  \forall i \neq j$. We already have one solution given by $|\alpha=\rho\ra$ as defined in the previous subsection by (\ref{alpha}). However, we still have the usual un-rescaled diagonal simplicity constraint which do not involve the Immirzi parameter. Then, as for the crossed simplicity constraints, the result is also disappointing
and we have that $\forall |\psi \ra, \, |\phi \ra \in \cH^\rho$:
\bes
\la \psi | \vJ^L_i\cdot \vJ^L_j -\f14 E_i^LE_j^R | \phi \ra &= & \la \psi | -\f12 F_{ij}^{L \dagger} F_{ij}^L| \phi \ra \nn \\
&=& \rho^2 \la \psi | -\f12 F_{ij}^{R} F_{ij}^{R \dagger}| \phi \ra \nn \\
&=&\rho^2 \la \psi | \vJ^R_i\cdot \vJ^R_j -\f14 (E_i^R+2)(E_j^R+2) | \phi \ra.
\ees
Thus, since $E_i^L=E_i^R$, we get at the leading order in $j$ that the "right" observables $(\vJ_i^R\cdot \vJ_j^R -E_iE_j) \sim |J_i| |J_j|(\cos \theta_{ij}^R-1)$ are rescaled by the proportionality coefficient $\rho^2$ with respect to the "left" observables $\vJ_i^L\cdot \vJ_j^L -E_iE_j) \sim |J_i| |J_j|(\cos \theta_{ij}^L-1)$ where $\theta_{ij}$ is the angle between the two vectors $\vJ_i$ and $\vJ_j$. However, these observables which are corrected observables compared to the scalar product observables, do not have any real interesting geometrical interpretations and it does not seem possible to extract the expected relation: $\la \vJ_i^L \cdot \vJ_j^L \ra= \rho^2 \la \vJ_i^R \cdot \vJ_j^R \ra$.

Here again, it seems that the main obstacle is imposing strongly the diagonal simplicity constraints. In the following section, we will show how to relax the diagonal simplicity constraints and solve weakly all the simplicity constraints using coherent states for an arbitrary value of the Immirzi parameter.

\section{Weakening the constraints and using the $\U(N)$ coherent states}

Until now we tried to solve the crossed simplicity constraints weakly whereas the diagonal simplicity constraints were imposed strongly. Indeed, in the previous section, we focused on the issue of the cross-diagonal simplicity constraints $\vJ_i^L\cdot \vJ_j^L-\vJ^R_i\cdot \vJ_j^R=0$ which have to be imposed weakly since they do not form a closed algebra. We defined some new sets of constraints $\{ \cC_{ij}, \, i<j \}$ or $\{ \cF_{ij} \}$ which allow to solve the cross simplicity constraints weakly and which are compatible with the diagonal simplicity constraints in such a way that  the sets of all constraints form a closed algebra and therefore can all be imposed strongly in a consistent way. Here, we propose to relax all simplicity constraints because there are in fact physically on an equal footing and there is no physical reason to deal with the diagonal simplicity constraints in a different way from the cross simplicity ones. The idea is to use coherent states to solve weakly all simplicity constraints in the semi-classical regime. We first go back to the usual $\SU(2)$ coherent states, then we will propose the $\U(N)$ coherent states that solve weakly all simplicity conditions for arbitrary Immirzi parameter.

\subsection{Back to $\SU(2)$ coherent intertwiners}

The $\SU(2)$ coherent intertwiners $||j_i, \hat{n}_i \ra_L\otimes ||j_i, \hat{n}_i \ra_R$ are currently used to solve the simplicity constraints.  The usual analysis has been recalled in section \ref{SUcoh}. It is interesting to notice that these intertwiners are strong solutions to the diagonal simplicity constraints and that  there does not seem to exist any  other exact equation strongly solved by these states  in order to weakly solve the cross diagonal simplicity constraints even in the semi-classical regime.

In fact, $||j_i, \hat{n}_i \ra\otimes ||j_i, \hat{n}_i \ra= \int_{\Spin4} dG \, G\, \triangleright \otimes_{i=1}^N |2j_i, \hat{n}_i \ra$ span a Hilbert space of intertwiners which is the Hilbert space of intertwiners symmetric under the exchange of the left and right part. We denote it $\cH_{\textrm{sym}}$.
This symmetric Hilbert space $\cH_{\textrm{sym}}$ is generated by applying the operators
$E_{ij}^L E_{ij}^R$, $F_{ij}^LF_{ij}^R$, $F^{L \dagger}_{ij}F_{ij}^{R \dagger}$ on the vacuum state $|0\ra$.
It is obvious that any state $\psi \in \cH_{\textrm{sym}}$ satisfies all the non-diagonal simplicity constraints $\la \vJ_i^L\cdot \vJ_j^L\ra=\la \vJ^R_i\cdot \vJ_j^R\ra$ in expectation values. $\cH_{\textrm{sym}}$ is even the largest Hilbert space such that all the matrix elements of the constraints vanish:
\be
\forall \, \psi, \, \phi \in \cH_{\textrm{sym}}, \quad \la \psi |\vJ_i^L\cdot \vJ_j^L-\vJ^R_i\cdot \vJ_j^R| \phi \ra=0.
\ee

However, this symmetry property of the states does not seem to be fundamental. Indeed, we have seen in  Section \ref{SU2coh} that there is a second sector solution to the cross simplicity constraints given by the states $||j_i, -\hat{n}_i \ra_L\otimes ||j_i, \hat{n}_i \ra_R$. These states are not symmetric anymore in the exchange of the left and right part but they clearly satisfy the simplicity constraints in expectation value. Moreover, the previous analysis is not generalizable to the case of Euclidean gravity with a finite Immirzi parameter $\gamma$: the cross simplicity constraints become  $\vJ_i^L\cdot \vJ_{j}^L=\rho^2 \vJ_i^R\cdot \vJ_{j}^R$ and thus, the symmetric intertwiners cannot be used  to solved them weakly anymore. The resolution done in \ref{SU2coh} is not generalizable when the Immirzi parameter is taken into account; usually the diagonal simplicity constraints are imposed strongly and the quadratic cross simplicity constraints are replaced by linear constraints $\vJ_i^L=\pm \rho \vJ_i^R$ which are then used to construct a so-called Master constraint in order to solve weakly the off-diagonal simplicity constraints \cite{EPRL}.

We will now see that it is in fact possible to keep the standard quadratic simplicity constraints and to solve weakly all the simplicity constraints for any finite value of the Immirzi parameter.

\subsection{The final proposal: using $\U(N)$ coherent states}

Following the coherent state approach to solving the simplicity constraints, we propose to use the $\U(N)$ coherent states instead of the usual $\SU(2)$ coherent intertwiners. As we have already reviewed earlier, a $\U(N)$ coherent state $|J, \{z_k\}\ra$ is labeled by the total area $J$ and the $N$ spinors $z_k$ which define the semi-classical geometry underlying the intertwiner state. Now, considering $\Spin(4)$-intertwiners, we consider tensor products of $\U(N)$ coherent states for both the left and right sectors, that is $|J_L, \{z^L_k\}\ra\otimes |J_R, \{z^R_k\} \ra$. We would like to relax all simplicity constraints. Since we also relax the diagonal simplicity constraints, we do not require the matching of the total areas of the left and right sectors and we work with a priori two different $\U(N)$ representations, $J_L\neq J_R$. Then, the simplicity constraints impose that the classical geometry of the left and right intertwiners  are the same up to an overall scale. This will translate into relations between the spinors of the left and right sectors, $z^L_k$ and $z^R_k$.

Let us start by recalling the norm of the $\U(N)$ coherent states and the expectation values (normalized by the norm) of the geometric observables on them:
\bes
&&\la J, \{z_k\}\,|\,J, \{z_k\}\ra = A(z)^{2J},\qquad
A(z)\,=\,\f12\sum_k \la z_k|z_k\ra \\
&&\la E_{ij}\ra
=
J\,\f{\la z_i|z_j\ra}{A(z)},\qquad\forall i,j \nn\\
&&\la \vJ_i\cdot\vJ_j\ra
=
\f14 \f{J^2}{A(z)^2}\,\vV(z_i)\cdot\vV(z_j)+
\f{J}{8\,A(z)^2} \left(\vV(z_i)\cdot\vV(z_j)-3|\vV(z_i)|\,|\vV(z_j)|\right),\qquad \forall i\ne j, \nn
\ees
%
where we have implicitly assumed that the spinors $z_k$ satisfy the closure conditions. In case they do not close, the formulas above still hold up to replacing $A(z)$ by the the determinant $\sqrt{\det\, X(z)}$ as explained in the previous sections.

From these expressions, two things are clear. First, the total area label $J$ is simply a scale factor, it does not affect further the details of the classical geometry determined by the spinor labels. Thus, it appears that the ratio of the total area of the left and right sectors defines directly the Immirzi parameter $\rho=\f{J_L}{J_R}$. Second, if we want to match up to an overall factor the expectation values of the scalar product $\la \vJ_i\cdot\vJ_j\ra$ of the left and right sectors, it is clear that we have to require that the 3-vectors $\vec{V}(z_k)$ are the same up to a sign for the left and right sectors. Thus we distinguish two classes of solutions, which correspond to the two regimes, standard (s) and dual ($\star$), of simplicity constraints:
\begin{enumerate}

\item We require ${z^L_k= z^R_k}$ and consider the tensor product $|J_L, \{z_k\}\ra\otimes |J_R, \{z_k\} \ra$. This means that $\vec{V}(z_k^L)=\vec{V}(z_k^R)$. This corresponds to the {\bf standard simplicity regime (s)}. At leading order in the total area $J_{L,R}$, we have the equality of the expectation values of the scalar product observables:
    \be
    \la \vJ_i^L\cdot\vJ_j^L\ra \,\sim\, \rho^2\,\la \vJ_i^R\cdot\vJ_j^R\ra,
    \qquad \rho=\f{J_L}{J_R}.
    \ee
    Moreover, we also have the exact  equality of the expectation values of the $\u(N)$ generators:
    \be
    \la E_{ij}^L\ra =\rho  \la E_{ij}^R\ra.
    \ee
    There is still a $\U(N)$ action on the set of coherent states $|J_L, \{z_k\}\ra\otimes |J_R, \{z_k\} \ra$. Indeed the diagonal action $(u,u)$ acts simultaneously on the two sets of spinors, $(z_k,z_k)\arr ((u\,z)_k,(u\,z)_k)$. Thus these are still coherent states {\it \`a la} Perelomov.

\item We require {\bf $z^L_k= \varsigma z^R_k$} and consider the tensor product $|J_L, \{z_k\}\ra\otimes |J_R, \{\varsigma z_k\} \ra$. This means that $\vec{V}(z_k^L)=-\vec{V}(z_k^R)$ and corresponds to the {\bf dual simplicity regime ($\star$)}. At leading order in the total area $J_{L,R}$, we still have the equality of the expectation values of the scalar product observables:
    \be
    \la \vJ_i^L\cdot\vJ_j^L\ra \,\sim\, \rho^2\,\la \vJ_i^R\cdot\vJ_j^R\ra,
    \qquad \rho=\f{J_L}{J_R}.
    \ee
    However the equality of the expectation values of the $\u(N)$ generators is slightly modified due to the fact that the $\varsigma$ map is anti-unitary. Indeed, taking into account that $\la \varsigma z_i|\varsigma z_j \ra=\la z_j|z_i \ra=\overline{\la z_j|z_i \ra}$, we now have:
    \be
    \la E_{ij}^L\ra =\rho  \la E_{ji}^R\ra =\rho  \overline{\la E_{ij}^R\ra}\,.
    \ee
    The  $\U(N)$ action which is consistent with this set of coherent states $|J_L, \{z_k\}\ra\otimes |J_R, \{\varsigma z_k\} \ra$ is  the diagonal action $(u,\bar{u})$ which is actually generated by our $\u(N)$ simplicity condition $\cC_{ij}$ and which  acts simultaneously on the two sets of spinors as
    $(u,\bar{u})\vartriangleright(z_k,\varsigma z_k)\,=\, ((u\,z)_k,\varsigma(u\,z)_k)$.
    Thus these are also coherent states {\it \`a la} Perelomov.

\end{enumerate}

Therefore, just like when using coherent intertwiners to solve weakly the simplicity constraints, we can clearly implement the two regimes of simplicity for the intertwiners. However, there are clear advantages of this new approach over the usual one. First, there are no big difference in the properties of the $\U(N)$ coherent states corresponding to the two sectors. Second, the $E_{ij}$ observables allow to easily distinguish the two sectors. Third, we have $\U(N)$ actions in both cases which allow consistently deform these intertwiners, thus endowing them with a true structure of coherent states and not mere semi-classical states.

\medskip

For the moment, we have managed to solve weakly both diagonal and cross simplicity constraints using the coherent states $|J_L, \{z_k^L\}\ra\otimes |J_R, \{z_k^R\} \ra$ with $z_k^L=z_k^R$ or $z_k^L=\varsigma z_k^R$. This provides solutions to the simplicity constraints for values of the Immirzi parameter corresponding to the ratio $\rho= J_L/J_R$. This parameter still takes discrete values. However, since we have decided to relax the diagonal simplicity constraints and thus {\it not} require an exact match between the individual spins $j_i^{L,R}$ of the left and right sectors, we can further relax our implicit assumption that the total area need to be fixed. Then we would only require a matching of the total areas of the left and right sectors in expectation value and the parameter $\rho= \la J_L\ra/\la J_R\ra$ will be allowed to take any (positive) real value.

To implement this, we come back to the $F$-constraints considered earlier in section \ref{Fconstraints} and in section \ref{immparam?}~:
\be
F_{ij}^L= \rho F_{ij}^R.
\ee
These constraints were not compatible with the diagonal simplicity constraints. However, since we have decided to relax these diagonal simplicity constraints, we can neglect them and impose the $F$-constraints strongly.
We can easily solve these constraints since we know how to diagonalize the annihilation operators $F_{ij}$. Indeed, a generic solution will be given by the tensor product of $\beta$-states:
\be
|\beta_L, \{z_k^L\}\ra\otimes |\beta_R, \{z_k^R\} \ra,
\qquad
\textrm{with}\quad
\beta_L=\rho\beta_R
\quad\textrm{and}\quad
z_k^L=z_k^R.
\ee
We remind the definition of the $\beta$-states as superpositions of coherent states for different values of the total area:
\be
|\beta,\{z_k\}\ra
\,=\,
\sum_J \f{\beta^{2J}}{\sqrt{J!(J+1)!}}\,|J,\{z_k\}\ra,
\ee
which satisfy the eigenvalue equation:
$$
F_{ij}\,|\beta,\{z_k\}\ra\,=\,
\beta^2Z_{ij}
\,|\beta,\{z_k\}\ra,\qquad
\textrm{with}\quad
Z_{ij}=(z^0_iz^1_j-z^1_iz^0_j).
$$
Once again, we can easily compute the norm of these states, as well as the expectation values of the geometric observables~:
\bes
&&\la \beta, \{z_k\}\,|\,\beta, \{z_k\}\ra = \f{I_1(2x)}{x},\qquad
\textrm{with}\quad x\,=\,|\beta|^2A(z), \\
&&\la E_{ij}\ra
=
\f{xI_2(2x)}{I_1(2x)}\,\f{\la z_i|z_j\ra}{A(z)},\qquad\forall i,j \nn\\
&&\la \vJ_i\cdot\vJ_j\ra
=
\f14 \f{\vV(z_i)\cdot\vV(z_j)}{A(z)^2}\,\f{x\left(\f32 I_2(2x)+xI_3(2x)\right)}{I_1(2x)}
-\f38\f{|\vV(z_i)|\,|\vV(z_j)|}{A(z)^2}\,\f{xI_2(2x)}{I_1(2x)},\qquad \forall i\ne j, \nn
\ees
where the $I_n$'s are the modified Bessel functions of the first kind and the parameter $x=|\beta|^2A(z)$ depends very simply on the label $\beta$.
For large values of $x$, i.e for large area $A(z)$ or large value of $\beta$ (this is more or less the same since the label $\beta$ can be entirely absorbed as a overall rescaling of the spinors $z_k$ in the definition of the $\beta$-states), these expressions simplify at leading order and we get~:
\bes
&&\la E_{ij}\ra
\sim
x\,\f{\la z_i|z_j\ra}{A(z)},\qquad\forall i,j \\
&&\la \vJ_i\cdot\vJ_j\ra
\sim
\f{x^2} 4 \f{\vV(z_i)\cdot\vV(z_j)}{A(z)^2}
,\qquad \forall i\ne j. \nn
\ees
Thus, considering tensor product states $|\rho \beta, \{z_k\}\ra\otimes |\beta, \{z_k\} \ra$ with $\beta_L=\rho\beta_R$ and $z_k^L=z_k^R$, we obtain exact solutions to the $F$-constraints $(F_{ij}^L-\rho F_{ij}^R)\,|\psi\ra=0$. And these solutions satisfy weakly the simplicity conditions at leading order in the semi-classical limit, $\la \vJ_i^L\cdot\vJ_j^L\ra\sim\rho^2\la \vJ_i^R\cdot\vJ_j^R\ra$ and $\la E_{ij}^L\ra\sim\rho\la E_{ij}^R\ra$.

We proceed similarly with the other sectors and consider tensor product states $|\rho \beta, \{z_k\}\ra\otimes |\beta, \{\varsigma z_k\} \ra$, with $\beta_L=\rho\beta_R$ and $z_k^R=\varsigma z_k^L$.  These solutions satisfy weakly the simplicity conditions at leading order in the semi-classical limit. Indeed, we have obviously $\la \vJ_i^L\cdot\vJ_j^L\ra\sim\rho^2\la \vJ_i^R\cdot\vJ_j^R\ra$, but $\la E_{ij}^L\ra\sim\rho\,\overline{\la E_{ij}\ra}$. However, the main difference is that we have not been able to identify a set of constraints as the $F$-constraints which would characterize these tensor states. Indeed, looking at the action of the $F_{ij}$ operators, we get:
\bes
&&F_{ij}^L\,|\rho \beta, \{z_k\}\ra\otimes |\beta, \{\varsigma z_k\} \ra
\,=\,
\rho\beta\,Z_{ij}\,|\rho \beta, \{z_k\}\ra\otimes |\beta, \{\varsigma z_k\} \ra,\nn\\
&&F_{ij}^R\,|\rho \beta, \{z_k\}\ra\otimes |\beta, \{\varsigma z_k\} \ra
\,=\,
\beta\,\overline{Z_{ij}}\,|\rho \beta, \{z_k\}\ra\otimes |\beta, \{\varsigma z_k\} \ra,\nn
\ees
and we actually don't know any operator which would act anti-holomorphically on states $|\beta, \{z_k\}\ra$ so as to produce the value $\overline{Z_{ij}}$. This is very similar to what happens when solving the simplicity constraints using the standard coherent intertwiners: the coherent intertwiner span a subspace in the standard regime (s) while they still span the whole Hilbert space of intertwiners in the dual regime ($\star$). However, our approach still has two very interesting advantages: the $\U(N)$ action on our solution states and the straightforward inclusion of the Immirzi parameter in our framework as a simple scale factor.

\medskip

We would like to finish this last section with a remark on the phase of the spinors. Indeed, the matching of the expectation values of the scalar product observables of the left and right sectors only requires a matching of the 3-vectors $\vec{V}(z^L_k)=\pm\vec{V}(z^R_k)$ with the sign depending on whether we are in the standard regime or the dual regime. In order to impose these equalities, we have required that $z^R_k=z^L_k$ or that $z^R_k=\varsigma z^L_k$. However, the 3-vector $\vec{V}(z)$ only determines the spinor $z$ up to a global phase, $z\arr e^{i\theta}\,z$. We can thus multiply any of the $2N$ spinors  $z^L_k$ and $z^R_k$ by arbitrary phases without affecting the expectation values $\la \vJ^L_i\cdot\vJ^L_j \ra$ and $\la \vJ^R_i\cdot\vJ^R_j \ra$. Therefore, we can consider generally coherent states $|\rho J, \{e^{i\theta_k^L}z_k\}\ra\otimes |J, \{e^{i\theta_k^R}z_k\} \ra$ with arbitrary phases  $\theta_k^L,\theta_k^R$. These tensor products will still solve weakly the quadratic simplicity constraints on the scalar product operators.
The expectation values of the $\u(N)$ generators $\la E_{ij}^{L,R}\ra$ are nevertheless sensitive to these phases and are equal only up a phase. Since the geometry of the 3-vectors $\vV(z_k^{L,R})$, and thus the geometry of the intertwiner, do not depend on the phases of the spinors, it is natural to wonder about their physical/mathematical relevance.

The answer proposed in \cite{UN3} is that these phases are relevant to the spin network construction when we glue intertwiners together. Indeed, following the interpretation of loop quantum gravity in term of discrete twisted geometries \cite{twisted}, these phases (or more precisely the relative phase between two intertwiners glued along an edge) encode the extrinsic curvature at the discrete level. In our context, having these freedom in shifting these phase without affecting the intrinsic geometry of the intertwiner (defined in term of the 3-vectors) should allow to glue these $\U(N)$ coherent intertwiners in a consistent way without interfering with the simplicity constraints.

\part{Spinfoam models and the semi-classical limit} \label{part3}

Spin foam models define transition amplitudes between quantum states of geometry through state-sum models which can be understood as discrete space-time geometries. They are supposed to describe the quantum space-time structure at the Planck scale. One of the main issues is then to extract some semi-classical information from the formalism and to show its relation to the more standard perturbative approach to the quantization of general relativity based on quantum field theory. In particular one would hope to understand why such perturbative approach fails.  The quantum gravity semi-classical limit analysis amounts to proving that we can recover general relativity in a large scale (or low energy) regime of the spinfoam models and to showing  how to compute the quantum corrections to the classical dynamics of the gravitational field. We will detail in Chapter \ref{groupInt6j}, Chapter \ref{recursion6j} and Chapter \ref{4dphysical} three projects relevant to tackle this open problem but let us first introduce the framework.
\smallskip

 In the spin foam context, 
 a proposal for reconstructing the graviton propagator in this discrete setting from correlations between geometrical observables such as the areas of elementary surfaces \cite{graviton1, graviton2} has been developed. This proposal allows to extract  semi-classical correlations at large scales, which we hope to compare with the perturbative calculations performed in quantum general relativity treated as a quantum field theory.
 %
 %
 The main ingredients of these ``spin foam graviton propagator" calculations are 
\begin{enumerate}
  \item[1.]   a suitable boundary state peaked on a classical 3-geometry, 
  \item[2.] the spin foam amplitudes for the bulk geometry, 
  \item[3.]  the relevant observables probing the space-time geometry. 
\end{enumerate}

%
Since the original proposal, there have been a lot of works developing this line of research. They  mainly focused on creating the new mathematical tools needed for these semi-classical calculations and on using these computations as a criteria to select spinfoam models with a correct semi-classical behavior and as a result to discriminate some of  them (see e.g \cite{alesci}). All these developments hint towards the fact that the graviton propagator in spin foam models lead back to Newton's law for gravity at large scales while being regularized at the Planck scale. This behavior has been confirmed by numerical simulations in the simplest cases \cite{numerics1,numerics2}.\smallskip

However, up to now, the most explicit calculations have been done at the leading order (in the scale parameter) and for the simplest space-time triangulation (a single tetrahedron in 3d and a single 4-simplex in 4d).  In order to make the link with the standard quantum field theory perturbative expansion, we need to do  calculations with physical boundary states, \ie states that solve the Hamiltonian constraint on the boundary. Moreover, we need  to be able to push these calculations further and calculate the correlations both at higher order (``loop corrections") and  more refined triangulations (smoother boundary state). 
The three key elements to improve our understanding of the spinfoam graviton are   therefore to
\begin{enumerate}
\item[(i)]  determine the exact correlations when considering  \textit{physical} boundary states;
\item[(ii)] determine the higher order  corrections to the correlations (the leading order of the correlations gives the classical propagator of the graviton) by studying the asymptotic behavior of the vertex amplitude;
\item[(iii)]  determine the behavior of the  correlations by considering different (physical) boundary states. 
\end{enumerate}

We have focused on (i)  in the context of 4d gravity.
 Up to now, most of the  recent works have focused on building quantum coherent states with good semi-classical properties. One important issue is the requirement that the states are physical \ie solve the Hamiltonian constraint on the 3d boundary.  The requirement of working with a physical state should allow us to determine explicitly the width of the gaussian state defining the boundary state (see \cite{physical} for 3d gravity).   This width is relevant in the context of the geometrical correlations because it enters the exact numerical factor in front of these correlations. Therefore, if we want to have the \textit{exact} correlations and not only their scaling properties, we need a definite prediction of that width.
\smallskip 

We have addressed the issue of  (ii)  in the context of 3d gravity.  Considering a single tetrahedron, the solution of this issue requires understanding the corrections to the asymptotical behavior of the spin foam vertex amplitude associated to a single tetrahedron. We recall that this is the basic building block of of a 3d spin foam model, since a spin foam can then be constructed by gluing these spin foam vertices  in order to describe the whole quantum space-time. In the Ponzano-Regge model (see Chapter \ref{PRmodel}), the spinfoam vertex is given by the \sj-symbol from the recoupling theory of the representations of $\SU(2)$.  Therefore, a necessary step towards providing explicit formulae or procedures to compute all orders of the perturbative expansion (in term of the length scale) of the graviton correlations in the Ponzano-Regge model is to compute all orders of the asymptotic expansion of the \sj-symbol. 
\smallskip

Let us now detail the plan of this last part. We will first give a quick review of the \textit{spin foam graviton propagator} framework in the next chapter.
In  Chapter \ref{3dtoymodel}, we will define explicitely the 3d "spin foam graviton" for the simplest possible setting given by the 3d toy model introduced in \cite{3dtoymodel1, 3dtoymodel2}. In Chapters \ref{groupInt6j} and \ref{recursion6j}, we will present two different methods to develop the asymptotic of the \sj-symbol: either from a brute-force calculation based on the explicit formula of the \sj-symbol in term of factorials \cite{article1}, or from recursion relations for the \sj-symbol \cite{article2}. Finally, in  Chapter \ref{4dphysical}, we will show that it is possible to define a physical boundary state for a spin foam model for 4d quantum gravity. These results were published in \cite{article5} in which  the consequences of the physical state requirement for the Euclidean Barrett-Crane model for the simplest case of a space-time triangulation constructed from a single 4-simplex have been investigated.

\chapter{The graviton propagator}

Let us  describe the spin foam framework for deriving the graviton propagator from correlations between area observables. We recall that the kinematical Hilbert space of quantum geometry states for loop quantum gravity  is spanned by spin network states $|\psi\ra=|\gamma, j_l, i_n\ra$ where   $\gamma$ is a graph, $j_l$ is a ``spin" labeling an irreducible representation of the gauge group $G$ associated to the link $l$ of the graph, and $i_n$ is associated to the node $n$ of $\gamma$ and labels  intertwiners. 
\\
The gauge group $G$ will be specified when we  consider a specific spin foam model. Note that we changed the notations: a spin network is now denoted by $\psi$ whereas it was  called $s$ in the previous parts.  In this part $s$ will represent the specific spin network associated to the boundary graph of a 4-simplex. 
\smallskip

We now consider a 4d space-time region $\mathcal{M}$  with a 3d boundary $\Sigma$. The spin network state $|\psi\ra$ defines the quantum state of geometry of the boundary $\Sigma$, the spin foam amplitude $K[\psi]$ (\ref{propaSF}) defines the dynamical probability amplitude of that state and is supposed to contain the whole dynamical content of quantum gravity.

We consider in this part the case where the boundary $\Sigma$ is connected and the kernel $K$ is then defined as a function of only one boundary spin network  $\psi=(\gamma, j_l, i_n)$.
\be
K[\psi]=\sum_{\mathcal{F}|_{\partial\mathcal{F}=\psi }} \mathcal{A}_\mathcal{F}[\psi],
\ee
where we used the same compact notation as in (\ref{propaSFcompact}). We recall that the sum is over  spin foams $\mathcal{F}=(C,c)$ such that  their two-complexes are compatible with the graph $\gamma$ and their colorings $c$ are compatible with the representations and intertwiners $(j_l, i_n)$.
We start by considering a semi-classical spin network functional $\Psi_{q}[\psi]$ peaked on a classical 3d metric $q$ for the boundary $\Sigma$. We further require that this boundary state $\Psi_{q}[\psi]$ induces a space-time structure in the bulk peaked around the flat Minkowski metric. In particular, this normally fixes the classical boundary $q$ to be the 3-metric induced on $\Sigma$ by the Minkowski metric on ${\cal M}$. Then we construct correlations $W$ between the metric fluctuations for the chosen boundary state $\Psi_{q}$.
\be \label{twopointSF}
W^{abcd}(x,y;q)=\sum_{\psi}K[\psi] \la \psi | h^{ab}(x) h^{cd}(y) |\psi \ra \Psi_{q}[\psi]
\,=\,{\textrm Tr}\, \left[K \,h^{ab}(x) h^{cd}(y)\,\Psi_q\right]\,,
\ee
where the trace is taken over the Hilbert of spin networks. Here $x$ and $y$ are two points localized on the boundary $\Sigma$. In our discrete spin network setting, they are usually determined as nodes of the graph $\gamma$ underlying the spin network state $\Psi_q$. The metric fluctuations $h^{ab}(x) h^{cd}(y)$ are usually constructed as geometrical quantities. There are two basic types of such geometric observables in the discretized geometry setting of spinfoams: the diagonal component of the metric tensor can be interpreted as areas and the off-diagonal components as dihedral angles between simplices. More details on the 3d case are given in the next chapter \ref{3dtoymodel}.

Finally, this formula defines the 2-point function for the gravitational field in the spinfoam framework. It can be considered as the equivalent of the standard 2-point function of the conventional quantum field theory  framework, which defines the graviton propagator\footnote{In the following formula, $T$ stands for the standard time-ordering prescription.} \cite{graviton1}. 
\be \label{2point}
W_{\mu \nu \rho \sigma} (x, y)= \la 0 | T\{ h_{\mu \nu} (x) h_{\rho \sigma} (y) \}| 0\ra.
\ee

 \paragraph{Physical observables of the theory.}

The goal of the proposal for computing a "spin foam graviton propagator" is to probe the geometry induced by the spin foam amplitudes through calculating correlations between geometric observables. Thus, we first need to identify appropriate observables. As we already mentioned it, geometrical observables such as areas or dihedral angles appear in the metric fluctuations $h^{ab}(x) h^{cd}(y)$ which enter into the 2-points function definition (\ref{twopointSF}). That is to consider correlations between different components of the metric amounts to defining the correlations between areas and volumes (which can be defined from area and dihedral angles variables \cite{bianca-simone}) at different points on the boundary $\Sigma$ defined by the labels $(j_l, i_n)$. Indeed, the data $(j_l, i_n)$ living on the boundary graph $\gamma$ encodes the geometrical information of the boundary $\Sigma$: the representation $j_l$ gives the area of the triangle $\Delta \in \Delta_\partial$ dual to the link $l \in \Delta_\partial^\ast$ and the intertwiner $i_n$ describes the shape and the volume of the tetrahedron of the triangulation $\Delta_\partial$ dual to the node $n \in \Delta^\ast_\partial$. These boundary representations and intertwiners are thus the typical geometrical observables that we consider for the "spin foam graviton propagator".

\paragraph{The boundary state.}
In the spin foam setting,  $\Psi_{q}[\psi]$ is a semi-classical state peaked on both intrinsic and extrinsic geometry \cite{graviton1}. It is the extrinsic data that determines the 4-metric induced in the bulk.
Considering boundary states, while most of the recent work has focused on building quantum coherent states with good semi-classical properties, one important issue is the requirement that the states are physical i.e solve the Hamiltonian constraint on the 3d boundary. This ``physical state" criteria can be entirely formulated in term of a compatibility equation between the boundary state and the spinfoam bulk amplitude by the two following conditions: 
\be \label{conditionsSF} \left\{ \tabl{l}{
\sum_\psi |\Psi_q[\psi]|^2=1 \\
\sum_\psi K[\psi] \Psi_q[\psi]=1
} \right.
\ee
The first condition is the normalization of the boundary state, while the second condition translates truthfully the requirement to work with a physical state and can be considered as the ``Wheeler-deWitt" condition.

The consequences of this last condition were investigated in the framework of the Ponzano-Regge spinfoam model for 3d quantum gravity, more particularly in a toy model where the 3d space-time is triangulated by a single tetrahedron \cite{3dtoymodel1, 3dtoymodel2}. This 3d toy model is introduced in  Chapter \ref{3dtoymodel}. 

 We  studied the consequences of both conditions (\ref{conditionsSF}) on $\Psi_q$ in the context of  the Barrett-Crane model \cite{article5}. These results are presented in  Chapter \ref{4dphysical}.
 
\paragraph{The kernel.}
The main spinfoam models used to define the bulk amplitudes are the Barrett-Crane model (see Chapter \ref{BCmodel}), which exists in both its Euclidean version \cite{BC,carlo-alej} and its Lorentzian counterpart \cite{BC2,bcl_carlo}, and the more recent EPRL-FK models (see Chapter \ref{EPRLFK} and \cite{EPR,EPRL,FK,LS, coh1})
and their generalizations \cite{eprl_jerzy, polish} for 4d gravity. Details concerning the Barrett-Crane kernel are given in the Chapter \ref{4dphysical}.

\chapter{The asymptotic expansion in 3d} 

Most of the explicit calculations in the "spin foam graviton" framework have been done at the leading order (in the scale parameter) and for the simplest space-time triangulation (\ie a single tetrahedron in 3d and a single 4-simplex in 4d). The graviton propagator in spin foam models seems to lead back to Newton's law for gravity at large scales while being regularized at the Planck scale. This behavior has been confirmed by numerical simulations in the simplest cases \cite{numerics1,numerics2}. In order to make the link with the standard quantum field theory perturbative expansion, we now need to be able to push these calculations further and calculate the correlations both at higher order (``loop corrections") and for more refined triangulations (smoother boundary state). In both works \cite{article1, article2} presented respectively in Chapter \ref{groupInt6j} and Chapter \ref{recursion6j}, we focus on the first aspect: the leading order of the correlations gives the classical propagator of the graviton and we would like to compute the higher order (quantum) corrections.
Following the lines of \cite{3dtoymodel2,valentin,4d}, this requires understanding the corrections to the asymptotical behavior of the spinfoam vertex amplitude, which is the amplitude associated to a single tetrahedron in 3d quantum gravity or to a single 4-simplex in 4d models. 

In this chapter, we focus on 3d quantum gravity although 3d general relativity has no local degrees of freedom. Indeed, the 2-point function of the linearised quantum theory (\ref{2point}) is a well defined quantity that can be evaluated once a gauge-fixing is chosen. However, this quantity is a pure gauge and the quantum theory does not properly  have  a propagating graviton \cite{hooft}. It provides nevertheless a nice laboratory to test the ideas which have been proposed in the 4d case since the 3d quantum gravity model is much simpler than the 4d one. 

The canonical framework to compute 3d correlation in quantum gravity is reviewed in the next section \label{3dtoymodel}; we will see that the structure of this "spin foam graviton" framework is particularly clear in 3d. Then the study of the \sj-symbol is presented in Section \ref{methods6j},  as well as Chapters\ref{groupInt6j} and   \ref{recursion6j}.

\section{The canonical framework for 3d correlation in gravity/ The boundary states and the kernel for 3d correlation in gravity} \label{3dtoymodel}

We present in this Section the simplest  setting -- given by a 3d toy model introduced in \cite{3dtoymodel1, 3dtoymodel2} --  to illustrate the "spin foam graviton propagator" framework and to point out the importance to study the asymptotic behavior of the \sj-symbol. In spite of the simplicity of the model, we will see that the framework developed here, has rather generic features for computing graviton propagator/correlations in non-perturbative quantum gravity from spin foam amplitudes. 
\smallskip

\begin{figure}[ht]
\begin{center}
\includegraphics[width=4cm]{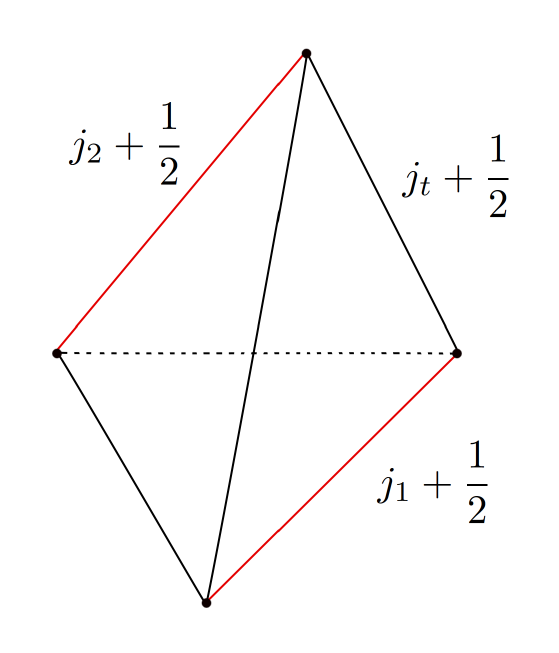}
\end{center}
\caption{The dynamical tetrahedron as evolution between two hyperplanes containing $e_1$ and $e_2$. The labels give the physical lengths: $l_1=j_1+1/2$, $l_2=j_2+1/2$ and we assume to have measured the time $T=(j_t+1/2)/\sqrt{2}$. ($j_1, j_2, j_t \in \N/2$)} \label{toytetra}
\end{figure}

We consider a triangulation consisting of a single tetrahedron embedded in flat 3d Euclidean space-time: we are interested in the correlations of length fluctuations of the bottom edge and of the top edge of the tetrahedron (see Fig. \ref{toytetra}). In order to define transition amplitudes in a background independent context for a certain region of space-time, we perform a perturbative expansion with respect to the geometry of the boundary. It is the classical geometry which will act as a background for the perturbative expansion. This classical geometry in the case of a single tetrahedron is defined by its edge lengths and its dihedral angles since these quantities specify respectively the intrinsic and extrinsic curvatures of the boundary. 
\\
More precisely,   given a background that we will precise in the following,  to compute the correlations between fluctuations around the background of the length of the bottom edge $l_1\equiv j_1+1/2$ and the top edge $l_2\equiv j_2+1/2$ of the tetrahedron represented in Fig. \ref{toytetra}, we consider the situation where the edge lengths of the four other edges are fixed to the unique value $j_t+1/2$. In this setting, which is referred to as the time-gauge setting, the four imposed bulk edge length can be interpreted as representing the time measured between two planes containing $l_1$ and $l_2$ given by $T=(j_t+1/2)/\sqrt{2}$.
The background around which we study the fluctuations is introduced by the state $|\psi\ra$ which will peak $j_1$ and $j_2$ around a given value $j_0$. Then to study the perturbative expansion of the 2-point function, we choose a Coulomb-like gauge and for simplicity we also take $j_t=j_0$  and our starting point becomes
\bes \label{propa3d}
W^{1122}&= &\f{1}{\cN} \sum_{j_1, j_2} \cO_{j_1}(j_0) \Psi_{e_1}(j_1) \cO_{j_2}(j_0) \Psi_{e_1}(j_1) \left\{ \tabl{ccc}{j_1& j_0 & j_0 \\
j_2 &j_0 & j_0} \right\} \\
\textrm{with } && \cO_{j}(j_0)\equiv \f{1}{d_{j_0}^2}\left( d_j^2-d_{j_0}^2\right) \quad \textrm{ and } \quad \Psi_{e}(j)\equiv \exp \left(-\f{\alpha}{2} (\delta_{d_{j}})^2+ i \theta\f{d_j}{2} \right) \nn
\ees
where $\cN$ is the normalization factor given by the same sum without the observable insertions $\cO_{j_e}(j_0)$. Moreover, $d_j\equiv 2j+1=l/2$ and $\delta_{d_{j}}=d_j-d_{j_0}$ and we recall that the $\{6j\}$ symbol is the Ponzano-Regge kernel (see Chapter \ref{PRmodel}).  To define the boundary state $\Psi$, we used the Gaussian ansatz. $\Psi$ peaks the geometry around the equilateral tetrahedra which all edges are equal to $l_0=j_0+1/2=d_{j_{0}}/2$ and all dihedral angles (defined as the angles between the external normal to the triangles) have the same value $\theta= \arccos(-1/3)$.  $W_{1122}$ then measures the correlations between length fluctuations for the edges $e_1$ and $e_2$ for the tetrahedron, and it can be interpreted as the 2-point function for gravity, contracted along the directions of $e_1$ and $e_2$ \cite{3dtoymodel1}.
\\
Thus, the pertubative expansion of this propagator needs the knowledge of the expansion of the $\{6j\}$-symbol. The well-known asymptotics of the $\{6j\}$ symbol is given by \cite{PR, 6jsaddle1, 6jsaddle3, 6jsaddle2}
\be
\left\{ \tabl{ccc}{j_1&j_0&j_0\\
j_2&j_0&j_0 } \right\}
\sim \frac{\cos\left(S_R[j_e]+\pi/4\right)}{\sqrt{12 \pi V(j_1, j_2, j_0)}},
\ee
where $V(j_1, j_2, j_0)=\f{1}{12}\sqrt{4l_0^2l_1^2l_2^2-l_1^2l_2^4-l_1^4l_2^2}$ is the volume of the tetrahedron and $S_R[j_e]$ is the Regge action
\be
S_R[j_e]= \sum_e (j_e+\f12) \theta_e(j_e),
\ee
where $\theta_e$ are dihedral angles. The Regge action is a discretized version of general relativity, which captures the non-linearity of the theory \cite{regge}. In the pertubative expansion of the propagator (\ref{propa3d}), there are therefore two sources of correction: contributions coming from higher orders in the expansion of the Regge action and contributions coming from higher orders in the expansion of the $\{6j\}$ symbol. We focus now on the second aspect and we propose in the next section different methods to study the asymptotic expansion of the $\{6j\}$ symbol.


\section{How to study the \sj-symbol?} \label{methods6j}

They are three basic ways to compute the leading order asymptotics of the \sj-symbol and to show its relation to the Regge action for 3d gravity.
\begin{itemize}
\item {\it Recursion relations}~\cite{SG1, SG2}. Using the invariance of the \sj-symbol under Pachner moves (Biedenharn-Elliott identity) or directly its definition as a recoupling coefficient, one can derive a recursion relation for the \sj-symbol. This recursion formula is actually very useful for numerical computations, but it can also be approximated at large spins by a (second order) differential equation. One then derive the asymptotics from a WKB approximation.

\item {\it Integral formula}~\cite{freidel,6jsaddle3}. One can write the square of the \sj-symbol as an integral over four copies of $\SU(2)$. In the large spin regime, we can use saddle point techniques and one derives the right asymptotics after a careful analysis of non-degenerate and degenerate configurations for the saddle points. This is the technique used to derive the asymptotics of the Barrett-Crane and EPR-FK vertex amplitudes.

\item {\it Brute-force approximation}~\cite{razvan}. One can start from the explicit algebraic formula of the \sj-symbol as a sum over some products of factorials. Using the Stirling formula and after lengthy calculations, we approximate the sum by an integral and use saddle point techniques again which lead to the same asymptotics.

\end{itemize}
We would like to mention  aslo  the more sophisticated and rigorous proof of the asymptotics by Roberts \cite{6jsaddle1} based on geometric quantization, but he also uses an integral formula and the saddle point method.

\chapter{Group integral techniques and the asymptotic expansion of the \sj} \label{groupInt6j}
The method is based on the explicit algebraic formula of the \sj-symbol as a sum over some products of factorials. Using the Stirling formula, we approximate the sum by an integral and use saddle point techniques. These results were published in \cite{article1}

\section{The \sj-symbol and the Racah's single sum formula}

The \sj-symbol is the basic building block of the Ponzano-Regge model which is a state sum model for 3d Euclidean gravity formulated as a $\SU(2)$ gauge theory. The Ponzano-Regge model is defined over a triangulation of space-time: we build the 3d space-time manifold from tetrahedra glued together along their respective triangles and edges. We assign an irreducible representation (irreps) of $\SU(2)$ to each edge $e$ of the triangulation. These irreps are labeled by a half-integer $j_e\in\N/2$, the spin, and the dimension of the corresponding representation space is given by $d_{j_{e}}=2j_{e}+1$. Each tetrahedron of the triangulation has six edges labeled by six spins $j_{e_1},..,j_{e_6}$ and we associate it with the corresponding \sj-symbol, which is the unique (non-trivial) $\SU(2)$ invariant built from these six representations. It is giving by combining four normalized Clebsh-Gordan coefficients corresponding to the four triangles of the tetrahedron. Finally, the Ponzano-Regge amplitude for a given colored triangulation is simply given by the product of the \sj-symbols associated to all its tetrahedra.

\begin{figure}[ht]
\begin{center}
\includegraphics[width=4cm]{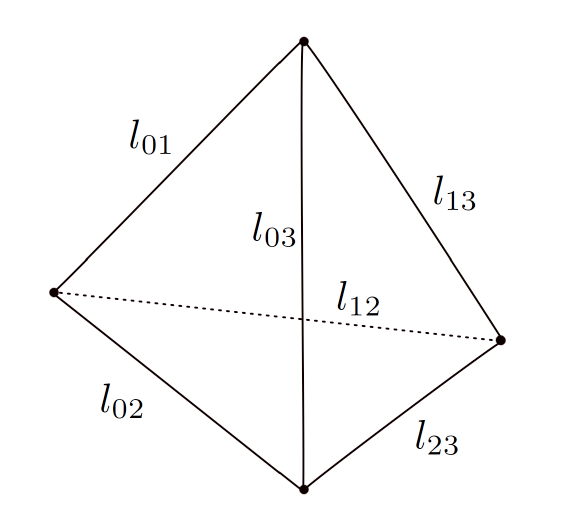}
\caption{A single tetrahedron: the edge lengths are given by $l_{IJ}=\frac{d_{j_{IJ}}}{2}$.}  \label{fig_tetrahedron}
\end{center}
\end{figure}

Looking more closely at a single tetrahedron, we label its four triangles by $I=0,..,3$. Then each of its six edges is labeled by the couple of triangles to which it belongs, $(IJ)$ with $0\leq I<J \leq 3$. To each edge is attached a $\SU(2)$ irrep of spin $j_{IJ}$, which defines the length of that edge $j_{IJ}+\frac{1}{2}=\frac{d_{j_{IJ}}}{2}$ (see Fig. \ref{fig_tetrahedron}).
There are several ways of expressing the \sj-symbol. The basic formula is the Racah's single sum formula which expresses the \sj-symbol as a sum over some products of factorials (see Appendix \ref{6j}). This is our starting point as in \cite{razvan}~:
\be
\label{Racah6j}
\left\{
\begin{array}{ccc}
j_{01}&j_{02}&j_{03}\\
j_{23}&j_{13}&j_{12}
\end{array}
\right\}
\,=
\sqrt{\Delta(j_{01}, j_{02}, j_{03}) \Delta(j_{23}, j_{02}, j_{12}) \Delta(j_{23}, j_{13}, j_{03}) \Delta(j_{01}, j_{13}, j_{12}) }
\sum_{\textrm{max } v_I}^{\textrm{min } p_J} \frac{(-1)^t (t+1)!}{\prod_I(t-v_I)! \prod_J (p_J-t)!}
\ee
where the $v_I$ and $p_i$ are given by the following sums:
$$
\forall K=0..3,\quad v_K= \displaystyle{\sum_{I \neq K}} j_{IK},\qquad
\forall k=1..3,\quad p_k= \displaystyle{\sum_{i \neq 0, k}} (j_{0i} + j_{ki}).
$$
The factors $\Delta(j_{01}, j_{02}, j_{03})$ are weights associated to each triangle and are defined by:
$$
\Delta(j_{01}, j_{02}, j_{03})= \frac{(j_{01}+j_{02}-j_{03})!(j_{01}-j_{02}+j_{03})!(-j_{01}+j_{02}+j_{03})!}{(j_{01}+j_{02}+j_{03}+1)!}.
$$
From this point, in all sums and products throughout this paper, capital indices $K$ will run from $0$ to $3$ and lower-cases indices $k$ will run from $1$ to $3$.

We are interested in the large spin expansion of the \sj-symbol when scaling all the spins homogeneously. Actually we will scale the lengths $d_{j_{IJ}}/2$ instead of the spins $j_{IJ}$ because the structure of the expansion will be simpler (we expect an alternation of cosines and sines without any mixing up at all orders as in \cite{valentin}) and the geometrical interpretation (when possible) is expected to be simpler. Then we rescale all $d_{j_{IJ}}$ by $\lambda d_{j_{IJ}}$ in (\ref{Racah6j}), which is equivalent to changing $j_{IJ}=\f{d_{j_{IJ}}}{2}-\f12$ to $\lambda\f{d_{j_{IJ}}}{2}-\f12$. This gives:
\equa{ \label{Racah_dj}
\tabl{l}{
\left\{ \tabl{lll}{\lambda d_{j_{01}}/2-1/2&\lambda d_{j_{02}}/2-1/2&\lambda d_{j_{03}}/2-1/2\\
\lambda d_{j_{23}}/2-1/2&\lambda d_{j_{13}}/2-1/2&\lambda d_{j_{12}}/2-1/2\\} \right\}= \\
\\
\;\;\;\;\;\sqrt{\Delta(\lambda d_{j_{01}}, \lambda d_{j_{02}}, \lambda d_{j_{03}}) \Delta(\lambda d_{j_{23}}, \lambda d_{j_{02}}, \lambda d_{j_{12}}) \Delta(\lambda d_{j_{23}}, \lambda d_{j_{13}}, \lambda d_{j_{03}}) \Delta(\lambda d_{j_{01}},\lambda d_{j_{13}}, \lambda d_{j_{12}}) }\\
\\
\;\;\;\;\; \displaystyle{\sum_{\lambda \,\textrm{max } \tilde{v}_I-\frac{3}{2}}^{\lambda \,\textrm{min } \tilde{p}_j-2}} (-1)^t \frac{(t+1)!}{\prod_i(t-\lambda \tilde{v}_I+\frac{3}{2})! \prod_j (\lambda \tilde{p}_j-t-2)!}
}}
with the new conventions:
$$
\tilde{v}_K= \sum_{I \neq K} \frac{d_{j_{IK}}}{2},\qquad
\tilde{p}_k= \sum_{i \neq 0, k} \frac{(d_{j_{0i}} + d_{j_{ki}})}{2},
$$
$$
\Delta(\lambda d_{j_{01}}, \lambda d_{j_{02}}, \lambda d_{j_{03}})= \frac{\left(\frac{\lambda }{2}(d_{j_{01}}+d_{j_{02}}-d_{j_{03}})-\frac{1}{2}\right)!\left(\frac{\lambda }{2}(d_{j_{01}}-d_{j_{02}}+d_{j_{03}})-
\frac{1}{2}\right)!\left(\frac{\lambda }{2}(-d_{j_{01}}+d_{j_{02}}+d_{j_{03}})-
\frac{1}{2}\right)!}{\left(\frac{\lambda }{2}(d_{j_{01}}+d_{j_{02}}+d_{j_{03}})-\frac{1}{2}\right)!}.
$$
The quantity $\tilde{v}_K$ gives the perimeter of the triangle $K$ while the $\tilde{p}_k$'s are the perimeters of (non-planar) quadrilaterals.

\section{Perturbative expansion of the 6j-symbol}
In this section, we will give a procedure to obtain the full perturbative expansion of the \sj-symbol in term of the length scale $\lambda$ and we compute explicitly the leading order (Ponzano-Regge formulae) then the next-to-leading order analytically.

\subsection{General procedure}
We give all the necessary formulae to obtain the Ponzano-Regge corrections at any order. But calculations are only performed explicitly at the next-to-leading order for a generic \sj-symbol. We start from equation (\ref{Racah_dj}).

\paragraph{First approximation: factorials.}

The factorial can be expanded in a series:
\equa{\label{fact}
 n!= \sqrt{2 \pi n}\left(\frac{n}{e}\right)^n\left(1 + \frac{1}{12n}+ \frac{1}{288n^2}- \frac{139}{51840n^3}-\frac{571}{2488320n^4} + \cdots \right)
}
In equation (\ref{Racah_dj}), there are factorials of the form: $n!$, $ (n+1/2)!$ and $(n-1/2)!$, which are rigourously defined through Euler's $\Gamma$ function. From (\ref{fact}) we deduce asymptotic expansions for $(n+1/2)!$ and $(n-1/2)!$ (see the details in appendix \ref{factorials}). In order to get the next-to-leading order (NLO) in the $1/\lambda$ expansion of the \sj-symbol, we replace the factorials in equation (\ref{Racah_dj}) by their respective asymptotic expansion:
\equa{ \tabl{l}{
n! \sim \sqrt{2 \pi }e^{(n+\frac{1}{2})\ln(n)-n}\left(1 + \frac{1}{12n}\right)\\
\\
(n+\frac{1}{2})! \sim \sqrt{2 \pi} e^{(n+1)\ln(n)-n}\left(1+ \frac{11}{24n}  \right) \\
\\
(n-\frac{1}{2})! \sim \sqrt{2 \pi} e^{n\ln(n)-n}\left(1- \frac{1}{24n}  \right). \\
}}
Then, equation (\ref{Racah_dj}) reads at first order as:
\equa{\label{6japprox1}
\left\{ \tabl{lll}{\lambda d_{j_{01}}/2-1/2&\lambda d_{j_{02}}/2-1/2&\lambda d_{j_{03}}/2-1/2\\
\lambda d_{j_{23}}/2-1/2&\lambda d_{j_{13}}/2-1/2&\lambda d_{j_{12}}/2-1/2\\} \right\}=
\frac{1}{2\pi} e^{\frac{\lambda }{2}h(d_{j_{IJ}})}
 \left(1 -\frac{1}{24\lambda } H(d_{j_{IJ}}) + O \left(\frac{1}{\lambda^2 }\right)\right) \,\Sigma\,.
}
The first factor is given by:
\bes
\label{termsh}
&& h(d_{j_{IJ}})= \sum_{I<J}d_{j_{IJ}}h_{d_{j_{IJ}}}\\
&&\textrm{with}\quad
h_{d_{j_{IJ}}}= \frac{1}{2} \ln \left(
\frac{(d_{j_{IJ}}-d_{j_{IK}}+d_{j_{IL}})(d_{j_{IJ}}+d_{j_{IK}}-d_{j_{IL}})(d_{j_{IJ}}-d_{j_{JK}}+d_{j_{JL}})
(d_{j_{IJ}}+d_{j_{JK}}-d_{j_{JL}})}{(d_{j_{IJ}}+d_{j_{IK}}+d_{j_{IL}})(-d_{j_{IJ}}+d_{j_{IK}}+d_{j_{IL}})
(d_{j_{IJ}}+d_{j_{JK}}+d_{j_{JL}})(-d_{j_{IJ}}+d_{j_{JK}}+d_{j_{JL}})}\right), \nonumber
\ees
where $(KL)$ is the opposite side to $(IJ)$, that is $K \neq L$ and $K,L \neq I,J$.
The second factor is due to the NLO of the factorials:
\be
\label{termsH}
H(d_{j_{IJ}})=2\displaystyle{\sum_{j,K}} \frac{1}{\tilde{p}_j-\tilde{v}_K} - 2\displaystyle{\sum_K} \frac{1}{\tilde{v}_K}= \displaystyle{\sum_I}\left[\frac{-r^I+\sum_{K \neq I}r^I_K}{2A_I}\right],
\ee
where $A_I$ is the area of triangle $I$, $r^I$ is the radius of the incircle of triangle $I$ and $r^I_K$  is the radius of the excircle to the triangle $I$ tangent to the side $d_{j_{IK}}$ of the triangle $I$.
Finally, $\Sigma$ is a Riemann sum:
\be
\label{termsSigma}
\Sigma=\frac{1}{\lambda ^2}\displaystyle{\sum_{x=\textrm{max } \tilde{v}_I/2}^{\textrm{min } \tilde{p}_j/2}} e^{F(x)} \left( 1 -\frac{1}{12\lambda }G(x) + O\left(\frac{1}{\lambda^2 }\right)\right) e^{\lambda f(x)}
\ee
with the pre-factor and the action given by:
\bes
f(x)&=& i \pi x+ x \ln(x) - \displaystyle{\sum_K}(x-\tilde{v}_K) \ln(x-\tilde{v}_K) - \displaystyle{\sum_j}(\tilde{p}_j-x) \ln(\tilde{p}_j-x), \nonumber\\
F(x)&=& \frac{1}{2} \ln \left( \frac{x^3 \prod_j(\tilde{p}_j-x)^3}{\prod_K (x-\tilde{v}_K)^4} \right),\\
G(x)&=&-\frac{13}{x}+\frac{47}{2} \displaystyle{\sum_K}\frac{1}{x-\tilde{v}_K} +13\sum_j \frac{1}{\tilde{p}_j-x}. \nonumber
\ees
The details of the computation are given in Appendix \ref{stirling}.

\medskip

\paragraph{Second approximation: Riemann sum.}

The second approximation consists in replacing the Riemann sum $\Sigma$ of (\ref{6japprox1}) by an integral. One $k^{-1}$ factor of $\Sigma$ plays the role of $dx$. We can then rewrite equation (\ref{6japprox1}) as:
\be
\{6j\}\sim\frac{1}{2\pi}  \left(1 -\frac{1}{24\lambda} H(d_{j_{IJ}}) + O \left(\frac{1}{\lambda}\right)\right) e^{\frac{\lambda}{2}h(d_{j_{IJ}})} \frac{1}{\lambda}\displaystyle{\int_{\textrm{max } \f{\tilde{v}_I}{2}}^{\textrm{min } \f{\tilde{p}_j}{2}}} dx \, e^{F(x)} \left( 1 -\frac{1}{12\lambda}G(x) + O\left(\frac{1}{\lambda^2}\right)\right) e^{\lambda f(x)}.
\ee
%
This approximation does not generate any corrections at leading order and at first order. It will nevertheless enter at second order in terms in $1/\lambda^2$.

\paragraph{Third approximation: saddle point approximation.}
We have to study an integral of the form $I=\int_a^b dx g(x) e^{\lambda f(x)}$ where $\lambda$ is a large parameter. The asymptotic expansion of such an integral is given by contributions around the stationary points of the action $f$ which are  points, denoted $x_0$, of the complex plane such that $f^{\prime}(x_0)=0$. We expand the action $f(x)$ and the function $g(x)$ around the stationary points $x_0$ in term of $\delta x=x-x_0$:
$$
f(x)=\displaystyle{\sum_{j=0}^\infty}\frac{f(x_0)^{(j)}}{j!}(\delta x)^j=f(x_0)+\frac{f^{\prime \prime }(x_0)}{2}(\delta x)^2+ f_{x_0}^{>2}(\delta x)  \quad \textrm{and}\quad g(x)=\displaystyle{\sum_{j=0}^\infty}\frac{g(x_0)^{(j)}}{j!}(\delta x)^j=g(\delta x).
$$
We then expand $K(\delta x)=g(\delta x)e^{kf_{x_0}^{>2}(\delta x)}$ in power of $\delta x$. Following the standard stationary phase approximation, we extend the integration domain to the whole $\R$.  The integrals are then ``generalized Gaussians" which can easily be computed. We  group the resulting terms according to their dependence on $1/\lambda$, being careful because of the function $g(x)$ which  depends on $1/\lambda$. We recall that $g(x)$  was obtained by replacing the factorials in (\ref{Racah_dj}) by their series expansion and we write $g(x)$ under the general form:
$$
g(x)=\displaystyle{\sum_{i=1}^\infty}\frac{g_i(x)}{i!\,\lambda^i}.
 $$
Then the complete perturbative expansion of $I$ can be written as:
\equa{\label{complete}
I= \displaystyle{\sum_{x_0}} e^{\lambda f(x_0)}\sqrt{\frac{2\pi}{-f^{\prime \prime}(x_0)\lambda}} \left(1+\displaystyle{\sum_{n=1}^\infty}\frac{1}{\lambda^n} \left[ \displaystyle{\sum_{p=0}^{n-1}} \tilde{N}_p\frac{(2p-1)!!}{(-f^{\prime \prime}(x_0))^p} + \displaystyle{\sum_{p=0}^{2n}}N_p \frac{(2n+2p-1)!!}{(-f^{\prime \prime}(x_0))^{n+p}} \right] \right)
}
where \equa{ \tabl{l}{
\tilde{N}_p=\displaystyle{\sum_{i=1}^{E[\frac{2p}{3}]}} \frac{1}{i!(n-p+i)!} \displaystyle{\sum_{l_1 \cdots l_i=3}^{E[\frac{2p}{i}]}} \frac{g_{n-p+i}^{(2p-\sum_{j=1}^il_j)}(x_0)}{(2p-\sum_{j=1}^il_j)!} \displaystyle{\prod_{j=1}^i} \frac{f^{l_j}(x_0)}{(l_j)!}\\
\\
N_0=\frac{g_0^{(2n)}(x_0)}{(2n)!}+\displaystyle{\sum_{i=1}^{E[\frac{2n}{3}]}} \frac{1}{(i!)^2} \displaystyle{\sum_{l_1 \cdots l_i=3}^{E[\frac{2n}{i}]}} \frac{g_{i}^{(2n-\sum_{j=1}^il_j)}(x_0)}{(2p-\sum_{j=1}^il_j)!} \displaystyle{\prod_{j=1}^i} \frac{f^{l_j}(x_0)}{(l_j)!}\\
\\
N_p=\displaystyle{\sum_{i=p}^{E[\frac{2(p+n)}{3}]}} \frac{1}{i!(i-p)!} \displaystyle{\sum_{l_1 \cdots l_i=3}^{E[\frac{2(n+p)}{i}]}} \frac{g_{i-p}^{(2(n+p)-\sum_{j=1}^il_j)}(x_0)}{(2(n+p)-\sum_{j=1}^il_j)!} \displaystyle{\prod_{j=1}^i} \frac{f^{l_j}(x_0)}{(l_j)!} \textrm{ for } p\geq1\\
}}
The details of the computation are given in Appendix \ref{saddlepoint}. From this expansion and adjusting the first approximation
to get the proper dependence on $\lambda$ for $g$ and the pre-factors, it is possible to compute analytically the whole asymptotic expansion of the \sj-symbol.

Here to get  explicitly the next-to-leading order of the \sj-symbol asymptotic expansion, we only  need the next-to-leading order of the $1/\lambda$ expansion of $I$, so we cut the previous formulae at $n=1$, then
$$
I \sim  \displaystyle{\sum_{x_0}} e^{\lambda f(x_0)}\sqrt{\frac{2\pi}{-f^{\prime \prime}(x_0)\lambda }} \left(1+\frac{1}{\lambda }\left(\tilde{N}_0+ \frac{N_0}{-f^{\prime \prime }(x_0)}+\frac{3N_1}{(-f^{\prime \prime }(x_0))^2}+\frac{15N_2}{(-f^{\prime \prime }(x_0))^3}\right) \right)
$$
with the expansion coefficients given by
$$
\tilde{N}_0=g_1(x_0), \quad
N_0=\frac{g_0^{\prime \prime }(x_0)}{2}, \quad N_1=\frac{f^{(3)}(x_0)g_0^\prime(x_0)}{3!}+\frac{f^{(4)}(x_0)g_0(x_0)}{4!}, \quad N_2=\frac{g_0(x_0)}{2}\left(\frac{f^{(3)}(x_0)}{3!}\right)^2.
$$
We recall that $g(x)=e^{F(x)}\left(1-\frac{G(x)}{12\lambda }\right)$; that is: $g_0(x)=e^{F(x)}$ and $g_1(x)=-\frac{G(x)}{12}e^{F(x)}.$  Finally, we obtain the approximation:
\equa{\tabl{ll}{\label{intNLO}
I \sim \displaystyle{\sum_{x_0}} &\sqrt{\frac{2\pi}{-f^{\prime \prime}(x_0)\lambda} }\, e^{F(x_0)+\lambda f(x_0)} \\
&\left[1 +
\frac{1}{\lambda } \left(- \frac{G(x_0)}{12}- \frac{F^{\prime \prime}(x_0)+(F^{\prime}(x_0))^2}{2 f^{\prime \prime}(x_0)}+\frac{f^{(4)}(x_0)+4f^{(3)}(x_0)F^\prime(x_0)}{8(f^{\prime \prime}(x_0))^2}- \frac{5(f^{(3)}(x_0))^2}{24(f^{\prime \prime}(x_0))^3} \right) + O \left(\frac{1}{\lambda^2} \right) \right]
}}
This gives us the following expression for the asymptotic expansion of the \sj-symbol at second order:
\equa{\label{NLO1}
\tabl{l}{
\left\{ \tabl{lll}{\lambda d_{j_{01}}/2-1/2&\lambda d_{j_{02}}/2-1/2&\lambda d_{j_{03}}/2-1/2\\
\lambda d_{j_{23}}/2-1/2&\lambda d_{j_{13}}/2-1/2&\lambda d_{j_{12}}/2-1/2\\} \right\} \\
\\
\;\;\;\;\;\;\;\; \sim \displaystyle{\sum_{x_0}} \sqrt{\frac{1}{-f^{\prime \prime}(x_0)2\pi \lambda^{3}} } \exp\left(F(x_0)+\lambda f(x_0)\right) \exp\left(\displaystyle{\sum_{I<J}}\frac{\lambda d_{j_{IJ}}}{2}h_{d_{j_{IJ}}}\right) \\
\\
\;\;\;\;\;\;\;\;\;\;\; \left[1 +
\frac{1}{\lambda} \left(- \frac{H(j_{IJ})}{24}-\frac{G(x_0)}{12}- \frac{F^{\prime \prime}(x_0)+(F^{\prime}(x_0))^2}{2 f^{\prime \prime}(x_0)}+\frac{f^{(4)}(x_0)+4f^{(3)}(x_0)F^\prime(x_0)}{8(f^{\prime \prime}(x_0))^2}- \frac{5(f^{(3)}(x_0))^2}{24(f^{\prime \prime}(x_0))^3} \right) + O \left(\frac{1}{\lambda^2} \right) \right]
}}
where $x_0$ are the stationary points of the phase, i.e. $f^\prime(x_0)=0$. The next step is to identify these stationary points.

\subsection{Contributions of the stationary points}

The phase $f(x)$ is an analytical function given by:
\equa{
f(x)= i \pi x+ x \ln(x) - \displaystyle{\sum_K}\left(x-\frac{\tilde{v}_K}{2}\right) \ln\left(x-\frac{\tilde{v}_K}{2}\right) - \displaystyle{\sum_j}\left(\frac{\tilde{p}_j}{2}-x\right) \ln\left(\frac{\tilde{p}_j}{2}-x\right)
}
therefore the stationary points $x_0$ satisfy the following equation as shown in \cite{razvan}:
 \equa{ \label{stationaryequa}
  f^\prime(x)= i\pi + \ln(x)-\sum \ln\left(x-\tilde{v}_K/2\right) + \sum \ln\left(\tilde{p}_j/2-x\right)=0
  }
  which is equivalent to
  \equa{\label{stationaryequa2}
  x \prod_j (p_j-x)=-\prod_K(x-v_K)
  }
The previous equation reduces to a quadratic equation $Ax^2-Bx+C=0$ with
 \equa{
 \tabl{l}{
A=-\displaystyle{\sum_{j<l}}\tilde{p}_k \tilde{p}_l+\displaystyle{\sum_{K<L}}\tilde{v}_K \tilde{v}_L=\frac{1}{2}\left(\displaystyle{\sum_{\tabl{c}{I<J, K<L,\\ (I,J) \neq (K,L)}}d_{j_{IJ}}d_{j_{KL}}}\right)\\
B=-\tilde{p}_1\tilde{p}_2\tilde{p}_3+\displaystyle{\sum_{I<J<K}}\tilde{v}_I\tilde{v}_J\tilde{v}_K=\frac{1}{4}\left[\left(\displaystyle{\sum_{\tabl{c}{I<J, K<L,\\ (I,J) \neq (K,L)}}d_{j_{IJ}}d_{j_{KL}}}\right)\left(\displaystyle{\sum_{I<J}}d_{j_{IJ}}\right)+\displaystyle{\sum_J}\left(\displaystyle{\prod_{K \neq J}}d_{j_{JK}}\right)\right]\\
C=\displaystyle{\prod_K}v_K\\
}}
As shown in \cite{razvan}, the discriminant $\Delta= -(B^2-4AC)$ is given in terms of the $d_{j_{IJ}}$ by:
\equa{\tabl{ll}{
\Delta&=\frac{1}{16}\left[ \displaystyle{\sum_{\tabl{c}{I<J,\\ K<L,\\ (I,J)\neq (K,L)}}}d_{j_{IJ}}d_{j_{KL}}\left(\displaystyle{\sum_{\tabl{c}{M<N,\\ (M,N)\neq (I,J),\\(M,N)\neq (K,L)}}}d_{j_{MN}}^2-d_{j_{IJ}}^2-d_{j_{KL}}^2\right)- \displaystyle{\sum_K}\displaystyle{\prod_{L\neq K}}d_{j_{KL}}^2\right] \\
&\\
&=2 \begin{vmatrix}
0 &\left(\frac{d_{j_{23}}}{2}\right)^2 & \left(\frac{d_{j_{13}}}{2}\right)^2&\left(\frac{d_{j_{12}}}{2}\right)^2& 1 \\
\left(\frac{d_{j_{23}}}{2}\right)^2 &0&\left(\frac{d_{j_{03}}}{2}\right)^2&\left(\frac{d_{j_{02}}}{2}\right)^2 & 1 \\
\left(\frac{d_{j_{13}}}{2}\right)^2 &\left(\frac{d_{j_{03}}}{2}\right)^2  & 0&\left(\frac{d_{j_{01}}}{2}\right)^2&1 \\
\left(\frac{d_{j_{12}}}{2}\right)^2 & \left(\frac{d_{j_{02}}}{2}\right)^2 &\left(\frac{d_{j_{01}}}{2}\right)^2&0&1\\
1&1&1&1&0\\
\end{vmatrix} =2^4(3!)^2V^2
}}
where $V$ is the volume of the tetrahedron of edge length $d_{j_{IJ}}/2$. In the following we will focus on the case where $\Delta>0$, i.e. $V^2>0$, which corresponds to tetrahedra in flat Euclidean space. The other case $\Delta<0$ corresponds to tetrahedra admitting an embedding in the 2+1d Minkowski space-time. And so, we get two stationary points:
\equa{ \label{statiopoints}
x_\pm=\frac{B\pm i \sqrt{\Delta}}{2A}
}
The geometrical interpretation of the stationary points is not clear yet. We have shown that $\Delta$ is related to the volume of the tetrahedron. $B$ and $A$ are also related to invariant of the tetrahedron:
$$B=\sum_I \frac{v_I}{2} A + 24V \cot \theta$$
where we recall that $v_I$ is the perimeter of the triangle I of the tetrahedron. The angle $\theta$ is the Brocard angle of the tetrahedron. Indeed, $\frac{d_{j_{01}}}{2} \frac{d_{j_{02}}}{2}\frac{d_{j_{03}}}{2} ::\frac{d_{j_{03}}}{2} \frac{d_{j_{23}}}{2}\frac{d_{j_{13}}}{2} :: \frac{d_{j_{12}}}{2} \frac{d_{j_{02}}}{2}\frac{d_{j_{23}}}{2}  :: \frac{d_{j_{01}}}{2} \frac{d_{j_{12}}}{2}\frac{d_{j_{13}}}{2} $ are the barycentric coordinates of the second Lemoine point of the tetrahedron denoted $L$. This point is such that the distance from $L$ to the face $I$ of the tetrahedron is equal to $R_I \tan \theta$ where $R_I$ is the radius of the circumscribed circle of the triangle $I$ and $\theta$ is then defined by $\displaystyle{\sum_J}\left(\displaystyle{\prod_{K \neq J}}\frac{d_{j_{JK}}}{2}\right)= 12V \cot \theta$.

The geometrical significance of the stationary points still has to be understood. However, we can now give the explicit form of the leading order and of the next to leading order of the \sj-symbol.

\medskip

\textit{Leading order.}
We first focus on the leading order and on the $x_+$ contribution. This analysis has already been done in \cite{razvan} and we just recall the main steps and give the notations:

 $f(x_+)= \displaystyle{\sum_{I<J}}\frac{d_{j_{IJ}}}{2}f_{d_{j_{IJ}}}$ where
\equa{ \label{functionf}
\tabl{l}{
f_{d_{j_{0i}}}= \ln\left[\frac{(x_+-\tilde{v}_0)(x_+-\tilde{v}_i)}{\prod_{j\neq i}(\tilde{p}_j-x_+)}\right] \textrm{ for } i, j \in \{1, \cdots, 3\} \\
\\
f_{d_{j_{ik}}}=\ln\left[\frac{(x_+-\tilde{v}_k)(x_+-\tilde{v}_i)}{(\tilde{p}_k-x_+)(\tilde{p}_i-x_+)}\right] \textrm{ for } i, k \in \{1, \cdots, 3\} \\
}}

The second derivative of $f$ is given by:
$$\tabl{ll}{-f^{\prime \prime}(x_+)&= \sum_K \frac{1}{x_+-\tilde{v}_K}+\sum_j\frac{1}{\tilde{p}_j-x_+}-\frac{1}{x_+}\\
&\\
&=-i \sqrt{\Delta}\exp(-\ln(x_+\prod_j(\tilde{p}_j-x_+)))
}$$
where we have used the equation (\ref{stationaryequa2}) which gives $x_+\prod_j (\tilde{p}_j-x_+)=-\prod_K(x_+-\tilde{v}_K)$. In the same way, we can simplify $F(x_+)=-\frac{1}{2} \ln\left( x_+ \prod_j \left( \tilde{p}_j-x_+ \right) \right)$. The exponential piece of $f^{\prime \prime}(x_+)$ and $e^{F(x_+)}$ compensate and we get:
$$ \frac{1}{\sqrt{-f^{\prime \prime}(x_+)}} e^{F(x_+)}= \frac{1}{\sqrt{-i\sqrt{\Delta}}}$$

Collecting these different results yields  the following contribution of the $x_+$ stationary point:
\equa{\tabl{l}{
 \sqrt{\frac{1}{-f^{\prime \prime}(x_+)2\pi \lambda^{3}} } \exp\left(F(x_+)+\lambda f(x_+)\right) \exp\left(\displaystyle{\sum_{I<J}}\frac{\lambda d_{j_{IJ}}}{2}h_{d_{j_{IJ}}}\right)=\\
 \;\;\;\;\;\frac{1}{\sqrt{2\pi \lambda^3\sqrt{\Delta}}} \exp\left[ i \frac{\pi}{4}+ \sum_{IJ}(\lambda d_{j_{IJ}}/2)(h_{d_{j_{IJ}}}+f_{d_{j_{IJ}}})\right]
}}

The same analysis for the $x_-$ contribution yields the same contribution as the previous one with an opposite phase:
\equa{\tabl{l}{
 \sqrt{\frac{1}{-f^{\prime \prime}(x_-)2\pi \lambda^{3}} } \exp\left(F(x_-)+\lambda f(x_-)\right) \exp\left(\displaystyle{\sum_{I<J}}\frac{\lambda d_{j_{IJ}}}{2}h_{d_{j_{IJ}}}\right)=\\
 \;\;\;\;\;\frac{1}{\sqrt{2\pi \lambda^3\sqrt{\Delta}}} \exp\left[ -i \frac{\pi}{4}+ \sum_{IJ}(\lambda d_{j_{IJ}}/2)(h_{d_{j_{IJ}}}+\overline{f_{d_{j_{IJ}}}})\right]}}

We must now compute $f_{d_{j_{IJ}}}$ which is a complex logarithm. We recall that the principal value of the logarithm is defined by $\textrm{Log} z := \ln |z| +i \textrm{Arg} z$. Therefore, we have to compute $\Im(f_{d_{j_{IJ}}})=\theta_{IJ}$.  From (\ref{functionf}), we can write that:
\equa{ \label{angle}
\tabl{l}{
\theta_{0i}= \textrm{Arg} (x_+-\tilde{v}_0) +\textrm{Arg} (x_+-\tilde{v}_i) -  \displaystyle{\sum_{j \neq i}}\textrm{Arg} (\tilde{p}_j-x_+) \\
\theta_{ik}= \textrm{Arg} (x_+-\tilde{v}_k) +\textrm{Arg} (x_+-\tilde{v}_i) -  \textrm{Arg} (\tilde{p}_k-x_+)- \textrm{Arg} (\tilde{p}_i-x_+) \\
}}
The analysis done in \cite{razvan} shows that $\theta_{IJ}$ can be identified as the (exterior) dihedral angles of the tetrahedron. Moreover,
\equa{\tabl{l}{
\Re(f_{d_{j_{0i}}})=\ln \left| \frac{(x_+-\tilde{v}_0)(x_+-\tilde{x}_i)}{\prod_{j\neq i} (\tilde{p}_j-x_+)} \right|\\
\\
\Re(f_{d_{j_{ik}}})=\ln \left| \frac{(x_+-\tilde{v}_k)(x_+-\tilde{v}_i)}{(\tilde{p}_i-x_+)(\tilde{p}_k-x_+)} \right|
}}
A tedious (but interesting) computation shows that:
\equa{
\Re(f_{d_{j_{IJ}}}) + h_{d_{j_{IJ}}}=0
}
Then, summing the contributions of $x_+$ and $x_-$ we get the leading order of the 6j-symbol:
\equa{\label{LO}
\left\{ \tabl{lll}{\lambda d_{j_{01}}/2-1/2&\lambda d_{j_{02}}/2-1/2&\lambda d_{j_{03}}/2-1/2\\
\lambda d_{j_{23}}/2-1/2&\lambda d_{j_{13}}/2-1/2&\lambda d_{j_{12}}/2-1/2\\} \right\}
 \stackrel{\textrm{L.O.}}{\sim}\sqrt{ \frac{1}{12\pi \lambda^3 V}} \cos\left[  \frac{\pi}{4}+S_R \right]
}
where $S_R= \sum_{I<J}\frac{\lambda d_{j_{IJ}}}{2}\theta_{IJ}$ is the Regge action. This is the well-known limit given by Ponzano and Regge \cite{PR} and which has justified their state sum model for 3d Euclidean gravity where the \sj-symbol is the spinfoam amplitude for a single tetrahedron. 

\medskip

\begin{figure}[ht]
\includegraphics[width=5cm]{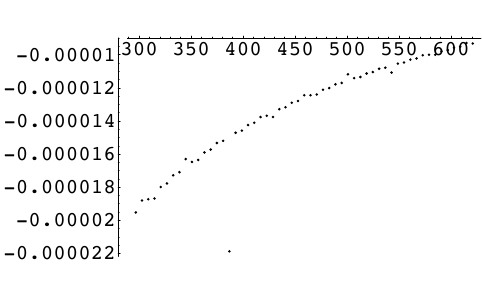} \;\;\;\;\;\;\;\;\;\;\;\;\;\;\;\;\;\;\;\;\;\;
\includegraphics[width=5cm]{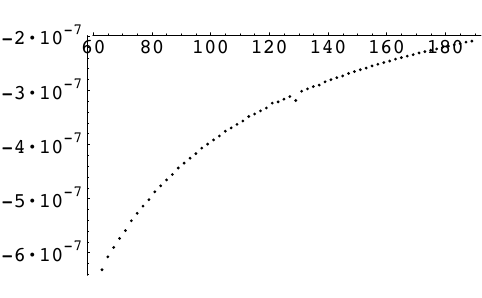}
\caption{Plots of the difference $\delta_{NLO}$ between the \sj-symbol and its analytical approximation up to NLO. On the left, we look at the \sj-symbol for $d_1=5\lambda, d_2=7\lambda, d_3=9\lambda, d_4=7\lambda, d_5=9\lambda, d_6=9\lambda$ with the x-coordinate standing for $3\lambda$. On the right, we've plotted the case $d_1=15\lambda, d_2=17\lambda, d_3=19\lambda,d_4=19\lambda, d_5=21\lambda, d_6=17\lambda$ with $\lambda$ running from 60 to 200.} \label{geneplot}
\end{figure}

\textit{Next to leading order.}
The next-to-leading order is then given by the term in $\frac{1}{\lambda^{5/2}}$ in equation (\ref{NLO1}). Using equations (\ref{termsH}-\ref{termsh}-\ref{termsSigma}), we rewrite the leading order in terms of $x_\pm, \tilde{v}_I, \tilde{p}_j$ and $\Delta$:
\equa{\label{NLO}
\frac{1}{\sqrt{48\pi \lambda^5 V}}\left\{ A(x_+, \tilde{v}_I, \tilde{p}_j,\Delta) e^{i(S_R+\frac{\pi}{4})}+A(x_-, \tilde{v}_I, \tilde{p}_j,\Delta) e^{-i(S_R+\frac{\pi}{4})}\right\}
}
where
\equa{\tabl{l}{
A(x_+, \tilde{v}_I, \tilde{p}_j,\Delta)=- \frac{H(d_{j_{IJ}})}{24}+\frac{1}{24i \sqrt{\Delta}\Delta \prod_I (x_+-\tilde{v}_I)} [-\Delta^2-3i (\displaystyle{\sum_K} \displaystyle{\prod_{L\neq K}} (x_+-\tilde{v}_L) ) \Delta \sqrt{\Delta}
\\ \;\;\; \;\;\;\;\;\;+( 9\displaystyle{\sum_K} \displaystyle{\prod_{L\neq K}} (x_+-\tilde{v}_L)^2 +6 \displaystyle{\prod_I} (x_+-\tilde{v}_I) \displaystyle{\sum_{K<L}} (x_+-\tilde{v}_K)(x_+-\tilde{v}_L))\Delta\\
\;\;\;\;\;\;\;\;\; -6i( -\displaystyle{\prod_j}(\tilde{p}_j -x_+)^3-\displaystyle{\sum_K}\displaystyle{\prod_{L\neq K}}(x_+-\tilde{v}_L)^3 +\displaystyle{\sum_j} x_+^3\displaystyle{ \prod_{l\neq j}} (\tilde{p}_l-x_+)^3\\
\;\;\;\;\;\;\;\;\;\;\;\;\;\;\;\;\;\;-(\displaystyle{\sum_K} \displaystyle{\prod_{L\neq K}} (x_+-\tilde{v}_L))(-\displaystyle{\prod_j}(\tilde{p}_j -x_+)^2+\displaystyle{\sum_K}\displaystyle{\prod_{L\neq K}}(x_+-\tilde{v}_L)^2 -\displaystyle{\sum_j} x_+^2\displaystyle{ \prod_{l\neq j}} (\tilde{p}_l-x_+)^2) )\sqrt{\Delta}\\
\;\;\;\;\;\;\;\;\; -5(-\displaystyle{\prod_j}(\tilde{p}_j -x_+)^2+\displaystyle{\sum_K}\displaystyle{\prod_{L\neq K}}(x_+-\tilde{v}_L)^2 -\displaystyle{\sum_j} x_+^2\displaystyle{ \prod_{l\neq j}} (\tilde{p}_l-x_+)^2)^2  ]
}}
Since $x_\pm$ are conjugated to each other, we obviously have $A(x_+)=\overline{A(x_-)}$.
Moreover, numerical computations shows that $\Re(A(x_\pm, \tilde{v}_I, \tilde{p}_j,\Delta))=0$, and in particular $A(x_+)=-{A(x_-)}$. This is a priori a non-trivial result to obtain from the previous formulas. Nevertheless, we tested it numerically for various choices of spin and it always turned out true. Thus we believe that there should be a way to show it analytically.
We can then give an explicit formula of the NLO of the \sj-symbol:
\be
\label{NLOsin}
\{6j\}\underset{\lambda\arr\infty}{\sim}
\{6j\}_{NLO}=
\frac{1}{\sqrt{ 12\pi \lambda^3 V}} \cos\left[  \frac{\pi}{4}+S_R \right]
-\frac{1}{\sqrt{12\pi \lambda^5 V}}\Im(A(x_+,\tilde{v}_I, \tilde{p}_j,\Delta)) \sin(S_R+\pi/4).
\ee
This result is confirmed by numerical simulations. The plots in Fig. \ref{geneplot} represent numerical simulations of the \sj-symbol minus its approximation given above (\ref{NLOsin}). Moreover, to enhance the comparison, we have multiplied by $\lambda^{5/2}$ to see how the coefficient of the next to leading order is approached and we have divided by $\cos (S_R+\pi/4)$ (oscillations of the next-to-next-to-leading order) to suppress the oscillations; that is we have plotted:
\be
\delta_{NLO}\,\equiv\,\lambda^{5/2}\,\f{\{6j\}-\{6j\}_{NLO}}{\cos (S_R+\pi/4)}.
\ee
As expected, the numerical simulations show that this rescaled difference $\delta_{NLO}$ goes to 0 as $1/\lambda$ when $\lambda$ goes to $\infty$. Moreover, the data for $\delta_{NLO}$ without any oscillation suggest that we correctly divided by $\cos (S_R+\pi/4)$ and thus the NNLO of the\sj-symbol should oscillate in $\cos (S_R+\pi/4)$. Therefore, this strongly suggest that the asymptotic expansion of the \sj-symbol in term of the length scale $\lambda$ is given by an alternative of cosines and sinus at each order. We strongly underline that this is true because we have rescaled the edge lengths $d_{j_{IJ}}$. If we had instead rescaled the spins $j_{IJ}$ as usually done, we would have found an oscillatory behavior controlled by a mixing of $\cos$ and $\sin$ at each order (as shown explicitly for the case of the isosceles tetrahedron in \cite{valentin}). This suggests that the $d_{j_{IJ}}$ are indeed the right parameter to consider when studying the semi-classical behavior of the \sj-symbol.

The only thing left to do in the present analysis is to provide the NLO coefficient $\Im(A(x_+))$ with a geometrical interpretation and to show rigourously that $\Re(A(x_+))$ vanishes.

Finally, we rewrite the approximation up to NLO of the \sj-symbol in a slightly different manner:
\be
\{6j\}\sim
\frac{1}{\sqrt{ 12\pi \lambda^3 V}} \cos\left[  \frac{\pi}{4}+S_R + \f1\lambda\Im(A(x_+)) +O\left(\f1{\lambda^2}\right)\right].
\ee
This shows that the next-to-leading corrections to the \sj-symbol can be directly considered as corrections to the Regge action for (3d) gravity:
$$
S_R^{corrected}\,\equiv\,S_R + \f1\lambda\Im(A(x_+)).
$$
We point out that an expansion in $1/\lambda$ with alternating $\cos$ and $\sin$ could be similarly re-absorbed as corrections to the Regge action. This would define in the spinfoam framework the quantum gravity corrections to classical 3d gravity due to the fundamental discreteness of the theory. Such correction would enter the gravitational correlations (of the ``graviton propagator" type) at second order as suggested in  \cite{josh}.


\section{Some particular cases}

\subsection{The equilateral tetrahedron}

\begin{figure}[ht]
\begin{center}
\includegraphics[width=5cm]{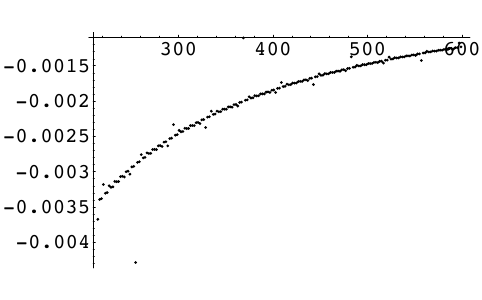}
\caption{Difference between the equilateral \sj-symbol and the analytical result (\ref{NLOequa}). The x-axis stands for $d$ and $d$ goes from 200 to 600.} \label{equiplot}
\end{center}
\end{figure}

For the equilateral tetrahedron, all the edges have the same length: that is $\forall I,J,\,\, d_{j_{IJ}}=d$. The tetrahedron with edge length $d/2$ has a volume $V=(d/2)^3\,\sqrt{2}/12$ and has all equal dihedral angles $\theta=\arccos(-1/3)$.
In this case, the expressions greatly simplify. For instance, the stationary points are $x_\pm= \frac{11 \pm i\sqrt{\frac{1}{2}}}{6}d$. Equations (\ref{LO}) and (\ref{NLO})reduce to:
\equa{\label{NLOequa}
\{6j\}^{\textrm{NLO}}_{\textrm{equi}} =\frac{2^{5/4}}{\sqrt{\pi d^3}} \cos\left(S_R+\frac{\pi}{4}\right)-\frac{31}{72\sqrt{\sqrt{2}\pi d^5}}\sin\left(S_R+\frac{\pi}{4}\right)
}
where the Regge action is $S_R=3d\theta$. The result was already obtained in \cite{valentin}. We confirm it by numerical simulations. The plot in fig.\ref{equiplot} gives the equilateral \sj-symbol minus its NLO approximation (\ref{NLOequa}). Like for the previous plots, we have multiplied by $\lambda^{5/2}$ to see how the coefficient of the next to leading order is approached and we have divided by $\cos (S_R+\pi/4)$ (oscillations of the next to next to leading order) to suppress the oscillations.

\subsection{The isosceles thetrahedron}

\begin{figure}[ht]
\begin{center}
\includegraphics[width=4cm]{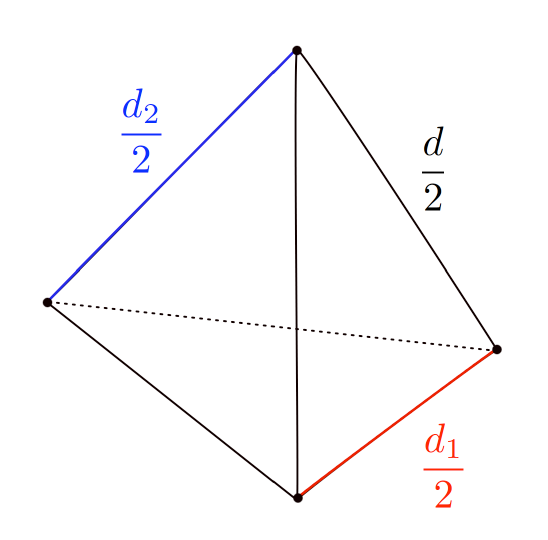}
\caption{The isosceles tetrahedron} \label{fig_isotetra}
\end{center}
\end{figure}

We now consider an isosceles tetrahedron that is a tetrahedron which has two opposite edges of length equal to $\frac{d_1}{2}$ and $\frac{d_2}{2}$ and the remaining four edges of the same length equal to $\frac{d}{2}$ (see Fig. \ref{fig_isotetra}). The volume of the tetrahedron is:
$$
V^2=\frac{1}{2^8 (3!)^2} d_1^2d_2^2\left( 4{d}^{2}-d_1^{2}-d_2^{2} \right),
$$
and the dihedral angles are:
$$
\theta= \arccos \left( \frac{-d_1d_2}{\sqrt{4d^2-d_1^2}\sqrt{4d^2-d_2^2}}\right),\qquad \theta_{1,2}=2\arccos \left( \frac{d_{2,1}}{\sqrt{4d^2-d_{1,2}^2}}\right).
$$
Once again, equations (\ref{LO}) and (\ref{NLO}) simplify and we get~:
\equa{ \label{isoequa}
\{6j\}_{\textrm{NLO}}^{(\textrm{iso})}= \frac{1}{\sqrt{12\pi V\lambda^3}}\, \cos \left(S_R +\frac{\pi}{4}\right) -  \frac{F(d,d_1,d_2)}{24V \lambda\sqrt{12\pi V\lambda^3}}\,  \sin \left(S_R +\frac{\pi}{4}\right)
}
where
$
F(d,d_1,d_2)=\frac {768{d}^{6}(d^2-d_1^{2}-d_2^{2})+736{d}^{4}d_1^{2}d_2^{2}
+240d^{4}(d_1^{4}+d_2^{4})-176{d}^{2}d_1^2d_2^{2}(d_1^2+d_2^2)-24{d}^{2}(d_1^{6}+d_2^{6})+10d_1^2 d_2^{2}(d_1^4+d_2^4)+25 d_1^{4}d_2^{4}}
{96\left( 4d^2-d_1^2 \right)\left(4d^2-d_2^2 \right)\left( 4{d}^{2}-d_1^{
2}-d_2^{2} \right)},
$
and the Regge action $S_R=2d \theta+\frac{d_1}{2}\theta_1+\frac{d_2}{2}\theta_2$.  Let us point out that the volume increases as $\lambda^3$ while $F$ goes as $\lambda^2$, so that the NLO scales properly as $\lambda^{-5/2}$.

\begin{figure}[ht]
\includegraphics[width=5cm]{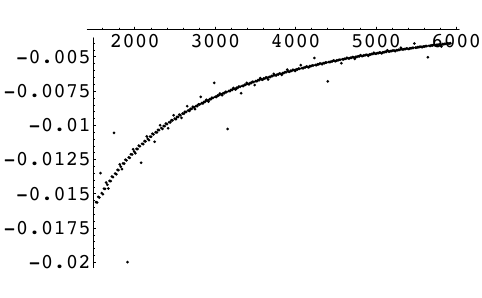} \;\;\;\;\;\;\;\;\;\;\;\;\;\;
\includegraphics[width=5cm]{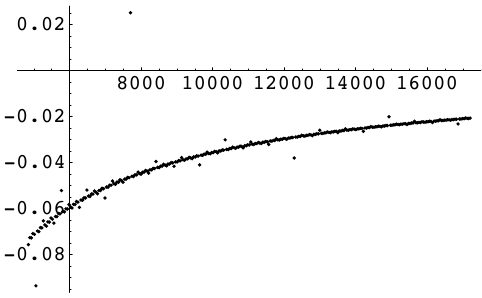}
\caption{Differences between isosceles \sj-symbols and their analytical approximation (\ref{isoequa}). The x-axis stands for $d$ with $\lambda$ goes from 200 to 600. On the left hand side, we consider isosceles tetrahedra with $d_1=3\lambda, d_2=3\lambda, d=7\lambda$. On the right hand side, we've plotted the case $d_1=9\lambda, d_2=3\lambda, d=21\lambda$.} \label{isoplot}
\end{figure}

This reproduces the result previously obtained in \cite{valentin}. We can easily check that this reduces to the previous equilateral case when $d_1=d_2=d$ and we further confirm it by numerical simulations. The plots in Fig. \ref{isoplot}  represents numerical simulations of an isosceles \sj-symbol minus the analytical formula (\ref{isoequa}). Like for the previous plot, we have multiplied the data by $\lambda^{5/2}$ to see how the coefficient of the NNLO order is approached and we have divided by $\cos (S_R+\pi/4)$ (NNLO oscillations) to suppress the oscillations.
Finally, the geometrical interpretation of the term $F(d,d_1,d_2)$ remains to be understood. If we can't provide it with a geometrical meaning, there is little hope to interpret the NLO coefficient $\Im(A(x_+))$ in the generic case. Nevertheless, we give a more compact expression for the denominator of $F$:
\be
96\left( 4d^2-d_1^2 \right)\left(4d^2-d_2^2 \right)\left( 4{d}^{2}-d_1^{
2}-d_2^{2} \right)=
\,
96^3\f{V^2}{\cos^2\theta}.
\ee
We still need to express the numerator of $F$ in term of geometrical objects. For instance, we could express it in term of $d^2$, $(4d^2-d_1^2)(4d^2-d_2^2)$ and $(4{d}^{2}-d_1^{2}-d_2^{2})$, which would provide a formula in term of the volume and the dihedral angles. Nevertheless, we haven't been able to find such a useful rewriting of this NLO coefficient.

\chapter{Recursion relations and the asymptotic expansion of the \sj} \label{recursion6j}
An alternative way to proceed in order to probe the asymptotic behavior and the induced corrections of the correlations  is to use the exact recursion relations satisfied by the \sj-symbol. It is this method we present in this Chapter. Let us point out that recursion relations satisfied by other spin foam amplitudes have been investigated in \cite{recursion}.
The following results have been published in \cite{article2}.
\section{Exact and Approximate Recursion Relation for the Isosceles Tetrahedron}

We will focus on the isosceles tetrahedron, which is relevant for the computations of geometrical correlations in the simplest non-trivial toy model in 3d quantum gravity \cite{3dtoymodel1}. Such  tetrahedron has four of its edges of equal length with the two remaining opposite edges of arbitrary length. The corresponding isosceles \sj-symbol is:
$$
\{a,b\}_J\equiv\sixj{a&J&J}{b&J&J},
$$
where $J\in \N/2$ and $a,b$ are integers smaller than $2J$ (to satisfy the triangular inequality). The associated tetrahedron has edge lengths $l_j=d_j/2$ for $j=a,b,J$, where $d_j=2j+1$ is the dimension of the $\SU(2)$ representation of spin $j$.
The volume $V$ of the tetrahedron is given by the simple formula:
\be \label{volume}
V_J(a,b)=\f1{12}l_al_b\sqrt{4l_J^2-\left(l_a^2+l_b^2\right)},
\ee
while the (exterior) dihedral angles $\theta_j$  can also be written in term of the edge lengths (see e.g. \cite{valentin} for more details):
\be \label{angles}
\cos\theta_a\,=\,
-\f{4l_J^2-l_a^2-2l_b^2}{4l_J^2-l_a^2},\quad
\cos\theta_b\,=\,
-\f{4l_J^2-l_b^2-2l_a^2}{4l_J^2-l_b^2},\quad
\cos\theta_J\,=\,
\f{-l_al_b}{\sqrt{4l_J^2-l_a^2}\sqrt{4l_J^2-l_b^2}}.
\ee
The general recursion relation for the \sj-symbol given by Schulten and Gordon in \cite{SG1,SG2} simplifies in this specific isosceles case:
\be \label{exactrecursion}
(l_a+\frac{1}{2})\left[4l_J^2-(l_a+\frac{1}{2})^2\right]\{a+1,b\}_J
-2l_a\left[(4l_J^2-l_a^2)\cos(\theta_a)+\frac{1}{4}\right]\{a,b\}_J
+(l_a-\frac{1}{2})\left[4l_J^2-(l_a-\frac{1}{2})^2\right]\{a-1,b\}_J =0
\ee
In the asymptotic regime, we know (analytically and numerically) the behavior of the \sj-symbol at the leading order:
\be
\label{LO}
\{a,b\}_J \,\sim\,
\{a,b\}_J^{LO}\,\equiv
\f1{\sqrt{12\pi V}}\,\cos\left(
l_a\theta_a +l_b\theta_b +4l_J\theta_J +\f\pi4
\right),
\ee
which is actually valid under the assumption that the tetrahedron with edge lengths $l_a,l_b,l_J$ exists (else the generic asymptotics can be expressed in term of Airy functions). The oscillatory phase is given by the Regge action $S_R=l_a\theta_a +l_b\theta_b +4l_J\theta_J$.
Using the obvious trigonometric identity $\cos((n+1)\phi)+\cos((n-1)\phi)=\,2\cos\phi\,\cos n\phi$, we can write an exact recursion relation for the leading order of the \sj-symbol:
\be \label{exactrecurLO}
\sqrt{V_J(a+1,b)}\,\{a+1,b\}_J^{LO}
-2\cos\theta_a \,\sqrt{V_J(a,b)}\,\{a,b\}_J^{LO}
+\sqrt{V_J(a-1,b)}\,\{a-1,b\}_J^{LO}=0.
\ee
A similar recursion relation holds for $b$-shifts and also $J$-shifts.

The most natural idea is to compare this recursion relation for the leading order to the previous  equation on the exact \sj-symbol to see how to use them to extract the next-to-leading correction to the asymptotic behavior. We can first  find the link between the leading order of equation (\ref{exactrecursion}) and the leading order of equation (\ref{exactrecurLO}). Both equations can be written under the same form at the leading order:
\equa{
\{a+1,b\}_J -2\cos \theta_a \{a,b\}_J +\{a-1,b\}_J \approx 0,
}
which turns into a simple second order differential equation in the large spin limit.
Then the next-to-leading order of the equation (\ref{exactrecursion}):
\equa{\tabl{ll}{
\sqrt{V_J(a,b)}\left(1+ \f{1}{2l_a} \left(1-\f{2l_a^2}{4l_J^2-l_a^2}\right)\right)&\{a+1,b\}_J-2\cos \theta_a \sqrt{V_J(a,b)}\{a,b\}_J \\
&+\sqrt{V_J(a,b)}\left(1- \f{1}{2l_a} (1-\f{2l_a^2}{4l_J^2-l_a^2})\right)\{a-1,b\}_J \approx 0\\
}}
will have to be compared to an recursion relation for the next-to-leading order of the \sj-symbol.

\section{Pushing to the Next-to-Leading Order}

We are interested in the asymptotic expansion  of the \sj-symbol. It was shown in previous works \cite{valentin,article1} that  $l_j$ seems to be the right parameter to consider when studying the semi-classical behavior of the \sj-symbol. So from now we write:
$$
\sixj{a&J&J}{b&J&J}\equiv \{l_a, l_b\}_{l_J}.
$$
Notice that shifting $a$ by $\pm 1$ is equivalent to shifting the edge length $l_a=a+1/2$ by $\pm 1$. We rescale now $l_j$ by $\lambda l_j$ and we replace the exact \sj-symbol by a series in $1/\lambda$ alternating cosines and sinus of the Regge action (shifted by $\pi/4$) in the previous equation (\ref{exactrecursion}). The fact that there is no mixing up of cosines and sinus at all order was show in \cite{valentin}. More precisely, we write the \sj-symbol asymptotic expansion under the form:
%
\equa{\label{6jNNNLO}
\tabl{ll}{
\{\lambda l_a, \lambda l_b\}_{\lambda l_J}=\frac{1}{\lambda^{3/2}D(l_a,l_b,l_J)}[\cos(\lambda S_R+\pi/4)+ &\frac{F^{(1)}(l_a,l_b,l_J)}{\lambda}\sin(\lambda S_R+\pi/4)+\frac{G^{(1)}(l_a,l_b,l_J)}{\lambda}\cos(\lambda S_R+\pi/4))\\
&+ \frac{F^{(2)}(l_a,l_b,l_J)}{\lambda^2}\cos(\lambda S_R+\pi/4)+\frac{G^{(2)}(l_a,l_b,l_J)}{\lambda^2}\sin(\lambda S_R+\pi/4))\\
&+ \frac{F^{(3)}(l_a,l_b,l_J)}{\lambda^3}\sin(\lambda S_R+\pi/4)+\frac{G^{(3)}(l_a,l_b,l_J)}{\lambda^3}\cos(S_R+\pi/4)\\
&+ \frac{F^{(4)}(l_a,l_b,l_J)}{\lambda^4}\cos(\lambda S_R+\pi/4)+\frac{G^{(4)}(l_a,l_b,l_J)}{\lambda^4}\sin(S_R+\pi/4) + O(\lambda^{-5})],
}}
where the pre-factor denominator $D(l_a,l_b,l_J)$ is given by the square-root of the tetrahedron volume as in equation \Ref{LO}.
To study the asymptotics, it is convenient to factorize the whole equation (\ref{exactrecursion}) by $\lambda^{3/2}$. We then write $\{ l_a \pm 1/ \lambda, l_b \}_{l_J}$ for $\{\lambda l_a \pm 1, \lambda l_b \}_{\lambda l_J}$. We also factorize the coefficients of the recursion relation. We start by defining $C(l_j)=l_a(4l_J^2-l_a^2)=\frac{16(A(la,l_J,l_J))^2}{l_a}$ where $A(a,b,c)=\frac{1}{4}\sqrt{(a+b+c)(a+b-c)(a-b+c)(-a+b+c)}$ is the area of the triangle of edge lengths given by $a$, $b$ and $c$.
The coefficient which appears in front of $\{l_a \pm 1/\lambda, l_b\}_{l_J}$ becomes $C(l_a\pm1/(2\lambda),l_b,l_J)=(l_a\pm 1/(2\lambda))(4l_J^2-(l_a\pm1/(2\lambda))$, where we underline that the shift is $\pm1/(2\lambda)$ and not simply $\pm1/\lambda$.  We expand $C(l_a\pm1/(2\lambda),l_b,l_J)$ in term of derivatives:
$$
C(l_a\pm1/(2\lambda),l_b,l_J)= \sum_n \frac{1}{n!}\frac{1}{(2\lambda)^n}\frac{\partial^nC}{\partial l_a^n}
$$
with \equa{\label{coefequa}
\left\{
\tabl{l}{C=l_a(4l_J^2-l_a^2) \\
\f{\pd C}{\pd l_a}=4l_J^2-3l_a^2\\
\f{ \pd^2 C}{\pd l_a^2}=-6l_a^2\\
\f{ \pd^3 C}{\pd l_a^3}=-6\\
\f{\pd^nC}{\pd l_a^n}=0 \textrm{ for } n\geq 4}
\right.
}
Then to express  $\{ l_a \pm 1/\lambda, l_b\}_{ l_J}$ we need to expand $D(l_a\pm 1/\lambda)$, $F^{(i)}(l_a\pm 1/\lambda)$, and $G^{(i)}(l_a\pm 1/\lambda)$: $(i\in \{1\cdots 4\})$
 \equa{\label{coef6j}
 \left\{
 \tabl{l}{
D(l_a\pm 1/\lambda)= D\pm \f{1}{\lambda} \frac{\partial D}{\pd l_a}+\f{1}{2\lambda^2}\f{\pd^2 D}{\pd l_a^2}\pm \f{1}{3!\lambda^3}\f{\pd^3D}{\pd l_a^3} +\f{1}{4!\lambda^4}\f{\pd^4D}{\pd l_a^4}\\
F^{(i)}(l_a\pm 1/\lambda)=\sum_{k=0}^{4-i}(-1)^{k}\f{1}{k!\lambda^k}\f{\pd^k F^{(i)}}{\pd l_a^k} \; \\
G^{(i)}(l_a\pm 1/\lambda)=\sum_{k=0}^{4-i}(-1)^{k}\f{1}{k!\lambda^k}\f{\pd^k G^{(i)}}{\pd l_a^k} \\
}
\right.}
$F^{(1)}(l_j)$ was computed in a previous paper \cite{valentin,article1}. It was also suggested that the asymptotic expansion of the \sj-symbol in term of the length  scale $\lambda$ is given by an alternative of cosines and sinus at each order, so we expect that $G^{(i)}(l_j)=0$ for $\forall i\ge 1$. Finally, we also need to expand the Regge action $\lambda S_R(l_a\pm \frac{1}{\lambda})$, remembering that $\theta_j=\theta_j(l_a)$~:
 \equa{ \label{action}
\lambda S_R(l_a\pm \frac{1}{\lambda})= \lambda S_R +\displaystyle{\sum_{k=0}^{4}}\f{(-1)^{k+1}}{(k+1)!\lambda^k}\f{\pd^k \theta_a}{\pd l_a^k}
}
 with $$\left\{ \tabl{l}{ \f {\partial \theta_a}{\partial l_a}= \f{-2l_a l_b}{(4l_J^2-l_a^2)\sqrt{4l_J^2-l_a^2-l_b^2}} \\
\\
\f{\pd^2\theta_a}{\pd l_a^2}= - \f{2l_b(4l_J^2l_a^2-2l_a^4-l_a^2l_b^2+16l_J^4-4l_b^2l_J^2)}{(4l_J^2-l_a^2)^2[4l_J^2-l_a^2-l_b^2]^{3/2}}\\
\\
\f{\pd^3 \theta_a}{\pd l_a^3}=-\f{2l_al_b(24l_b^4l_J^2+40l_J^2l_a^2l_b^2-12l_J^2l_a^4+5l_a^4l_b^2-192l_J^4l_a^2-240l_J^4l_b^2+2l_a^2l_b^4+6l_a^6+576l_J^6)}{(4l_J^2-l_a^2)^3[4l_J^2-l_a^2-l_b^2]^{5/2}}  \\
\\
\f{\pd^4 \theta_a}{\pd l_a^4}= \f{1}{(4l_J^2-la^2)^4(4l_J^2-l_a^2-l_b^2)^{7/2}}(6(8l_a^{10}+8l_a^8l_b^2+152l_J^2l_a^6l_b^2-720l_J^4l_a^6+7l_a^6l_b^4+3520l_J^6l_a^4-1472l_J^4l_a^4l_b^2+2l_a^4l_b^6\\
\quad \quad \quad+140l_a^4l_b^4l_J^2-560l_a^2l_b^4l_J^4-3840l_J^8l_a^2+2432l_J^6l_a^2l_b^2+48l_a^2l_J^2l_b^6+2048l_J^8l_b^2-448l_J^6l_b^4-3072l_J^{10}+32l_b^6l_J^4)l_b)
}\right. $$
We can now write an asymptotic recursion equation from equations (\ref{exactrecursion}), (\ref{coefequa}), (\ref{6jNNNLO}), (\ref{coef6j}) and (\ref{action}) in terms of $\lambda$ neglecting terms of order $O(\lambda^{-4})$ and smaller, assuming that $\lambda$ is large. This leads to a couple of equations at each order, one for the $\cos$-oscillations and one for the term in $\sin$:
\begin{itemize}
\item The first equation is given by the terms of order $\lambda^{0}$ and it is trivially satisfied $(0=0)$ since we have already written  the leading order of the \sj-symbol proportional to $\cos(S_R+\f\pi4)$ (the Ponzano-Regge asymptotic formulae).
\item The second equation is given by the terms of order $\lambda^{-1}$:
\equa{ \label{equa0}
\left( \f{1}{2C}\f{\pd C}{\pd l_a}-\f1D\f{\pd D}{\pd l_a}\right)\sin(\theta_a) +\f12 \f{\pd \theta_a}{\pd l_a} \cos(\theta_a)=0
}
which can be rewritten as a differential equation for $D$:
\equa{ \label{equa1}
\frac{\partial \ln D}{\partial l_a}=\frac{1}{2} \left[ \frac{\partial \theta_a}{\partial l_a} \frac{\cos \theta_a }{\sin \theta_a} + \frac{\partial \ln C}{\partial l_a} \right].
}
This allows to determine $D$: $\ln D= \f12 \ln(C\sin(\theta_a))+ K$, which simplifies into $D=K\sqrt{ l_a l_b \sqrt{4l_J^2-l_a^2-l_b^2}}$ where $K$ is a constant factor. Thus this second equation shows that $D$ is correctly proportional to the square-root of the volume $V$ of the isosceles tetrahedron. To determine the normalization constant $K$ (as well as $G^{(1)}$), the orthonormality property of \sj-coefficients can be employed: $\sum_a 4l_a \sqrt{l_bl_{b^\prime}} \{a,b\}_J \{a,b^\prime \}_J= \delta_{b b^\prime }$ and we get the $K=\sqrt{12\pi}$. The details are given in the next section.

\item The third equation is given by the terms of order $\lambda^{-2}$ and which are proportional to $\cos (S_R + \f\pi4)$
\equa{ \tabl{ll}{ \label{equa2}
 \f{\pd F^{(1)}}{\pd l_a}=\f{l_a}{4C \sin \theta_a} + &\left( \f{1}{2C}\f{\pd C}{\pd l_a } -\f1D\f{\pd D}{\pd l_a}\right) \f12 \f{\pd \theta_a}{\pd l_a}+\f16 \f{\pd^2 \theta_a}{\pd l_a^2}\\
 &+ \f{\cos \theta_a}{\sin \theta_a} \left(\f{1}{2D}\f{\pd^2 D}{\pd l_a^2}-\f{1}{8C}\f{\pd^2 C}{\pd l_a^2}+\f1D\f{\pd D}{\pd l_a}\left(\f{1}{2C}\f{\pd C}{\pd l_a}-\f1D \f{\pd D}{\pd l_a}+(\f12\f{\pd \theta_a}{\pd l_a})^2 \right) \right)
}}
where we used the fact that $\left(\f{1}{2C}\f{\pd C}{\pd l_a}-\f1D\f{\pd D}{\pd l_a}\right)\sin(\theta_a) +\f12 \f{\pd \theta_a}{\pd l_a} \cos(\theta_a)=0$ (eqn. \Ref{equa1}) to remove all the terms proportional to $F^{(1)}$ itself.
The first term of the right-hand side of the equation (\ref{equa2}) comes from the variation of the coefficient in front of $\{a,b\}_J$ in the recursion equation (\ref{exactrecursion}). The terms with a derivative of $C$ with respect to $l_a$ come from the coefficients in front of $\{a\pm 1, b\}_J$ and $\{a,b\}_J$. The variation of $C$ with respect to $l_a$ is given by the variation of the areas of the triangles of the tetrahedron. From eqn.(\ref{equa1}), we relate it to the variations of $D$ (the volume) and to the variations of the dihedral angle $\theta_a$: $\f1C \f{\pd C}{\pd l_a}=\f2D \f{\pd D}{\pd l_a} -\f{\cos \theta_a}{\sin \theta_a}\f{\pd \theta_a}{\pd l_a}$. The terms with a derivative of $D$ with respect to $l_a$ come from the variation of the leading order of the asymptotic of the \sj-symbol and the terms with a derivative of the dihedral angle $\theta_a$ come from the variations of the Regge action $S_R$. We can now compute the derivative of $F^{(1)}$ with respect of $l_a$ (equation (\ref{equa2})) in terms of $l_a$, $l_b$ and $l_J$ the edge lengths of the tetrahedron:
\equa{\tabl{ll}{
 \f{\pd F^{(1)}}{\pd l_a}=-&\f{1}{48(l_a^2(-4l_J^2+l_a^2)^2(4l_J^2-l_a^2-l_b^2)^{(5/2)}l_b)}(-32l_b^6l_J^2l_a^2+10l_b^6l_a^4+96l_b^6l_J^4-960l_J^6l_b^4+15l_a^6l_b^4+400l_J^4l_b^4l_a^2-100l_J^2l_b^4l_a^4\\
 &-168l_J^2l_a^6l_b^2-1664l_J^6l_b^2l_a^2+20l_a^8l_b^2+576l_J^4l_b^2l_a^4+3072l_J^8l_b^2-3072l_J^{10}+48l_J^4l_a^6+2304l_J^8l_a^2-576l_J^6l_a^4)
 }}
 and then easily integrate this equation over $l_a$:
\equa{\tabl{ll}{
F^{(1)}(l_j)&=\\
&-\frac {768{l}_J^{6}(l_J^2-l_a^2-l_b^2)+736{l}_J^{4}l_a^{2}l_b^{2}+240l_J
^{4}(l_a^{4}+l_b^{4})-176{l}_J^{2} l_a^
{2}l_b^{2}(l_a^2+l_b^2)-24{l}_J^{2}( l_a^{6}+l_b^{6})+10l_a^{2} l_b^
{2}(l_a^4+l_b^4)+25 l_a^{4}l_b^{4}}
{ 24\left( 4l_J^2-l^2_b \right)\left( 2l_J^2-l_a^2 \right)\left( 4{l}_J^{2}-l_b^{
2}-l_a^{2} \right)^{3/2}l_al_b }  +Z(l_b, l_J)
}}
The integration constant $Z(l_b, l_J)$ can be determined using the symmetry properties of the \sj-symbol: symmetry of the isosceles \sj-symbol with respect to $l_a$ and $l_b$, coupling of $l_a$, $l_b$ and $l_J$ by this isosceles \sj-symbol and homogeneity of $F^{(1)}$ ($[F^{(1)}]=l_j^{-1}$) imply that $Z(l_b,l_J)=0$. Then this gives us the same result as in the previous paper \cite{article1}. Moreover, using the definitions of the tetrahedron volume (\ref{volume}) and of the dihedral angles (\ref{angles}), we can express $F^{(1)}$ in terms of some geometrical characteristics of the tetrahedron:
\equa{ \label{geometricNLO}
F^{(1)}=-\f{\cos \theta_J\left(3(12V)^8-(12V)^4l_a^4l_b^4\left(3(l_a^2-l_b)^2+2l_a^2l_b^2\right)-l_a^{12}l_b^{12} \right)+6l_a^{12}l_b^{12}}{48(12V)^3l_a^8l_b^8}
}
\item The fourth equation is given by the terms of order $\lambda^{-2}$ and which are proportional to $\sin (S_R + \f\pi4)$. It is the same equation as the previous one for $G^{(1)}$  but the right-hand side is now equal to zero (homogenous equation). That is we simply get that
\equa{
 \f{\pd G^{(1)}}{\pd l_a}= 0
}
so $G^{(1)}=Z(l_b,l_J)$ is just a constant of integration. Once again the symmetry properties of the \sj-symbol implies that $G^{(1)}=0$.
\item The next equation is given by the terms of order $\lambda^{-3}$ and which are proportional to $\sin (S_R + \f\pi4)$. We get an equation for the first derivative of $F^{(2)}$ with respect to $l_a$
\equa{ \tabl{l}{\label{equa5}
\f{\pd F^{(2)}(l_j)}{\pd l_a}= \f{\cos \theta_a}{2 \sin \theta_a} \f{\pd^2F^{(1)}}{\pd l_a^2} - \left(\f{1}{\sin^2\theta_a}\f{\pd \theta_a}{\pd l_a} + F^{(1)}\right) \f{\pd F^{(1)}}{\pd l_a} \\
\;\;\;+ \f{\cos \theta_a}{\sin \theta_a} \left[  -\f{1}{4!}\f{\pd^3 \theta_a}{\pd l_a^3}+\left(\f1D \f{\pd D}{\pd l_a}\f{1}{2C}\f{\pd C}{\pd l_a} + \f{1}{2D}\f{\pd^2D}{\pd l_a^2}-\left(\f1D\f{\pd D}{\pd l_a}\right)^2-\f{1}{8C}\f{\pd^2C}{\pd l_a^2}+ \f{1}{3!}\left(\f12\f{\pd \theta_a}{\pd l_a}\right)^2\right)\f12 \f{\pd \theta_a}{\pd l_a} +\left(\f1D\f{\pd D}{\pd l_a}-\f{1}{2C}\f{\pd C}{\pd l_a}\right) \f{1}{3!}\f{\pd^2\theta_a}{\pd l_a^2}  \right] \\
\;\;\; +\f{1}{2C}\f{\pd C}{\pd l_a}\f{1}{2D}\f{\pd^2 D}{\pd l^2_a}+\f{1}{8C}\f{\pd^2 C}{\pd l^2_a}\f{1}{D}\f{\pd D}{\pd l_a} +\left( \f{1}{2C}\f{\pd C}{\pd l_a}-\f{1}{D}\f{\pd D}{\pd l_a}\right) \f12 \left(\f{1}{2}\f{\pd \theta_a}{\pd l_a}\right)^2 +\f{1}{3!}\f{\pd^2 \theta_a}{\pd l^2_a}\f{1}{2}\f{\pd \theta_a}{\pd l_a} +\left(\f{1}{D}\f{\pd D}{\pd l_a}\right)^3-\f{1}{D}\f{\pd D}{\pd l_a}\f{1}{D}\f{\pd^2 D}{\pd l^2_a}+ \f{1}{3!D}\f{\pd^3 D}{\pd l^3_a}\\
\;\;\; -\f{1}{8\cdot 3! C}\f{\pd^3 C}{\pd l^3_a} -\f{1}{2C}\f{\pd C}{\pd l_a}\left(\f{1}{D}\f{\pd D}{\pd l_a}\right)^2
}}
We recall that $D$ is proportional to the square root of the tetrahedron volume, $C$ can be expressed in terms of the volume $V$ and the sinus of the dihedral angle $\theta_a$ (see equation (\ref{equa1})). To integrate this equation, we first express explicitly~\footnotemark it in terms of $l_a$, $l_b$ and $l_J$,
\footnotetext{$\f{\pd F^{(2)}}{\pd l_a}(l_j)=-\f{1}{2304\left((4l_J^2-l_b^2)l_a^3(4l_J^2-l_a^2)^3(4l_J^2-l_a^2+l_b^2)^4l_b^2\right)}(-1604l_a^8l_b^8l_J^2+1250816l_a^4l_J^8l_b^6-207104l_a^6l_b^6l_J^6+31904l_a^{10}l_J^4l_b^4
 -169344l_a^4l_b^8l_J^6-3920l_a^6l_b^{10}l_J^2+24992l_a^6l_b^8l_J^4-46848l_a^{10}l_J^6l_b^2-7129088l_a^4l_J^{10}l_b^4+1770496l_a^6l_J^8l_b^4+16832l_a^4l_b^{10}l_J^4
+34368l_a^8l_J^4l_b^6-6816l_a^{12}l_J^4l_b^2-278912l_a^8l_J^6l_b^4+14524416l_a^2l_J^{12}l_b^4+486144l_a^2l_J^8l_b^8-560l_b^{12}l_a^4l_J^2-43776l_a^2l_b^{10}l_J^6-3317760la^2lJ^{10}l_b^6
+22241280l_a^4l_J^{12}l_b^2+794l_a^{12}l_b^6-6955008l_a^6l_J^{10}l_b^2+2801664l_J^{12}l_b^6+672l_a^{14}l_b^2l_J^2-26542080l_a^4l_J^{14}-451584l_J^{10}l_b^8+46080l_J^8l_b^{10}
-2304l_b^{12}l_J^6-21233664l_J^{18}+37158912l_a^2l_J^{16}+1072128l_a^8l_J^8l_b^2-1528la^{12}l_b^4l_J^2-10911744l_J^{14}l_b^4-35979264l_a^2l_J^{14}l_b^2+1728l_b^{12}l_J^4l_a^2
+9953280l_a^6l_J^{12}+228096l_a^{10}l_J^8-10368l_a^{12}l_J^6-2073600l_a^8l_J^10+27l_a^{10}l_b^8+23592960l_J^{16}l_b^2+400l_a^8l_b^{10}-88l_a^{14}l_b^4-8144l_a^{10}l_b^6l_J^2+100l_b^{12}l_a^6)$}
and then deduce $F^{(2)}$:
\equa{ \tabl{ll}{ \label{NNLO}
F^{(2)}(l_j)&=\f{-1}{4608\left((4l_J^2-l_a^2)^2(4l_J^2-l_b^2)^2(4l_J^2-l_a^2-l_b^2)^3l_a^2l_b^2\right)}(-2359296l_a^2l_J^{10}l_b^4-224512l_a^6l_J^6l_b^4+100l_a^{12}l_b^4+576l_J^4l_b^{12}+112896l_J^8l_b^8\\
&+2727936l_a^4l_J^{12}+5308416l_J^{16}+212l_b^{10}l_a^6-5898240l_a^2l_J^{14}-11520l_a^{10}l_J^6+941056l_a^4l_J^8l_b^4+31584l_a^8l_J^4l_b^4\\
&-2416l_a^4l_b^{10}l_J^2
-79872l_a^8l_J^6l_b^2-480l_b^{12}l_J^2l_a^2-7040l_a^8l_b^6l_J^2-2416l_a^{10}l_J^2l_b^4+100l_a^4l_b^{12}+212l_a^{10}l_b^6\\
&+2727936l_J^{12}l_b^4-700416l_J^{10}l_b^6-5898240l_J^{14}
l_b^2-11520l_J^6l_b^{10}-700416l_a^6l_J^{10}+609l_a^8l_b^8+112896l_a^8l_J^8\\
&+576l_a^{12}l_J^4-2359296l_a^4l_J^{10}l_b^2+528384l_a^6l_J^8l_b^2+5849088l_a^2l_J^{12}l_b^2-79872l_a^2l_b^8l_J^6+31584l_a^4l_b^8l_J^4-7040l_a^6l_b^8l_J^2\\
&-224512l_a^4l_b^6l_J^6+58816l_a^6l_b^6l_J^4+8640l_a^{10}l_J^4l_b^2+8640l_b^{10}l_J^4l_a^2-480l_a^{12}l_J^2l_b^2+528384l_a^2l_J^8l_b^6)
}}
which is the only result with the required symmetries\footnote{If the result is not symmetric after integration, a non-null integration constant has to be added and  its determination can be done using the symmetry properties of the \sj-symbol. Indeed, we have $\f{\pd F^{(2)}}{\pd l_a}= H(l_a, l_b, l_J)$ so by integration over $l_a$, $F^{(2)}(l_j)=h(l_a, l_b, l_J)+Z(l_b, l_J)$. Moreover by symmetry, we must have $\f{\pd F^{(2)}}{\pd l_b}= H(l_a=l_b, l_b=l_a, l_J)$ and then integrating over $l_b$, we obtain a second expression for $F^{(2)}$: $F^{(2)}(l_j)=h(l_a=l_b, l_b=l_a, l_J)+Z(l_a, l_J)$ which implies that the constant of integration satisfies $Z(l_b,l_J)-Z(l_a,l_J)=h(l_a=l_b, l_b=l_a, l_J)- h(l_a, l_b, l_J)$. This equation allows to determine $Z$ and to get (\ref{NNLO}).}. The geometrical meaning of this function does not seem obvious. Nevertheless, we can give a more compact expression for the denominator of $F^{(2)}$:
\equa{\label{denoF2}
(4l_J^2-l_a^2)^2(4l_J^2-l_b^2)^2(4l_J^2-l_a^2-l_b^2)^3l_a^2l_b^2= \f{(12V)^6}{\cos^4\theta_J}.
}
\item The next equation comes from the terms of order $\lambda^{-3}$ which are proportional to $\cos(S_R+\f\pi4)$:
\equa{
\f{\pd G^{(2)}}{\pd l_a}(l_j)=0
}
which implies once again that $G^{(2)}=Z(l_b,l_J)$ is a constant of integration. Then the symmetry properties of the \sj-symbol implies $G^{(2)}(l_j)=0$.
\end{itemize}

\begin{figure}[ht]
\begin{center}
\includegraphics[width=4cm]{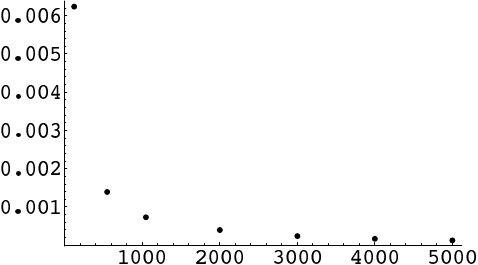}
\caption{Difference between the equilateral \sj-symbol and the analytical result (\ref{equaNNLO}). The x-axis stands for $l=d/2$ and $d$ goes from 100 to 5000. The error decreases as expected as $l^{-1}$ confirming our asymptotic formula.} \label{plotequiNNLO}
\end{center}
\end{figure}

We can now give the asymptotic expansion of an isosceles \sj-symbol until the next to next to leading order (NNLO):
\equa{ \label{isoNNLO}
\{ l_a, l_b\}^{\textrm{NNLO}}_{l_J}= \f{1}{\sqrt{12\pi V_{l_J}(l_a,l_a)}} \left[\cos (S_R +\f\pi4)+F^{(1)}(l_j) \sin(S_R+\f\pi4)+F^{(2)}(l_j) \cos(S_R+\f\pi4) \right]
}
where the expression for $F^{(1)}$ and $F^{(2)}$ are given by equations (\ref{geometricNLO}) and (\ref{NNLO}). This result seems to confirm that the expansion of the \sj-symbol is a series alternating cosines and sinus of the Regge action (shift by $\f\pi4$). In the case of an equilateral tetrahedron, all the edges have the same length, that is $l_a=l_b=l_J=l$ and $V= \f{\sqrt{2}}{12} l^3$. Then equation (\ref{isoNNLO}) reduces to:
\equa{\label{equaNNLO}
\{6j\}^{\textrm{NNLO}}_{\textrm{equi}}= \f{1}{\sqrt{\pi l^3\sqrt{2}}} \cos(S_R+\f\pi4)-\f{ 31}{72\,2^{1/4}\,2^{5/2}\sqrt{\pi l^5}}\sin(S_R +\f\pi4)-\f{45673}{20736}\f{1}{\,2^{1/4}\,2^{4}\sqrt{\pi l^7}}\cos(S_R+\f\pi4)
}
where the Regge action is given by $S_R=6 l \theta$ and $\theta=\theta_a=\theta_b=\theta_J= \arccos(-1/3)$.
This result is confirmed by numerical simulations. The plot in Fig. \ref{plotequiNNLO} represents numerical simulations of the equilateral \sj-symbol minus its approximation (\ref{equaNNLO}). Moreover, to enhance the comparison, we have multiplied by $l^{7/2}$ to see how the coefficient of the NNLO is approached and we have divided by $\sin(S_R+\f\pi4)$ (oscillations of the next to next to next to leading order) to suppress the oscillations. This gives an error that decreases as expected as $l^{-1}$.

\begin{itemize}
\item The next two equations come from the terms of order $\lambda^{-4}$. The equation for $G^{(3)}$ is the same as the one for $G^{(1)}$ and $G^{(2)}$, that is: $\f{\pd G^{(3)}}{\pd l_a}=0$. Using the same arguments of symmetry we deduce that $G^{(3)}=0$. This confirms our expectation of a series alternating cosines and sines in the asymptotic of the \sj-symbol:
\equa{\label{asympt}
 \{\lambda l_a, \lambda l_b\}_{\lambda l_J}=\frac{1}{\lambda^{3/2}D(l_a,l_b,l_J)}\left[\cos(\lambda S_R+\f\pi4)+ \displaystyle{\sum_{k=1}^\infty}\frac{F^{(k)}(l_a,l_b,l_J)}{\lambda^k}\cos(\lambda S_R+\f\pi4+\eps(k)\f\pi2)\right],
}
where $\eps(k)=-1$ when $k$ is odd and $\eps(k)=0$ when $k$ is even.
We already have an expression for $F^{(1)}$ and $F^{(2)}$. The equation for $F^{(3)}$ is the second equation of order $\lambda^{-4}$ and gives its first derivative with respect to $l_a$ in terms of $l_a$, $l_b$ and $l_J$. It is straightforward (though lengthy) to integrate it over $l_a$ and get the expression\footnotemark of $F^{(3)}$ in terms of $l_a$, $l_b$ and $l_J$. In the equilateral case ($l_a=l_b=l_J=l$), the formula reduces to:
\equa{
F^{(3)}=\frac{28833535}{17915904}\frac{1}{2^{9/2}2^{1/4}l^3}
}
\footnotetext{$F^{(3)}(l_j)=-\f{1}{3317760(l_a^3(l_a-2l_J)(l_a+2l_J)(4l_J^2-l_b^2)^3(4l_J^2-l_a^2-l_b^2)^(9/2)l_b^3(4l_J^2-l_a^2)^2)}(-6965741568l_a^6l_J^{10}l_b^8+1069728l_a^6l_b^{16}l_J^2+28270080l_a^{16}l_J^6l_b^2-743040l_J^4l_a^2l_b^{18}+1750503260160l_J^{20}l_a^2l_b^2+379404800l_a^{12}l_J^6l_b^6+788705280l_a^{10}l_J^{10}l_b^4-33547200l_a^6
l_b^{14}l_J^4-98578608l_a^{12}l_b^8l_J^4-81032970240l_a^8l_J^{14}l_b^2+379404800l_a^6l_b^{12}l_J^6-389214720l_J^8l_a^2l_b^{14}-323813376000l_a^6l_J^{18}-156036l_a^8l_b^{16}-22176l_J^2l_a^4l_b^{18}-3824640l_a^{16}l_J^4l_b^4-28408l_a^{18}l_b^6-31104000l_a^{16}l_J^8-1262270545920l_J^{18}l_a^2l_b^4-1262270545920l_J^{18}l_a^4l_b^2+461814497280l_J^{16}l_a^2l_b^6+1069728l_a^{16}l_J^2l_b^6+74144841728l_a^6l_J^{12}l_b^6+824520278016l_J^{16}l_a^4l_b^4+461814497280l_a^6l_J^{16}l_b^2+788705280l_J^{10}l_a^4l_b^{10}-267541217280l_a^6l_J^{14}l_b^4+115174656l_J^6l_a^4l_b^{14}-999532800l_J^8l_a^4l_b^{12}+68611276800l_a^8l_J^{16}-3824640l_J^4l_a^4l_b^{16}-1242078720l_a^{10}l_J^8l_b^6+28270080l_J^6l_a^2l_b^{16}+38483472384l_J^{12}l_a^4l_b^8+2157096960l_a^{12}l_J^{10}l_b^2-1242078720l_a^6l_b^{10}l_J^8+509607936000l_J^{24}-33547200l_a^{14}l_b^6l_J^4+1222041600l_J^{12}l_a^2l_b^{10}+11119152l_a^{12}l_b^{10}l_J^2-999532800l_a^{12}l_J^8l_b^4+783522201600l_J^{20}l_a^4+2157096960l_J^{10}l_a^2l_b^{12}-28408l_a^6l_b^{18}-389214720l_a^{14}l_J^8l_b^2-783601920l_a^8l_J^8l_b^8-98578608l_a^8l_J^4l_b^{12}-6965741568l_a^8l_J^{10}l_b^6-323813376000l_J^{18}l_b^6+1222041600l_a^{10}l_J^{12}l_b^2+115174656l_a^{14}l_J^6l_b^4-743040l_a^{18}l_J^4l_b^2-22176l_a^{18}l_J^2l_b^4-81032970240l_J^{14}l_a^2l_b^8+7201368l_a^{14}l_b^8l_J^2-114011616l_a^{10}l_b^{10}l_J^4-985242009600l_J^{22}l_b^2+568565760l_a^{10}l_b^8l_J^6-1401753600l_a^{12}l_J^{12}-267541217280l_J^{14}l_a^4l_b^6+568565760l_a^8l_J^6l_b^{10}+11119152l_a^{10}l_b^{12}l_J^2+783522201600l_J^20l_b^4-342523l_a^{12}l_b^{12}+38483472384l_a^8l_J^{12}l_b^4-3981312000l_a^{10}l_J^{14}-156036l_a^{16}l_b^8-511602l_a^{10}l_b^{14}+344217600l_b^{14}l_J^{10}-1401753600l_b^{12}l_J^{12}+1036800l_b^{18}l_J^6+344217600l_a^{14}l_J^{10}+68611276800l_J^{16}l_b^8-3981312000l_J^{14}l_b^{10}-511602l_a^{14}l_b^{10}-31104000l_b^{16}l_J^8-985242009600l_J^{22}l_a^2+1036800l_a^{18}l_J^6+7201368l_a^8l_b^{14}l_J^2)$}
Therefore, we have the expression of the asymptotic expansion of the equilateral \sj-symbol up to the next-to-next-to-next to leading order (NNNLO):
\bes
\{6j\}^{\textrm{NNNLO}}_{\textrm{equi}}&= \f{1}{2^{1/4}\,\sqrt{\pi l^3}} &
\left[ \cos(S_R+\f\pi4)-\f{31}{72\,2^{5/2}\,l}\sin(S_R +\f\pi4)\right.\nn\\
&&-\left.\f{45673}{20736\,2^4\,l^2}\cos(S_R+\f\pi4) +\frac{28833535}{17915904\,2^{9/2}\,l^3}\sin(S_R+\f\pi4)\right].
\ees
We check this result numerically by computing it using Mathematica for two values of spins $j=50$ and $j=100$. More precisely, we computed the renormalized error $\f{(\{6j\}_{\textrm{equi}}-\{6j\}^{\textrm{NNNLO}}_{\textrm{equi}})l^{9/2}}{\cos(S_R+\f\pi4)}$ and we got the expected $1/\lambda $-behavior. However for $l> 100$, Mathematica is not accurate enough and the numerical errors are too important to get exploitable results.
In the general isosceles case, the expression of $F^{(3)}$ is quite complicated and its geometrical interpretation remains to be understood. Nevertheless, we can again give as before a more compact formula for the denominator of $F^{(3)}$:
\equa{\label{denoF3}
\textrm{denominator}_{F^{(3)}}=3317760 (4l_J^2-l_a^2-l_b^2)^{9/2}l_a^3l_b^3(4l_j^2-l_a^2)^3(4l_J^2-l_b^2)^3= 30(48)^3\f{(12V_J(a,b))^9}{\cos^6\theta_J}
}
From this equation, the equation giving the denominator of $F^{(2)}$ and remembering that the denominator of $F^{(1)}$ can be written under a similar form: denominator$_{F^{(1)}}=48\frac{(12V)^3}{\cos^2\theta_J}$, we can conjecture that:
\equa{
\textrm{denominator}_{F^{(k)}} \propto \frac{(12V_J(a,b))^{3k}}{(\cos \theta_J)^{2k}}
}
where $F^{(k)}$ are the terms appearing in the asymptotic expansion of the \sj-symbol (\ref{asympt}). And consequently, the numerator of $F^{(k)}$ is a polynomial in $l_j$ of degree $8k$.
\end{itemize}
So, using the recursion relation for the isosceles \sj-symbol as well as its symmetry properties, we have computed explicitly the asymptotic expansion of the isosceles \sj-symbol to the fourth order up to an overall factor $K$ (this integration constant $K$ comes from the integration of the first equation (\ref{equa1})). The well-known value $K=\sqrt{12 \pi}$ which already appears in the Ponzano-Regge formula  can be obtained easily using the unitary property of the \sj-symbol, as we show in the next section.  The equilateral case has been checked against numerical calculations. This method using the recursion relation is fairly easy to implement. It requires integrating a rational fraction at each level and does not involve neither Riemann sum nor saddle point analysis.
Moreover, since the coefficient $C$ of the recursion relation (\ref{exactrecursion}) is a polynomial of degree 3; $\f{\pd^n C}{\pd l_a^n}=0$ for $n \geq 4$. Therefore, we expect to get a stable relation for the first derivative of $F^{(k)}$ and $G^{(k)}$ with respect to $l_a$ for  $k \geq 3$. On one hand, this allows to prove that $G^{(k)}$ always vanishes; and on the other hand, it should provide a systematic method to extract $F^{(k)}$ for arbitrary order $k$.


We conclude this section with a general remark on the asymptotic expansion of the \sj-symbol. In the context of 3d quantum gravity, it is often argued that the leading order of the \sj-symbol is a $\cos(S)$ instead of a complex phase $\exp(+iS)$, thus reflecting that the path integral is invariant under a change of (local) orientation (see e.g.\cite{laurent}). This obviously neglects the $+\pi/4$ shifts, which can be considered as a quantum effect (like an ordering ambiguity). However, in the light of the present expansion, it is clear that we have terms of the type $\sin(S)$ beyond the leading order and such terms are not invariant under the change $S\arr -S$. This means that the role of this symmetry in the spinfoam path integral should be more subtle than originally thought.

\section{Consequences of the unitary property of the \sj-symbol}

The orthogonality property of the \sj-symbols states that:
\equa{
\sum_{l_a} 4l_a \sqrt{l_b l_{b^{\prime}}} \{l_a, l_b\}_{l_J} \{l_a, l_{b^{\prime}}\}_{l_J}=\delta_{l_b l_{b^{\prime}}}
}
This relation corresponds to the unitarity of the evolution in the Ponzano-Regge 3d quantum gravity.
We want to use this property to determine the constant of the leading order of the \sj-symbol. From the recursion relation we have shown that $\{l_a, l_b\}^{\textrm{LO}}_{l_J}= \f{K}{\sqrt{V_J(a,b)}} \cos(S_R+\f\pi4)$; but $K$ is still undetermined. For large spin and for $l_b \approx l_{b^\prime}$, we can approximate the unitary property at the leading order in $(l_b-l_{b^\prime})$ by:
\equa{\label{unitary}
 \int_0^\infty dl_a 4 l_a l_b \f{K^2}{V_J(a,b)}\cos(S_R(l_a, l_b)+\f\pi4) \cos(S_R(l_a, l_{b^\prime})+\f\pi4)\approx \delta(l_b-l_{b^\prime } ).
}
The product of the cosines can be simplified at leading order:
\bes
\cos(S_R(l_a, l_b)+\f\pi4) \cos(S_R(l_a, l_{b^\prime})+\f\pi4)
&=&
\f12\left[\cos(S_R(l_a, l_b)+S_R(l_a, l_{b^\prime})+\f\pi2)+\cos(S_R(l_a, l_b)-S_R(l_a, l_{b^\prime}))\right] \nn\\
&\sim&
\f12\left[\cos(2S_R(l_a, l_b)+\f\pi2)+\cos((l_b-l_{b^\prime})\theta_b)\right], \nn
\ees
%
where the dihedral angle $\theta_b=\arccos \left( -\f{4l_J^2-l_b^2-2l_a^2}{4l_J^2-l_b^2}\right)$ is considered as a function of the length $l_a$. We do a saddle point approximation. The first term oscillates and its integral is exponentially suppressed. And we are left with the second term, which should satisfy the following equation:
\equa{
 \int_{-\infty}^\infty dl_a  l_a l_b \f{K^2}{V_J(a,b)}\cos((l_b-l_{b^\prime }) \theta_b) \approx \delta(l_b-l_{b^\prime } )
}
We recall that:
$$
\f{1}{2\pi} \int_{-\infty}^\infty dl_a \cos(l_a(l_b-l_{b^\prime }))= \delta(l_b-l_{b^\prime })
$$
therefore we can conclude that
\equa{
l_al_b \f{K^2}{V}=\f{1}{2\pi} \left| \f{\pd \theta_b}{\pd l_a} \right|.
}
$\theta_b$ and $l_b$ are so conjugate variables and $K$ comes from the Jacobian of the change of variables between $l_a$ and $\theta_b$.
Computing the derivative of the dihedral angle gives:
\be
\f{\pd \theta_b}{\pd l_a}= \f{-2}{\sqrt{4l_J^2-l_a^2-l_b^2}}=\f{-l_al_b}{6V_J(a,b)}
\quad
\Rightarrow
\quad
K=\f{1}{\sqrt{12\pi}}.
\ee
Moreover, pushing the approximation of the unitary property to the next to leading order in $(l_b-l_{b^\prime})$ and using the next to leading order of the \sj-symbol shows that $G^{(1)}=0$. This was already shown in the previous part using the recursion relation and the symmetry properties of the \sj-symbol and comes as a confirmation.

\section{``Ward-Takahashi identities" for the spinfoam graviton propagator}

We are interested in the two-point function in 3d quantum gravity for the simplest triangulation given by a single tetrahedron. This provides the first order of the ``spinfoam graviton propagator" in 3d quantum gravity.

Considering the isosceles tetrahedron, we focus on the correlations between the two representations $a$ and $b$:
\be
\la \cO(a) \tcO(b)\ra_{\psi_J}=
\f{1}{Z} \,\sum_{a,b}\psi_J(a)\psi_J(b)\cO(a) \tcO(b) \{a,b\}_J,
\qquad
Z\equiv\,\sum_{a,b}\psi_J(a)\psi_J(b)\{a,b\}_J,
\ee
where $\psi_J(j)$ is the boundary state, which depends also on the bulk length scale $J$, and $\cO,\tcO$ are the observables whose correlation we are studying.

Now, inserting a recursion relation with shifts on $a$, $b$ or $J$ in the sum over the representation labels $\sum_{a,b}$ leads to equations relating the expectation values of different observables. We distinguish two cases: when the state $\psi_J$ does not change or when the length scale $J$ also varies.

\subsection{Relating Observables}

Inserting the recursion relation on $a$-shifts in the definition of the correlation function, we obtain the following exact identity:
\be\tabl{ll}{
\la \f{\psi_J(a-1)}{\psi_J(a)}\cO(a-1)\tcO(b)(l_a-\f12)(4l_J^2-(l_a-\f12)^2)\ra_\psi&-
\la \cO(a)\tcO(b)2l_a(2\cos \theta_a (4l_J^2-l_a^2)+\f14)\ra_\psi\\
&+ \la \f{\psi_J(a+1)}{\psi_J(a)}\cO(a+1)\tcO(b)(l_a+\f12)(4l_J^2-(l_a+\f12)^2)\ra_\psi
=0.
}\ee
We call this a Ward identity for our spinfoam correlation.
If the observable diverges at $a=0$, more precisely if it contains terms in $1/a$ or in $1/(a+1)$, then we need to take into account extra boundary terms in this equation corresponding to contributions at $a=0$. But all observables usually considered are regular in this sense.

Then one can choose different sets of observables $\cO$ and $\tcO$ and one gets different identities on the correlation functions of the spinfoam model. 
 For example, taking $\cO(a)=l_a$, we get:
$$ \tabl{ll}{
\la \f{\psi_J(a-1)}{\psi_J(a)}\tcO(b)(l_a-1)(l_a-1/2)(4l_J^2-(l_a-1/2)^2)\ra_\psi&-
\la \tcO(b)(2\cos \theta_a l_a^2(4l_J^2-l_a^2)+l_a^2/2)\ra_\psi\\
&+\la \f{\psi_J(a+1)}{\psi_J(a)}\tcO(b)(l_a+1)(l_a+1/2)(4l_J^2-(l_a+1/2)^2)\ra_\psi
=0.
}$$
We recall that the area of the triangle of edge lengths given by $l_a$, $l_J$, $l_J$ is equal to $A(l_a, l_J)= \f14 l_a\sqrt{4l_J^2-l_a^2}$; then $(l_a\pm1)(l_a \pm 1/2)(4l_J^2-(l_a\pm 1/2)^2)=16[A^2(l_a\pm1/2,l_J)\pm \f{A^2(l_a\pm1/2,l_J)}{2(l_a\pm 1/2)}]$, therefore we can rewrite the previous equation as  an equation between correlation functions of the observable $\tcO(b)$ and different observables proportional to the square of the triangle area $A(l_a, l_J)$:
$$ \tabl{ll}{
\la \f{\psi_J(a-1)}{\psi_J(a)}[A^2(l_a-1/2,l_J)-\f{A^2(l_a-1/2,l_J)}{2(l_a- 1/2)}]\tcO(b)\ra_\psi&-
\la(2\cos \theta_aA^2(l_a,l_J)+l_a^2/2)\tcO(b)\ra_\psi\\
&+\la \f{\psi_J(a+1)}{\psi_J(a)}[A^2(l_a+1/2,l_J)+\f{A^2(l_a+1/2,l_J)}{2(l_a+ 1/2)}]\tcO(b)\ra_\psi
=0.
}$$
The standard choice of boundary is a phased Gaussian \cite{graviton1, 3dtoymodel1, physical}:
\be
\psi_J(j) \,\sim\, e^{i2l_j\vtheta} e^{-2\alpha\f{(l_j-l_J)^2}{l_J}},
\ee
where $\vtheta$ is a fixed angle defining a posteriori the external curvature of the boundary and $\alpha$ is an arbitrary real positive number (which can be fixed by the requirement of a physical state \cite{physical}).
In this case, we can compute explicitly the ratios $\psi(a\pm1)/\psi(a)$ entering the Ward identity:
$$
\f{\psi_J(a\pm 1)}{\psi_J(a)}= e^{\pm i2\vtheta} e^{\mp 4 \alpha \f{l_a-l_J}{l_J}} e^{-\f{2\alpha}{l_J}}
$$
Of course, this ratios does not depend on $b$; therefore if the observable $\tcO(b)=1$, then the dependence on $b$ only appears in one correlation function through the cosine of the dihedral angle $\theta_a$.
%
As another example, we consider $\cO(a)=l_a^{-1}$ and $\tcO(b)=\f{4l_J^2-l_b^2}{(2l_J)^{4}}$, then:
$$\tabl{ll}{
\la \f{\psi_J(a-1)}{\psi_J(a)}\, \f{l_a-1/2}{l_a-1}\,\f{4l_J^2-(l_a-1/2)^2}{4l_J^2}\, \f{4l_J^2-l_b^2}{4l_J^2} \ra_\psi &-2\la \cos\theta_a\f{4l_J^2-l_a^2}{4l_J^2}\, \f{4l_J^2-l_b^2}{4l_J^2} + \f{1}{16l_J^2}\, \f{4l_J^2-l_b^2}{4l_J^2} \ra_\psi \\ &+ \la \f{\psi_J(a+1)}{\psi_J(l_a)} \,\f{l_a+1/2}{l_a+1}\, \f{(4l_J^2-(l_a+1/2)^2}{4l_J^2}\, \f{4l_J^2-l_b^2}{4l_J^2}\ra_\psi=0
}$$
which can be approximated by:
$$
\la e^{- i2\vtheta} e^{ 4 \alpha \f{l_a-l_J}{l_J}}\,\Delta((l_a-1/2)^2)\Delta(l_b^2) \ra_\psi -2e^{\f{2\alpha}{l_J}}
\la \cos\theta_a\Delta(l_a^2)\Delta(l_b^2)+ \f{1}{16l_J^2}\Delta(l_b^2) \ra_\psi + \la e^{ i2\vtheta} e^{-4 \alpha \f{l_a-l_J}{l_J}} \,\Delta((l_a+1/2)^2)\Delta(l_b^2)\ra_\psi \approx 0
$$
where $\Delta(l_j^2)=\f{l_j^2-4l_J^2}{4l_J^2}$.

\subsection{Rescaling the Tetrahedron}

We can now vary also the length scale $l_J$.  First let's notice that in the same way we wrote an exact recursion relation for the leading order of the isosceles  \sj-symbol shifting the representation $a$ (equation (\ref{exactrecurLO})), we can write a similar exact recursion relation for the leading order of the \sj-symbol shifting the label $J$; that is
\equa{
\sqrt{V_{J+1}(a,b)} \{a,b\}^{\textrm{LO}}_{J+1}-2\cos(4\theta_J)\sqrt{V_{J}(a,b)} \{a,b\}^{\textrm{LO}}_{J}+\sqrt{V_{J-1}(a,b)} \{a,b\}^{\textrm{LO}}_{J-1}=0
}
Inserting this recursion relation on $J-$shifts in the definition correlation function, we obtain the following identity:
\bes
\la \sqrt{V_{J+1}(a,b)} \f{\psi_J(a)\psi_J(b)}{\psi_{J+1}(a)\psi_{J+1}(b)} \cO(a) \tcO(b) \ra_\psi
&+\la \sqrt{V_{J-1}(a,b)} \f{\psi_J(a)\psi_J(b)}{\psi_{J-1}(a)\psi_{J-1}(b)} \cO(a) \tcO(b) \ra_\psi&\nn\\
&-2\la \cos(4\theta_J) \sqrt{V_{J}(a,b)} \cO(a) \tcO(b) \ra_\psi &
=0
\ees
The correlation functions appearing in this equation are in fact approximation. We are allowed to use the leading order of the \sj-symbol because the boundary state used picks the function on large $j_0$. And for the same reason, we can expand $\sqrt{V_{J\pm1}(a,b)}$ and the ratios $\f{\psi_J(a)\psi_J(b)}{\psi_{J\pm1}(a)\psi_{J\pm1}(b)}$:
\equa{\tabl{l}{
\la \sqrt{V_{J}(a,b)}\left(1-\f{2l_J}{4l_J^2-l_a^2-l_b^2}\right) e^{-4\alpha\f{(2l_J-(l_a+l_b))}{l_J}[1+\f{3l_J-2(l_a+l_b)}{2l_J(2l_J-l_a-l_b)}]} \cO(a) \tcO(b) \ra_\psi  -2\la \cos(4\theta_J) \sqrt{V_{J}(a,b)} \cO(a) \tcO(b) \ra_\psi \\
\quad \quad \quad \quad \quad \quad+\la \sqrt{V_{J}(a,b)}\left(1+\f{2l_J}{4l_J^2-l_a^2-l_b^2}\right) e^{4\alpha\f{(2l_J-(l_a+l_b))}{l_J}[1-\f{3l_J-2(l_a+l_b)}{2l_J(2l_J-l_a-l_b)}]} \cO(a) \tcO(b) \ra_\psi \approx 0.
}}
We hope that such equation will turn out useful to study the asymptotic properties of the correlations function as the length scale $J$ grows large, but we leave this for future investigation.

\chapter{ Physical boundary state for the quantum 4-simplex} \label{4dphysical}
As we already mentioned it, $\Psi_q$  in (\ref{twopointSF}) needs to be a physical boundary state.
In this Chapter, we work in the context of 4d gravity and we present results published in \cite{article5} where we investigate the consequences of the physical state requirement (\ref{conditionsSF}) for the Euclidean Barrett-Crane model (see Chapter \ref{BCmodel}) for the simplest case of a space-time triangulation constructed from a single 4-simplex. In this context, we show that this requirement fixes uniquely the width of the quantum boundary state (in term of the classical data) similarly to what happens in the 3d toy model.

We recall that the Barrett-Crane vertex amplitude is given by
\be
A_v=\tj\,=\,
\int_{\SU(2)} [dg_a]^{\otimes 5}\,
\prod_{a<b}\chi_{j_{ab}}(g_ag_b^{-1}),
\ee
where $a,b=1,\cdots, 5$ label the nodes of the dual graph -- a pentahedral graph -- of the 4-simplex. Therfore the propagator kernel 
 \be
 K[s]=\mathcal{A}[d_{j_{ab}}]= \prod_{a<b} (d_{j_{ab}})^2 \prod_{a<b}\left(A_e(d_{j_{ab}})\right)^{1/2}\tj
 \ee
is independent of the intertwiners and we  can rewrite the two conditions (\ref{conditionsSF}) as\footnote{Choose factorizable ansatz, product state $\psi(j_1,..,j_{10})=\prod_{i=1}^{10} \phi_i(j_i)$. Then equilateral ansatz, $\phi_i(j)=\phi(j)$ for all triangles $i=1..10$. Then following \cite{anal1}, we introduce the Fourier transform of the state:
\be
f(g)=\sum_j \phi(j) \chi_j(g).
\ee
Then the two conditions for a physical state simply translates to:
\be
\int dg\, |f(g)|^2=1,
\ee
\be
\int [dg_m]^5\,\prod_{a<b} f(g_mg_n^{-1})=
\int [d\theta_{mn}]^{10}\delta(\det [\cos\theta_{mn}])
\prod_{m<n}f(\theta_{mn})\,=\,1.
\ee
}:
\be \label{norm}
\sum_{j_{ab}} |\psi_q(d_{j_{ab}})|^2=1 \quad \textrm{ for the normalized condition,}
\ee
\be \label{dyna}
\sum_{j_{ab}} K(d_{j_{ab}})\psi(d_{j_{ab}})=1 \quad \textrm{ for the "Wheeler-deWitt" condition,}
\ee
with the propagator kernel a function of the dimensions $d_{j_{ab}}$ only
\be \label{kernel}
K[j_{ab}]= \mu \left(\prod_{a<b}d_{j_{ab}}\right)^{\sigma}\tj
\ee
with $\mu$ and $\sigma$ undetermined coefficients. The edge amplitude $A_e$ is taken into account in this factor $\mu \left(\prod_{a<b}d_{j_{ab}}\right)^{\sigma}$.
\\
For more generic spinfoam models such as the EPRL-FK models (see Chapter \ref{EPRLFK}), we need to take into account the intertwiners too in the definition of the spinfoam vertex and of the physical states.

In the following, we will  work on the Barrett-Crane model in the large spin limit. Thus,  in the propagator kernel formula (\ref{kernel}), we consider the asymptotic formula of the $\tj$-symbol 
\be 
\tj \, \sim \, P(d_{j_{ab}}) \cos \left( \f{1}{l_P^2}S_R[d_{j_{ab}}]\right) + D(d_{j_{ab}})
\ee
where we recall that $S_R[d_{j_{ab}}]= \sum_{a<b}d_{j_{ab}}\theta_{ab}$ and the function $P(d_{j_{ab}})$ is a slowly varying factor, that grows as $\zeta^{-9/2}$ when scaling all triangle areas $d_{j_{ab}}$ by $\zeta$. In the large $d_{j_0}$ limit, we can thus write $P(d_{j_{ab}}=d_{j_0})=\f{P}{d_{j_0}^{9/2}}$, where $P$ is a constant, that we can choose positive. $D(d_{j_{ab}})$ is a contribution coming from degenerate configurations of the 4-simplex. This is a non-oscillating term which has no geometrical meaning scaling like $1/\zeta^2$. However, this sick term is negligible in most computations we are interested in, such as in the computation of  (\ref{twopointSF}) (see \cite{graviton1, graviton2}) or as we will see in the computation of (\ref{conditionsSF}), because it will not match the boundary data induced by $\Psi_q$ which peaks the asymptotic around the non-degenerate semi-classical configuration.
\\
%
We now focus on the issue of the semi-classical boundary state. The function $\Psi_q$ should describe the boundary value of the gravitational field on the boundary 4-simplex. We thus consider a state peaked on the geometry of a regular 4-simplex: $q=(d_{j_0}, \Theta)$. The simplest possibility is to choose a Gaussian peaked on theses values:
\be \label{gaussian}
\Psi_{\q}(d_{j_{ab}})= e^{-\sum_{a<b,c<d} \alpha_{cd}^{ab}(d_{j_{ab}}-d_{j_0})(d_{j_{cd}}-d_{j_0})+i\sum_{a<b} \Theta \, d_{j_{ab}}}
\ee
The phase of this semi-classical state determines where the state is peaked in the conjugate variables: $\Theta$ is the variable conjugate to the spin $j_0$ and it codes the extrinsic geometry of the boundary. $\alpha_{cd}^{ab}$ is a given ten by ten matrix. It depends on $d_{j_0}$ in such a way that the relative uncertainties of area and angle on this state become small in the large $d_{j_0}$ limit, namely:
\be
\f{\la \Psi_q|\Delta d_{j_{ab}}| \Psi_q\ra}{\la\Psi_q| d_{j_{ab}}| \Psi_q\ra} \rightarrow 0, \qquad \qquad \f{\la\Psi_q|\Delta \theta_{ab}| \Psi_q\ra}{\la\Psi_q| \theta_{ab}| \Psi_q\ra} \rightarrow 0 \qquad \qquad \forall \,a<b
\ee
Assuming  that the matrix elements $\alpha_{cd}^{ab} \sim \alpha d_{j_0}^{-n}$ in the large spin limit with $\alpha$ which does not scale with $d_{j_0}$, the fluctuation determined by the gaussian state (\ref{gaussian}) are of the order:
\be \label{uncertainties}
\f{\la\Psi_q|\Delta d_{j_{ab}}| \Psi_q \ra}{\la\Psi_q| d_{j_{ab}}| \Psi_q \ra} \sim \f{d_{j_0}^{n/2-1}}{\sqrt{\alpha}}, \qquad \qquad \f{\la\Psi_q|\Delta \theta_{ab}| \Psi_q \ra}{\la \Psi_q| \theta_{ab}| \psi_q \ra} \sim d_{j_0}^{-n/2}\sqrt{\alpha}
\ee
which restricts $n \in  ] 0, 2[$. \\
In the following, we thus focus on a Gaussian state as boundary state such as the matrix elements of the ten by ten matrix $\alpha^{ab}_{cd}$ are given by
\be
\alpha^{ab}_{cd} \sim \f{\alpha}{d_{j_0}^n}= \f{a+ib}{d_{j_0}^n} \quad \textrm{ with } n \in ]0,2[ \textrm{ and } a, b \in \R
\ee
in the large spin limit. We now study the consequences of the two conditions (\ref{norm}) and (\ref{dyna}) on this boundary state. That is the aim is now to determine the consequences on $\alpha_{cd}^{ab}$ of the requirement that our boundary state is a physical state.

\section{Semi-Classical States: the Decoupled Gaussian Ansatz}
We first start by considering an additional ansatz for the boundary state. We take a factorized boundary state:
\be
\Psi_q[j_{ab}]= \prod_{a<b}\phi(j_{ab})
\ee
where each $\phi(j_{ab})$ is peaked around the background value $q=(d_{j_0}, \Theta)$ which corresponds to an equilateral 4-simplex.
%
In this simplest case, the ten by ten matrix $\alpha_{cd}^{ab}$ reduces to a diagonal matrix $\alpha \id_{10\times 10}$. 
Such a boundary state has been used in  \cite{numeric2} since it is up to now the only setting in which numerical simulations can be performed. However, this assumption has not been tested yet. Could such a decoupled gaussian state capture a true physical state? Could it satisfy conditions (\ref{norm}) and (\ref{dyna})?

Two choices for a factorized boundary state have so far appeared in the literature: \begin{itemize}
\item  the Gaussian state \cite{graviton, numeric1}, where each factor is given by,
\be \label{gaussianf}
\phi(j)= e^{-\alpha (d_{j}-d_{j_0})^2}e^{i\Theta d_{j}}
\ee
where $\alpha$ is a complex number.
\item The Bessel-based state \cite{simone1, graviton3d}, where each factor is given by:
\be \label{bessel}
\phi_B(j)= \f{I_{|d_j-d_{j_0}|}(\f{d_{j_0}}{\alpha})-I_{d_j+d_{j_0}}(\f{d_{j_0}}{\alpha})}{\sqrt{I_0(\f{2d_{j_0}}{\alpha})-I_{2d_{j_0}+1}(\f{2d_{j_0}}{\alpha})}} \cos\left(d_j\Theta\right)
\ee
\end{itemize}
In the large spin limit regime, we focus only on the Gaussian ansatz. Indeed, in the large spin limit the Bessel part of (\ref{bessel}) reduces to the Gaussian in (\ref{gaussianf}). Therefore, at the leading order, the only difference between (\ref{gaussianf}) and (\ref{bessel}) is in the phase: the phase in (\ref{gaussianf}) is complex whereas the phase in (\ref{bessel}) is real. A gaussian state with a real phase would lead to the same results as a gaussian state with a complex phase. The interested reader can find details concerning the case of a gaussian state with a real phase in appendix \ref{real} and we now tackle the issue of defining a physical state coming from a decoupled gaussian state with a complex phase.
Therefore, we consider a boundary state of the form,
\be \label{Factorizedgaussian}
\Psi_{\q}[j_{ab}]= \f{1}{\mathcal{N}}\prod_{a<b}e^{-\alpha (d_{j_{ab}}-d_{j_0})^2}e^{-i\Theta d_{j_{ab}}}
\ee
with $\mathcal{N}$ the normalization constant and $\alpha \in \C$. 
We now want to test this assumption using conditions (\ref{norm}) and (\ref{dyna}). These conditions lead to the two following equations on $\mathcal{N}$ and $\alpha$: 
\be \label{normapprox}
1=\sum_{j_{ab} }|\Psi_{\q}(j_{ab})|^2= \f{1}{(\mathcal{N}^2)^{10}}\sum_{\{j_{ab}\}} e^{-2\Re(\alpha)\sum_{a<b}(d_{j_{ab}}-d_{j_0})^2}
\ee
and,
\be \label{WWapprox}
1=\sum_{j_{ab} }K[d_{j_{ab}}] \Psi_{\q}(j_{ab})\simeq \frac{\mu}{\mathcal{N}^{10}}\sum_{\{j_{ab}\}}(\prod_{a<b}d_{j_{ab}})^\sigma P(d_{j_{ab}}) \sum_{\epsilon=\pm1} e^{-\Re(\alpha)\sum_{a<b}(d_{j_{ab}}-d_{j_0})^2+i\sum_{a<b}[d_{j_{ab}}(\epsilon \theta_{ab}-\Theta)-\Im(\alpha)(d_{j_{ab}}-d_{j_0})^2]}
\ee
The first equation corresponds to the normalization condition (\ref{norm}) for $\Psi_{\q}[j_{ab}]$. The second equation is the "Wheeler-deWitt" condition for $\Psi_{\q}[j_{ab}]$ in the simple case $K[j_{ab}]= \mu \left(\prod_{a<b}d_{j_{ab}}\right)^{\sigma}\tj$.  Solving them allow to determine uniquely $\mathcal{N}$ and $\alpha$ in term of the coefficients $\mu$ and $\rho$. 
The analysis is done in the large spin limit regime.
In equation (\ref{WWapprox}), we have already used the asymptotic formulae of the $\tj$-symbol. Moreover, we also replace $\alpha$ in (\ref{normapprox}) and (\ref{WWapprox}) by its asymptotic expression
\be 
\alpha \sim \f{a+ib}{d_{j_0}^n}
\ee
with $a,\, b \in \R$ and $n$ restricted to belonging to $ ]0,2[$ in order that the asymptotic behavior of the  relative uncertainties of the area and angle on this state is correct.  Then solving the two obtained equations requires to distinguish three cases with respect to the value of the power $n \in ]0,2[$.
The final result can be stated as:
\begin{proposition} The requirement on a factorized Gaussian state (\ref{Factorizedgaussian}) to be a physical state fixes the width of the Gaussian $\alpha$ for certain values of the power $n$:
\begin{enumerate} 
\item $0<n<1$: the width of the Gaussian $\alpha$ is uniquely determined and the coefficient $\sigma$ is restricted. More precisely, \be \label{result1} \left\{ \tabl{l}{
\alpha \in \R_+, \\
a= \f{(\mu P)^{2/5} \pi}{2}, \; b=0\\
\sigma= \f14\left( \f95-n \right) \textrm{ and } \sigma>\f15,
} \right.
\ee
\item $1<n<2$: there is no solution in this case.
\item $n=1$: A solution exists for
$$\sigma=\f15$$
and then, the value of $\alpha$ can be determined graphically. For example,
$$
a\simeq 0.1 \textrm{ and } b\simeq1.9 \; \textrm{ for } P=\mu=1.
$$
\end{enumerate}
\end{proposition}
Before proving this result, let us comment on the third case. This last case should be the most natural since the semi-classical state is then peaked on the same way on $\Theta$ encoding the extrinsic geometry of the boundary and on $d_{j_0}=A_0$ encoding the intrinsic geometry, and it is in fact not at all transparent. This leads to the conclusion that the choice of a decoupled gaussian state to define a physical state might be to simple and should be modified. A new proposition is given in the next subsection. Let us now give the proof of the results stated above.

\begin{proof}
In the large spin limit, the summation in equations (\ref{normapprox}) and (\ref{WWapprox}) can then be approximated with an integral  and we can write:
\be \label{intNorm}
1\simeq \f{1}{2\mathcal{N}^2} \int d(d_{j})e^{-2\Re(\alpha)(d_{j}-d_{j_0})^2}
\ee
and
\be \label{intDyna}
1\simeq \frac{\mu}{(2\mathcal{N})^{10}}\int \prod_{a<b}d[d_{j_{ab}}] (\prod_{a<b}d_{j_{ab}})^\sigma P(d_{j_{ab}}) \sum_{\epsilon=\pm1} e^{-\Re(\alpha)\sum_{a<b}(d_{j_{ab}}-d_{j_0})^2+i\sum_{a<b}[d_{j_{ab}}(\epsilon \theta_{ab}-\Theta)-\Im(\alpha)(d_{j_{ab}}-d_{j_0})^2]}
\ee
The first integral (\ref{intNorm}) is just a Gaussian integral, which can been integrated directly:
\be \label{equa1}
\mathcal{N}^2=\f{1}{2}\sqrt{\f{\pi}{2\Re(\alpha)}}
\ee
 given a first relation at the leading ordrer between $\mathcal{N}$ and $\alpha$ .
 \\
 To evaluate the second integral (\ref{intDyna}) in the large spin limit, we first notice that the Gaussian implies
 $$\delta d_{j_{ab}}=d_{j_{ab}}-d_{j_0}\ll1\qquad \forall a<b,
 $$
  thus we can expand the Regge action around $d_{j_0}$:
 \be
 S_R[d_{j_{ab}}]=\sum_{a<b}d_{j_{ab}} \theta_{ab}(d_{j}) \simeq \sum_{a<b} d_{j_0} \Theta +\sum_{a<b} \f{\pp S_R}{\pp d_{j_{ab}}}|_{d_{j}=d_{j_0}} \delta d_{j_{ab}} +\f12\sum_{a<b, c<d} \f{\pp^2 S_R}{\pp d_{j_{ab}}d_{j_{cd}}}|_{d_{j}=d_{j_0}} \delta d_{j_{ab}} \delta d_{j_{cd}}
 \ee
The Schafli identity implies that
$$\f{\pp S_R}{\pp d_{j_{ab}}}|_{d_{j}=d_{j_0}} =\theta_{ab}(d_j)|_{d_{j}=d_{j_0}}= \Theta$$
since in the equilateral 4-simplex, all the dihedral angles are equal to $\Theta$. And we introduce the Hessian,
\be \label{hessian}
N^{ab}_{cd}= \f{\pp^2 S_R}{\pp d_{j_{ab}}d_{j_{cd}}}|_{d_{j}=d_{j_0}}=\f{\pp \theta_{ab}}{\pp d_{j_{cd}}}|_{d_{j}=d_{j_0}},
\ee
 then:
\be
 S_R[d_{j_{ab}}] \simeq 10d_{j_0} \Theta + \Theta \sum_{a<b} \delta d_{j_{ab}} +\f12 \sum_{a<b, c<d} N^{ab}_{cd}  \delta d_{j_{ab}} \delta d_{j_{cd}}=\Theta \sum_{a<b}  d_{j_{ab}} +\f12 \sum_{a<b, c<d} N^{ab}_{cd}  \delta d_{j_{ab}} \delta d_{j_{cd}}.
 \ee
We replace $ S_R[d_{j_{ab}}]$ by  this expansion in (\ref{intDyna}):
\be \tabl{ll}{
1 \simeq \frac{\mu}{(2\mathcal{N})^{10}}\int \prod_{a<b}d[d_{j_{ab}}] (\prod_{a<b}d_{j_{ab}})^\sigma P(d_{j_{ab}}) &[ e^{-2i\Theta\sum_{a<b}d_{j_{ab}}-\f{i}{2}\sum_{a<b,c<d}N^{ab}_{cd}  \delta d_{j_{ab}} \delta d_{j_{cd}}-\alpha \sum_{a<b}\delta d_{j_{ab}}^2}\\
&+ e^{\f{i}{2}\sum_{a<b,c<d}N^{ab}_{cd}  \delta d_{j_{ab}} \delta d_{j_{cd}}-\alpha \sum_{a<b}\delta d_{j_{ab}}^2}]
}\ee
The first exponential is a rapidly oscillating term in $d_{j_{ab}}$ which will vanish when we perform the integration over $d_{j_{ab}}$ so we only have to consider the second term. Moreover, at the leading order in $\delta d_{j_{ab}}$ we can replace $(\prod_{a<b}d_{j_{ab}})^\sigma P(d_{j_{ab}}) $  by $d_{j_{0}}^{10\sigma} P(d_{j_{0}})$ in the integral (\ref{intDyna}). And we recall that  $P(d_{j_{ab}})$ grows as $\zeta^{-9/2}$ when scaling all triangle areas $d_{j_{ab}}$ by $\zeta$, so we can write $P(d_{j_{ab}}=d_{j_0})=\f{P}{d_{j_0}^{9/2}}$ in the large $d_{j_0}$ limit, where $P$ is a constant, that we can choose positive. Therefore, once again we have to integrate a Gaussian integral:
\be \label{WWapprox2}
1\simeq \frac{\mu(d_{j_{0}})^{(10\sigma-9/2)} P}{(2\mathcal{N})^{10}}\int \prod_{a<b}d[\delta d_{j_{ab}}]  e^{-\delta d_{j_{ab}}M^{ab}_{cd}\delta d_{j_{cd}}}
\ee
where $M$ is a ten by ten matrix defined by
\be
M^{ab}_{cd}=\alpha \delta_{cd}^{ab} -iN^{ab}_{cd}\qquad a<b, \; c<d,
\ee
with  $\delta_{cd}^{ab}=1$ if the two couples of indices are the same and it vanishes otherwise;  $N^{ab}_{cd}=\f{\pp \theta_{ab}}{\pp d_{j_{cd}}}|_{d_{j}=d_{j_0}}$ was explicitly computed in \cite{graviton}\footnote{
$$N^{ab}_{cd}=\f{\sqrt{3}}{4\sqrt{5}d_{j_0}}\left(\tabl{cccccccccc}{
-4 & 7/2 & 7/2 & 7/2& 7/2 & 7/2 & 7/2 & -9 & -9 & -9\\
7/2 & -4 & 7/2 & 7/2 & 7/2 &-9 &-9 & 7/2 & 7/2& -9 \\
7/2 & 7/2 & -4 & 7/2 & -9 & 7/2 & -9 & 7/2 & -9 & 7/2\\
7/2 & 7/2 & 7/2 & -4 & -9 & -9 & 7/2 & -9 & 7/2 & 7/2\\
7/2 & 7/2 & -9 & -9 & -4 & 7/2 & 7/2 & 7/2 & 7/2 & -9 \\
7/2 & -9 & 7/2 & -9 & 7/2 & -4 & 7/2 & 7/2 & -9 & 7/2\\
7/2 & -9 & -9 & 7/2 & 7/2 & 7/2 & -4 & -9 & 7/2 & 7/2\\
-9 & 7/2 & 7/2 & -9 & 7/2 & 7/2 & -9 & -4 & 7/2 & 7/2 \\
-9 & 7/2 & -9 & 7/2 & 7/2 & -9 & 7/2 &7/2 & -4 & 7/2 \\
-9 & -9 & 7/2 & 7/2 & -9 & 7/2 & 7/2 & 7/2 & 7/2 & -4} \right)
$$}:
\be
N^{ab}_{cd}=\f{\tilde{N}^{ab}_{cd}}{d_{j_0}}
\ee
where $\tilde{N}$ is a ten by ten real symmetric constant matrix with all coefficients independent of $d_{j_0}$. In particular, $\tilde{N}^{ab}_{ab}=-\sqrt{\f35} \, \;\forall a<b$. Therefore, $M$ is a symmetric matrix with all its diagonal coefficients equal to $\alpha+i \sqrt{\f35}\f{1}{d_{j_0}}$.

At this stage we have to distiguish the three cases mentioned above. Assuming that $\alpha$ is of the form $\alpha=\f{a+ib}{d_{j_0}^n}$, with $a, b \in \R$, $n\in ]0,2[$. The three cases with respect to the power $n \in ]0,2[$ are: \begin{enumerate}
\item $0<n<1$ corresponds to $\alpha \gg \f{1}{d_{j_0}}$. In this case, we could negligible the terms of the order $\f{1}{d_{j_0}}$ with respect to $\alpha$. 
\item $1<n<2$ corresponds to $ \alpha \ll \f{1}{d_{j_0}}$. In this case, $\alpha$ will be negligible with respect to the terms of the order $\f{1}{d_{j_0}}$.
\item $n=1$ corresponds to $\alpha \sim \f{1}{d_{j_0}}$. This case should be the most natural case since it peaks in the same way the triangle areas of the 4-simplex around the background value $A_0=d_{j_0}$ and the dihedral angles around the background value $\Theta$.
\end{enumerate}
Let us now separately study the three cases to solve equation (\ref{WWapprox2}).
\begin{enumerate}
\item {\bf The first case $0<n<1$} is the easiest one.
\\
Indeed in this case, the matrix $M\sim \f{a+ib}{d_{j_0}^n}+ \f{\tilde{N}}{d_{j_0}}$  can be approximated by a  ten by ten diagonal matrix: $M \simeq \alpha \id=(a+ib) \id$ since $\alpha \sim  \f{a+ib}{d_{j_0}^n} \gg \f{1}{d_{j_0}}$ and the ten integrals are then decoupled. Thus, we just have to compute a one-dimensional Gaussian integral:
\be \label{equa2-3}
\mathcal{N}= \f{(\mu P)^{1/10}}{2} (d_{j_{0}})^{(\sigma-9/20)} \int d[\delta d_j] e^{-\alpha (\delta d_{j})^2}= \f{(\mu P)^{1/10}}{2} d_{j_{0}}^{(\sigma-9/20)} \sqrt{\f{\pi}{\alpha}}
\ee
which gives a second equation for $\mathcal{N}$ and $\alpha$. Therefore using equation (\ref{equa1}), we obtain the following equation on $\alpha$:
\be
\f{1}{2}\sqrt{\f{\pi}{2\Re(\alpha)}}= \f{(\mu P)^{1/5}}{4} d_{j_{0}}^{(2\sigma-9/10)}\f{\pi}{\alpha}
\ee
Finally, expressing $\alpha$ under the form $\f{a+ib}{d_{j_0}^n }$, $0<n<1$), we get that:
\be
\left\{ \tabl{l}{
\alpha \in \R_+, \\
\sigma= \f14\left( \f95-n \right) \textrm{ and } \sigma>\f15, \\
a= \f{(\mu P)^{2/5} \pi}{2}, \quad b=0.
} \right.
\ee
In this case, $\alpha$ is real and positive. Furthermore, we get a condition on the normalization factor of the spinfoam vertex $\sigma > 1/5$.

\item {\bf The second case $1<n<2$} implies that $ \alpha \sim  \f{a+ib}{d_{j_0}^n} \ll \f{1}{d_{j_0}}$.
\\
In this case, the matrix $M$ reduces to $M=-iN$. We have to integrate: $\int d[\delta d_{j_{ab}}] \exp( i \sum \delta d_{j_{ab}} N^{ab}_{cd} \delta d_{j_{cd}})$. Recall that for a $m \times m $ symmetric invertible matrix $A$ with signature $\sigma(A)$ we have:
\be
\int_{\R^m}[dX_i] \exp \left[ i\left( \sum_{i,j} X_i A_{ij}X_j\right)\right]= e^{i\sigma(A) \f{\pi}{4}}\sqrt{\f{\pi^m}{|\det(A)|}}
\ee
In our case, $N$ is real, symmetric and its signature, which is the difference between the number of positive eigenvalues and the number of negative eigenvalues, is equal to $-2$. Therefore, $e^{i\sigma(N)\f{\pi}{4}}=-i$ and our Gaussian integral is an imaginary number which is not compatible with the first equation on $\mathcal{N}$ (\ref{equa1}). So, we cannot have $ \alpha = \f{a}{d_{j_0}^n }$ with $1<n<2$.

\item {\bf The third case $n=1$} corresponds to $\alpha=\f{a+ib}{d_{j_0}}$.%
\\
 Then the determinant of $M$ is a complex number: its real part  and its imaginary part are polynomials of degree 10 and of arguments $a$ and $b$. We thus obtain a complex equation for $\mathcal{N}$ and $\alpha$:
\be
1\simeq \frac{\mu (d_{j_{0}})^{(10\sigma-9/2)} P}{(2\mathcal{N})^{10}}\int \prod_{a<b}d[\delta d_{j_{ab}}]  e^{-\delta d_{j_{ab}}M^{ab}_{cd}\delta d_{j_{cd}}}=\frac{\mu (d_{j_{0}})^{(10\sigma-9/2)} P}{(2\mathcal{N})^{10}} \sqrt{\frac{\pi^{10}}{ \det(M)}}
\ee
Combining this equation with the first equation (\ref{equa1}) that we already have on $\mathcal{N}$ and $\Re(\alpha)=a$ we get a complex equation on $a$ and $b$:
\be \label{equa2-1}
2^5 \left(\f{d_{j_0}\pi}{2a}\right)^{5/2}-\mu (d_{j_{0}})^{(10\sigma-9/2+5)} P \sqrt{\frac{\pi^{10}}{ \det(\tilde{M}(a,b))}}=0
\ee
with $\tilde{M}=d_{j_0}M$. This equation implies a condition on the normalization factor of the spinfoam vertex: $$
\sigma=\f15.
$$
 We can then solve numerically the previous equation by plotting a 3d graph representing the square of the norm of the complex number given by the left-hand side of the previous equation (\ref{equa2-1}) in terms of $a$ and $b$ using Maple. For example, for $P=\mu=1$, the value on the surface is null for $$
 a\simeq 0.1\quad \textrm{ and }  \quad b \simeq 1.9.
 $$
  Therefore, there exists a specific value $\alpha$ for which $\Psi_{\q}$ is a physical state. However, whereas this case for which $\Psi_{\q}$ is peaked in the same way around the intrinsic and extrinsic geometry $\q$ should be the most natural, it is quite complicated. 
\end{enumerate} 
\end{proof}
The alternative is then to work with a more complication semi-classical state. Indeed although a factorized gaussian wave-packet  is up to now the only one which has allowed to perform numerical simulations, the previous analysis seems to show that it is too simple to catch all the features of a physical state. We show in the next section that  a tensorial Gaussian state is more adapted to describe a physical state. Indeed we will see that such a state allows to compensate the imaginary part which comes from the second derivative of the Regge action and given by the matrix $iN$ and consequently to simplify the resolution of this third case studied above.

\section{The Coupled Gaussian Ansatz}
Our new assumption is to consider a boundary state of the form:
\be \label{coupledGaussian}
\psi_q[d_{j_{ab}}]= \f{1}{\mathcal{N}}e^{-\sum_{a<b,c<d}\alpha_{cd}^{ab} (d_{j_{ab}}-d_{j_0})(d_{j_{cd}}-d_{j_0})}e^{i\Theta \sum_{a<b}d_{j_{ab}}}
\ee
where $\alpha$ is now a ten by ten complex matrix. And we choose
\be \label{coupledWidth}
\alpha_{cd}^{ab}= \beta(d_{j_0}) \delta^{ab}_{cd} + iN_{cd}^{ab}
\ee
 where $\beta \in \R$ and the imaginary part of $\alpha$ is  now  the conjugate variables of the dihedral angles of the tetrahedron in the semi-classical regime introduced in (\ref{hessian}):$$
 N^{ab}_{cd}=\f{\pp \theta_{ab}}{\pp d_{j_{cd}}}|_{d_{j}=d_{j_0}}.
 $$
$N^{ab}_{cd}$ is the Hessian of $S_R$ and we will see that it is this choice which allows to simplify the construction of a physical semi-classical state peaked in the same way on the extrinsic and extrinsic geometry of the 3d boundary. We recall that $N$ depends on $d_{j_0}$ such that $N=\f{\tilde{N}}{d_{j_0}}$ with $\tilde{N}$ a matrix with constant coefficients. 
\begin{proposition}
For $0<n\leq 1,$ the width $\beta$ (the real part of the matrix $\alpha$; see (\ref{coupledWidth})) of the Gaussian state (\ref{coupledGaussian}) is uniquely defined and the coefficient $\sigma$ is restricted. More precisely,
\be \label{resultcoupled1} \left\{ \tabl{l}{
\beta \in \R_+, \\
a= \f{(\mu P)^{2/5} \pi}{2}, \; b=0\\
\sigma= \f14\left( \f95-n \right) \textrm{ and } \sigma\geq\f15,
} \right.
\ee
Therefore, the case $n=1$ appears now in the continuity of the case $0<n<1$. This can be considered as an improvement compared to the previous case of a factorized Gaussian semi-classical state. However, the case $1<n<2$, which admited no solution in the factorized Gaussian state, has not been solved yet. The resolution of this case needs the knowledge of the next to leading order of the asymptotic expansion of the $\{10j\}$-symbol.
\end{proposition}
\begin{proof}
Regarding the asymptotic behavior of $\beta$, the real part of the matrix $\alpha$, we distinguish the three same cases as in the decoupled gaussian state case. The result of the first case is not modified. Indeed, in the first case,\begin{enumerate}
\item $\beta \sim \f{a}{d_{j_0}^n}$  with $0<n<1$, therefore, $\beta \gg \f{1}{d_{j_0}}$ and the matrix $M$ can be approximated by a ten by ten diagonal matrix $M\simeq \beta \, \id$ and the state is again factorized and we obtain the same result as in (\ref{result1}):
\be \label{result2} \left\{ \tabl{l}{
\beta \in \R_+ \\
a= \f{(\mu P)^{2/5} \pi}{2}\\
\sigma= \f14\left( \f95-n \right) \textrm{ and } \sigma> \f15
} \right.
\ee
\item The second case is when $]0,2[$ in $\beta \sim \f{a}{d_{j_0}^n}$ ($a\in \R$). Since $\beta \ll \f{1}{d_{j_0}}$ in this case, the real part of $\alpha$ is negligible compare to its imaginary part.  Consequently since this imaginary part compensates the second derivative of the Regge action in the exponential of the second condition (\ref{dyna}), we should go to the next order in $\delta d_j$. To compute this expansion in $\delta d_j$ we need to know the next to leading order of the asymptotic expansion of the $\tj$-symbol. We can thus say nothing now in this case for the moment.
\item The third case, $\beta \sim \f{a}{d_{j_0}}$ ($a \in \R$), is greatly simplified compared to the equivalent case for the decoupled gaussian wave-packet. Indeed, applying condition (\ref{norm}) and condition (\ref{dyna}) to this state, we obtain the two following equations for $\mathcal{N}$ and $a$:
\be \left\{ \tabl{l}{
1\simeq \f{1}{2^{10}\mathcal{N}^2}\left(\int d[\delta d_{j}] e^{-2\f{a}{d_{j_0}}(\delta d_{j})^2} \right)^{10}= \f{1}{2^{10}\mathcal{N}^2} \left(\frac{\pi d_{j_0}}{2a} \right)^5\\
1 \simeq \f{d_{j_0}^{10\sigma-9/2}\mu P}{2^{10}\mathcal{N}}\left(\int d[\delta d_j] e^{-\f{a}{d_{j_0}}(\delta d_j)^2}\right)^{10}=  \f{d_{j_0}^{10\sigma-9/2}\mu P}{2^{10}\mathcal{N}}\left(\frac{\pi d_{j_0}}{a} \right)^5
} \right.
\ee
where we use to compute the second equation the same analysis as in the factorized gaussian state case.  The key element which simplifies everything is that the second derivative of the Regge action added to the imaginary part of $\alpha$ is null since $\alpha$ imaginary part is the Hessian. This system of equations can then be simplified in a single equation for $a$:
\be
\f{1}{2^{10}}\left( \f{\pi d_{j_0}}{2a} \right)^5 =\f{d_{j_0}^{(20\sigma-9)}(\mu P)^2}{2^{20}} \left( \f{\pi d_{j_0}}{a}\right)^{10}
\ee
which implies a condition on the normalization factor of the $\tj$-symbol in the spinfoam vertex:
\be
\sigma= \f{1}{5}
\ee
 and the unique solution for $a$:
 \be
 a= \f{\pi (\mu P)^{2/5}}{2}.
 \ee
 Therefore, we now get the same value of $a$ in this case as in the first case. Moreover, $\sigma$ is now uniquely determined since  $n=1$.
\end{enumerate}
\end{proof}
This analysis on the consequences of conditions (\ref{norm}) and (\ref{dyna}) on a coupled gaussian wave-packet (\ref{coupledGaussian}) 
is therefore more convicting on the ability of such a state to capture a true physical rather than a too simple state defined  as a factorized gaussian state. Indeed, the width $\alpha$ of the coupled gaussian state is now uniquely determined by the requirement of a normalized physical state.

Let us recall that in all our calculations presented in this Chapter, we used $K_1[s]$ the lowest order term in the expansion of the propagator kernel $K[s]$ in the group field theory coupling constant\footnote{ The lowest order term in the expansion of the propagator kernel $K[s]$ in the group filed coupling constant $\lambda$ is in fact $K_0$ but we are expecting no contribution of zero order to the sum coming from $K_0$ \cite{graviton2}.} $\lambda$  -- $s$ symbolizes the 4-simplex boundary graph. The total propagator kernel $K[s]$ is given by  $K[s]=\sum_{V=1}^{\infty} \lambda^{V}K_V[s]$. This operator should describe a unitary "evolution". In the spin foam framework the notion of evolution is not well-defined, 
however we can require the following normalization condition on $K[s]$
\be
K[s]\bar{K}[s]=(K[s])^2=1
\ee
that describes the process of creation of a  4-simplex starting from a null 4-volume following by the  annihilation of this 4-simplex into a null 4-volume again. This condition also says that $K$ is by definition a physical state. Using the $\lambda$ expansion of  $K[s]=\sum_{V=1}^{\infty} \lambda^{V}K_V[s]$, we can expand this normalization condition in power of $\lambda$
\be
\sum_{V=1}^\infty \lambda^{V}\sum_{V_1, V_2/V_1+V_2=V+1}K_{V_1}K_{V_2}=1
\ee
which simplifies  at the leading order in $\lambda$ in
\be
\lambda^2K_1[s]^2=1
\ee
meaning that approximatively $\lambda K_1[s]$ is a physical state and  introducing a new constraint on $\lambda K_1[s]=  \lambda\mu \prod_{a<b}d_{j_{ab}}^{\sigma}\tj$. It is this new constraint which allows to determine $\mu$:
\be
(\lambda\mu)^2=\f{1}{\left(\prod_{a<b}d_{j_{ab}}\right)^{2\sigma}\tj^2}
\ee
where $\sigma \geq 1/5$ in order that $\mu$ is finite. This restriction on the domain of validity of $\sigma$ is consistent with the results obtained previously.
\\
Since up to now we have only considered the lowest order term in the expansion of $K[s]$ in  $\lambda$, we have in fact fixed  the simplest bulk triangulation and taken into account only a finite number of degrees of freedom.
The next step would be, keeping the same boundary spin network $s$, to explicitly write the next terms: $K_2$, $K_3$,... of the $\lambda$-expansion of $K[s]=\sum_{V=1}^{\infty} \lambda^{V}K_V[s]$. This is not obvious.  A priori,  $K_4$ could be determined from $K_1$ by doing a 5-1 move. Otherwise, from an effective field theory point of view we can  write $K[s]=K_1[s]+ k[s]$ where $k$ takes into account all the possible counter-terms and therefore all the possible bulk geometries for a given boundary spin network. This is equivalent to look at a coarse-grained lattice in which all vertices can be considered as shrunk to a single effective vertex. $k[s]$ is then the weight associated to this effective vertex. In this weight $k[s]$, we are expected to get a term proportional to the Barrett-Crane spin foam vertex amplitude, $\epsilon \tj$, which would come from the contribution of the flat 4-simplex. Moreover, an additional contribution of the form $\rho  \int dG\, P(G) \,\Gamma_{10j}(G)$  should come from the non-flat 4-simplices where $P(G)$ is a factor term and $\Gamma_{10j}$ such that $\Gamma_{10j}(\id)=\tj$ takes into account the curvature. Otherwise, other terms should certainly appear in a more precise analysis. The issue would be then to determine which terms of $k[s]$ contribute in the requirement of a physical state for $\psi_q$; that is, we should understand how the condition $\sum_{j_{ab}} K[d_{j_{ab}}] \psi(d_{j_{ab}})=1$ is explicitly modified by the term $k[d_{j_{ab}}]$ of $K[d_{j_{ab}}] $. We would expect that the requirement for the boundary state to be a physical state selects the bulk triangulation. It is the choice of the phase which seems to be important. The choice of  a different phase for $\Psi_q$ should select another bulk triangulation $\Gamma$ which could not be flat anymore.

To conclude this chapter, let us summarize. In the context of 4d gravity we have investigated the consequences of the physical state requirement (\ref{conditionsSF}) for the Euclidean Barrett-Crane model for the simplest case of a space-time triangulation constructed from a single 4-simplex. In this context, we have shown that this requirement fixes uniquely the width of the quantum boundary state (in term of the classical data) similarly to what happens in the 3d toy model.  The results presented above are therefore relevant in order to perform numerical computations in the context of the spin foam graviton propagator framework. Moreover, an important conclusion of this analysis is that a too simple boundary state, such as a factorized Gaussian state, does not easily capture a true physical state and that a better ansatz is to consider a more complicated state, such as a coupled gaussian state.

\chapter{Conclusion}

In this last Chapter, we will recall some of the key-issues of the loop quantum gravity/spinfoam approaches and we will review  the new results we have obtained while addressing these issues. We will also point out some problems which have been left open and  some new questions that have arisen.

\medskip
 
Our research has  focused on the loop quantum gravity and  spin foam frameworks \cite{book-carlo, book-thomas}.  Loop quantum gravity  has been established   as a background independent and non perturbative quantization of general relativity, through the canonical quantization scheme. 
  The loop quantum gravity kinematical aspects  are well  under control and   its building blocks  are the so-called \textit{spin network states}. These provide a basis for the kinematical Hilbert space and diagonalize some geometric operators, such as  the area operator.  
  On the other side, the loop quantum gravity dynamical aspects are encoded in the Hamiltonian constraint and the physical states solving this constraint, \ie the kernel of the constraint. They are still to be completely understood.   As a tentative answer to this issue, the spin foam framework was  introduced in order to provide a history formalism for loop quantum gravity, thus defining dynamics and transition amplitudes between spin network states.
Spin foam models can also be naturally interpreted as a form of  path integral approach to quantum gravity, the covariant approach, as opposed to the canonical approach which relies in splitting space-time into ``space" and ``time". The spin foam framework has been the starting point of all our work.

\medskip

Let us recall the main issues we discussed. 
\begin{enumerate}
\item[\textit{(1.)}] This  spin foam  framework in 4d is based on the following observation: 4d general relativity can be seen as  a topological theory (\ie with non-local degrees of freedom) plus constraints (which reintroduce local degrees of freedom). The constraints are  directly imposed at the quantum level and the key-issue is to understand how to implement these constraints to obtain a consistent quantum gravity  model.
\item[\textit{(2.)}] The  precise link between spin foam models and  loop quantum gravity is still missing in 4d (for the 3d case, see \cite{ale-karim}).  Spin foam models and loop quantum gravity are different approaches which use different methods and lead to different results. For example, a discrepancy comes from the fact that   most of the spin foam models for 4d gravity have been constructed as discretized path integral for constrained BF field theories with the Lorentz group $\SL(2,\C)$ as gauge group. Consequently, their boundary are resulting $\SL(2,\C)$-invariant spin network states while the kinematical Hilbert space of loop quantum gravity is spanned by $\SU(2)$ spin networks.
\item[\textit{(3.)}] The semi-classical limit is another important open issue in the loop quantum gravity/spin foam approaches. It is very important to understand the semi-classical features of the quantum gravity firstly to check that we can really recover gravity in the classical regime and secondly to calculate the corrections due to  quantum gravitational effects in the semi-classical regime.
\end{enumerate}


\medskip

In Chapter \ref{UNChap}, we addressed the issue \textit{(1.)} in the context of Euclidean 4d gravity from an original perspective \cite{article3} which uses  the recently developed $\U(N)$ framework.  We recall that in the spin foam quantization procedure, the simplicity constraints which turn the SO(4) BF theory into 4d Euclidean gravity theory, are discretized and have to be imposed on the $\Spin(4)$-intertwiners from which are built the quantum states of geometry and the spinfoam transition amplitudes. The issue to implement the simplicity constraints without freezing too many local degrees of freedom comes from the fact that they do not form a closed algebra at the discrete level and cannot be imposed strongly.
\smallskip

We revisited the implementation of the discrete simplicity constraints \cite{article3} using the $\U(N)$ framework initially developed for $\SU(2)$-intertwiners  in \cite{UN1, UN2, UN3}. Based on the Schwinger representation of the $\su(2)$ Lie algebra in term of a couple of harmonic oscilaltors, this framework introduces a new set of $\SU(2)$-invariant operators acting on the space of $\SU(2)$-intertwiners. These operators act on pairs of legs $(i,j)$ of the intertwiners: $E_{ij}$ generates $\U(N)$ transformations that deform the shape of the intertwiner, while $F_{ij}$ and $F\dag_{ij}$ act as annihilation and creation operators consistent with the $\U(N)$-action. The key result of this approach is that these $\SU(2)$-invariant observables form a closed algebra.

In the spinfoam context, we deal with the $\Spin(4)$-intertwiners. Using the decomposition of $\Spin(4)=\SU(2)_L\times\SU(2)_R$ in left and right sectors, we now have invariant operators acting on both sectors $E_{ij}^{L,R},\, F_{ij}^{L,R}, \, F_{ij}^{L,R}{}\dag$ which can be used to investigate how to impose the simplicity constraints. More precisely, the idea we developed has been to recast the discrete simplicity constraints in term of observables -- defined in term of the $\U(N)$ operators -- that form a closed algebra. At the end of the day, it allowed us to propose a set of $\U(N)$ coherent states that solve the simplicity constraints weakly at large scale for arbitrary values of the Immirzi parameter.

More precisely, we  have reviewed the $\U(N)$ coherent states introduced in \cite{UN3}.   For a $N$-valent $\SU(2)$-intertwiner, they are labeled by  the total area $J=\sum_i j_i$ and  a set of $N$ spinors $z_k$. These coherent states $|J, \{z_k\}\ra$ form a over-complete basis for the space of $\SU(2)$-intertwiners at fixed area $J$ and are simply related to the Livine-Speziale coherent intertwiners \cite{coh1}. Moreover, we gave explicitly the action of the $\SU(2)$ invariant operators on these $\U(N)$ coherent states as differential operators. As such, we  have  firstly completed the analysis of the $\U(N)$ framework for $\SU(2)$-intertwiners initiated in \cite{UN1, UN2, UN3}.


Secondly, we have applied these new $\U(N)$ tools to the analysis of the simplicity constraints for $\Spin(4)$-intertwiners. The simplicity constraints couple the left and right sectors of the intertwiners. We have focused in re-expressing them in term of the $E,F,F\dag$ operators of the $\U(N)$ formalism. Following the usual approach, we have always distinguished the diagonal simplicity constraints from the cross simplicity constraints. The diagonal constraints act on single legs of the intertwiner and require that the $\Spin(4)$-representation living on a leg $i$ be simple i.e that the left and right spins are equal $j_i^L=j_i^R$ (or $j_i^L=\rho j_i^R$ for a non-trivial Immirzi parameter). These diagonal constraints are always imposed strongly on the intertwiner states. On the other hand, the cross simplicity constraints deal with pairs of legs and are standardly solved weakly in the most recent spinfoam models \ie only in expectation value (with minimal uncertainty). We started by showing that the discrete simplicity constraints which do not form a closed algebra can be replaced by  a new set of constraints $\{\cC_{ij}\}$ which forms a $\u(N)$-algebra. These new $\u(N)$ simplicity constraints are very simply constructed in terms of the $E$-operators. We also explored other possibilities of constraint operators based on the operators $F$ and $F\dag$. In the end, it appeared that distinguishing the diagonal constraints from the cross constraints and imposing the first strongly while solving the later only weakly always lead to difficulties. Thus, in the last part of our work, we proposed to put all (diagonal and cross) simplicity constraints on the same footing and to solve all of them at once in a weak way. This led us to introduce constraints $F_{ij}^L-F_{ij}^R=0$ involving only annihilation operators. These constraints can be considered as the holomorphic constraints of the Gupta-Bleuler quantization procedure. Solving them in term of $\U(N)$ coherent states provideed us with weak solutions to all simplicity constraints, for arbitrary values of the Immirzi parameter.
\medskip

The next important question to explore is how to generalize this framework to  the Lorentzian case in order to check whether it is also possible to construct coherent states which could solve all simplicity constraints with an arbitrary Immirzi parameter. Another issue is to understand how to glue these $\U(N)$ coherent intertwiners consistently into spin network states in order to generalize our analysis to triangulations formed of an arbitrary number of polyhedra glued together. Finally, we hope that the introduction of these $\U(N)$ coherent states as a basis of the boundary physical Hilbert space of spinfoam model could help to understand the symmetries of the spinfoam amplitudes and their behavior under (discrete) deformations or diffeomorphisms.

\bigskip

In Chapter \ref{ProjectChap}, we have addressed the issue \textit{(2.)} in the context of the Lorentz 4d quantum gravity investigating the correspondence between the $\SU(2)$ spin network states of the canonical loop quantum gravity framework and the projected spin networks arising in spin foam models \cite{article4}. 
We first introduced the projection map from projected cylindrical functions down to $\SU(2)$ cylindrical functions. Reversely, we have studied the lifting maps allowing to inverse this projection map and raise $\SU(2)$ spin network to projected spin networks on $\SL(2,\C)$. We have obtained a whole family of such lifting maps, parameterized by the Immirzi parameter. In this way, we established an isomorphism between the space of $\SU(2)$ spin networks and the space of proper projected cylindrical functions at fixed Immirzi parameter. We have also shown that allowing the Immirzi parameter to run through all possible real values, we sweep the whole space of proper projected cylindrical functions. Finally, we have analyzed the differences between the two scalar products respectively for $\SU(2)$ functionals and $\SL(2,\C)$ functionals, and we have explained how to modify the lifting maps so as to ensure that these two scalar products match exactly.
\medskip

This work hints towards considering that the most useful perspective would be to compare $\SU(2)$ spin networks to projected spin networks and not directly to $\SL(2,\C)$ spin networks as was done in recent work on bridging between the EPRL-FK spinfoam models and the canonical approach \cite{eprl_jerzy}.
\\
Physically, $\SL(2,\C)$ spin networks erase all data about the time-normal field, which is actually instrumental in properly implementing  the simplicity constraints. Mathematically, both $\SU(2)$ spin networks and projected spin networks involve $\SU(2)$ intertwiners, which allows for a direct map between the two Hilbert spaces. Therefore, we propose to use consistently projected spin networks as boundary states for the EPRL-FK spinfoam models and we hope that the present work will be useful in order to consistently translate loop quantum gravity's dynamics into spinfoam amplitudes.
\\
We would like to also point out that our projected cylindrical functions obtained through a lift of $\SU(2)$ spin networks look similar to the recently introduced ``holomorphic" spin network functionals introduced to study the semi-classical behavior of the EPRL-FK spin amplitudes \cite{claudio,claudio2}. We think that this is an issue worth studying in more details.
\\
Finally, we hope that the relation between $\SU(2)$ spin networks and projected functionals which we uncovered will trigger more interest in studying the structure of the space of projected spin networks. More particularly, we would like to put emphasis on two issues. First, it would be interesting to understand the geometrical interpretation of un-proper projected spin networks \ie states carrying two different spins  per edge $j_e^s\ne j_e^t$ (when the spin along an edge is different at its source vertex and at its target vertex). Then, it would be interesting to investigate the coarse-graining of projected cylindrical functions and see if we can construct a projective limit \`a la Ashtekar-Lewandowski as was done in loop quantum gravity \cite{ALmeasure}. Such techniques have failed up to now when applied to spin network states for non-compact gauge groups such as the Lorentz group $\SL(2,\C)$. Nevertheless, we believe that this could be different when dealing with projected spin networks due to their effective $\SU(2)$ gauge invariance and their mapping into $\SU(2)$ spin networks.

\bigskip

In both Chapters \ref{groupInt6j} and \ref{recursion6j}, we have focused on the issue \textit{(3.)} in the context of 3d gravity.
In order to better understand the low-energy regime interpretation of the spin foam model, we investigated in Chapter \ref{groupInt6j} the asymptotical behavior of the \sj-symbol which is relevant in the construction of spin foam amplitudes. Starting from its expression as a (finite) sum over (half-)integers of algebraic combinations of factorials, we followed the footsteps of \cite{razvan} and  showed that one can derive systematically the corrections to the leading order formula at any order \cite{article1}. The method relied on three steps. First, we used the Stirling formula (with the appropriate corrections) to approximate the factorials. Second, we considered the sum as a Riemann sum and approximate it by an integral (over the real line). Finally, we performed a saddle point approximation to compute the behavior of the \sj-symbol for (homogeneously) large spins.

Using this framework, we showed that we recover an oscillating leading order (LO) with frequency given by the Regge action as is well-known and was already proved in \cite{razvan}. Then we computed analytically the next-to-leading (NLO) corrections. The formula that we obtained is explicit, although not compact, and we could not interpret it geometrically in a clear way. Nevertheless, we performed two simple checks. First, we checked that our complicated formula reduced to the known expression for the NLO for isosceles tetrahedra \cite{valentin}. Second, we checked it numerically in various cases and found a perfect fit. These numerical simulations also confirmed that the NLO is a $\f\pi2$-phase shift with respect to the LO (the NLO is given by a $\sin$ instead of a $\cos$) and that the NNLO is back in phase with the LO (again a $\cos$), which confirmed our expectation of an alternating asymptotical series in $\cos+\f1{j}\sin+\f1{j^2}\cos+\f1{j^3}\sin+\dots$.
\\ We point out that we computed in details the corrections due to the Stirling formula and to the saddle point approximation. However we did not study the Riemann sum approximation. It does not contribute to the LO and NLO. It will only enter at the level of the NNLO.
\medskip

This work, mainly technical, can be applied to the computation of gravitational correlations for 3d quantum gravity following \cite{graviton1, 3dtoymodel2, valentin, 4d}. It will enter the quantum corrections to the propagator/correlations at second order, as was shown in \cite{3dtoymodel2}. Indeed, the first order corrections are derived from the path integral of the Regge action, while the deviations from the Regge action as computed here enter at second order (as two-loop corrections). From this perspective, this NLO of the \sj-symbol describes the leading order deviation of quantum gravity with respect to the classical gravity.

Beyond the technicality of these results, our purpose was to show that computing such corrections is indeed possible (although it does lead to complicated expressions) and that similar methods could be used for 4d spinfoam gravity. Although these methods allow straightforward (but lengthy) analytical calculations, which might be handled by a computer program, their drawback is the loss of the  geometrical meaning of the expressions obtained. An alternative way to proceed is to use the exact recursion relations satisfied by the \sj-symbol (see \cite{SG1, SG2}) and other spinfoam amplitudes (see \cite{recursion}) to probe the asymptotic behavior and the induced corrections of the correlations. 
\medskip

In Chapter \ref{recursion6j}, we have actually  studied this alternative method  in the context of 3d gravity \cite{article2}.
We have used the recursion relation satisfied by the \sj-symbol to study the structure of its asymptotical expansion for large spins. The exact recursion relation allowed us to compute explicitely the asymptotical approximation of the isosceles \sj-symbol up to fourth order. This confirmed previous results \cite{valentin,article1} and introduced techniques allowing further systematic analytical calculations of the corrections to the behavior of the\sj-symbol at large spins. However a clear and simple geometrical interpretation of the polynomials appearing in this expansion is still missing. 

This work using recursion relation is useful in particular for the study of large scale correlations in the spinfoam model for 3d quantum gravity. In this context, the recursion relation allowed us to write equations satisfied by the spinfoam correlations similar to the Ward identities of standard quantum field theory. Such recursion techniques have been shown to be further applied to the study of 4d spinfoam amplitudes and the resulting spinfoam graviton propagator \cite{recursion}.

\bigskip

A framework to address  issue \textit{(3.)} is the spin foam graviton proposal \cite{graviton1}. The graviton propagator is based on the extraction of some semi-classical correlations at large scale thanks to a suitable boundary state peaked on a given classical 3-geometry. One issue which has not been explored up to now in the context of 4d gravity  is that this boundary state  should be physical, \ie solve the Hamiltonian constraint on the 3d boundary.
We addressed this issue in Chapter \ref{4dphysical} and showed that it is possible to determine a physical semi-classical state for the Barrett-Crane model in the case of the simplest triangulation given by a 4-simplex \cite{article5}. The requirement of a normalized physical state determines uniquely the Gaussian width. This analysis showed also that we cannot take a too simple semi-classical state. Indeed, we have seen that a decoupled gaussian state which is peaked on the geometry of an equilateral 4-simplex does not seem to be able to capture a true physical state. Instead, a coupled gaussian state appears as defining more naturally a physical semi-classical state. 

\medskip

Our analysis seemed to indicate that the requirement for the boundary state to be a physical state selects the bulk triangulation.  However, a detailed investigation dealing with more refined boundary states is required to confirm this intuition and to really understand the dominant terms in the computation of the two-point function of the gravitational field.
\\
Another step would be to deal with the EPRL-FK models instead of the Barrett-Crane model. Since the EPRL-FK vertex has also for asymptotic the exponential of the Regge action, we  expect to get similar results in the restricted case of a single 4-simplex. However, the behavior under renormalization should be quite different since the space of intertwiners in the EPRL-FK model introduces new degrees of freedom in the 3d space geometry compared to the Barrett-Crane geometry which is describe by a unique intertwiner and  seems to have too many degrees of freedom frozen. These degrees of freedom will have to be taken into account when gluing 4-simplices together.

\bigskip

Some key-issues of the spin foam framework for quantum gravity have been tackled here and some relevant results proposed. There have been many techniques developed to build and to study the spin foam amplitudes but there are still a lot of issues to clear up and understand, much more than what has already be done! \\
 For example, one issue not really taken into account yet concerns the issue of the renormalization in the spin foam models. Indeed, an important aspect to keep in mind is that it is fundamental to develop a framework with the appropriate tools to study the coarse-graining of spin foam amplitudes in order to truly define the continuum limit of spinfoam models and their semi-classical limit. And beyond this, we need to identify a family of models parametrized by a finite number of parameters. That is, we need to understand what are the physical relevant coupling constants such as it has been done in standard quantum field theory and to identify a family of spin foam models stable under coarse-graining. \\
 To be continued....

\appendix
\chapter{Quick Overview of $\SL(2,\C)$ Representations} \label{appendix2}

The Plancherel decomposition formula for $\SL(2,\C)$ states that $L^2$ functions with respect to the Haar measure on $\SL(2,\C)$ uniquely decompose in term of the matrix elements of the group element in the unitary irreducible  representations of $\SL(2,\C)$ of the principal series. Such irreducible representation (irreps) are labeled by a couple of numbers $(n,\rho)$, where $n\in\N/2$ is a half-integer and $\rho\in\R$ a real number.
There also exists a supplementary series of unitary irreps, labeled by a single real number bounded by 1 in modulus, but they do not enter the Plancherel decomposition.
Then the Plancherel formula for a function $f\in L^2(\SL(2,\C))$ reads:
\be
f(G)=\f1{8\pi^4}\sum_n\int \mu(n,\rho) d\rho\, \tr\,\left[F(n,\rho)D^{(n,\rho)}(G)\right],
\ee
where the Fourier components $F(n,\rho)$ are matrices in the Hilbert space of the representation $(n,\rho)$ and are obtained by the reverse formula:
\be
F(n,\rho)=\int dG\,f(G)D^{(n,\rho)}(G^{-1}).
\ee
The measure of integration over the representation labels $\mu(n,\rho) d\rho \,\equiv(n^2+\rho^2)d\rho$ is called by the Plancherel measure. This Plancherel decomposition relies on the fact that the matrix elements $D^{(n,\rho)}(G)$ form an orthogonal basis for the Hilbert space $L^2(\SL(2,\C))$.


It will be useful for later to have the explicit action of the Lorentz generators in each $(n, \rho)$ representation. The relevant basis for us is the $\SU(2)$ basis obtained by decomposing the $\SL(2,\C)$ representation into $\SU(2)$ irreducible representations. Indeed, one can show that the $(n, \rho)$ representation decomposes onto all $\SU(2)$ irreps with spin $j$ bounded below by the half-integer $n$, thus implying that the Hilbert space of the $(n, \rho)$ representation is the direct sum of the Hilbert spaces corresponding to these $\SU(2)$  irreps:
\be
R^{(n,\rho)}=\bigoplus_{j\in n+\N} V^j.
\ee
Let us point out that we have chosen the canonical $\SU(2)$ subgroup, which stabilizes the 4-vector $\om$ or equivalently the identity matrix $\Om=\id$ (as explained previously). Then we give the action of the $\su(2)$-rotation generators $\vJ$ and boost generators $\vK$ in the standard basis for $\SU(2)$-representations in term of the spin $j$ and the magnetic momentum $m$, diagonalizing the rotation operator $J^3$:
\bes
J^3\,|j, m\ra &=& m|j,m\ra,  \label{actionJ}\\
J^+\,|j,m\ra&=& \sqrt{(j-m)(j+m+1)}\,|j,m+1\ra, \nn \\
J^-\,|j,m\ra &=&\sqrt{(j+m)(j-m+1)}\,|j,m-1\ra, \nn \\
K^3\,|j,m\ra&=&- \alpha_j \sqrt{j^2-m^2}\, |j-1,m\ra -\beta_jm \, |j,m\ra + \alpha_{j+1} \sqrt{(j+1)^2-m^2}\, |j+1,m\ra, \label{actionK}\\
K^+\,|j,m\ra&=&-\alpha_j \sqrt{(j-m)(j-m-1)}\, |j-1,m+1\ra -\beta_j \sqrt{(j-m)(j+m+1)} \, |j, m+1\ra \nn \\
&&  -\alpha_{j+1} \sqrt{(j+m+1)(j+m+2)}\,|j+1,m+1\ra, \nn \\
K^-\,|j,m\ra&= & \alpha_j \sqrt{(j+m)(j+m-1)}\, |j-1,m-1\ra- \beta_j \sqrt{(j+m)(j-m+1)} \, |j,m-1 \ra \nn \\
&& +\alpha_{j+1} \sqrt{(j-m+1)(j-m+2)}\,|j+1,m-1\ra , \nn
\ees
where the coefficients defining the action of the boost generators are given by:
\be
\label{actionbeta}
\beta_j=\f{ n\rho}{j(j+1)},
\qquad \alpha_j= \f{i}{j} \sqrt{\f{(j^2-n^2)(j^2+\rho^2)}{4j^2-1}}.
\ee
It is straightforward to check that this postulated action satisfied as expected the $\SL(2,\C)$ commutation relations~\footnotemark.
\footnotetext{
The  commutation relation of the $\SL(2,\C)$ Lie algebra are:
\bes
&&[J^+,J^3]=-J^+, \quad[J^-, J^3]=J^-, \quad [J^+, J^-]=2J^3 \nn \\
&& [J^+,K^+]=[J^-, K^-]= [J^3, K^3]=0, \quad [J^+, K^-]=-[J^-, K^+]=2K^3, \nn \\
&&[J^+, K^3]=-K^+, \quad [J^-,K^3]=K^-, \quad [K^+, J^3]=-K^+, \quad [K^-, J^3]=K^-, \nn\\
&&[K^+,K^3]=J^+, \quad[K^-, K^3]=-J^-, \quad [K^+, K^-]=-2J^3. \nn
\ees
}
Moreover, since the coefficient $\alpha_n=0$ vanishes for $j=n$, it is also clear that the truncation to spins $j\ge n$ is self-consistent. On the other hand, it is obvious that the coefficients $\alpha_j$ for $j>n$ will never vanish, thus there is no upper bound on the spin $j$. This is consistent with the fact that a unitary representation of $\SL(2,\C)$ necessarily has an infinite dimension.

From this action, we can check that the $\SU(2)$ Casimir operator has the usual value $\vJ^2=j(j+1)$. We can also compute the values of the two Casimir operators of $\SL(2,\C)$:
\be
C_1=\vec{K}^2-\vJ^2=\rho^2-n^2+1,
\qquad
C_2= \vJ \cdot \vec{K}=2n \rho.
\ee

Finally, we introduce the characters of the $\SL(2,\C)$ representations, $\Theta^{(n,\rho)}(G)\,\equiv\,\tr\,D^{(n,\rho)}(G)$. It is easy to evaluate it on $\SU(2)$ group elements since we know the decomposition of the $\SL(2,\C)$ representation into $\SU(2)$ representations~\footnotemark:
\be
\forall g\in\SU(2),\,
\Theta^{(n,\rho)}(g)
\,=\,
\sum_{j\in n+\N} \chi^j(g)
\,=\,
\sum_{j\in n+\N} \f{\sin(2j+1)\theta}{\sin\theta}
\,=\,
\f{\cos 2n\theta}{2\sin^2\theta},
\ee
where $\theta$ is the class angle of the group element $g$, i.e meaning that $g$ is conjugate to the diagonal matrix with entries $[e^{i\theta},e^{-i\theta}]$.
\footnotetext{
The character formula is straightforwardly generalizable to the whole $\SL(2,\C)$ group. Indeed, all group elements are conjugated to a diagonal matrix. Then we can evaluate the character on such matrices (see e.g. \cite{noncompact}):
$$
\Theta_{(n,\rho)}
\mat{cc}{e^{\lambda+i\theta} & 0 \\ 0&e^{-\lambda-i\theta}}
\,=\,
\f{e^{i\rho\lambda}e^{i2n\theta}+e^{-i\rho\lambda}e^{-i2n\theta}}{|e^{\lambda+i\theta}-e^{-\lambda-i\theta}|^2}.
$$
}

Now that we have quickly reviewed these basic facts on $\SL(2,\C)$ unitary representations and the Plancherel decomposition, we are ready to introduce the basis of projected spin networks for our Hilbert space $H$ of Lorentz invariant cylindrical functions.

\chapter{Coherent States for the Harmonic Oscillator} \label{cohHO}

Let us review the standard definition of coherent states for a single harmonic oscillator, defined by its creation and annihilation operators satisfying the commutation relation $[a,a\dag]=1$. The standard basis is defined by the number of quanta:
\be
a|n\ra=\,\sqrt{n}\,|n-1\ra,\qquad
a\dag|n\ra=\,\sqrt{n+1}\,|n+1\ra,\qquad
a\dag a|n\ra\,=\,n|n\ra.
\ee
Coherent states are defined through a sum over the standard basis:
\be
|z\ra= \sum_n \f{z^n}{\sqrt{n!}}\,|n\ra
= \sum_n \f{z^n}{\sqrt{n!}}\,\f{(a\dag)^n}{\sqrt{n!}}\,|0\ra
= e^{z\,a\dag}\,|0\ra.
\ee
This definition is not normalized, but we can easily compute its norm and define normalized states:
\be
\la z|z\ra= e^{|z|^2},\qquad
|z\ra_N\,\equiv\, e^{-\f{|z|^2}{2}}\,|z\ra.
\ee
The action of the $a,a\dag$ operators can be derived directly from the definition of the coherent states as series. The coherent states diagonalize the annihilation operator $a$ while the creation operator $a\dag$ acts as a derivation:
\be
a\,|z\ra=z\,|z\ra,\quad
a\dag\,|z\ra
=\sum_{n\ge1} n\f{z^{n-1}}{\sqrt{n!}}\,|n\ra
=\pp_z\,|z\ra.
\ee
This action can be straightforwardly on the normalized coherent states. Then we get a anti-holomorphic shift in the $a\dag$ action:
\be
a\,|z\ra_N=z\,|z\ra_N,\quad
a\dag\,|z\ra_N
=\pp_z\,e^{-\f{|z|^2}{2}}\,|z\ra
=\left(\pp_z-\f{\bar{z}}{2}\right)\,|z\ra_N.
\ee
The coherent states naturally provides an over-complete basis and a new decomposition of the identity:
\bes
\int \f{d^2z}{\pi}\, |z\ra_N{}_N\la z|
&=&
\int \f{d^2z}{\pi}\, e^{-|z|^2}|z\ra\la z|
\,=\,
\sum_{m,n}\f{|m\ra\la n|}{\sqrt{m!}\sqrt{n!}}\,\int \f{d^2z}{\pi}\, e^{-|z|^2}\bar{z}^nz^m \nn\\
&=&
\sum_{m,n}\f{|m\ra\la n|}{\sqrt{m!}\sqrt{n!}}
\int_0^{+\infty} dr\,e^{-r^2}r^{m+n+1} \int_0^{2\pi} \f{d\theta}{\pi}\, e^{i(m-n)\theta}\nn\\
&=&
\sum_{n}\f{2}{n!}|n\ra\la n|\,
\int_0^{+\infty} dr\,e^{-r^2}r^{2n+1}
\,=\, \id.
\ees
%
%
We can also check explicitly that the action of $a\dag$ on coherent states is correctly given by the adjoint of the action of the annihilation operator $a$:
\bes
\int [d^2z d^2w]\,\overline{\phi(z)}\psi(w) \la z|a\dag w\ra
&=&
-\int [d^2z d^2w]\,\overline{\phi(z)}\pp_w\psi(w) \la z| w\ra
\,=\,
\int [d^2z d^2w]\,\overline{\phi(z)}\psi(w)\pp_w\left(e^{\bar{z}w}\right)\nn\\
&=&
\int [d^2z d^2w]\,\bar{z}\overline{\phi(z)}\psi(w) \la z| w\ra
\,=\,
\int [d^2z d^2w]\,\overline{\phi(z)}\psi(w) \la az| w\ra.
\ees
Finally, these coherent states transform consistently under the $\U(1)$-action generated by the number of quanta operator $a\dag a$~:
\be
e^{i\tau a\dag a}\,|z\ra
\,=\,
\sum_n \f{z^n}{\sqrt{n!}}\,e^{i\tau n}\,|n\ra
\,=\,
|e^{i\tau}z\ra.
\ee

\chapter{Commutation Relations Of the $E,F,F\dag$ Action on Coherent States}

The commutation relations between these $F$, $F^\dagger$ and $E$ operators acting on the U(N) coherent states are straightforward to check:
\bes
\left[ E_{ij}, E_{kl}\right] |J, \{z_q\}\ra
&= & \delta^z_{kl}\left( \delta^z_{ij}\left( |J,\{z_k\} \ra \right) \right) - \delta^z_{ij}\left( \delta^z_{kl}\left( |J,\{z_k\} \ra \right) \right)\nn\\
&=&\left( \delta_{kj} \delta^z_{il}- \delta_{il} \delta^z_{kj} \right) |J,\{z_q\} \ra= \left( \delta_{jk}E_{il}- \delta_{il} E_{kj} \right) |J,\{z_q\}\ra ,
\ees
\bes
\left[E_{ij}, F_{kl} \right] |J, \{z_q\} \ra
&=&\sqrt{J(J+1)}\left( Z_{kl} \delta^z_{ij} \left( |J-1,\{z_q\} \ra \right) - \delta^z_{ij}\left( Z_{kl} |J-1, \{z_q\}\ra \right)\right)\nn\\
&=&\sqrt{J(J+1)} \left(\delta_{il} Z_{jk}- \delta_{ik}Z_{jl} \right) |J-1,\{z_q\}\ra \nn \\
&=& \left(\delta_{il} F_{jk}- \delta_{ik}F_{jl} \right) |J,\{z_q\}\ra ,
\ees
\bes
\left[E_{ij}, F_{kl}^\dagger \right] |J,\{z_q\}\ra
&= &\f{1}{\sqrt{(J+1)(J+2)}} \left( \Delta^z_{kl} \left( \delta_{ij}^z (|J+1,\{z_q\}\ra)\right)- \delta^z_{ij} \left( \Delta^z_{kl} (|J+1,\{z_q\} \ra ) \right) \right) \nn \\
&=&\f{1}{\sqrt{(J+1)(J+2)}}\left( \delta_{kj} \Delta^z_{il}-\delta_{lj} \Delta^z_{ik}\right) |J+1,\{z_q\}\ra \nn \\
&= &\left(\delta_{kj} F_{il}^\dagger- \delta_{lj}F_{ik}^\dagger \right) |J,\{z_q\}\ra ,
\ees
\bes
\left[ F_{ij}, F^\dagger_{kl}\right]|J,\{z_q\}\ra
&= &\Delta^z_{kl}\left(Z_{ij} |J,\{z_q\}\ra \right)- Z_{ij} \Delta_{kl}^z\left( |J,\{z_q\} \ra \right)\nn\\
&=& \left(\delta_{ki} \delta^z_{lj}-\delta_{kj} \delta^z_{li} -\delta_{li} \delta^z_{kj}+ \delta_{lj} \delta^z_{ki} + \Delta^z_{kl}(Z_{ij}) \right)|J,\{z_q\}\ra \nn \\
&=& \left(\delta_{ki} E_{lj}-\delta_{kj} E_{li} -\delta_{li} E_{kj}+ \delta_{lj} E_{ki} + 2(\delta_{ki}\delta_{lj}-\delta_{li}\delta_{kj}) \right)|J,\{z_q\}\ra .
\ees

\chapter{Norm of the $\U(N)$-invariant state: $|J\ra$} \label{norm_App}

Following the previous work done in \cite{2vertex}, we compute the norm of the $\U(N)$-invariant state $|J \ra= (f^\dagger)^J |0\ra$ where we have introduced the operator:
\be
f^\dagger= \sum_{kl} F_{kl}^{L \dagger } F_{kl}^{R \dagger}.
\ee
The norm of $|J\ra$ is then given by:
\be
\la 0 | f^J( f^\dagger)^J |0\ra
\ee
with $f= \sum_{kl} F_{kl}^{L } F_{kl}^{R }$. We need to determine the action of $f$ on $|J\ra$: $f |J\ra= f (f^\dagger)^J |0\ra$.
In order to compute this action of $f$ on $|J\ra$, we calculate the commutator between $f$ and $f^\dagger$:
\be
[f, f^\dagger]= 4 e \left(\f{E^L+E^R}{2} \right) + 4 \left( \f12 (E^L+E^R)^2+(E^L+E^R)(2N-1)+2N(N-1) \right)
\ee
where we have defined:
\be
e= \sum_{kl} E_{kl}^L E_{kl}^R
\ee
and we recall that $E^{L/R}=\sum_i E_i^{L/R}$. In our case of interest $E^L=E^R=E$. Moreover, to compute this commutator we used the fact that on the intertwiner space ($\vJ^L=\vJ^R=0$), the $F$ and $F^\dagger$ operators satisfy additional quadratic constraints:
\be
\sum_k \left(F^{L/R}_{ki}\right)^ \dagger F_{kj}^{L/R} = E_{ij}^{L/R} \left( \f{E^{L/R}}{2}+1\right), \qquad \sum_k F^{L/R}_{kj} \left(F_{ki}^{L/R}\right)^\dagger = (E_{ij}^{L/R}+2 \delta_{ij}) \left( \f{E^{L/R}}{2}+N-1\right).
\ee
We also need to compute the commutator between $e$ and $f^\dagger$. We use the fact that the $E$ and $F^\dagger$ also satisfy quadratic constraints on the intertwiner space:
\be
\sum_k \left(F^{L/R}_{ik}\right)^\dagger E_{jk}^{L/R} =  \left(F^{L/R}_{ij}\right)^\dagger \f{E^{L/R}}{2}, \qquad \sum_k E_{jk}^{L/R} \left(F^{L/R}_{ik}\right)^\dagger =  \left(F^{L/R}_{ij}\right)^\dagger \left( \f{E^{L/R}}{2} +N-1 \right),
\ee
then,
\be
[e, f^\dagger]= 2 f^\dagger \left(\f{E^L+E^R}{2}+N-1 \right)
\ee
where once again, $E^L$ and $E^R$ can be replaced by $E$. Moreover, this total area operator is clearly diagonal in the basis $|J\ra$:
\be
E |J\ra = 2J |J\ra
\ee
We can then deduce the action of $f$ on $|J \ra$:
\bes
f(f^\dagger)^J|0\ra&=&\sum_{k=0}^{J-1}\left\{4(2(J-1-k)+N) (f^\dagger)^k e \, (f^\dagger)^{J-1-k} |0\ra + 16 (2 (J-1-k)^2 +(J-1-k)(2N-1)) (f^\dagger)^{J-1}|0\ra \right\} \nn\\
&& \qquad \qquad+8 (J-1)N(N-1) (f^\dagger)^{J-1}|0 \ra \nn \\
&=&[\sum_{k=0}^{J-1}\left\{ 8(2(J-1-k)+N)(J-1-k)(J+N-k-3)+ 16 (2 (J-1-k)^2 +(J-1-k)(2N-1))\right\}  \nn \\
&& \qquad \qquad +8 (J-1)N(N-1)] |J-1\ra \nn \\
&=& 4J(J+1)(N+J-1)(N+J-2) |J -1 \ra
\ees
Using this action, we compute the norm of the state $|J \ra$ by recursion:
\be
\la J | J\ra = \la 0 | f^{J-1} f |J\ra= 4J(J+1)(N+J-1)(N+J-2) \la J-1 | J-1 \ra
\ee
which leads us to the scalar product:
\be
\la J | J\ra= 2^{2J} J! (J+1)! \f{(N+J-1)!(N+J-2)!}{(N-1)!(N-2)!}
\ee

\chapter{The \sj-symbol - recoupling theory}
\label{6j}

The \sj-symbol is a real number and it is obtained by combining four normalized Clebsh-Gordan coefficients along the six edges of a tetrahedron, with edge lengths given by $j_{IJ}+\frac{1}{2}=\frac{d_{j_{IJ}}}{2}$ ($0\leq I<J \leq 3$).
\begin{figure}[ht]
\begin{center}
\includegraphics[width=3cm]{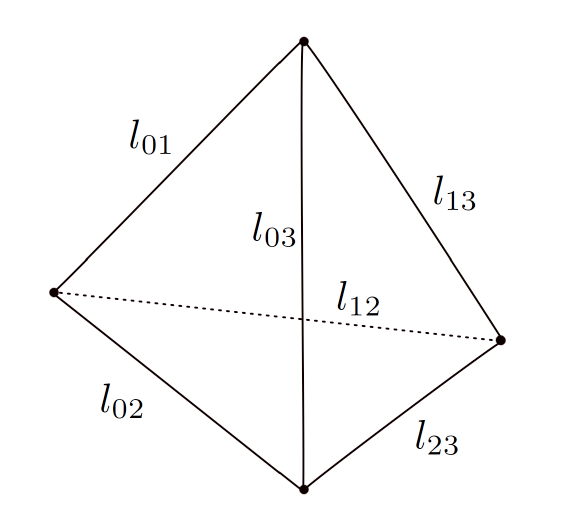}
\caption{Tetrahedron: edge lengths are given by $l_{IJ}=\frac{d_{j_{IJ}}}{2}$}
\end{center}
\end{figure}
We usually express the 6j-symbol in term of  the Wigner 3j-symbols~:
\equa{ \tabl{ll}{
 \left\{ \tabl{lll}{j_{01}&j_{02}&j_{03}\\
j_{23}&j_{13}&j_{12}\\} \right\}=& \sum_\alpha (-1)^{j_{01}+j_{03}+j_{01}-\alpha_{01}-\alpha_{03}-\alpha_{01}} \left( \tabl{lll}{j_{01}&j_{12}&j_{13}\\
\alpha_{01}&\alpha_{12}&-\alpha_{13}\\} \right)\\
& \left( \tabl{lll}{j_{13}&j_{23}&j_{03}\\
\alpha_{13}&\alpha_{23}&\alpha_{03}\\} \right) \left( \tabl{lll}{j_{03}&j_{02}&j_{01}\\
\alpha_{03}&\alpha_{02}&-\alpha_{01}\\} \right) \left( \tabl{lll}{j_{02}&j_{23}&j_{12}\\
\alpha_{02}&\alpha_{23}&\alpha_{12}\\} \right).
}}
The Wigner 3j symbols are very simply related to the Clebsh-Gordan coefficients $<j_{01} j_{12} \alpha_{01} \alpha_{12} | j_{13} \alpha_{13}>$ by:
$$<j_{01} j_{12} \alpha_{01} \alpha_{12} | j_{13} \alpha_{13}>= (-1)^{j_{01}-j_{12}+\alpha_{13}}(2j_{13}+1/2)^{1/2} \left( \tabl{lll}{j_{01}&j_{12}&j_{13}\\
\alpha_{01}&\alpha_{12}&-\alpha_{13}\\} \right),$$
And Racah gave a general formulae for the Clebsh-Gordan coefficient:
$$\tabl{ll}{ <j_{01} j_{12} \alpha_{01} \alpha_{12} | j_{13} \alpha_{13}>&= \delta( \alpha_{01}+ \alpha_{12}, \alpha_{13}) \Delta(j_{01}j_{12}j_{13}) \\
&\\
& \sqrt{(2j_{13}+1)(j_{01}+\alpha_{01})!(j_{01}-\alpha_{01})!(j_{12}+\alpha_{12})!(j_{12}-\alpha_{12})!(j_{13}+\alpha_{13})!(j_{13}-\alpha_{13})!}\\
&\\
&\sum_\mu \frac{(-1)^\mu}{(j_{01}-\alpha_{01}-\mu)!(j_{13}-j_{12}+\alpha_{01}+\alpha)!(j_{12}+\alpha_{12}-\mu)! (j_{13}-j_{01}-\alpha_{12}+\alpha)! \mu!(j_{01}+j_{12}-j_{13}-\mu)!}
}$$
where $\Delta(j_{01}, j_{12}, j_{13})= \frac{(j_{01}+j_{12}-j_{13})!(j_{01}-j_{12}+j_{13})!(-j_{01}+j_{12}+j_{13})!}{(j_{01}+j_{12}+j_{13}+1)!}$
From these, Racah gave a tensorial formulae for the 6j-symbol, the Racah's single sum formulae:
\equa{
\tabl{ll}{
\left\{ \tabl{lll}{j_{01}&j_{02}&j_{03}\\
j_{23}&j_{13}&j_{12}\\} \right\}= &\sqrt{\Delta(j_{01}, j_{02}, j_{03}) \Delta(j_{23}, j_{02}, j_{12}) \Delta(j_{23}, j_{13}, j_{03}) \Delta(j_{01}, j_{13}, j_{12}) }\\
& \displaystyle{\sum_{\textrm{max } v_I}^{\textrm{min } p_j}} (-1)^t \frac{(t+1)!}{\prod_i(t-v_I)! \prod_j (p_j-t)!}
}}
with $ v_K= \displaystyle{\sum_{I \neq K}} j_{IK} \; \forall K \in \{ 0, \cdots, 3\}$ and  $p_k= \displaystyle{\sum_{i \neq 0, k}} (j_{0i} + j_{ki}) \; \forall k \in \{1 \cdots 3 \}$.
%

\chapter{Factorials} \label{factorials}

The factorial $n!$ is defined for a positive integer $n$ as:
$$
n! \equiv n(n-1) \cdots 2\cdot 1=\Gamma (n+1),
$$
where $\Gamma(n)$ is the gamma function for integers $n$. This definition is generalized to non-integer values. Using the identities for the $\Gamma$ function, we write explicitly the values for half-integers:
$$
(-\frac{1}{2})!=\sqrt{\pi}, \qquad
(\frac{1}{2})!= \frac{\sqrt{\pi}}{2},\qquad
(n-\frac{1}{2})!=\frac{\sqrt{\pi}}{2^n}(2n-1)!! =\frac{\sqrt{\pi}(2n)!}{2^{2n}n!},\qquad
(n+\frac{1}{2})!=\frac{\sqrt{\pi}}{2^{n+1}}(2n+1)!!=\frac{\sqrt{\pi}(2n+1)!}{2^{2n+1}n!},$$
where $n!!$ is the double factorial~:
$$
n!! \equiv \left\{ \tabl{l}{
n \cdot (n-2) \cdots 5\cdot 3 \cdot 1 \;\; \textrm{ if } n>0 \textrm{ odd,} \\
n \cdot (n-2) \cdots 6\cdot 4 \cdot 2 \;\; \textrm{ if } n>0 \textrm{ even,} \\
1 \;\; \textrm{ if } n= -1 \textrm{ or } 0.
} \right .
$$
Using the asymptotic expansion of a large factorial $n! \sim \sqrt{2 \pi n}\left(\frac{n}{e}\right)^n\left(1 + \frac{1}{12n}=\frac{1}{288n^3}-\frac{139}{51840n^3}-\frac{571}{2488320n^4}\cdots \right)$, we can get an asymptotic expansion for:
\equa{\tabl{l}{
(n+1/2)! \sim \sqrt{2 \pi} e^{(n+1)\ln(n)-n} \left(1+\frac{1}{2n}\right) \left(1+ \frac{11}{12(2n)} + \frac{1}{288(2n)^2}- \frac{139}{51840(2n)^3}-\frac{571}{2488320(2n)^4} + \cdots \right) \\
\;\;\;\;\;\;\;\;\;\;\;\;\;\;\;\;\;\;\;\;\;\;\;\;\left(1 - \frac{1}{12n}- \frac{1}{288n^2}+ \frac{139}{51840n^3}+\frac{571}{2488320n^4} - \cdots \right),\\
\\
(n-\frac{1}{2})! \sim \sqrt{2 \pi} e^{n\ln(n)-n}\left(1+ \frac{11}{12(2n)} + \frac{1}{288(2n)^2}- \frac{139}{51840(2n)^3}-\frac{571}{2488320(2n)^4} + \cdots \right) \\
\;\;\;\;\;\;\;\;\;\;\;\;\;\;\;\;\;\;\;\;\;\;\;\;\left(1 - \frac{1}{12n}- \frac{1}{288n^2}+ \frac{139}{51840n^3}+\frac{571}{2488320n^4} - \cdots \right),\\
}}
or more simply, at the next-to-leading order:
\equa{\tabl{l}{
n! \sim \sqrt{2 \pi n}\left(\frac{n}{e}\right)^n\left(1 + \frac{1}{12n}\right),\\
\\
(n+\frac{1}{2})! \sim \sqrt{2 \pi} e^{(n+1)\ln(n)-n}\left(1+ \frac{11}{24n}  \right), \\
\\
(n-\frac{1}{2})! \sim \sqrt{2 \pi} e^{n\ln(n)-n}\left(1- \frac{1}{24n}  \right). \\
}}

\chapter{First approximation in the "brute force" asymptotic expansion of the \sj} 
\label{stirling}

The first approximation in the "brute force" asymptotic expansion of the \sj-symbol is to replace the factorials by the next to leading order of the Stirling formula. \\
In this section, all computations are done at the next-to-leading order.
We replace the factorials in equation (\ref{Racah_dj}) by their respective asymptotic expansion. 

\begin{itemize}
\item Then, a typical triangle coefficient:
$$\Delta(\lambda d_{j_{01}}, \lambda d_{j_{02}}, \lambda d_{j_{03}})= \frac{\left(\frac{\lambda }{2}(d_{j_{01}}+d_{j_{02}}-d_{j_{03}})-\frac{1}{2}\right)!\left(\frac{\lambda }{2}(d_{j_{01}}-d_{j_{02}}+d_{j_{03}})-\frac{1}{2}\right)!\left(\frac{\lambda }{2}(-d_{j_{01}}+d_{j_{02}}+d_{j_{03}})-\frac{1}{2}\right)!}{\left(\frac{\lambda }{2}(d_{j_{01}}+d_{j_{02}}+d_{j_{03}})-\frac{1}{2}\right)!}$$
will be
$$
\tabl{ll}{
\Delta(\lambda d_{j_{01}}, \lambda d_{j_{02}}, \lambda d_{j_{03}})&=2\pi [ e^{-\frac{\lambda }{2}(d_{j_{01}}+d_{j_{02}}+d_{j_{03}})\ln[\frac{\lambda }{2}(d_{j_{01}}+d_{j_{02}}+d_{j_{03}})]+\frac{\lambda }{2}(d_{j_{01}}+d_{j_{02}}+d_{j_{03}})}\left(1+\frac{1}{12\lambda (d_{j_{01}}+d_{j_{02}}+d_{j_{03}})}\right)\\
&\\
& \;\;\; e^{\frac{\lambda }{2}(d_{j_{01}}+d_{j_{02}}-d_{j_{03}})\ln[\frac{\lambda }{2}(d_{j_{01}}+d_{j_{02}}-d_{j_{03}})]-\frac{\lambda }{2}(d_{j_{01}}+d_{j_{02}}-d_{j_{03}})}\left(1-\frac{1}{12\lambda (d_{j_{01}}+d_{j_{02}}-d_{j_{03}})}\right)\\
&\\
& \;\;\; e^{\frac{\lambda }{2}(d_{j_{01}}-d_{j_{02}}+d_{j_{03}})\ln[\frac{\lambda }{2}(d_{j_{01}}-d_{j_{02}}+d_{j_{03}})]-\frac{\lambda }{2}(d_{j_{01}}-d_{j_{02}}+d_{j_{03}})}\left(1-\frac{1}{12\lambda (d_{j_{01}}-d_{j_{02}}+d_{j_{03}})}\right)\\
&\\
& \;\;\; e^{\frac{\lambda }{2}(-d_{j_{01}}+d_{j_{02}}+d_{j_{03}})\ln[\frac{\lambda }{2}(-d_{j_{01}}+d_{j_{02}}+d_{j_{03}})]-\frac{\lambda }{2}(-d_{j_{01}}+d_{j_{02}}+d_{j_{03}})}\left(1-\frac{1}{12\lambda (-d_{j_{01}}+d_{j_{02}}+d_{j_{03}})}\right)\\
}
$$
which simplifies
\equa{\tabl{ll}{
\Delta(\lambda d_{j_{01}}, \lambda d_{j_{02}}, \lambda d_{j_{03}})&=2\pi e^{\frac{\lambda }{2}[(-d_{j_{01}}+d_{j_{02}}+d_{j_{03}})\ln(-d_{j_{01}}+d_{j_{02}}+d_{j_{03}})+(d_{j_{01}}-d_{j_{02}}+d_{j_{03}})\ln(d_{j_{01}}-d_{j_{02}}+d_{j_{03}})]}\\
\\
&e^{-\frac{\lambda }{2}[(d_{j_{01}}+d_{j_{02}}-d_{j_{03}})\ln(d_{j_{01}}+d_{j_{02}}-d_{j_{03}})+(d_{j_{01}}+d_{j_{02}}+d_{j_{03}})\ln(d_{j_{01}}+d_{j_{02}}+d_{j_{03}})]}\\
\\
&[1-\frac{1}{12\lambda }(\frac{1}{-d_{j_{01}}+d_{j_{02}}+d_{j_{03}}}+\frac{1}{d_{j_{01}}-d-{j_{02}}+d_{j_{03}}}+\frac{1}{d_{j_{01}}+d_{j_{02}}-d-{j_{03}}} -\frac{1}{d_{j_{01}}+d_{j_{02}}+d_{j_{03}}})].
}}
The factor $\sqrt{\Delta(\lambda d_{j_{01}}, \lambda d_{j_{02}}, \lambda d_{j_{03}}) \Delta(\lambda d_{j_{23}}, \lambda d_{j_{02}}, \lambda d_{j_{12}}) \Delta(\lambda d_{j_{23}}, \lambda d_{j_{13}}, \lambda d_{j_{03}}) \Delta(\lambda d_{j_{01}},\lambda d_{j_{13}}, \lambda d_{j_{12}}) }$ in equation (\ref{Racah_dj}) can then easily be put into the form:
\equa{
(2\pi)^2e^{\frac{\lambda }{2}h(d_{j_{IJ}})}\left( 1-\frac{1}{24\lambda }H(d_{j_{IJ}}) \right)
}
where \equa{
\tabl{l}{
h(d_{j_{IJ}})= \displaystyle{\sum_{I<J}}d_{j_{IJ}}h_{d_{j_{IJ}}} \\
\textrm{with } h_{d_{j_{IJ}}}= \frac{1}{2} \ln \left( \frac{(d_{j_{IJ}}-d_{j_{IK}}+d_{j_{IL}})(d_{j_{IJ}}+d_{j_{IK}}-d_{j_{IL}})(d_{j_{IJ}}-d_{j_{JK}}+d_{j_{JL}})(d_{j_{IJ}}+d_{j_{JK}}-d-{j_{JL}})}{(d-{j_{IJ}}+d_{j_{IK}}+d_{j_{IL}})(-d_{j_{IJ}}+d_{j_{IK}}+d_{j_{IL}})(d_{j_{IJ}}+d_{j_{JK}}+d_{j_{JL}})(-d_{j_{IJ}}+d_{j_{JK}}+d_{j_{JL}})}\right)\\
 \;\;\; K \neq L \textrm{ and } K,L \neq I,J\\
H(d_{j_{IJ}})=2\displaystyle{\sum_{j,K}} \frac{1}{\tilde{p}_j-\tilde{v}_K} -2 \displaystyle{\sum_K} \frac{1}{\tilde{v}_K} \textrm{ where } K \in \{0, \cdots, 3 \} \textrm{ and } j \in \{1,\cdots, 3\} \\
}}
and we recall that  $ \tilde{v}_K= \displaystyle{\sum_{I \neq K}}\frac{ d_{j_{IK}}}{2} \; \forall K \in \{ 0, \cdots, 3\}, \; \tilde{p}_k= \displaystyle{\sum_{i \neq 0, k}} \frac{(d_{j_{0i}} + d_{j_{ki}})}{2} \; \forall k \in \{1 \cdots 3 \}$.
\item We now replace the factorials in the sum of (\ref{Racah_dj}) by their approximations and we change of variables: $t=\lambda x$:
\equa{\tabl{ll}{
\Sigma(\lambda d_{j_{IJ}})&= \displaystyle{\sum_{x=\textrm{max }\tilde{ v}_I}^{\textrm{min } \tilde{p}_j}} (-1)^{\lambda x} \frac{(\lambda x+1)(\lambda x)! \prod_j(\lambda (\tilde{p_j}-x)) \prod_j (\lambda (\tilde{p}_j-x)-1)}{\prod_I (\lambda (x-\tilde{v}_I)+3/2)(\lambda (x-\tilde{v}_I)+1/2)!\prod_j(\lambda (\tilde{p}_j-x))!}\\
&= \frac{1}{(2\pi)^3} \displaystyle{\sum_{x=\textrm{max } v_I}^{\textrm{min } p_j}} e^{G_1(x)}G_2(x)
}}
where \equa{\tabl{ll}{
G_1(x)=& i\pi \lambda x + 3 \ln \lambda  + \ln x +2 \displaystyle{\sum_j} \ln (\tilde{p}_j-x) -\displaystyle{\sum_I} \ln (x- \tilde{v}_I)+(\lambda x+1/2)(\ln x + \ln \lambda ) -\lambda x+ \displaystyle{\sum_I} \lambda (x-\tilde{v}_I) \\
&- \displaystyle{\sum_I} (\lambda (x-\tilde{v}_I)+1)(\ln \lambda  + \ln (x-\tilde{v}_I))-\displaystyle{\sum_j} (\lambda (\tilde{p}_j-x)+1/2)(\ln \lambda  +\ln(\tilde{p}_j-x)) +\displaystyle{\sum_j} \lambda (\tilde{p}_j-x)
 }}
which can be simplified using the fact that $\sum_I \tilde{v}_I= \sum_j \tilde{p}_j$:
\equa{
G_1(x)= -2\ln \lambda +\frac{1}{2}\ln\frac{x^3\prod_j(\tilde{p}_j-x)^3}{\prod_I(x-\tilde{v}_I)^4} +\lambda \left[ i\pi x  +x \ln x - \sum_I (x-v_I)  \ln (x -v_I) - \sum_j (p_j-x)\ln(p_j-x) \right]
}
and
\equa{\tabl{l}{
G_2(x)= \frac{1+\frac{1}{12\lambda x}}{(1+\frac{3}{2\lambda (x-\tilde{v}_I)}\prod_I(1+\frac{11}{24\lambda (x-\tilde{v}_I)})\prod_j(1+\frac{1}{12\lambda (p_j-x)})}\\
\;\;\;\;\;\;= 1-\frac{1}{\lambda }\left(- \frac{13}{12x}+ \displaystyle{\sum_I}\frac{47}{24(x-v_I)} + \sum_j \frac{13}{12(p_j-x)} +O\left( \frac{1}{\lambda }\right)\right)
}}
Moreover,
\equa{
e^{G_1(x)}=\frac{1}{\lambda ^2}e^{F(x)+\lambda f(x)}}
where \equa{\tabl{l}{
f(x)= i \pi x+ x \ln(x) - \displaystyle{\sum_I}(x-v_K) \ln(x-v_I) - \displaystyle{\sum_j}(p_j-x) \ln(p_j-x) \\
F(x)= \frac{1}{2} \ln \left( \frac{x^3\prod_j(p_j-x)^3}{\prod_I (x-v_I)^4 } \right).
}}
Then the sum can be approximated by:
\equa{
\Sigma(\lambda d_{j_{IJ}})=\frac{1}{(2\pi)^3\lambda ^2}\displaystyle{\sum_{x=\textrm{max } v_I}^{\textrm{min } p_j}} e^{\lambda f(x)+F(x)} \left( 1 -\frac{1}{12\lambda }G(x) + O\left(\frac{1}{\lambda }\right)\right) e^{\lambda f(x)}
}
where \equa{
G(x)=-\frac{13}{x}+ \displaystyle{\sum_K}\frac{47}{24(x-v_K)} + \sum_j \frac{13}{p_j-x}. \\
}
\end{itemize}


\chapter{Third approximation in the "brute force" asymptotic expansion of the \sj} \label{saddlepoint} 

The third approximation in the "brute force" asymptotic expansion of the \sj-symbol is done using the stationary phase method. \\
We are interested in the $1/\lambda  $ expansion of the integral:
$$I=\displaystyle{\int_{\textrm{max } \tilde{v}_I/2}^{\textrm{min } \tilde{p}_j/2}} dx e^{F(x)} \left( 1 -\frac{1}{12\lambda }G(x) + O\left(\frac{1}{\lambda }\right)\right) e^{\lambda f(x)}.$$
We do not give here the proof of the whole expansion (equation (\ref{complete})) because of the heavy formalism but we directly prove the next to leading order formula (equation (\ref{intNLO})); the procedure is the same but the computations are easier.
The asymptotic expansion of such an integral is given by contributions around the stationary points of the phase denoted $x_0$. We expand the phase $f(x)$ around the stationary points $x_0$ at fourth order and the function $g(x)=e^{F(x)}\left( 1- \frac{1}{12\lambda }G(x) \right)$ at second order and we extend the integration to infinity.
\equa{
\tabl{ll}{
I \sim \displaystyle{\sum_{x_0}} \displaystyle{ \int_{-\infty}^{+ \infty} }& d(\delta x) \left(g(x_0) + g^\prime(x_0) \delta x + \frac{1}{2} g^{\prime \prime}(x_0) (\delta x)^2 \right) e^{\lambda \left( f(x_0)+ \frac{1}{2}f^{\prime \prime}(x_0) (\delta x)^2\right)} \\
& \left(1+ \lambda  \left(\frac{1}{3!} f^{(3)}(x_0) (\delta x)^3 + \frac{1}{4!}f^{(4)}(x_0) (\delta x)^4\right) +\frac{\lambda ^2}{2} \left( \frac{1}{3!} f^{(3)}(x_0) (\delta x)^3 \right)^2  +O(\lambda ^2)\right)
}}
where in our case, $g(x)= e^{F(x)}\left( 1- \frac{1}{12\lambda }G(x) \right)$ and then the integration are "generalized" Gaussians:
\equa{
\tabl{ll}{
I \sim \displaystyle{\sum_{x_0}}  e^{F(x_0)+\lambda f(x_0)} &[ \left( 1 -\frac{1}{12\lambda }G(x_0) \right) \displaystyle{ \int_{-\infty}^{+ \infty} } d(\delta x) e^{-\lambda (\frac{-f^{\prime \prime}(x_0)}{2})(\delta x)^2} \\
&+\frac{1}{2}\left( (F^\prime(x_0))^2+ F^{\prime \prime} (x_0) \right) \displaystyle{ \int_{-\infty}^{+ \infty} }d(\delta x) (\delta x)^2 e^{-\lambda (\frac{-f^{\prime \prime}(x_0)}{2})(\delta x)^2} \\
&+ \lambda  \left(\frac{f^{(4)}(x_0)}{4!} + \frac{f^{(3)}(x_0)}{3!}F^\prime(x_0)\right) \displaystyle{ \int_{-\infty}^{+ \infty} }d(\delta x) (\delta x)^4 e^{-\lambda (\frac{-f^{\prime \prime}(x_0)}{2})(\delta x)^2} \\
&+ \frac{\lambda ^2}{2} \left( \frac{f^{(3)}(x_0)}{3!} \right)^2 \displaystyle{ \int_{-\infty}^{+ \infty} }d(\delta x) (\delta x)^6 e^{-\lambda (\frac{-f^{\prime \prime}(x_0)}{2})(\delta x)^2}+ O \left(\frac{1}{\lambda ^{3/2}} \right) ]
}}
which can easily be computed:
\equa{\tabl{ll}{
I \sim \displaystyle{\sum_{x_0}} &\sqrt{\frac{2\pi}{-f^{\prime \prime}(x_0)\lambda } } e^{F(x_0)+\lambda f(x_0)} \\
&\left[1 +
\frac{1}{\lambda } \left( -\frac{G(x_0)}{12}- \frac{F^{\prime \prime}(x_0)+(F^{\prime}(x_0))^2}{2 f^{\prime \prime}(x_0)}+\frac{f^{(4)}(x_0)+4f^{(3)}(x_0)F^\prime(x_0)}{8(f^{\prime \prime}(x_0))^2}- \frac{5(f^{(3)}(x_0))^2}{24(f^{\prime \prime}(x_0))^3} \right) + O \left(\frac{1}{\lambda } \right) \right]
}}
\chapter{Physical States with a Real Phase} \label{real}
We will discuss here the case already mentioned of the Bessel-based factorized boundary state:
\be
\psi_q[j_{ab}]= \prod_{a<b}\phi_B(j_{ab})
\ee
where
\be
\phi_B(j)= \f{I_{|d_j-d_{j_0}|}(\f{d_{j_0}}{\alpha})-I_{d_j+d_{j_0}}(\f{d_{j_0}}{\alpha})}{\sqrt{I_0(\f{2d_{j_0}}{\alpha})-I_{2d_{j_0}+1}(\f{2d_{j_0}}{\alpha})}} \cos\left(d_j\Theta\right)
\ee
We are interested in the large spin limit regime and its $d_{j_0} \rightarrow \infty$ limit behaves as a Gaussian peaked around $d_{j_0}$:
\be
\phi_B(j)\simeq \left( \f{\alpha}{ d_{j_0}\pi }\right)^{1/4} \exp\left[ -\f{\alpha}{2 d_{j_0}}(d_j-d_{j_0})^2\right] \cos (d_j \Theta)
\ee
The difference with the case studied previously of a Gaussian peaked around $d_{j_0}$ is the phase which is real here.
We recall that the factorized boundary state assumption has been made in order to perform numerical simulations and the choice of a real phase has been done in the work concerning the area correlator to turn it into an exact group integral \cite{anal1} and perform exact analytical computations.

We therefore consider now a factorized Gaussian state with a real phase:
\be
\psi_q[d_{j_{ab}}] \f{1}{\mathcal{N}}\prod_{a<b} \exp(-\alpha (d_{j_{ab}}-d_{j_0})^2) \cos(d_{j_0}\Theta)
\ee
 to see if the conditions necessary to have a physical state are modified.  Conditions (\ref{norm}) and (\ref{dyna}) give the two following equations for $\mathcal{N}$ and $\alpha$.
\be
1=\sum_{j_{ab} }|\psi(j_{ab})|^2= \f{1}{(\mathcal{N}^2)^{10}}\sum_{\{j_{ab}\}} e^{-2\Re(\alpha)\sum_{a<b}(d_{j_{ab}}-d_{j_0})^2} \prod_{a<b} \left(\f{1+ \cos(2d_{j_{ab}}\Theta)}{2} \right)
\ee
and,
\be
1=\sum_{j_{ab} }\tj \psi(j_{ab})\simeq \frac{1}{\mathcal{N}^{10}}\sum_{\{j_{ab}\}}(\prod_{a<b}d_{j_{ab}})^\sigma P(d_{j_{ab}}) \sum_{\epsilon=\pm1;\, \eta=\pm1} e^{-\Re(\alpha)\sum_{a<b}(d_{j_{ab}}-d_{j_0})^2+i\sum_{a<b}[d_{j_{ab}}(\epsilon \theta_{ab}+\eta \Theta)-\Im(\alpha)(d_{j_{ab}}-d_{j_0})^2]}
\ee
where we have already used the asymptotic formulae of the $\tj$-symbol. In the large spin limit, the summation in the two previous equations can then be approximated with an integral  and we can write:
\be \label{intNorm2}
1\simeq \f{1}{2\mathcal{N}^2} \int d(d_{j})e^{-2\Re(\alpha)(d_{j}-d_{j_0})^2} \left[\f12 +\f14 e^{i2d_j \Theta}+ \f14 e^{-i2d_j\Theta} \right]
\ee
and
\be \label{intDyna2}
1\simeq \frac{1}{(2\mathcal{N})^{10}}\int \prod_{a<b}d[d_{j_{ab}}] (\prod_{a<b}d_{j_{ab}})^\sigma P(d_{j_{ab}}) \sum_{\epsilon=\pm1;\, \eta=\pm} e^{-\Re(\alpha)\sum_{a<b}(d_{j_{ab}}-d_{j_0})^2+i\sum_{a<b}[d_{j_{ab}}(\epsilon \theta_{ab}+\eta\Theta)-\Im(\alpha)(d_{j_{ab}}-d_{j_0})^2]}
\ee
The first integral (\ref{intNorm2}) after some changes of variables is just three Gaussian integrals, which can been integrated directly:
\be
\mathcal{N}^2=\f{1}{4}\sqrt{\f{\pi}{2\Re(\alpha)}}\left[1+e^{-\f{\Theta^2}{2\Re(\alpha)}} \cos(d_{j_0} \Theta) \right]
\ee
but we recall that $\alpha \propto \f{1}{d_{j_0}^n}$ with $n \in ] 0,2 [$ from equations (\ref{uncertainties}) and therefore in the large spin limit the second term of the right-hand side of this relation vanishes and we obtain the same relation at the leading ordrer between $\mathcal{N}$ and $\alpha$ as in the case of a Gaussian state with an imaginary phase (up to a factor $1/2$). That is:
\be
\mathcal{N}^2=\f{1}{4}\sqrt{\f{\pi}{2\Re(\alpha)}}
\ee
 To evaluate the second integral (\ref{intDyna2}) in the large spin limit we expand as previously the Regge action around $d_{j_0}$:
\be
 S_R[d_{j_{ab}}] \simeq \Theta \sum_{a<b}  d_{j_{ab}} +\f12 \sum_{a<b, c<d} N^{ab}_{cd}  \delta d_{j_{ab}} \delta d_{j_{cd}}.
 \ee
 where  $$N^{ab}_{cd}= \f{\pp^2 S_R}{\pp d_{j_{ab}}d_{j_{cd}}}|_{d_{j}=d_{j_0}}=\f{\pp \theta_{ab}}{\pp d_{j_{cd}}}|_{d_{j}=d_{j_0}}=\f{\sqrt{3}}{4\sqrt{5}d_{j_0}}\left(\tabl{cccccccccc}{
-4 & 7/2 & 7/2 & 7/2& 7/2 & 7/2 & 7/2 & -9 & -9 & -9\\
7/2 & -4 & 7/2 & 7/2 & 7/2 &-9 &-9 & 7/2 & 7/2& -9 \\
7/2 & 7/2 & -4 & 7/2 & -9 & 7/2 & -9 & 7/2 & -9 & 7/2\\
7/2 & 7/2 & 7/2 & -4 & -9 & -9 & 7/2 & -9 & 7/2 & 7/2\\
7/2 & 7/2 & -9 & -9 & -4 & 7/2 & 7/2 & 7/2 & 7/2 & -9 \\
7/2 & -9 & 7/2 & -9 & 7/2 & -4 & 7/2 & 7/2 & -9 & 7/2\\
7/2 & -9 & -9 & 7/2 & 7/2 & 7/2 & -4 & -9 & 7/2 & 7/2\\
-9 & 7/2 & 7/2 & -9 & 7/2 & 7/2 & -9 & -4 & 7/2 & 7/2 \\
-9 & 7/2 & -9 & 7/2 & 7/2 & -9 & 7/2 &7/2 & -4 & 7/2 \\
-9 & -9 & 7/2 & 7/2 & -9 & 7/2 & 7/2 & 7/2 & 7/2 & -4} \right).$$
 And using once again the argument that rapidly oscillating term in $d_{j_{ab}}$ will vanish when performing the integration over $d_{j_{ab}}$, we have to consider only two terms in the sum (\ref{intDyna2}) which are two Gaussian integral:
\be
1\simeq \frac{(d_{j_{0}})^{(10\sigma-9/2)} P}{(2\mathcal{N})^{10}}\int \prod_{a<b}d[\delta d_{j_{ab}}] \left[ e^{-\delta d_{j_{ab}}M^{ab}_{cd}\delta d_{j_{cd}}} + e^{-\delta d_{j_{ab}}Q^{ab}_{cd}\delta d_{j_{cd}}}  \right]
\ee
where $M$  and $Q$ are both  ten by ten matrices defined by: $M^{ab}_{cd}=\alpha \delta_{cd}^{ab} -iN^{ab}_{cd}$ and $Q^{ab}_{cd}=\alpha \delta_{cd}^{ab} +iN^{ab}_{cd}$ with $a<b$, $c<d$ and $\delta_{cd}^{ab}=1$ if the two couples of indices are the same and it vanishes otherwise. Therefore, $M$ is a symmetric matrix with all its diagonal coefficients equal to $\alpha+i\sqrt{\f35}\f{1}{d_{j_0}}$ and $Q$ is a symmetric matrix with all its diagonal coefficients equal to $\alpha-i \sqrt{\f35}\f{1}{d_{j_0}}$. Then the analysis is the same as the one done in the case of a Gaussian state with an imaginary phase;  we have three case to consider:\begin{enumerate}
\item $\alpha \gg \f{1}{d_{j_0}}$, that is we consider here that $\alpha$ is proportional to $d_{j_0}^{-n}$ with $0<n<1$. We have then that $M \simeq \alpha Id$ and $Q\simeq \alpha Id$ and once again we write $\alpha=\f{a}{d_{j_0}^n}$, then we get the same result as in the case of the imaginary phase (up to some factors $2$ in $a$) (see equations \ref{result1}), that is: $\left\{ \tabl{l}{
\alpha \in \R \\
\sigma= \f14\left( \f95-n \right) \textrm{ and } \sigma>\f15 \\
a= 32P^4 \pi
} \right.$
\item $ \alpha \ll \f{1}{d_{j_0}}$, that is we consider that $\alpha$ is proportional to $d_{j_0}^{-n}$ with $1<n<2$. We have to integrate $\int d[\delta d_{j_{ab}}] \exp( i \sum \delta d_{j_{ab}} N^{ab}_{cd} \delta d_{j_{cd}})+ \exp(- i \sum \delta d_{j_{ab}} N^{ab}_{cd} \delta d_{j_{cd}})$. The signature of the matrix $N$ is equal to $-2$; therefore, $e^{i \sigma(N)\f{\pi}{4}}=-i=-e^{i \sigma(-N)\f{\pi}{4}}$ and then the previous integral is null so in this case we also cannot have $\alpha=\f{a}{d^n_{j_0}}$ with $1<n<2$.
\item $\alpha \sim \f{1}{d_{j_0}}$: it is the most natural case which peaks in the same way the triangle areas of the 4-simplex around the background value $A_0=d_{j_0}$ and the dihedral angles around the background value $\Theta$. But once again due to the fact we have chosen a factorized boundary state, this case is very complicated although the imaginary parts of the determinant of $M$ and of the determinant of $N$ will compensate.  This confirms that a factorized boundary state is not the most natural to capture a physical state. And with a real phase it is not even possible to consider a tensorial state which could compensate $iN$ and $-iN$ as we did in the case of the imaginary phase.

\end{enumerate}


\bibliographystyle{ieeetr}
\bibliography{biblio}
\end{document}